MASTER OF SCIENCE THESIS IN ASTROPHYSICS

May 16th, 2022

# STARBURSTS AT COSMIC DAWN

## Formation of Globular Clusters, Ultra-Faint Dwarfs, and Population III star clusters at $z > 6$

**Author:** OLOF NEBRIN

**Supervisor:** PROF. GARRELT MELLEMA

Stockholm University

Department of Astronomy

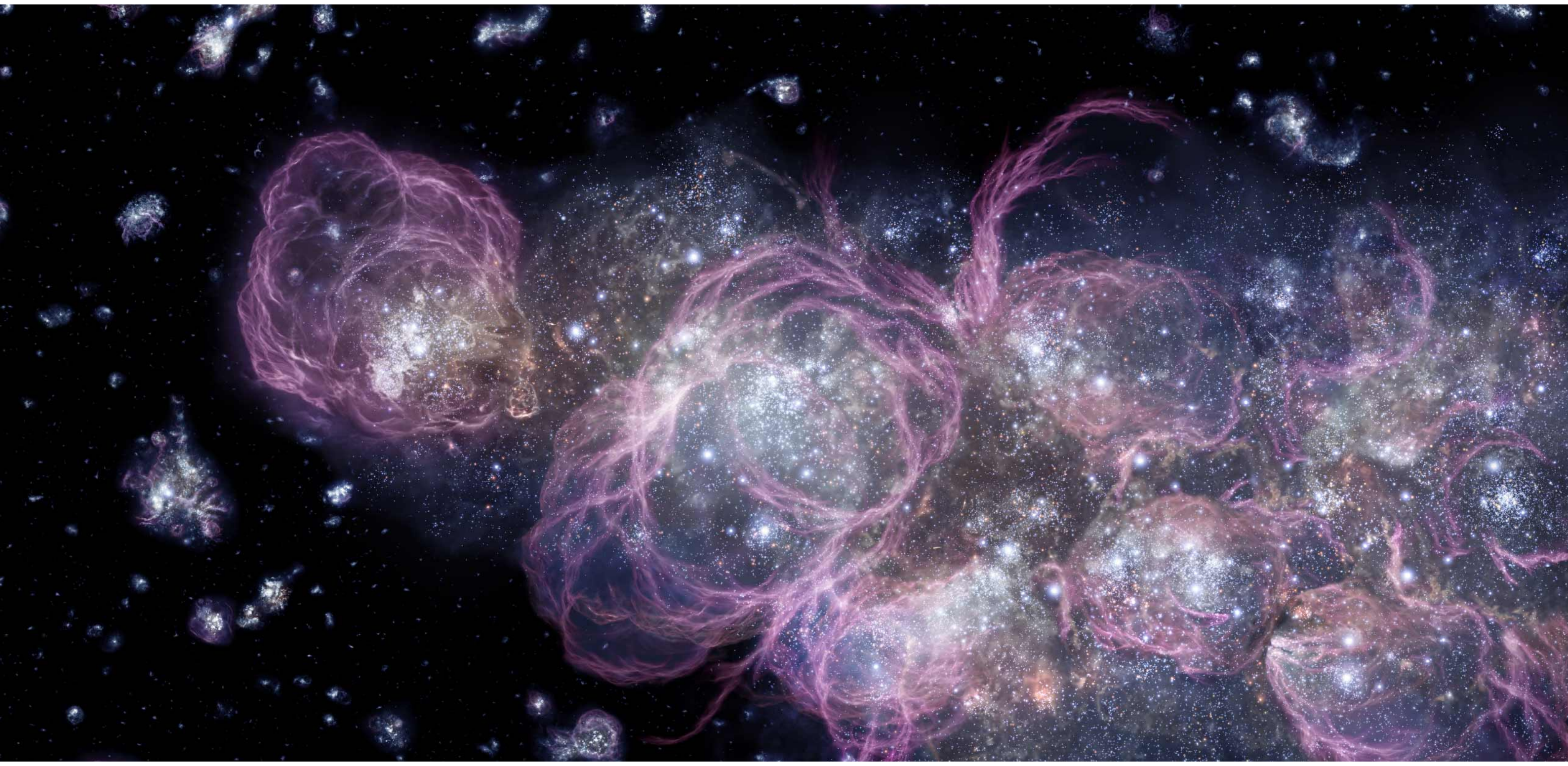

# Abstract

In the standard model of cosmology (ΛCDM) the first stars, star clusters, and galaxies are expected to have formed in short bursts of star formation in low-mass dark matter halos at high redshifts ($z \sim 6 - 30$). Up to this point, attempts to predict the properties and abundances of these luminous objects have made use of numerically expensive cosmological simulations. On top of being numerically expensive, these simulations often lack the required sub-parsec resolution needed to resolve the formation of compact star clusters and/or neglect possibly dominant stellar feedback processes. Motivated by this, I introduce ANAXAGORAS, as far as I know the most detailed analytical *ab initio* model of starbursts in low-mass halos to date. The model incorporates sub-models for gas cooling (including a new determination of the $H_2$-cooling threshold in minihalos), central gas accretion and disk formation (using a new self-similar solution), stellar feedback from radiation pressure (direct stellar radiation, Lyman-$\alpha$ scattering in H I, and multiple scattering of IR photons by dust), stellar winds, expanding H II regions, and (crudely) supernovae. The resulting star formation efficiency is used to predict the fraction of stars that remain gravitationally bound in a cluster following gas expulsion, and what fraction escape the central region of the halo, yet remain bound by the dark matter halo. I apply ANAXAGORAS to study star formation at $z > 6$ in satellite halos of the Milky Way using a halo merger tree code, as well as Population III (Pop III) star formation in minihalos.

For the Milky Way setup, hundreds of galaxies are predicted to form with luminosities ($L_V <$ few $\times 10^4$ $L_\odot$), half-mass radii ($\sim 10 - 200$ pc), mass-to-light ratios ($M/L_V \sim 100 -$ few $\times 10^3$ $M_\odot/L_\odot$), and ages ($13.18^{+0.29}_{-0.31}$ Gyrs) in good agreement with the observed local population of Ultra-Faint Dwarfs. This shows that ΛCDM is able to explain the properties of the faintest dwarf galaxies without fine-tuning. Furthermore, at least $\sim 40$ compact (initial half-mass radii $\sim 0.1 - 5$ pc), old ($13.27^{+0.21}_{-0.39}$ Gyrs) globular cluster (GC) candidates with initial stellar masses $10^5 - 10^6$ $M_\odot$ are predicted to form at the center of low-mass halos, and could survive to the present-day and explain at least a fraction of the observed metal-poor GCs. Their properties are consistent with recent candidates for GCs residing in dark matter halos. Thus, ANAXAGORAS lends support to the viability of the scenario of GC formation in minihalos. Finally, the formation of Population III (Pop III) stars in minihalos is studied, with the conclusion that if Pop III stars are not overly massive (25 $M_\odot$) between $\sim 1 - 30$ stars could form per minihalo at $z > 20$, with the number increasing to $\sim 10 - 500$ stars per minihalo at $z < 15$ as Lyman-Werner feedback delay star formation until halos reach larger masses. In the case where Pop III stars are more massive (140 $M_\odot$) most minihalos form just a single star. Due to self-shielding of $H_2$ in minihalos, I find that the cosmological Lyman-Werner background is insufficient to produce Pop III galaxies in atomic-cooling halos, with the implication that the number of massive Pop III galaxies/star clusters in the early Universe has been greatly overestimated in the literature that ignores self-shielding.

# Contents

**Cover image:** An artist's impression of bursty star formation in the early Universe ($z \gtrsim 6$) by Adolf Schaller/STSCI.

# Part I.
# Introduction

## 1. Cosmological parameters, common notation and abbreviations

Unless stated otherwise, in this thesis I will adopt $\Omega_\mathrm{M} = 0.315$, $\Omega_\mathrm{B} = 0.0493$, $\Omega_\Lambda = 0.685$, $h = 0.674$, $n_\mathrm{s} = 0.965$, and $\sigma_8 = 0.811$, consistent with the PLANCK 2018 results (Planck Collaboration *et al.*, 2020). The baryon fraction is denoted by $f_\mathrm{B} = \Omega_\mathrm{B}/\Omega_\mathrm{M} = 0.157$, and I will assume a primordial helium mass fraction of $Y = 0.247$ (see e.g. results and references in Cooke & Fumagalli, 2018). For these cosmological parameters, the mean matter density of the Universe at redshift $z$ is given by

$$\bar{\rho}_\mathrm{M} = \frac{3\Omega_\mathrm{M} H_0^2}{8\pi G}(1+z)^3 = 2.69 \times 10^{-27}\ \mathrm{g\ cm^{-3}} \left(\frac{1+z}{10}\right)^3 . \tag{1}$$

We also find a mean baryon density, $\bar{\rho}_\mathrm{B} = f_\mathrm{B}\bar{\rho}_\mathrm{M}$, of

$$\bar{\rho}_\mathrm{B} = 4.22 \times 10^{-28}\ \mathrm{g\ cm^{-3}} \left(\frac{1+z}{10}\right)^3 . \tag{2}$$

There are many quantities and equations in ANAXAGORAS, so for the convenience of the reader the notation used for some of the important quantities is reproduced here for easy reference:

- $M$, $R_\mathrm{vir}$, $T_\mathrm{vir}$: The (total, dark matter + baryon) mass, virial radius, and virial temperature of the halo, respectively (see Eq. 6). In numerical estimates, the following notation is commonly used: $M_\mathrm{x} \equiv M/10^\mathrm{x}\ \mathrm{M_\odot}$ and $T_\mathrm{vir,x} \equiv T_\mathrm{vir}/10^\mathrm{x}$ K.

- $T_\mathrm{h}$, $\mu_\mathrm{h}$, $c_\mathrm{s,h}$: The temperature, mean molecular weight, and (isothermal) sound speed of the gas in the halo (excluding the central disk), respectively. We often use the notation $T_\mathrm{h,x} \equiv T_\mathrm{h}/10^\mathrm{x}$ K and $\mu_\mathrm{h,x} \equiv \mu_\mathrm{h}/\mathrm{x}$ in numerical estimates.

- $T$, $\mu$, $c_\mathrm{s}$: The temperature, mean molecular weight, and (isothermal) sound speed of the gas in the disk, respectively. The notation $T_\mathrm{x} \equiv T/10^\mathrm{x}$ K and $\mu_\mathrm{x} \equiv \mu/\mathrm{x}$ will often be used in numerical estimates.

- $\lambda$: The (Bullock) spin parameter (see Eq. 85). The notation $\lambda_\mathrm{x} \equiv \lambda/\mathrm{x}$ will often be used in numerical estimates.

- $M_\mathrm{disk}$, $R_\mathrm{disk}$: The (total, stellar + gas) mass and radius of the central disk wherein stars form.

- $M_{\star,\mathrm{tot}}$; Total stellar mass formed in the disk before stellar feedback quenches further star formation.

- $M_\mathrm{dSph}$, $R_\mathrm{dSph}$: The predicted stellar mass (some fraction of $M_{\star,\mathrm{tot}}$) and half-mass radius of a resulting dwarf spheroidal galaxy (see Eq. 192).

- $M_{\star,\mathrm{cl}}$, $R_\mathrm{cl,f}$: The predicted stellar mass (some fraction of $M_{\star,\mathrm{tot}}$) and half-mass radius of a resulting gravitationally bound star cluster (see Eq. 181).

The following abbreviations are often used throughout the thesis:

- GC: Short for globular cluster.

- UFD: Short for Ultra-Faint Dwarf.

- dSph: Short for dwarf spheroidal.

- DM: Short for dark matter.
- CDM: Short for cold dark matter.
- CMB: Short for Cosmic Microwave Background.
- NFW: Short for Navarro-Frenk-White, referring to dark matter halos in the CDM paradigm with the Navarro-Frenk-White density profile (see Appendix 11 for an overview).
- IGM: Short for intergalactic medium.
- EoR: Short for Epoch of Reionization.
- MW: Short for Milky Way.
- LW: Short for Lyman-Werner, referring to Lyman-Werner photons ($10.2 - 13.6$ eV) that can photodissociate $H_2$ (see Section 3.1.1).
- Pop II: Short for Population II.
- Pop III: Short for Population III.

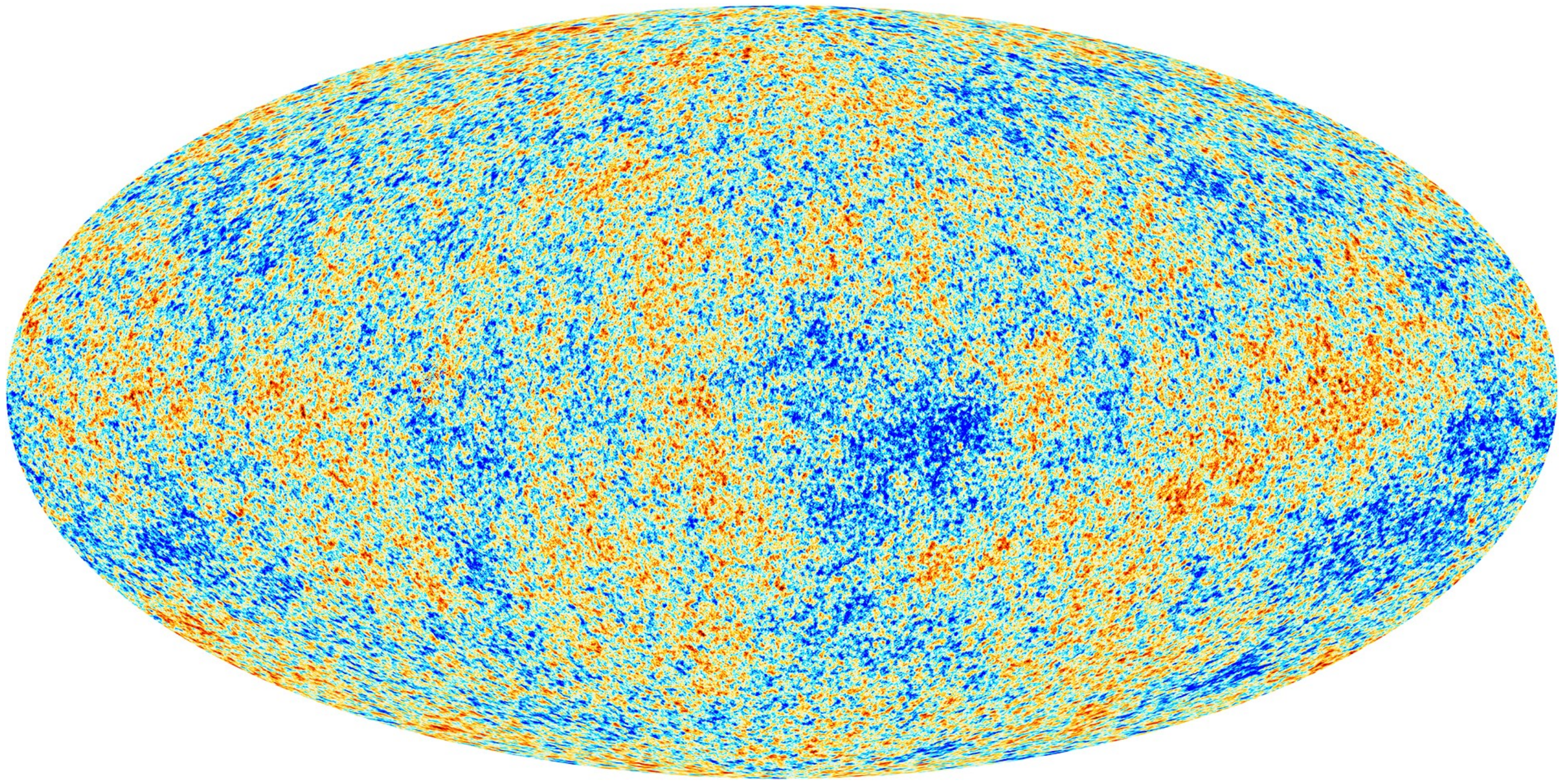

**Figure 1.** The Cosmic Microwave Background (CMB), as seen by the PLANCK satellite. The CMB has a sky-averaged present-day temperature of $T_{\mathrm{CMB},0} = 2.726$ K, but temperature fluctuations of order $\sim 10^{-5}$ has been observed (first by the COBE satellite, Smoot *et al.*, 1992). These are thought to be due to density fluctuations in the very early Universe that form the seeds for subsequent galaxy formation. **Credit:** ESA and the Planck Collaboration.

## 2. Structure formation in the early Universe

> *"If you wish to make an apple pie from scratch you must first invent the Universe."*
>
> — CARL SAGAN

Observations of the Cosmic Microwave Background (CMB) radiation indicate that the very early Universe ($z \sim 1000$) was very uniform, but not perfectly so (see Figure 1). From temperature fluctuations in the CMB one can infer the presence of small ($\sim 10^{-5}$) Gaussian density fluctuations, most likely produced by quantum fluctuations in the scalar field driving inflation, the extremely rapid accelerated expansion at even earlier times (e.g. Bardeen *et al.*, 1983; Guth & Pi, 1985; Peebles, 1993; Planck Collaboration *et al.*, 2020).

In regions of slightly higher density than the cosmic mean (Eq. 1) and negligible stabilizing pressure, the greater gravitational attraction would amplify the small ("linear") density fluctuation until it becomes of order unity ("non-linear"), and collapse to form a so-called halo. In the standard cold dark matter (CDM) paradigm (Peebles, 1982; Blumenthal *et al.*, 1984), dark matter (DM) behaves as a pressureless fluid. Due to the absence of any pressure that could counteract gravitational collapse, dark matter halos could therefore form with a wide range of masses, perhaps down to mass-scales as small as that of the Earth (Diemand *et al.*, 2005).[1] The standard model of cosmology ($\Lambda$CDM) assumes both CDM and the existence of dark energy that is indistinguishable from a cosmological constant $\Lambda$, which results in a late-time accelerated expansion of the Universe. The RMS linear density fluctuation $\sigma(M)$ at $z = 0$ on a mass-scale $M$ in $\Lambda$CDM, assuming a spectral index $n_s = 0.965$,[2] is approximately given by (see Appendix 2 for details):

$$
\begin{aligned}
\sigma(M) &\simeq \mathcal{N}\left\{\frac{36}{1+3\bar{M}} - \ln^3\left(\frac{\bar{M}}{1+\bar{M}}\right)\right\}^{1/2} \\
\mathcal{N} &= 0.0845\,\sigma_8 \\
\bar{M} &\equiv \frac{8M}{M_{\rm eq}} \\
M_{\rm eq} &= 2.4\times10^{17}\left(\frac{\Omega_{\rm M}h^2}{0.14}\right)^{-1/2}\left(\frac{1+z_{\rm eq}}{3400}\right)^{-3/2}\,\mathrm{M}_\odot\,,
\end{aligned}
\tag{3}
$$

where $1 + z_{\rm eq} \simeq 3400$ is the redshift of matter-radiation equality (Planck Collaboration *et al.*, 2020). Thus, it can be seen that the Universe contains larger density fluctuations on smaller scales (we have $\sigma \propto [36 - \ln^3 \bar{M}]^{1/2}$ on small scales). This coupled with the fact that the growth rate of linear CDM density fluctuations (on sub-horizon scales) is scale-independent, it follows that the earliest structures in $\Lambda$CDM are low-mass halos that subsequently grow in a hierarchical manner to form larger halos. Because of this, the first stars and galaxies are expected to form in the lowest mass dark matter halos wherein gas could accrete, cool efficiently, and collapse. Some important properties of halos at a given redshift can be estimated as follows. A spherical overdense region of radius $R$ in the early Universe, before dark energy domination $(1+z > (\Omega_\Lambda/\Omega_{\rm M})^{1/3} \simeq 1.3)$, evolves in the same manner as a closed matter-dominated Universe. Utilizing this, the so-called spherical collapse model can be used to describe the evolution of this region (see e.g. Peebles, 1993; Tegmark *et al.*, 1997; Peacock, 1999; Stiavelli, 2009):

$$
\begin{aligned}
R(z) &= \frac{1}{2}R_{\rm max}\,(1-\cos\theta) \\
\frac{t}{t_{\rm vir}} &= \frac{\theta-\sin\theta}{2\pi}\,,
\end{aligned}
\tag{4}
$$

where $R_{\rm max}$ is the maximum radius reached of the overdense region before it formally collapse to a point at time $t_{\rm vir}$. Since the scale factor in a matter-dominated Universe grows as $a \propto t^{2/3}$, the time can be related to redshift using $t/t_{\rm vir} = [(1 + z_{\rm vir})/(1 + z)]^{3/2}$. The time $t_{\rm vir}$ is known as the virialization time, since irregularities in the matter distribution would in reality lead to the formation of a virialized structure of non-zero size at that point. Using the virial theorem, one can show that the virialized halo will have a radius $R_{\rm vir} = \frac{1}{2}R_{\rm max}$ and an average density $\rho_{\rm vir} = \Delta_{\rm vir}\bar{\rho}_{\rm M}(z_{\rm vir})$, with $\Delta_{\rm vir} \simeq 18\pi^2$ at high redshifts (see e.g. Tegmark *et al.*, 2006, for more general expressions valid at all redshifts). The parametric solution of Eq. (4) for an overdense region enclosing a mass $M$ can be conveniently approximated to within $\sim 0.4\%$ as well as when the first galaxies formed. More recently, constraints both from high redshifts (e.g. due to the 21-cm signal from Cosmic Dawn) and low redshifts (e.g. the GD-1 tidal stream) has effectively ruled out these scenarios, instead favouring the CDM paradigm (e.g. Banik *et al.*, 2021; Nebrin *et al.*, 2019).

[1] In alternative models of dark matter like warm dark matter (WDM) or fuzzy dark matter (FDM), there is in effect a pressure that prevents clumping on small scales. This typically lead to the RMS linear overdensity $\sigma(M)$ reaching a maximum on mass-scales of $M \lesssim 10^{10}$ $\mathrm{M}_\odot$, usually motivated by a desire to suppress the believed predicted overabundance of dwarf galaxies in such halos by CDM. This would dramatically alter the abundance of low-mass halos in the local Universe,

[2] The latest results from Planck indicate a spectral index $n_s = 0.965 \pm 0.04$ (68% confidence interval) (Planck Collaboration *et al.*, 2020), and as noted earlier, we adopt $n_s = 0.965$. In practice, as shown in Appendix 2, this means that the convenient analytical fit in Eq. (3) has been calibrated against an exact integration that assumed $n_s = 0.965$.

by the following expression, which becomes exact in the limits $t \to 0$ and $t \to t_{\rm vir}$:

$$
\begin{aligned}
R(M,z) &\simeq \frac{1}{2} R_{\rm vir}(M, z_{\rm vir}) \left\{ 12 \sin\left(\frac{\pi t}{t_{\rm vir}}\right) \right\}^{2/3} \left\{ 1 - 0.237 \sin^{3/5}\left(\frac{\pi t}{t_{\rm vir}}\right) \right\} \\
\frac{t}{t_{\rm vir}} &= \left(\frac{1+z_{\rm vir}}{1+z}\right)^{3/2} .
\end{aligned} \tag{5}
$$

Following virialization at redshift $z_{\rm vir}$, a halo of mass $M$ would form. The resulting virial radius $R_{\rm vir}$, velocity $v_{\rm vir}$, and temperature $T_{\rm vir}$ (assuming a mean molecular weight $\mu_{\rm h}$ of halo gas) of the halo at redshift $z$ (dropping the subscript "vir" from now on) is given by (e.g. Barkana & Loeb, 2001)

$$
\begin{aligned}
R_{\rm vir} &= \left(\frac{3M}{4\pi \Delta_{\rm vir} \bar{\rho}_{\rm M}}\right)^{1/3} &&= 697 M_7^{1/3} \left(\frac{1+z}{10}\right)^{-1} \ {\rm pc} \\
v_{\rm vir} &= \left(\frac{GM}{R_{\rm vir}}\right)^{1/2} &&= 7.86\ M_7^{1/3} \left(\frac{1+z}{10}\right)^{1/2} \ {\rm km\ s^{-1}} \\
T_{\rm vir} &= 0.75 \times \frac{\mu_{\rm h} m_{\rm H} v_{\rm vir}^2}{2 k_{\rm B}} &&= 3370\ \mu_{\rm h,1.2} M_7^{2/3} \left(\frac{1+z}{10}\right) \ {\rm K} ,
\end{aligned} \tag{6}
$$

where I have used the correction factor 0.75 suggested by Fernandez *et al.* (2014) in the virial temperature $T_{\rm vir}$ for better agreement with simulations,[3] and assumed a virial overdensity $\Delta_{\rm vir} = 18\pi^2$, as noted earlier. The virial temperature is approximately the temperature the gas is expected to be shock-heated to as it falls into the halo during the second half of the evolution of the overdense region. It will prove convenient to also express the halo mass, the virial radius, and the virial velocity in terms of the virial temperature. Doing so using the above equations yield

$$
\begin{aligned}
M &= 5.11 \times 10^7 \, T_{\rm vir,4}^{3/2} \left(\frac{1+z}{10}\right)^{-3/2} \ {\rm M_\odot} \\
v_{\rm vir} &= 13.5 \, T_{\rm vir,4}^{1/2} \ {\rm km\ s^{-1}} \\
R_{\rm vir} &= 1.20 \, T_{\rm vir,4}^{1/2} \left(\frac{1+z}{10}\right)^{-3/2} \ {\rm kpc} .
\end{aligned} \tag{7}
$$

The epoch when the first stars and galaxies form is known as *Cosmic Dawn* (e.g. Pritchard & Loeb, 2012; Loeb & Furlanetto, 2013; Norman *et al.*, 2018; Nebrin *et al.*, 2019). From the spectrum of CDM density fluctuations (Eq. 3) and a careful consideration of gas cooling in halos with the properties laid out in Eq. (6) (Section 3.1), this is expected to occur at redshifts $z \sim 10-30$ in halos with masses $M \gtrsim 10^6$ $\rm M_\odot$ (e.g. Barkana & Loeb, 2001; Miralda-Escudé, 2003; Bromm, 2013; Schauer *et al.*, 2019, 2020).[4] The epoch of Cosmic Dawn precedes the *Epoch of Reionization*, when the first sources of ionizing photons reionized the intergalactic medium (IGM), which is observationally constrained to have been completed by $z \simeq 5-6$ (e.g. Bañados *et al.*, 2018; Zhu *et al.*, 2021).

[3] Fernandez *et al.* (2014) found that this correction factor gives the correct halo mass threshold for efficient Lyman-$\alpha$ cooling ($T_{\rm vir} > 10^4$ K).

[4] Although because of the Gaussian nature of the density fluctuations, extremely rare large fluctuations could potentially produce stars as early as $z \sim 65$ (Naoz *et al.*, 2006).

# Part II.
# Fossils of the early Universe

Observing the formation of the first stars and galaxies at high redshifts require very powerful telescopes, and may potentially even be out of reach of the recently launched James Webb Space Telescope (Weisz & Boylan-Kolchin, 2019). Another way to glean information about these objects is to study nearby relics or “fossils” from this era, often referred to as stellar archaeology or near-field cosmology (e.g. Bland-Hawthorn & Freeman, 2000; Frebel, 2010; Karlsson *et al.*, 2013; Weisz & Boylan-Kolchin, 2019). Below a very brief overview is given of three types of relics from the early Universe.

# 1. Globular Clusters

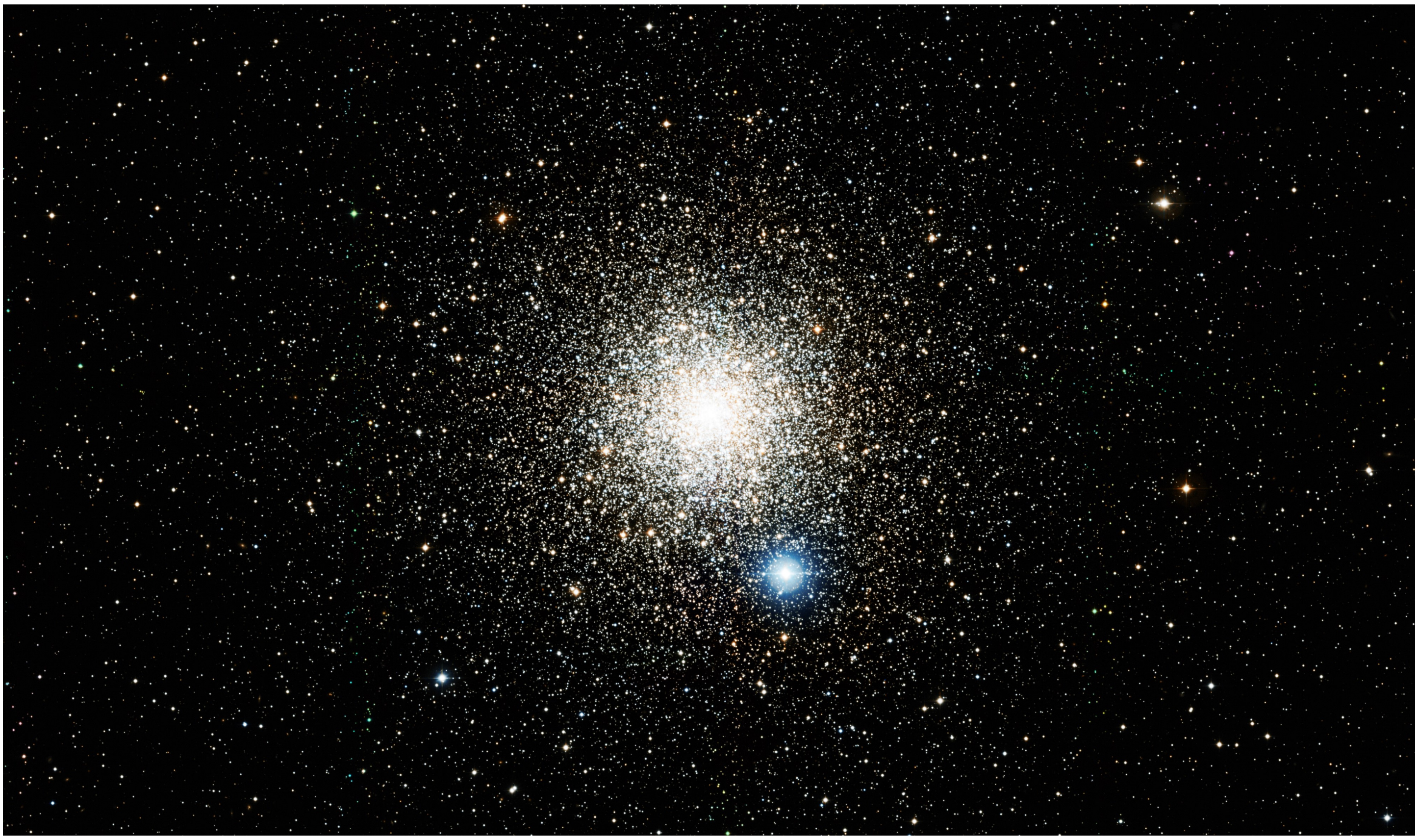

**Figure 2.** The globular cluster NGC 6752, located at a distance of $\simeq 4.1$ kpc from the Sun. It has a mass of $\simeq 2.8 \times 10^5$ $M_\odot$, a half-mass radius $\simeq 5.3$ pc, and a mean metallicity [Fe/H] $\simeq -1.58$ (Baumgardt, 2017; Baumgardt & Hilker, 2018; Bailin, 2019; Baumgardt *et al.*, 2020). A rising velocity dispersion has been observed in the outskirts of NGC 6752 — potential evidence of it having formed in a dark matter halo (Carlberg & Grillmair, 2021). **Credit:** ESO.

Globular clusters (GCs) are compact gravitationally bound clusters of stars (Figure 2).[5] There are 158 known GCs orbiting the Milky Way, with (median and 95% interval) masses $1.26^{+8.45}_{-1.23} \times 10^5$ $M_\odot$ and half-mass radii $4.86^{+22}_{-2.9}$ pc (Baumgardt, 2017; Baumgardt & Hilker, 2018; Baumgardt *et al.*, 2020). Populations of GCs are also observed around other galaxies, with the number and total mass of GCs around a given galaxy scaling approximately linearly with the halo mass of the host galaxy (e.g. Harris *et al.*, 2017; Forbes *et al.*, 2018a). They are often observed to have a bimodal metallicity distribution, and are therefore commonly divided into metal-rich ("red", [Fe/H] $> -1$) and metal-poor ("blue", [Fe/H] $< -1$) subpopulations (e.g. Brodie & Strader, 2006; Harris *et al.*, 2015). The blue GCs make up the majority of the total population in most galaxies, especially in low-mass halos. In particular, the observed mean fraction $f_{\rm blue}$ of GCs that are blue around a host galaxy in a halo of mass $M$ is approximately given by (Harris *et al.*, 2015)

$$f_{\rm blue} = 0.72 \left( \frac{M}{10^{12}\ M_\odot} \right)^{-0.07}, \quad (8)$$

valid for the range $10^{10}$ $M_\odot \lesssim M \lesssim 10^{13}$ $M_\odot$. Thus, for a Milky Way-like halo with $M \simeq 1.1 \times 10^{12}$ $M_\odot$ (Harris *et al.*, 2017; Rodriguez Wimberly *et al.*, 2021), blue GCs make up $\sim 72\%$ of the total GC population on average, or roughly $\sim 100$ in number. The spread in iron abundance among stars within a given GC is typically observed to very small (Bailin, 2019). Since the first core-collapse supernova is expected to explode after 3.4 Myrs (e.g. Hopkins *et al.*, 2018a) and enrich the gas out of which the stars are made with iron, this is indicative of most GCs having formed in short starbursts lasting only a few million years (Wirth *et al.*, 2021). It has proven hard to determine the ages of GCs accurately due to large systematic errors of $\sim 1$ Gyrs (e.g. O'Malley *et al.*, 2017; Forbes *et al.*, 2018b; Boylan-Kolchin & Weisz, 2021), but blue GCs tend to have greater estimated ages than red GCs (e.g. Kruijssen *et al.*, 2019b). In particular, the most metal-poor GCs ([Fe/H] $\lesssim -1.5$) have estimated ages of $12.5 - 13.5$ Gyrs (VandenBerg *et al.*, 2013; O'Malley *et al.*, 2017; Camargo & Minniti, 2019; Kruijssen *et al.*, 2019b; Jimenez *et al.*, 2019). Within the age uncertainty, this is consistent with the idea that at least a fraction of blue GCs formed at Cosmic Dawn, prior to the reionization of the intergalactic medium. Indeed, within the CDM paradigm of structure formation, the old ages, low metallicities, and spatial distribution of blue GCs have sparked much speculation that they could have formed in low-mass dark matter halos ($M \sim 10^6 - 10^9$ $\mathrm{M_\odot}$) at Cosmic Dawn (e.g. Peebles, 1984; Rosenblatt *et al.*, 1988; Moore *et al.*, 2006; Brodie & Strader, 2006; Griffen *et al.*, 2013; Trenti *et al.*, 2015; Creasey *et al.*, 2019). Recent high-resolution cosmological simulations have lent some theoretical support to this scenario, finding that GC-like objects form at the centers of some low-mass halos at high redshifts (Boley *et al.*, 2009; Kimm *et al.*, 2016; Ricotti *et al.*, 2016). However, as discussed in Section 1.1, these simulations lack one or more important stellar feedback process, and so cannot be taken as physically robust *ab initio* predictions of ΛCDM.

[5] This is in contrast to open star clusters that are typically not gravitationally bound.

One way of testing the dark matter scenario of GC formation observationally would be to detect the gravitational influence of dark matter on the stars in a GC. This is hard to do in practice because for reasonable dark matter density distributions the stars are expected to dominate gravitationally out to many half-light radii (Vitral & Boldrini, 2021). Furthermore, the dark matter halo in which a GC formed could be tidally stripped as it is accreted by the Milky Way (Mashchenko & Sills, 2005). Nevertheless, some preliminary evidence of dark matter in GCs has recently emerged from the study of stars in the outskirts of several GCs, where the potential presence of dark matter is more clearly manifested (Bianchini *et al.*, 2019; Boldrini & Vitral, 2021; Carlberg & Grillmair, 2021). Moreover, recent observed gravitationally lensed low-luminosity "galaxies" at redshifts $z \sim 6-8$ have estimated sizes $\sim 10$ pc and masses $\sim 10^6$ $\mathrm{M_\odot}$, similar to GCs or Ultra-Compact Dwarf galaxies (Bouwens *et al.*, 2021), consistent with the idea that some GC-like objects can form prior to reionization. These lines of evidence are only suggestive and preliminary, and there are also well-motivated scenarios of GC formation in the course of normal star formation in galaxies (e.g. Ma *et al.*, 2020), since in principle any (sufficiently massive) gas cloud with a large surface density ($\gtrsim 10^4$ $\mathrm{M_\odot}$ $\mathrm{pc^{-2}}$) is expected to resist stellar feedback, and form a gravitationally bound massive star cluster (e.g. Grudić *et al.*, 2018; Li *et al.*, 2019).[6] The reader is referred to Forbes *et al.* (2018b) for a recent review of GC formation scenarios. In this thesis we do not focus on making any assumptions regarding GC formation at high redshifts in low-mass halos. Instead, using an *ab initio* approach, we want to *test the idea* that ΛCDM *predicts* the formation of GCs in low-mass halos at Cosmic Dawn, as first hypothesized by Peebles (1984). If it is found that ΛCDM unambiguously predict the formation of GCs in low-mass halos, we could then hope to test our cosmological paradigm by studying observed GCs.

[6]However, Renaud (2020) has argued for some skepticism that GC formation is completely analogous to the formation of local Young Massive Clusters.

# 2. Ultra-Faint Dwarfs

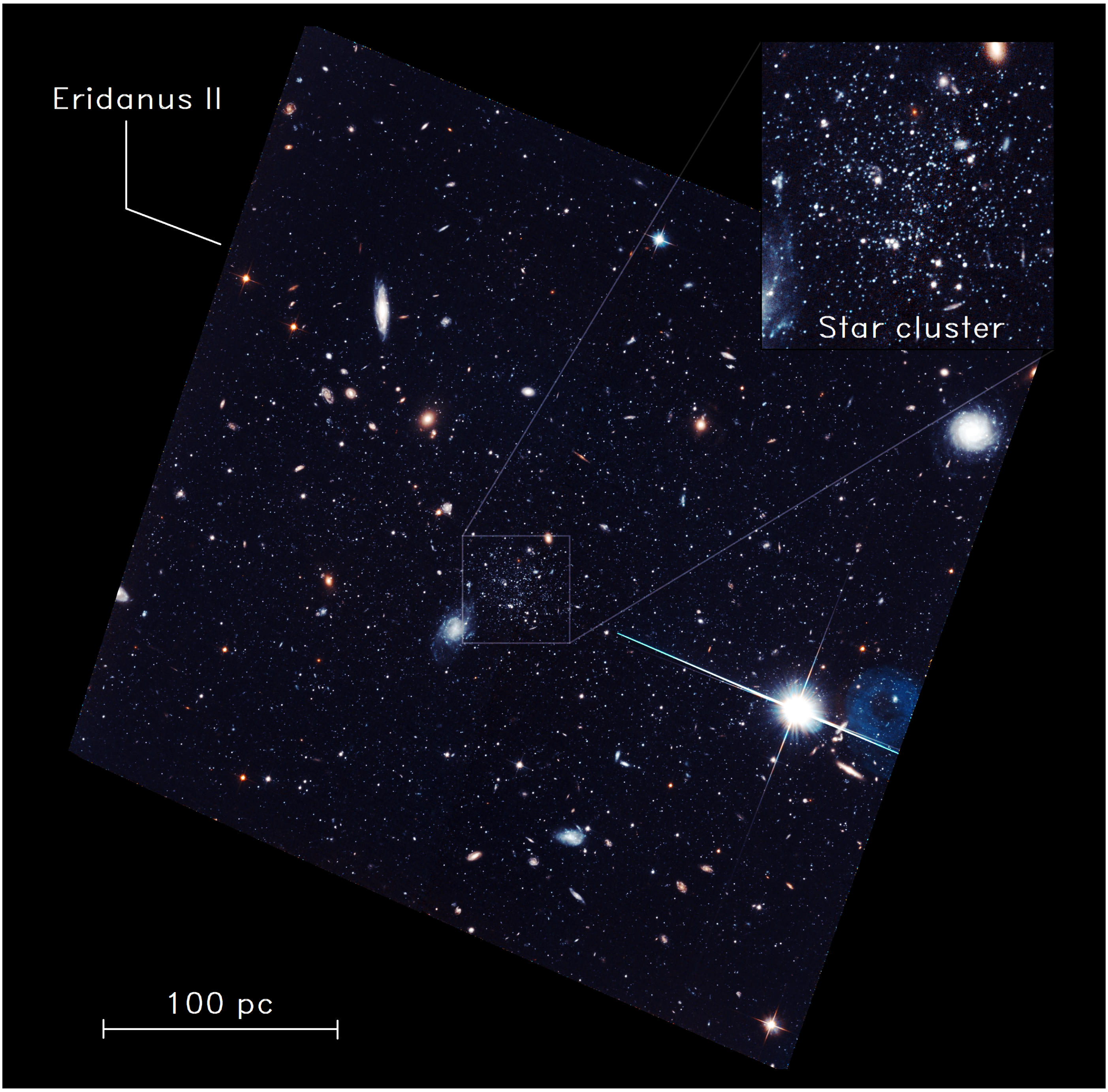

**Figure 3.** Eridanus II (Eri II), an Ultra-Faint Dwarf (UFD) galaxy situated near the virial radius of the Milky Way halo, as observed by Simon *et al.* (2020) using the Hubble Space Telescope. While it is hard to resolve any clear structure, most of the field stars in this image belong to Eri II. This UFD has a half-light radius of $R_{\rm h} = 246 \pm 17$ pc (Simon, 2019), a stellar mass of $M_\star \simeq 1.1 \times 10^5$ $\rm M_\odot$ (Gallart *et al.*, 2021), and a low mean metallicity of $[{\rm Fe/H}] = -2.38 \pm 0.13$ (Li *et al.*, 2017). The evidence suggests that it — like most UFDs — formed most of its stars prior to the Epoch of Reionization (e.g. Simon *et al.*, 2020; Gallart *et al.*, 2021), perhaps in a single short ($<$ 100 Myrs) burst (Gallart *et al.*, 2021). Not far from the center of Eri II is a star cluster, somewhat reminiscent of a faint GC or an extended cluster, with a stellar population similar to that of Eri II (Gallart *et al.*, 2021) — perhaps indicative of both originating at Cosmic Dawn. **Credit:** Figure adapted from Simon *et al.* (2020) with some added contrast and color enhancement for visual clarity, and added text.

As noted Section 2, in ΛCDM the first stars and galaxies are expected to form in low-mass halos at high redshifts. The shallow gravitational potential of these halos is expected to make stellar feedback very efficient, and star formation inefficient. After reionization ($z \simeq 6$) the intergalactic medium is heated to $\simeq 2 \times 10^4$ K, and so halos with peak virial temperatures significantly below this value would not continue to accrete gas and form new stars. Qualitatively, we therefore expect a relic population of very faint and old dwarf galaxies, that formed the majority of their stars at redshifts $z > 6$ (Bullock *et al.*, 2001b; Ricotti & Gnedin, 2005). In fact, this was seen as a possible solution to the so-called "missing satellites problem" of CDM (Klypin *et al.*, 1999; Moore *et al.*, 1999), where the large density fluctuations on small scales (Eq. 3) would otherwise seem to produce more dwarf galaxies in low-mass halos than observed (for a recent nice review, see Bullock & Boylan-Kolchin, 2017). Subsequent deep surveys able to detect very low surface brightness galaxies — starting with the Sloan Digital Sky Survey (SDSS) in 2005 — has confirmed this prediction, revealing a large population of so-called Ultra-Faint Dwarf (UFD) galaxies (see Simon, 2019, for an excellent review). These UFD galaxies have V-band luminosities $L_{\rm V} < 10^5$ $L_\odot$, corresponding to stellar masses of $\lesssim 10^5$ $M_\odot$, comparable to many GCs. Like blue GCs, the stars in UFDs are also very old, consistent with having formed most of their stars prior to reionization (e.g. Brown *et al.*, 2014; Simon *et al.*, 2020; Gallart *et al.*, 2021). Observed UFDs are very metal-poor, with mean iron abundances of $-3 \lesssim$ [Fe/H] $\lesssim -2$. Blue GCs on the other hand are observed to have a "metallicity floor" of [Fe/H] $\simeq -2.5$, which has prompted some attempts to explain this observation (e.g. Abe & Yajima, 2018). However, recent evidence has emerged of tidally disrupted GCs with [Fe/H] $\simeq -2.7$ and [Fe/H] $\simeq -3.4$ (Wan *et al.*, 2020; Martin *et al.*, 2022), more in line with the metallicities of observed UFDs. Thus, UFDs have similar stellar masses, ages, and metallicities as blue GCs.

A key feature that sets UFDs apart from blue GCs are their sizes. In particular, UFDs are quite extended for their luminosities, with the more luminous among them reaching half-light radii of $\sim$ few $\times$ 100 pc, and some fainter ones having half-light radii of just $\sim$ 20 pc (see the purple stars in Figure 19 for a plot of observed confirmed and candidate UFDs). It is because of their large half-light radii and small stellar masses — and so low surface brightness — that UFDs are so hard to detect (see Figure 3 for the example of Eridanus II). It is common practice to classify an object as a galaxy rather than star cluster if dark matter is required to explain the dynamics of the stars (Willman & Strader, 2012). By this definition, there are currently 21 spectroscopically confirmed UFDs with estimated velocity dispersions indicating a gravitationally dominant dark matter halo, with slightly more candidate UFDs that await confirmation (Simon, 2019). However, due to incomplete sky coverage in surveys and the faint nature of UFDs (which makes them very hard to detect), it is estimated that the actual total number of UFDs around the Milky Way is $\sim 100 - 300$ (Newton *et al.*, 2018; Nadler *et al.*, 2020).

Concurrent with the growing number of detections of UFDs, high-resolution cosmological simulations of increasing realism and complexity have started to resolve the formation of realistic UFD-like galaxies in low-mass halos at Cosmic Dawn (e.g. Ricotti & Gnedin, 2005; Ricotti *et al.*, 2016; Kimm *et al.*, 2017; Wheeler *et al.*, 2019; Agertz *et al.*, 2020; Applebaum *et al.*, 2021; Grand *et al.*, 2021). Intriguingly, the results of the very high-resolution ($\lesssim 1$ pc) simulation of Ricotti *et al.* (2016) even suggest a connection between UFD and GC formation at Cosmic Dawn: UFDs could form in star clusters that become gravitationally unbound as gas is expelled by feedback. The stars would then expand to greater radii in the halo until the enclosed mass of dark matter can keep the stars bound, leaving a faint extended UFD-like galaxy. This possibility is part of the motivation for studying both UFD and GC formation at Cosmic Dawn in this thesis.

# 3. Population III star clusters

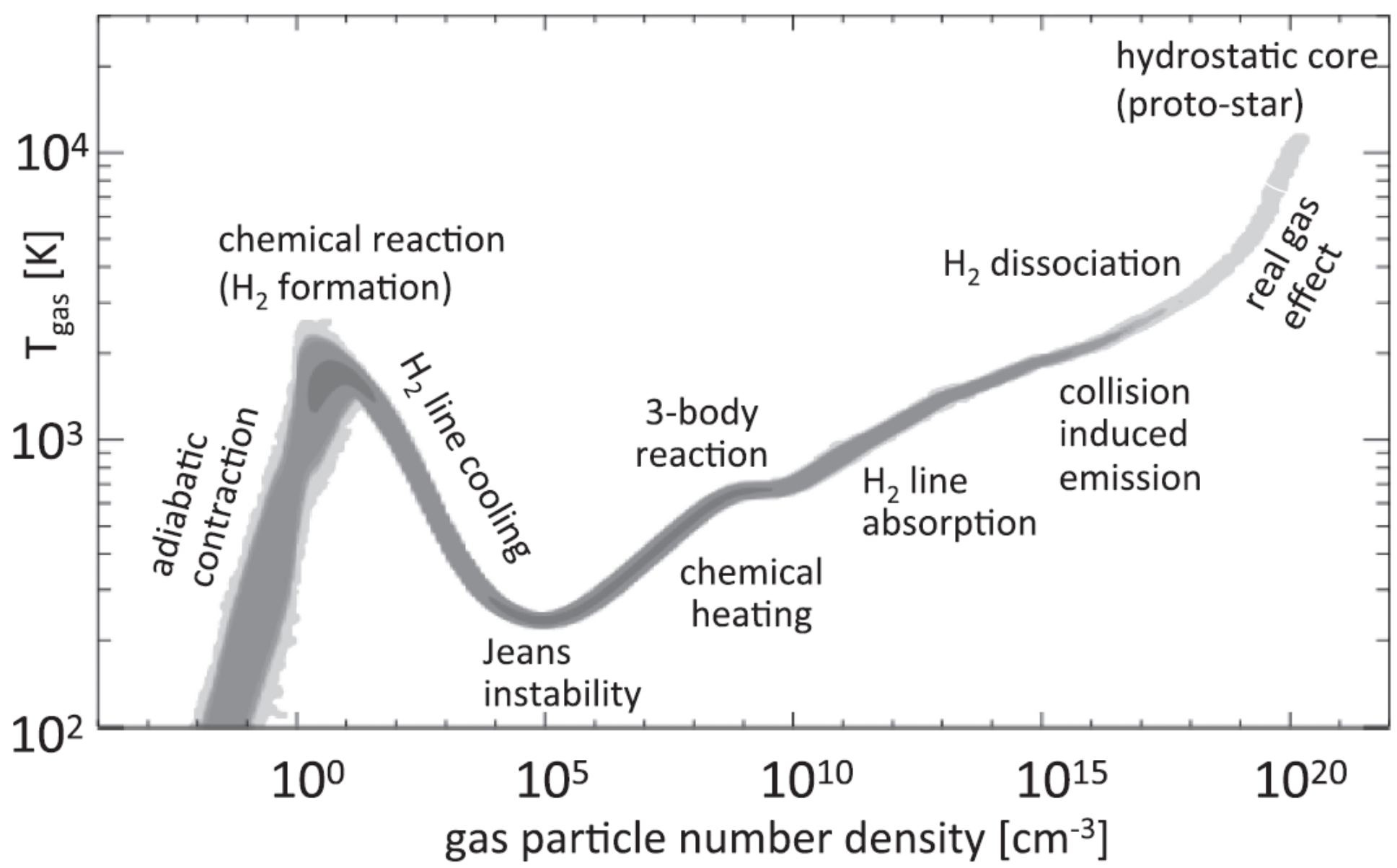

**Figure 4.** The temperature evolution of collapsing metal-free gas. Initially the gas temperature rise due to adiabatic heating until the temperature and density becomes high enough to trigger efficient $H_2$-cooling. At this point the gas cools down to $\sim 200$ K at a density of $n \sim 10^4$ cm$^{-3}$, after which the gas temperature start to rise again since the critical density has been reached above which the cooling rate scale as $n$ rather than $n^2$. **Credit:** Figure taken from Yoshida (2019).

The chemically pristine Universe left from the Big Bang was mainly composed of hydrogen and helium, with trace amounts of deuterium and lithium (e.g. Galli & Palla, 1998, 2013). Out of this primordial gas the first stars formed, usually denoted Population III (Pop III) stars (see e.g. Bromm, 2013; Yoshida, 2019, for reviews). The absence of important metal cooling agents like carbon or iron meant that the primordial gas mainly cooled via molecular hydrogen, $H_2$. This is expected to trigger cooling and gas collapse in halos with masses $M \gtrsim 10^6$ M$_\odot$ (e.g. Yoshida *et al.*, 2003; Schauer *et al.*, 2019), as will be discussed in great detail in Section 3.1. As the collapsing gas reaches higher densities, $H_2$-cooling (mainly via the rotational $J = 2 \rightarrow J = 0$ transition with $\Delta E/k_{\rm B} = 512$ K) push the gas temperature down to $T \simeq 200$ K at a density $n_{\rm H} \sim 10^4$ cm$^{-3}$ (roughly the critical density above which the cooling rate scales like $\propto n_{\rm H}$ rather than $\propto n_{\rm H}^2$). At higher densities the temperature is expected to rise, as shown in Figure 4. A temperature $T \simeq 200$ K is significantly higher than the temperature $T \sim 10$ K in dense present-day molecular clouds. Because of this, the characteristic mass of fragments (the Jeans or Bonnor-Ebert mass, $\propto T^{3/2} n_{\rm H}^{-1/2}$) and the gas accretion rate $\propto T^{3/2}$ (e.g. Shu, 1977) is expected to be higher. Taken together with the absence of dust around the protostar — which can heat the gas through gas-dust collisions if the photoheated dust is hotter than the gas (e.g. Sharda & Krumholz, 2022) — Pop III stars are expected to be more massive than present-day stars (Pop I and Pop II).

Mainly due to difficulties in modelling radiative feedback and numerical constraints (i.e. high resolution and long time integration are numerically expensive), the IMF of Pop III stars is still very uncertain. In some simulations Pop III stars form with masses $\sim 10 - 1000$ M$_\odot$ (Hirano *et al.*, 2014; Susa *et al.*, 2014; Hirano *et al.*, 2015; Hosokawa *et al.*, 2016), while other simulations predict Pop III stellar masses of $\lesssim$ few $\times$ 10 M$_\odot$ (Greif *et al.*, 2011; Latif *et al.*, 2022). Some preference for the latter range can be given for the following reasons:

- The recent state-of-the-art simulations of Latif *et al.* (2022) found Pop III stars with masses in the range $0.1 - 37$ $M_\odot$, with $\sim 65\%$ of Pop III stars having masses in the range $2 - 37$ $M_\odot$. As explained by these authors, the discrepancy with the results of earlier simulations suggesting much larger Pop III stellar masses mainly stem from Latif *et al.* (2022) including radiative feedback from *all* Pop III stars in a given minihalo, and not just the most massive Pop III star.

- While not a single Pop III star has been found in the local Universe — as expected if they are typically massive and hence short-lived — recently, ultra metal-poor stars have been found that showcase evidence of having been enriched by Pop III hypernovae (Placco *et al.*, 2021; Skúladóttir *et al.*, 2021; Yong *et al.*, 2021). The estimated masses of the Pop III progenitors fall in the range $20 - 30$ $M_\odot$.

- Sharda & Krumholz (2022) recently studied the characteristic stellar mass in gas of different metallicities using a detailed semi-analytical model, taking into account stellar feedback. At metallicities $Z/Z_\odot \lesssim 10^{-6}$ and high gas pressures ($nT = 10^8$ K cm$^{-3}$) these authors find that the characteristic stellar mass reaches $\sim 20 - 30$ $M_\odot$ (and drops rapidly to $\sim 0.3$ $M_\odot$ for $Z/Z_\odot \gtrsim 10^{-6}$, consistent with observed present-day stars). This is broadly consistent with both the simulation results of Latif *et al.* (2022) and the above mentioned indirect observational evidence of Pop III hypernovae.

Regardless of whether the characteristic mass of Pop III stars is $\sim 10$ $M_\odot$ or higher, the first stars would be more massive than present-day stars on average, and because of it, their stellar feedback (which comes mainly from massive stars) would be even more vigorous than in the local Universe. An interesting question to answer would therefore be whether massive Pop III star clusters (also known as Pop III galaxies) containing $> 100$ stars can form (e.g. Stiavelli & Trenti, 2010; Vikaeus *et al.*, 2021), or if, because of the strong stellar feedback, only $\sim 1 -$ few $\times 10$ Pop III stars form per halo (e.g. Xu *et al.*, 2016; Latif *et al.*, 2022). This is one of the questions studied in this thesis.

# Part III. ANAXAGORAS: A detailed model of starbursts in low-mass halos at Cosmic Dawn

## 1. Motivation

To conduct a realistic study of the formation of the first galaxies and star clusters in the Universe one need to satisfy two criteria:

1. Have high enough spatial resolution or detail in the simulation or analytical modelling to resolve parsec-sized star clusters and gas clouds, while still capturing the impact of the cosmological environment and initial conditions out of which these objects form.

2. Include all potentially important stellar feedback processes that have an impact on the star formation efficiency. This includes feedback from supernovae, photoionization, radiation pressure, and stellar winds.

In recent years researchers have developed increasingly sophisticated models and simulations to study the formation and evolution of the first star clusters and galaxies. Below we give a very short overview of state-of-the-art simulations and analytical models, and their shortcomings that motivate the development of Anaxagoras.

### 1.1. Cosmological zoom-in simulations

Recent advances in both computational power and astrophysical understanding has led to high-resolution cosmological simulations that zoom in on particular halos to resolve them with greater resolution than the external cosmological environment (see e.g. Cho, 2018; Faucher-Giguère, 2018b, for nice reviews). It is hoped that this will pave the way for extremely detailed *ab initio* simulations that are not fine-tuned to reproduce observations using calibrated subgrid models, and instead offering a robust test of our understanding of galaxy formation and cosmology. A prominent example of a detailed zoom-in cosmological simulation project is FIRE-2, with enough resolution to resolve giant molecular clouds (GMCs) in Milky Way-like galaxies (Hopkins *et al.*, 2018a). FIRE-2 can, without fine-tuning, produce realistic galaxy populations, including Ultra-Faint Dwarfs (Wheeler *et al.*, 2019). It has also recently been used to study the formation of globular clusters in galaxies, utilizing a resolution $\sim 1$ pc (Ma *et al.*, 2020). More relevant to the topic of this thesis, Kimm *et al.* (2016) and Ricotti *et al.* (2016) have used cosmological zoom-in simulations with sub-parsec resolution to study the formation of star clusters and faint galaxies in low-mass halos at Cosmic Dawn. The outcome of the simulation of Kimm *et al.* (2016) is shown in Figure 5. Both FIRE-2 and other recent state-of-the-art zoom-in simulations include detailed implementations of feedback from supernovae, photoionization, and, in some cases, (direct single-scattering) radiation pressure (Kimm *et al.*, 2016; Agertz *et al.*, 2020; Kimm *et al.*, 2017). However, among other things, these simulations do not take into account radiation pressure from resonant scattering of Lyman-$\alpha$ photons, which has only been included in a single idealized (rather than cosmological) galaxy simulation by Kimm *et al.* (2018). Both the results of Kimm *et al.* (2018) and analytical calculations suggest this to be a very important, possibly dominant, source of feedback in the low-metallicity gas clouds wherein the first star clusters and galaxies would form (Abe & Yajima, 2018; Kimm *et al.*, 2018; Tomaselli & Ferrara, 2021). Unfortunately, implementing Lyman-$\alpha$ scattering in galaxy formation simulations is prohibitively numerically expensive at the moment, compounding an issue already facing detailed zoom-in simulations in general. This prevents, among other things, a realistic numerical study of globular cluster formation in low-mass halos.

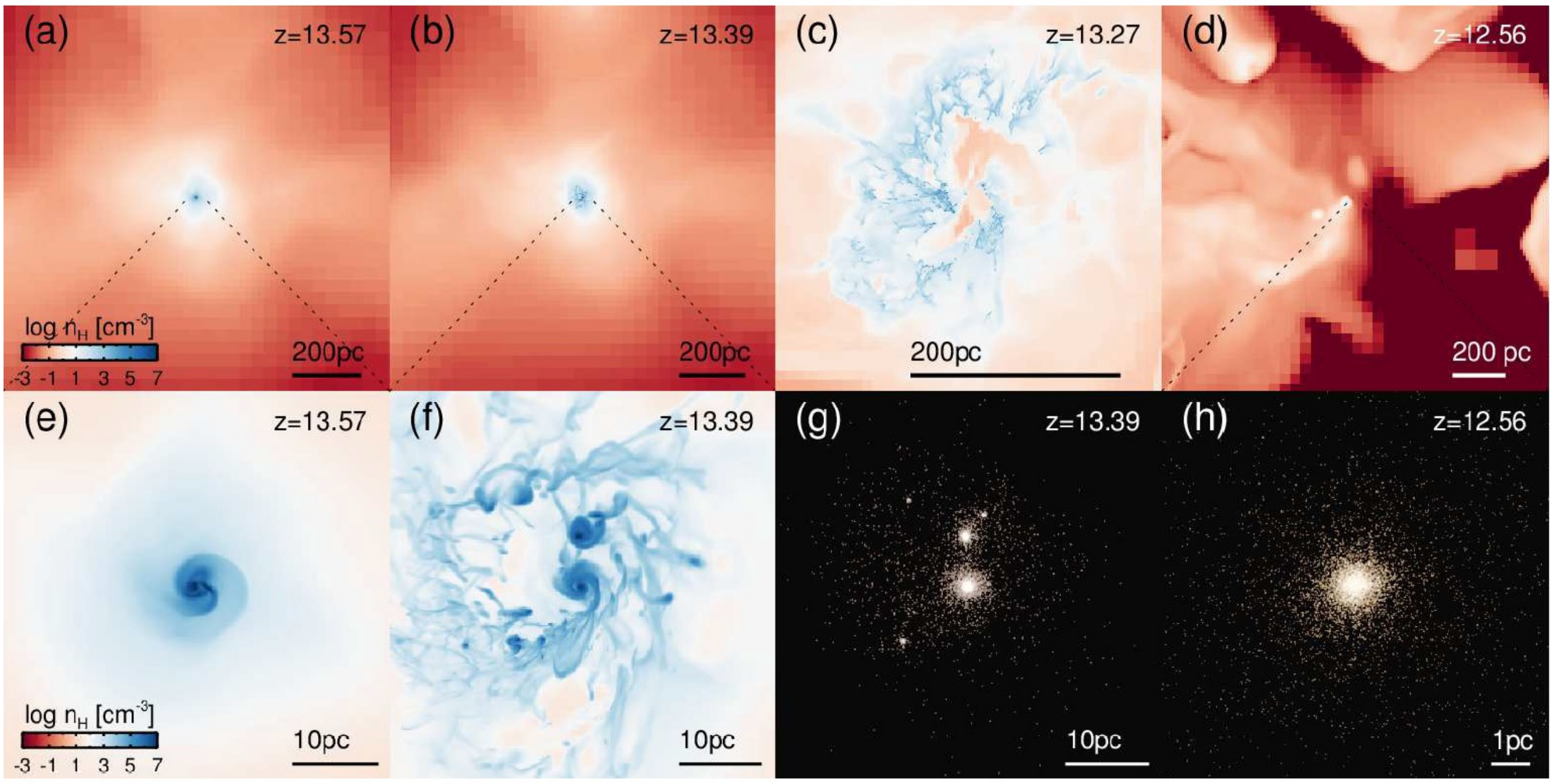

**Figure 5.** Snapshots from the high-resolution cosmological simulation of Kimm *et al.* (2016), showing the formation of a GC candidate with stellar mass $6 \times 10^5$ $M_\odot$ in a dense gaseous disk situated in a low-mass halo of mass $M \sim$ few $\times\, 10^7$ $M_\odot$ (with virial temperature $T_{\rm vir} \simeq 10^4$ K) at redshift $z \sim 13$. This state-of-the-art simulation include stellar feedback from supernovae (which can inject energy and momentum into the gas) and ionizing photons (which can heat the gas as well as exert radiation pressure). Due to the high density of the disk, the free-fall time-scale is short ($\lesssim 10^5$ yrs), and big swaths of gas can be rapidly converted into stars before it is expelled by supernovae from the halo (after several Myrs). Cooling by molecular hydrogen has been neglected in this simulation, which is why the halo first form stars when it reaches the atomic-cooling threshold $T_{\rm vir} \simeq 10^4$ K (see Section 3.1). **Credit:** Taken from Figure 1 in Kimm *et al.* (2016).

## 1.2. Analytical models

Analytical models have the advantage over simulations of not being numerically expensive, and so could yield predictions at a much faster rate, and could scan the parameter space (of initial conditions, importance of individual feedback processes, etc.) at a much faster pace. On the other hand, it is often very hard to find analytical solutions and build physically comprehensive models. For this reason, analytical models in the literature often make very crude approximations, guesses, or assumptions regarding star formation and feedback in low-mass halos in the early Universe. Perhaps the most similar model to the model developed in this thesis (Anaxagoras) is the cluster formation model of Devecchi & Volonteri (2009), further developed by Devecchi *et al.* (2010). These authors modelled the disk formation and subsequent star formation and black hole formation in low-mass halos at Cosmic Dawn. However, these models face several problems:

- *Neglect the effect of stellar feedback:* In the first model by Devecchi & Volonteri (2009), it is simply assumed that 25% of the gas in the disk is converted into a star cluster, thereby making no attempt at modelling stellar feedback at all. In the extended model of Devecchi *et al.* (2010) the star formation rate is taken to scale with the disk surface density using the empirical Schmidt-Kennicutt law (Schmidt, 1959; Kennicutt, 1998). While this is an improvement over the original model by Devecchi & Volonteri (2009), it is not clear that it is a physically motivated *ab initio* model which could be safely extrapolated to the extremely high surface densities, sub-parsec scales, very low metallicities, and short starburst time-scales ($\lesssim 10^7$ $M_\odot$) encountered in low-mass halos at Cosmic Dawn. Furthermore, this star formation recipe does not take into account pre-supernova feedback from radiation pressure, stellar winds, and photoionization heating.

- *Inaccurate modelling of disk formation:* The gas properties, in particular the gas surface density and the disk size, has a significant impact on the predicted star formation efficiency and half-mass radius of the luminous object formed. Contrary to the assumption of Devecchi *et al.* (2010), the specific angular momentum of the disk is not independent of the disk mass. Instead, both simulations and physically motivated arguments suggest that it scales approximately linearly with the disk mass (Bullock *et al.*, 2001a; Abel *et al.*, 2002; O'Shea & Norman, 2007, 2008; Meece *et al.*, 2014), as discussed in Section 4.2. This means that the angular momentum of the disk, and hence its size, would be overestimated at early times following disk formation.

For mainly these reasons, the analytical star formation models of Devecchi & Volonteri (2009) and Devecchi *et al.* (2010) do not satisfy the two criteria for a physically realistic model of star formation in low-mass halos at Cosmic Dawn. Other authors have developed more detailed analytical or semi-analytical models of star formation, but they cannot be applied for one reason or another to the case of starbursts in low-mass halos. For example, Rahner *et al.* (2017) developed a detailed semi-analytical model of feedback from radiation pressure, winds, supernovae, and photoionization. However, these authors do not attempt to model the star formation rate itself, nor are they concerned with star formation in low-mass halos at Cosmic Dawn (and so do not satisfy the first criteria above for a *relevant* physically realistic model). Furthermore, the implementation of radiation pressure in the model of Rahner *et al.* (2017) is only partially complete, since it does not include resonant scattering of Lyman-$\alpha$ photons, which, as noted earlier, is a potentially important feedback process in gas of low metallicity. Finally, other authors have developed models of star formation within gaseous disks in halos which do take into account several stellar feedback processes (e.g. Ostriker & Shetty, 2011; Faucher-Giguère *et al.*, 2013; Krumholz *et al.*, 2018), but these models assume a steady equilibrium inconsistent with the expected bursty star formation in low-mass halos in the early Universe (e.g. Kimm *et al.*, 2016; Ricotti *et al.*, 2016; Kimm *et al.*, 2017; Faucher-Giguère, 2018a), and do not, unlike Devecchi & Volonteri (2009) and Devecchi *et al.* (2010), model the formation of the disk itself. During the time of writing of this thesis, Furlanetto & Mirocha (2022) presented a simplified model of starbursts in low-mass halos at Cosmic Dawn. They emphasize, in accordance with the simulations of Kimm *et al.* (2016) (and the modelling in this thesis), that starbursts can be efficient if the star formation time-scale is shorter than the time delay to the first supernova explosion. However, their model is perhaps best seen as illustrative, since they do not include any model of pre-supernova feedback. In summary, no physically realistic (in the sense of satisfying the two criteria laid out above) analytical model of starbursts in low-mass halos exist at the moment.

## 1.3. Introducing Anaxagoras

Given the shortcomings of previous analytical models and simulations, one could either await (or take part in) the development of more detailed numerically expensive simulations, or try to develop a more detailed and comprehensive analytical model of starbursts in low-mass halos. In this thesis I take the latter approach. The philosophy underlying this endeavor could be traced back to the Enrico Fermi, who would famously solve complicated problems by splitting it up into sub-problems, and using quick order-of-magnitude estimates. In this case, uncertainties in each sub-problem often cancel out and yield surprisingly accurate final predictions. More analogous to ANAXAGORAS, even extremely complicated problems like the climate evolution of Earth-like planets — involving many intricate feedback processes — have been studied using semi-analytical models (e.g. Hart, 1978; Menou, 2015) that are significantly less numerically expensive than full 3D general circulation climate models (analogous to state-of-the-art cosmological simulations). Likewise, ANAXAGORAS aim to model starbursts in low-mass halos at Cosmic Dawn from first principles by splitting up the model into several sub-models:

- *The cooling of gas in low-mass halos, and subsequent disk formation:* Once the gas in a dark matter halo can cool efficiently, it is expected to lose pressure support, collapse, and form a disk. The threshold halo mass for efficient gas cooling is studied in detail in Section 3.1, where a new cooling threshold in minihalos is derived that is in better agreement with cosmological simulations than earlier estimates. If efficient gas cooling is possible, a new self-similar solution of gas collapse in dark matter halos, derived in Section 3.2, is employed to calculate the central gas accretion rate (i.e. the disk accretion rate). The time-dependent properties of the resulting disk are estimated from first principles in Section 4.

- *Star formation and stellar feedback:* From the calculated disk properties, the star formation rate is estimated from first principles taking into account the properties of the gaseous disk and a comprehensive model of pre-supernova feedback. This includes momentum feedback from radiation pressure (including single-scattering photons, resonant scattering of Lyman-$\alpha$ photons, and multiple scattering of IR photons by dust). The starburst is finally terminated either by expanding overpressurized H II regions (photoionization feedback) or supernova feedback, depending on which process is most efficient. *The model of stellar feedback is not tuned to reproduce observations.*

- *The formation of star clusters and faint dwarfs:* From the star formation efficiency (the fraction of the disk converted into stars) the fraction of stars that remain bound in a cluster is calculated (Section 5.6). The stars that do not remain bound can expand to greater radii within the dark matter halo and form a faint dwarf galaxy. ANAXAGORAS therefore provide predictions for the stellar masses (or luminosities) and half-mass radii of both star clusters and faint dwarfs produced in starbursts in low-mass halos at Cosmic Dawn.

To give the reader a better overview of the rather complicated structure of ANAXAGORAS, a flowchart of the model can be seen in Section 2. The flowchart can be helpful for the reader to see where a given sub-model, which can be convoluted in of itself, fit in the larger scheme of ANAXAGORAS.

## 2. Overview of Anaxagoras: A flowchart

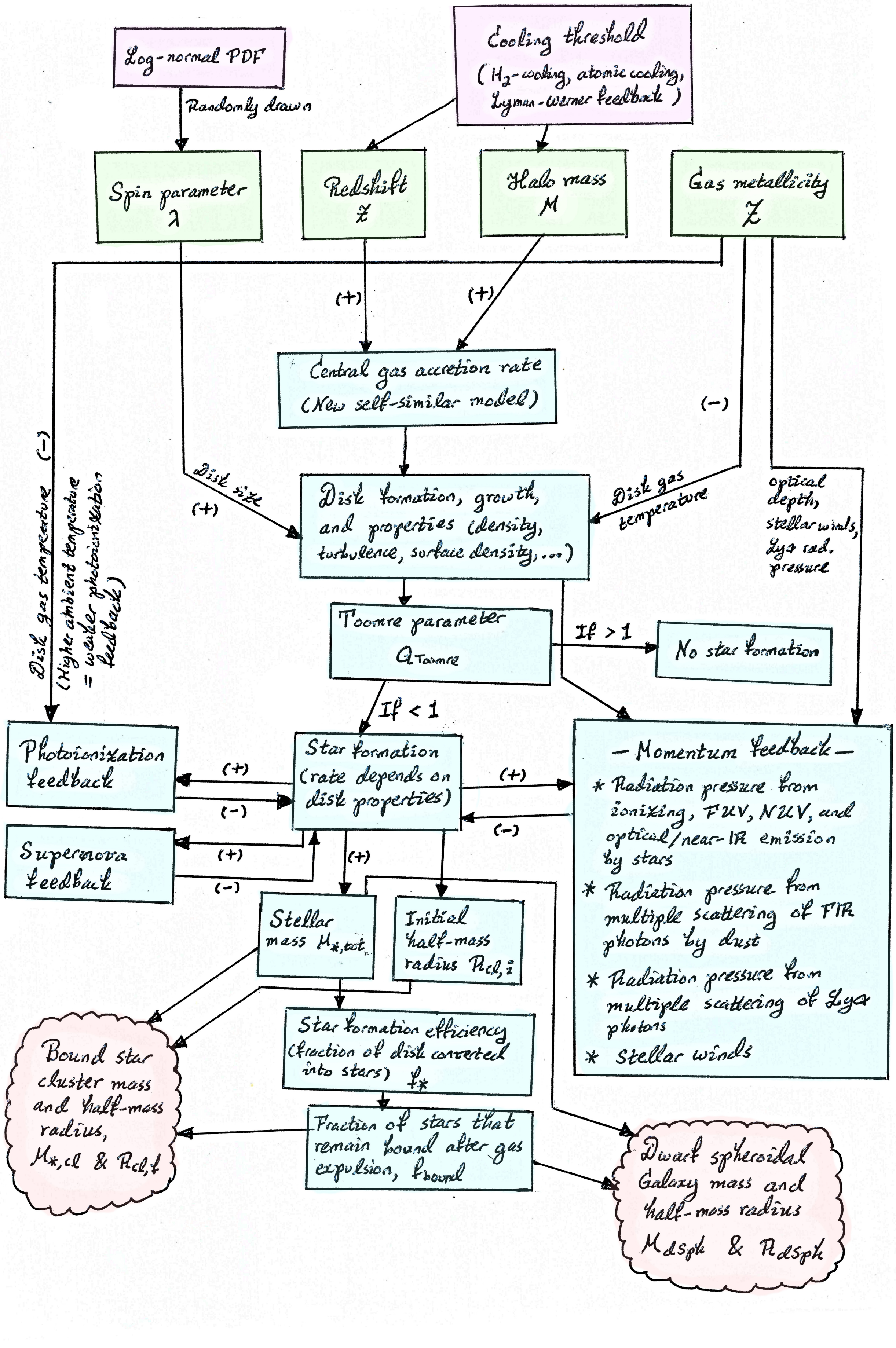

# 3. Cooling, collapse, and gas accretion in low-mass halos

## 3.1. Minimum halo mass for rapid gas cooling

In order for star formation to occur in a given halo containing gas, the gas must be able to cool efficiently so as to lose its pressure support and collapse to high densities. More specifically, the cooling time-scale $t_{\rm cool}$ should be shorter than the free-fall time-scale $t_{\rm ff}$ for gas in the halo in order for gas to undergo a near free-fall collapse to the center, fragment, and form stars (see e.g. Rees & Ostriker, 1977; Silk, 1977; White & Rees, 1978; Blumenthal *et al.*, 1984; Tegmark *et al.*, 1997; Tegmark & Rees, 1998; Nishi & Susa, 1999; Tegmark *et al.*, 2006; Bromm, 2013; Liu *et al.*, 2019). The two time-scales, assuming the gas in the halo has a temperature $T = T_{\rm vir}$ and central gas number density $n_{\rm core}$ just prior to efficient cooling, can be written as:

$$\begin{aligned} t_{\rm cool} &\equiv \frac{3 n_{\rm core} k_{\rm B} T_{\rm vir}/2}{\Lambda(T_{\rm vir}, n_{\rm core})} \\ t_{\rm ff} &\equiv \left(\frac{3\pi}{32 G \rho_{\rm tot,core}}\right)^{1/2}, \end{aligned} \tag{9}$$

where $\Lambda$ is the cooling rate per unit volume (in units of erg cm$^{-3}$ s$^{-1}$), and $\rho_{\rm tot,core}$ is the *total* (i.e. DM + baryon) matter density at the radius where the gas number density reaches its maximum value (i.e. $n_{\rm core}$). Cosmological hydrodynamical simulations and simple physical arguments indicate that, prior to efficient cooling, the baryons settle in a cored density profile (see e.g. Wise & Abel, 2007a; Visbal *et al.*, 2014a; Inayoshi *et al.*, 2015; Chon *et al.*, 2016). The constant-density core has a radius $R_{\rm core} \simeq 0.1\, R_{\rm vir}$, beyond which the density falls like $\rho_{\rm gas}(r) \propto r^{-2}$ (Visbal *et al.*, 2014a). For halos with masses $M \gtrsim$ few $\times\, 10^6$ M$_\odot$, Visbal *et al.* (2014a) found, using simulations, that the central gas number density $n_{\rm core} \equiv \rho_{\rm core}/\mu_{\rm h} m_{\rm H}$ (where $\mu_{\rm h}$ is the mean molecular weight of gas in the circumgalactic medium of the halo) in the absence of efficient radiative cooling has an upper limit:

$$n_{\rm core} < 5.4 \left(\frac{1+z}{10}\right)^3 \ {\rm cm^{-3}}\,. \tag{10}$$

Qualitatively, this could be interpreted to mean that the central gas density in atomic-cooling halos is proportional to the virial density of the halo, as one might expect.[7] Judging from their high-resolution simulations (see their figures 1-3), the central density for $M \gtrsim$ few $\times\, 10^6$ M$_\odot$ is typically smaller than the above upper limit by a factor of $\sim 0.6 - 0.7$. To be concrete we will therefore assume that the central gas number density in the high-mass (HM) limit is given by

$$n_{\rm core,HM} \simeq 3.5 \left(\frac{1+z}{10}\right)^3 \ {\rm cm^{-3}}\,, \tag{11}$$

which is a factor of 0.65 smaller than the upper bound in Eq. (10). In the low-mass (LM) limit ($M \lesssim$ few$\times$ $10^6$ M$_\odot$) Visbal *et al.* (2014a) showed that (in the absence of efficient radiative cooling) the observed central density in their simulations can be explained by assuming adiabatic contraction from IGM conditions (see also Tegmark *et al.*, 1997; Barkana & Loeb, 2001):

$$n_{\rm core,LM} \simeq \frac{\bar{\rho}_{\rm B}}{\mu_{\rm h} m_{\rm H}} \left(\frac{T_{\rm vir}}{T_{\rm IGM}}\right)^{1/(\gamma-1)}, \tag{12}$$

[7] In several early and simplified analytical models, the gas density was simply set to the virial density — even for minihalos (e.g. Tegmark *et al.*, 1997; Trenti & Stiavelli, 2009), in disagreement with the later simulation results of Visbal *et al.* (2014a) that found that the central density actually depend on $T_{\rm vir}$.

where $\gamma = 5/3$ is the adiabatic index,[8] and it is assumed that the central temperature in the halo ($T_\mathrm{h}$) is $T_\mathrm{h} \simeq T_\mathrm{vir}$. The IGM temperature (ignoring X-ray heating and photoionization) can be well fitted as a function of the scale factor $a = 1/(1+z)$ using the following formula (Tseliakhovich & Hirata, 2010; Liu *et al.*, 2019):[9]

$$T_\mathrm{IGM}(a) = \frac{T_\mathrm{CMB,0}}{a} \left\{ 1 + \frac{a/a_1}{1+(a_2/a)^{3/2}} \right\}^{-1}, \tag{13}$$

where $a_1 = 1/119$, $a_2 = 1/115$, and $T_\mathrm{CMB,0} = 2.726$ K is the present-day CMB temperature. This fitting formula gives the correct asymptotic scalings at high and low redshifts[10], along with the correct normalization. For the low redshifts of interest ($1+z \ll 115$), we get the adiabatic scaling $T_\mathrm{IGM}(z) \simeq 2.29\,[(1+z)/10]^2$ K. Upon using this in Eq. (12), along with Eqs. (6) and (2) for $T_\mathrm{vir}$, and $\bar{\rho}_\mathrm{B}$, respectively, we find (assuming $\mu_{1.2} \simeq 1$)

$$n_\mathrm{core,LM} \simeq 12\, M_7 \left(\frac{1+z}{10}\right)^{3/2}\ \mathrm{cm}^{-3}. \tag{14}$$

Following Visbal *et al.* (2014a) we take the central density in a given halo to then be given by the smaller of $n_\mathrm{core,LM}$ and $n_\mathrm{core,HM}$. Since the dividing line between the two limits occur at $M_7[(1+z)/10]^{-3/2} \simeq 0.292$, we can write this as:

$$n_\mathrm{core} \simeq \begin{cases} 3.5\,[(1+z)/10]^3\ \mathrm{cm}^{-3} & \text{for } M > 2.92\times10^6\,[(1+z)/10]^{3/2}\ \mathrm{M}_\odot \\ \\ 12\,M_7[(1+z)/10]^{3/2}\ \mathrm{cm}^{-3} & \text{for } M \leqslant 2.92\times10^6\,[(1+z)/10]^{3/2}\ \mathrm{M}_\odot \end{cases}. \tag{15}$$

A comparison between Eq. (15) and the results of Visbal *et al.* (2014a) is shown in Figure 6, showing a fair agreement between the median of their results and Eq. (15). Other authors have found that their cosmological simulations yield the same density and radius of the the constant-density core as Visbal *et al.* (2014a) (é.g. Chon *et al.*, 2016), which adds some confidence in our fit. On top of this, it should be stressed again that the scalings with redshift and halo mass in Eq. (15) have solid physical explanations, as discussed by Visbal *et al.* (2014a): In the low-mass limit the maximum gas density is set by adiabatic contraction, and for larger halo masses the gas density scales with the virial density of the halo. Thus, we conclude that Eq. (15) can be used for realistic estimates of the central gas density in halos prior to efficient radiative cooling and collapse.

[8]The gas is assumed to be monatomic here, which is a good approximation both during the contraction from the IGM as well as in the minihalo, since the fraction of gas in the form of $H_2$ is much smaller than unity.

[9]Tseliakhovich & Hirata (2010) derived this formula by fitting the result of detailed calculations using `RECFAST` (Seager *et al.*, 1999, 2000).

[10]At high redshifts $1+z \gtrsim 1000\,(\Omega_\mathrm{B}h)^{2/5} \sim 200$ the IGM temperature is approximately equal to the CMB temperature $T_\mathrm{CMB,0}/a$, since free electrons in the IGM are well coupled to the CMB by Compton scattering. As the residual electron fraction slowly decreases, and the CMB temperature decreases, the coupling between the IGM and the CMB temperatures becomes negligible. As a result, the IGM start to cool adiabatically, $T_\mathrm{IGM} \propto a^{-2}$. For more detailed discussion, see e.g. pp. 176-178 in Peebles (1993) or section 2.2 in Loeb & Furlanetto (2013).

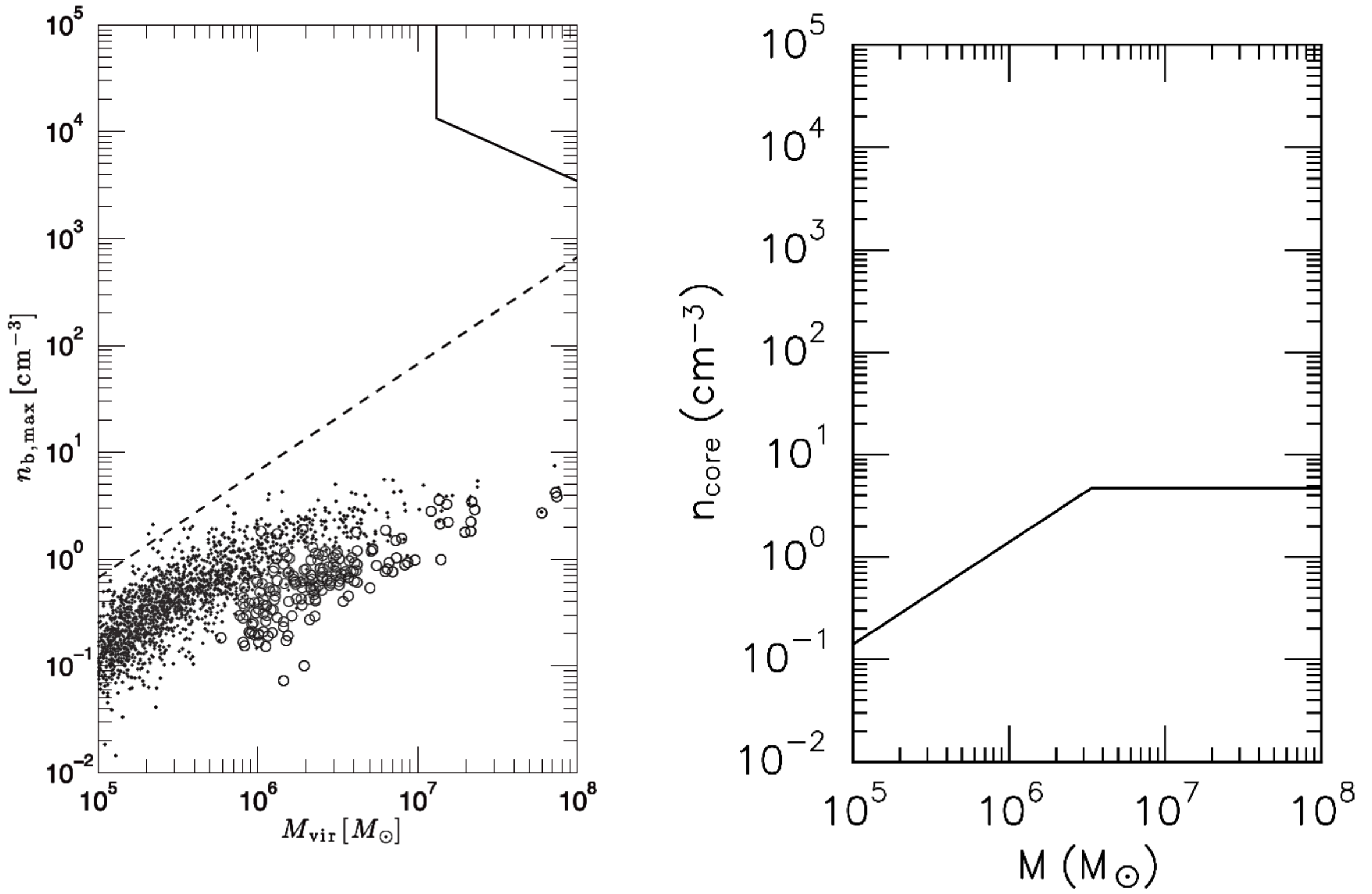

**Figure 6.** Comparison between the simulations of Visbal *et al.* (2014a) at $z = 10$ on the left, and Eq. (15) on the right for $n_{\rm core}$ ($n_{\rm b,max}$ in their notation) as a function of halo mass. The black dots in the plot by Visbal *et al.* (2014a) comes from their high-resolution simulations, while the circles come from their low-resolution runs. Their high-resolution simulations, which are more accurate, are most relevant for us. **Credit:** Left plot comes from Figure 2 in Visbal *et al.* (2014a).

### 3.1.1. $H_2$-cooling in minihalos in the absence of radiative feedback

> *"To make predictions for when the first objects formed and how big they were, you need to worry about something I hate: molecules."* — Max Tegmark

We start by considering minihalos with masses $M \leqslant 2.92 \times 10^6\,[(1+z)/10]^{3/2}$ M$_\odot$, and ignore the effect of radiative feedback on cooling, which will be considered separately. In this case the core density is given by $n_{\rm core} \simeq 0.12\,M_5[(1+z)/10]^{3/2}$ cm$^{-3}$. Using this in Eq. (9) for $t_{\rm cool}$, we find a cooling time-scale of:

$$t_{\rm cool} \simeq \frac{3.9 \times 10^{-15}\,{\rm erg\ cm^{-3}}}{\Lambda(T_{\rm vir}, n_{\rm core})}\,M_5^{5/3}\left(\frac{1+z}{10}\right)^{5/2}\,. \tag{16}$$

As shown in detail in Appendix 3, in metal-poor minihalos we can neglect metal line-cooling and focus on $H_2$-cooling, $\Lambda \simeq \Lambda_{\rm H_2}$. For temperatures 120 K $< T <$ 6400 K, Trenti & Stiavelli (2009) found that the $H_2$-cooling rate by Galli & Palla (1998) can be well-fitted by a power law in temperature (see also p. 17 in Stiavelli, 2009):[11]

$$\Lambda_{\rm H_2}(T, n) \simeq 3.98 \times 10^{-35}\,T^{3.4} n_{\rm H} n_{\rm H_2}\ {\rm erg\ cm^{-3}\ s^{-1}}\,, \tag{17}$$

[11] More accurate estimates of the $H_2$-cooling rate than provided by Galli & Palla (1998) exist in the literature, but for temperatures $T > 1000$ K the differences are quite small. See e.g. Figure 1 in Glover & Abel (2008).

where $n_{\rm H}$ and $n_{\rm H_2}$ are the hydrogen and molecular hydrogen number densities, respectively. This dependence on the product $n_{\rm H}n_{\rm H_2}$ comes from the fact that the $H_2$-cooling rate is dominated by collisional excitation of $H_2$ by H-atoms. To proceed we can write $n_{\rm H}n_{\rm H_2}$ in terms of $n$ (the total gas number density) by first defining the molecular hydrogen fraction $f_{\rm H_2} \equiv n_{\rm H_2}/n_{\rm H}$, so that $n_{\rm H}n_{\rm H_2} = f_{\rm H_2}n_{\rm H}^2$. Since[12] $n_{\rm H}/n = \mu(1-Y) = 0.926$ for primordial neutral gas ($\mu = 1.23$),[13] we then have $n_{\rm H}n_{\rm H_2} = 0.857\, f_{\rm H_2}n^2$, and so we get a cooling rate of:

$$\Lambda_{\rm H_2}(T,n) \simeq 3.41 \times 10^{-35}\, T^{3.4} f_{\rm H_2} n^2 \ {\rm erg\ cm^{-3}\ s^{-1}}\,. \tag{20}$$

Using $n_{\rm core} \simeq 0.12\, M_5[(1+z)/10]^{3/2}\ {\rm cm^{-3}}$ and $T_{\rm vir} \simeq 156\, M_5^{2/3}[(1+z)/10]$ K, we find the following $H_2$-cooling rate in the gaseous core of a minihalo:

$$\Lambda_{\rm H_2}(T_{\rm vir}, n_{\rm core}) \simeq 1.40 \times 10^{-29}\, M_5^{4.27} \left(\frac{1+z}{10}\right)^{6.4} f_{\rm H_2}\ {\rm erg\ cm^{-3}\ s^{-1}}\,. \tag{21}$$

Using this in Eq. (16) yields a cooling time-scale of

$$\begin{aligned} t_{\rm cool} &\simeq 2.8 \times 10^{14}\, M_5^{-2.60} \left(\frac{1+z}{10}\right)^{-3.9} f_{\rm H_2}^{-1}\ {\rm s} \\ &= 8.9 \times 10^{6}\, M_5^{-2.60} \left(\frac{1+z}{10}\right)^{-3.9} f_{\rm H_2}^{-1}\ {\rm yrs}\,. \end{aligned} \tag{22}$$

The difference in the value of $t_{\rm cool}$ and its scaling with halo mass and redshift[14] compared to Trenti & Stiavelli (2009) stems from their assumption that the gas density in the minihalo is simply $\propto \rho_{\rm vir} \propto (1+z)^3$, which as we saw earlier is an erroneous assumption since the central gas density of a minihalo is determined by adiabatic contraction, giving $n_{\rm core} \propto M(1+z)^{3/2}$. This, as we will ultimately see, will lead to a different redshift scaling of the $H_2$-cooling threshold halo mass. Next we need to estimate the free-fall time-scale in the gas core of the minihalo. Assuming a Navarro-Frenk-White (NFW, Navarro *et al.*, 1995, 1996, 1997) density profile for the dark matter halo we have[15]

$$\begin{aligned} \rho_{\rm DM}(r) &= \frac{\Delta_{\rm vir}\bar{\rho}_{\rm M}}{3f(c)} \frac{c^3}{(r/R_{\rm s})(1+r/R_{\rm s})^2} \\ &= \frac{4.78 \times 10^{-25}\, c^3\ {\rm g\ cm^{-3}}}{3f(c)(r/R_{\rm s})(1+r/R_{\rm s})^2} \left(\frac{1+z}{10}\right)^3\,, \end{aligned} \tag{23}$$

[12]The hydrogen number density is given by $n_{\rm H} = X\rho_{\rm B}/m_{\rm H}$ (with $X = 1-Y$ being the hydrogen mass fraction, assuming primordial gas), whereas $n = \rho_{\rm B}/\mu m_{\rm H}$, which yields $n_{\rm H}/n = \mu X = \mu(1-Y)$.

[13]For primoridal gas the mean particle mass is $\bar{m} = (n_{\rm H}m_{\rm H} + n_{\rm He}m_{\rm He})/n = (n_{\rm H}m_{\rm H} + n_{\rm He}4m_{\rm H})/n$, so that

$$\mu \equiv \frac{\bar{m}}{m_{\rm H}} = \frac{n_{\rm H} + 4n_{\rm He}}{n_{\rm H} + n_{\rm He}}\,. \tag{18}$$

Since $n_{\rm H} = X\rho_{\rm B}/m_{\rm H}$ and $n_{\rm He} = Y\rho_{\rm B}/4m_{\rm H}$ we get

$$\mu = \frac{X+Y}{X+Y/4}\,. \tag{19}$$

Since $X+Y=1$ for the primordial gas of interest, we get $\mu = (X+Y/4)^{-1} = 1.23$.

[14]Trenti & Stiavelli (2009) estimated a cooling time-scale of

$$t_{\rm cool} \simeq 6.2 \times 10^{14}\, M_5^{-1.6} \left(\frac{1+z}{10}\right)^{-5.4} f_{\rm H_2}^{-1}\ {\rm s}\,.$$

[15]See e.g. Mo *et al.* (1998) on how to rewrite the density profile in terms of the virial density of the halo. One difference is that we use $\Delta_{\rm vir} = 18\pi^2$ rather than $\Delta_{\rm vir} = 200$.

where $R_\mathrm{s} = R_\mathrm{vir}/c$ is the scale radius, $c$ is the halo concentration, and $f(c) \equiv \ln(1+c) - c/(1+c)$. For minihalos with masses $10^5\,\mathrm{M}_\odot \lesssim M \lesssim 10^6\,\mathrm{M}_\odot$ at high redshifts ($z \gtrsim 10$), the halo concentration is typically $c \sim 3$ (see e.g. Figure 7 in Correa *et al.*, 2015b). At the core radius $R_\mathrm{core} \simeq 0.1\,R_\mathrm{vir}$ we then find a DM density $\rho_\mathrm{DM,core} \equiv \rho_\mathrm{DM}(R_\mathrm{core})$ of

$$\rho_\mathrm{DM,core} \simeq 1.3 \times 10^{-23} \left(\frac{1+z}{10}\right)^3 \ \mathrm{g\ cm^{-3}}\,. \tag{24}$$

For comparison, the gas mass density $n_\mathrm{core}\mu_\mathrm{h} m_\mathrm{H}$ in the core of the minihalo is just $\simeq 2.5 \times 10^{-25}\,M_5[(1+z)/10]^{3/2}$ g cm$^{-3}$, indicating that the DM halo is gravitationally dominant. Thus, we can for neglect the baryons when calculating the free-fall time-scale in Eq. (9), giving us

$$\begin{aligned} t_\mathrm{ff} &\simeq 5.8 \times 10^{14} \left(\frac{1+z}{10}\right)^{-3/2} \ \mathrm{s} \\ &= 1.8 \times 10^{7} \left(\frac{1+z}{10}\right)^{-3/2} \ \mathrm{yrs}\,. \end{aligned} \tag{25}$$

Efficient cooling of the gas in the minihalo is possible if $t_\mathrm{cool} < t_\mathrm{ff}$, which upon using Eqs. (22) and (25) yields a condition on the required molecular hydrogen fraction:

$$f_{\mathrm{H}_2} > 0.48\,M_5^{-2.60} \left(\frac{1+z}{10}\right)^{-2.4} \equiv f_{\mathrm{H_2,req}}\,. \tag{26}$$

Only if $f_{\mathrm{H}_2} > f_{\mathrm{H_2,req}}$ will the gas be able to cool efficiently in the minihalo. The next task is therefore to check whether $f_{\mathrm{H}_2} > f_{\mathrm{H_2,req}}$ can be produced in the minihalo. Molecular hydrogen in chemically pristine gas is mainly formed in the following steps (e.g. McDowell, 1961; Peebles & Dicke, 1968; Hirasawa *et al.*, 1969; Hirasawa, 1969; Dalgarno & McCray, 1973; Tegmark *et al.*, 1997):

$$\begin{aligned} \mathrm{H} + e^- &\rightarrow \mathrm{H}^- + h\nu \qquad (\text{rate } k_2) \\ \mathrm{H} + \mathrm{H}^- &\rightarrow \mathrm{H}_2 + e^- \qquad (\text{rate } k_3)\,. \end{aligned} \tag{27}$$

The evolution of the number densities of electrons ($n_\mathrm{e}$), $\mathrm{H}^-$ ($n_{\mathrm{H}^-}$), and $\mathrm{H}_2$ ($n_{\mathrm{H}_2}$) are then governed by the following differential equations:

$$\begin{aligned} \dot{n}_\mathrm{e} &= -\,k_1 n_{\mathrm{H}^+} n_\mathrm{e} \\ \dot{n}_{\mathrm{H}^-} &= k_2 n_{\mathrm{H}^0} n_\mathrm{e} - k_3 n_{\mathrm{H}^0} n_{\mathrm{H}^-} \\ \dot{n}_{\mathrm{H}_2} &= k_3 n_{\mathrm{H}^0} n_{\mathrm{H}^-}\,, \end{aligned} \tag{28}$$

where $k_1$ is the recombination coefficient (collisional ionization can be neglected in this case). I will adopt the Case-B recombination coefficient, which is the relevant recombination coefficient in our case.[16] As discussed by Tegmark *et al.* (1997), $\mathrm{H}^-$ quickly reaches its equilibrium abundance $n_{\mathrm{H}^-,\mathrm{eq}} = k_2 n_\mathrm{e}/k_3$. The resulting $\mathrm{H}_2$ formation rate becomes:

$$\dot{n}_{\mathrm{H}_2} = k_2 n_{\mathrm{H}^0} n_\mathrm{e}\,. \tag{29}$$

[16]The Case-B recombination coefficient applies when a recombination to the ground state of hydrogen leads to a photon with energy 13.6 eV which then ionizes another hydrogen atom, leading to no effective change in the ionization state. In other words, the Case-B recombination coefficient should be used for a gas cloud that is optically thick at the Lyman limit. This is indeed the case for our minihalo: $n_\mathrm{core} \simeq 0.12\,M_5[(1+z)/10]^{3/2}$ cm$^{-3}$ and $R_\mathrm{core} \simeq 0.1\,R_\mathrm{vir} \simeq 15\,M_5^{1/3}[(1+z)/10]^{-1}$ pc yields an optical depth $\tau_0 \sim n_\mathrm{core}\sigma_0 R_\mathrm{core} \simeq 33\,M_5^{4/3}[(1+z)/10]^{1/2}$, where $\sigma_0 \simeq 6 \times 10^{-18}$ cm$^2$ is the photoionization cross-section at the Lyman limit.

TABLE 1. REACTION RATES

| Reaction | Rate $k$ adopted by Tegmark *et al.* (1997) (cm$^3$ s$^{-1}$) | Rate $k$ adopted in this thesis$^\dagger$ (cm$^3$ s$^{-1}$) |
|---|---|---|
| $H^+ + e^- \to H + h\nu$ | $k_1 \simeq 1.88 \times 10^{-10}\, T^{-0.64}$ | $k_1 \simeq 2.11 \times 10^{-10}\, T^{-0.72}$ |
| $H + e^- \to H^- + h\nu$ | $k_2 \simeq 1.83 \times 10^{-18}\, T^{0.88}$ | $k_2 \simeq 1.4 \times 10^{-18}\, T^{0.928}\, e^{-T/16200}$ |

$^\dagger$ For details on the fit to $k_1$, see Appendix 1. For $k_2$ we adopt the fit provided by Galli & Palla (1998).

At this point it is convenient to define $n_{\rm H} \equiv n_{\rm H^0} + n_{\rm H^+} + 2n_{\rm H_2}$, $x \equiv n_{\rm e}/n_{\rm H} \simeq n_{\rm H^+}/n_{\rm H}$, and (as before) $f_{\rm H_2} \equiv n_{\rm H_2}/n_{\rm H}$. This yields

$$1 = \frac{n_{\rm H^0}}{n_{\rm H}} + x + 2f_{\rm H_2}\,. \tag{30}$$

Thus, $\dot{f}_{\rm H_2} = k_2\,(1 - x - 2f_{\rm H_2})\,x n_{\rm H}$, and we get the following approximate differential equations for $x$ and $f_{\rm H_2}$ (Tegmark *et al.*, 1997):

$$\begin{aligned} \dot{x} &= -\,k_1 n_{\rm H} x^2 \\ \dot{f}_{\rm H_2} &= k_2 n_{\rm H} x\,, \end{aligned} \tag{31}$$

where it is assumed that $x + 2f \ll 1$ (as is the case for the largely neutral and atomic hydrogen in a minihalo prior to gas collapse). The rates $k_1$ and $k_2$ are given in Table 1, including the rates adopted by Tegmark *et al.* (1997). To get the recombination coefficient $k_1$ I have derived a power law fit that is in fair agreement with the more complicated fit by Draine (2011) (p. 139), and the theoretical calculation by Padmanabhan (2000) (p. 323).

In contrast, as shown in Appendix 1, the fit for $k_1$ adopted by Tegmark *et al.* (1997) (which they take from Hutchins, 1976) overestimates the (Case-B) recombination coefficient by $\sim 50\%$ at $T = 10^3$ K and $\sim 90\%$ at $T = 10^4$ K, ultimately leading to an underestimate of the abundance of $H_2$ that can be produced in the minihalo (since, as we shall soon see, $f_{\rm H_2} \propto 1/k_1$). Several later analytical models of the $H_2$-cooling threshold halo mass based on the reaction rates in Tegmark *et al.* (1997) — which have been applied in N-body simulations that tag minihalos for Pop III star formation, or to gauge the effect of Pop III star formation on the 21-cm signal — inherit this problem (e.g. Trenti & Stiavelli, 2009; Crosby *et al.*, 2013; Griffen *et al.*, 2018; Mebane *et al.*, 2018, 2020). The inaccurate value for $k_1$ used by Tegmark *et al.* (1997) can be traced to their adoption of the fit by Hutchins (1976), which in turn is derived from the Case-A (rather than Case-B) recombination coefficient by Spitzer (1956) (pp. 91-92). As for the reaction rate $k_2$, I adopt the fit from Galli & Palla (1998) which is in good agreement with the fit used by Tegmark *et al.* (1997) (taken from Hutchins, 1976) for $T \lesssim 3 \times 10^3$ K, but it is more accurate at higher temperatures (see e.g. the discussion in Glover, 2015).

Let us now proceed to estimate the molecular hydrogen fraction that can be attained in the minihalo by solving the differential equations in Eq. (31), following the derivation in Tegmark *et al.* (1997). The differential equation for $x$ can be separated and integrated to yield

$$x(t) = \frac{x_0}{1 + k_1 n_{\rm H} x_0 t}\,, \tag{32}$$

where $x_0$ is the initial ionization fraction. The resulting differential equation for $f_{\rm H_2}$ becomes

$$\dot{f}_{\rm H_2} = \frac{k_2 n_{\rm H} x_0}{1 + k_1 n_{\rm H} x_0 t}\,, \tag{33}$$

which can be integrated to give (compare with Eq. 17 in Tegmark *et al.*, 1997):

$$\begin{aligned} f_{\rm H_2}(t) &= \frac{k_2}{k_1}\ln(1 + k_1 n_{\rm H} x_0 t) \\ &\simeq 6.6\times 10^{-9}\, T^{1.648}\, e^{-T/16200}\ln(1 + k_1 n_{\rm H} x_0 t)\,, \end{aligned} \tag{34}$$

where I have assumed that the initial value of $f_{\rm H_2}$ is negligible.[17] Using $T \simeq T_{\rm vir} \simeq 156\, M_5^{2/3}[(1+z)/10]$ K then yields

$$f_{\rm H_2}(t) \simeq 2.7\times 10^{-5}\, M_5^{1.099}\left(\frac{1+z}{10}\right)^{1.648} e^{-0.00963\, M_5^{2/3}[(1+z)/10]}\ln(1 + k_1 n_{\rm H} x_0 t)\,. \tag{35}$$

Demanding that $f_{\rm H_2}(t)$ exceed the required $H_2$ abundance for efficient cooling in Eq. (26) gives us the following threshold halo mass for $H_2$-cooling:

$$M > 1.4\times 10^{6}\,\frac{e^{0.00260\, M_5^{2/3}[(1+z)/10]}}{\ln^{0.2703}(1 + k_1 n_{\rm H} x_0 t)}\left(\frac{1+z}{10}\right)^{-1.094} {\rm M}_\odot\,. \tag{36}$$

We still need to estimate the time $t$, which should be the time available for the molecular hydrogen fraction to grow in the halo. Tegmark *et al.* (1997) simply set $k_1 n_{\rm H} x_0 t \simeq 1$ on the basis of the logarithm only growing slowly at late times. However, here we want to use a more physically motivated prescription for better accuracy. We can estimate $t$ to be approximately equal to the growth time-scale of the gaseous core, $t_{\rm gr} \equiv M_{\rm core}/\dot{M}_{\rm core}$. This can be motivated as follows. For $t \gtrsim t_{\rm gr}$ we would have a large influx of new gas of primordial composition with negligible molecular hydrogen abundance, so that the average value in the gaseous core drops significantly, and the effective cooling rate averaged over the whole core falls off (see also Reed *et al.*, 2005, for a discussion of this point). Secondly, for $t \gtrsim t_{\rm gr}$ we expect the gas to be dynamically heated from mergers, which leads to a further drop (if not complete halt) to the net cooling rate. Only for times $t \lesssim t_{\rm gr}$ can we ignore all of this, which motivates our choice $t = t_{\rm gr}$. For a fixed redshift the core mass scales as $M_{\rm core} \propto n_{\rm core} R_{\rm vir}^3 \propto M^2$, and so $t_{\rm gr} = M/2\dot{M}$. Neistein *et al.* (2006) showed that the halo mass accretion rate can be estimated analytically using extended Press-Schechter theory (see also Correa *et al.*, 2015a; Salcido *et al.*, 2018):

$$\frac{\dot{M}}{M} = \sqrt{\frac{2}{\pi}}\frac{\delta_{\rm crit}}{\sqrt{S_q - S}}\frac{\dot{\mathfrak{D}}}{\mathfrak{D}^2}\,, \tag{37}$$

where $S \equiv \sigma^2(M)$, $S_q \equiv \sigma^2(M/q)$, and $\mathfrak{D}(z)$ is the growth factor of linear density perturbations. The factor $q$ is at most only very weakly dependent on the halo mass and cosmology (Neistein *et al.*, 2006; Correa *et al.*, 2015a), and it appears only within $S_q$, which is a weak function of $q$. Thus, we can treat it as a constant. For $10^5\,{\rm M}_\odot < M < 10^8\,{\rm M}_\odot$ — covering practically the whole halo mass range of interest in this thesis — Correa *et al.* (2015a) finds $2.32 < q < 2.55$ (see their Eqs. 22 and 23). To be concrete we can adopt $q = 2.38$, which is the value of $q$ at $M = 10^6\,{\rm M}_\odot$. As shown in Appendix 10, the resulting halo mass accretion rate becomes

$$\frac{\dot{M}}{M} \simeq \frac{3.04\times 10^{-15}}{\sigma_8\left|\ln \bar{M}\right|}\left(\frac{1+z}{10}\right)^{5/2} {\rm s}^{-1}\,, \tag{38}$$

[17] The initial abundance of $H_2$ in the IGM is expected to freeze out by $z \lesssim 40$ to a value of $f_{\rm H_2}(z \lesssim 40) \simeq 6\times 10^{-7}$ (Galli & Palla, 2013), which is much smaller than the required abundance for efficient cooling in Eq. (26), and hence can be neglected when estimating the cooling threshold.

using the analytical approximations of $\sigma(M)$ and $\mathfrak{D}(z)$ from Eqs. (3) and (326), respectively. The growth time-scale $t_{\rm gr} = M/2\dot{M}$ for the core is therefore:

$$t_{\rm gr} \simeq 1.6\times10^{14}\,\sigma_8 \left|\ln\bar{M}\right| \left(\frac{1+z}{10}\right)^{-5/2} \ {\rm s} \\ = 5.2\times10^{7}\,\sigma_8 \left|\ln\bar{M}\right| \left(\frac{1+z}{10}\right)^{-5/2} \ {\rm yrs}\,. \quad (39)$$

We can now use this to evaluate $k_1 n_{\rm H} x_0 t = k_1 n_{\rm H} x_0 t_{\rm gr}$ in Eq. (36). Using $n_{\rm H} = 0.926\,n_{\rm core} \simeq 1.1\,M_6[(1+z)/10]^{3/2}$ cm$^{-3}$ and the recombination coefficient from Table 1 (with $T = T_{\rm vir}$), we get $k_1 n_{\rm H} x_0 \simeq 2.3\times10^{-10}\,T_{\rm vir}^{-0.72} M_6[(1+z)/10]^{3/2} x_0$ s$^{-1}$, and so

$$k_1 n_{\rm H} x_0 t_{\rm gr} \simeq 3.7\times10^{4}\,\sigma_8 T_{\rm vir}^{-0.72} M_6 \left|\ln\bar{M}\right| \left(\frac{1+z}{10}\right)^{-1} x_0\,, \quad (40)$$

or, with $T_{\rm vir} \simeq 724\,M_6^{2/3}[(1+z)/10]$ K,

$$k_1 n_{\rm H} x_0 t_{\rm gr} \simeq 320\,\sigma_8 M_6^{0.52} \left|\ln\bar{M}\right| \left(\frac{1+z}{10}\right)^{-1.72} x_0\,. \quad (41)$$

The initial ionization fraction $x_0$ can be taken to be equal to the residual ionization fraction $x_{\rm res} \simeq 1.38\times10^{-5}\,\Omega_{\rm M}^{1/2}/(\Omega_{\rm B}h) = 2.33\times10^{-4}$ well after ($z \lesssim 50$) the Epoch of Recombination (see p. 13 in Stiavelli, 2009), and we can also use our standard value $\sigma_8 = 0.811$ from now on. Furthermore, we can approximate $\left|\ln\bar{M}\right|$ as a power law around $M = 10^6\,{\rm M}_\odot$ (at which $\left|\ln\bar{M}\right| = 24.1$),[18] giving us $\left|\ln\bar{M}\right| \simeq 24.1\,M_6^{-0.0415}$. Putting everything together yields

$$k_1 n_{\rm H} x_0 t_{\rm gr} \simeq 1.5\,M_6^{0.48} \left(\frac{1+z}{10}\right)^{-1.72}\,. \quad (42)$$

Thus, from Eq. (36) we find that rapid $H_2$-cooling (absent radiative feedback) is possible in minihalos with masses $M > M_{\rm th,H_2}$ where:

$$M_{\rm th,H_2} \simeq 1.4\times10^{6}\,\frac{e^{0.0120\,M_{\rm th,H_2,6}^{2/3}[(1+z)/10]}}{\ln^{0.2703}(1+1.5\,M_{\rm th,H_2,6}^{0.48}[(1+z)/10]^{-1.72})} \left(\frac{1+z}{10}\right)^{-1.094} {\rm M}_\odot\,. \quad (43)$$

This is a trancendental equation for $M_{\rm th,H_2}$, but we can make some progress by finding a very good approximate solution using iteration. As a first crude guess at a solution we can ignore the exponential, since the argument of the exponential is expected to be much smaller than unity. Furthermore, for $M_{\rm th,H_2,6} \sim \mathcal{O}(1)$ we expect that $\ln^{0.2703}(1+1.5\,M_{\rm th,H_2,6}^{0.48}[(1+z)/10]^{-1.72}) \sim \mathcal{O}(1)$. Thus, as a first guess we can take:

$$M_{\rm th,H_2}^{(1)} \simeq 1.4\times10^{6} \left(\frac{1+z}{10}\right)^{-1.094} {\rm M}_\odot\,. \quad (44)$$

We can then get an even better approximate solution by using this result on the right-hand side of Eq. (43) to get our final result:

[18]The power law index $\eta$ can be found by taking the logarithmic derivative of $\left|\ln\bar{M}\right|$ and evaluating it at $M = 10^6\,{\rm M}_\odot$:

$$\eta = \frac{1}{\left|\ln\bar{M}\right|}\frac{\partial\left|\ln\bar{M}\right|}{\partial\ln\bar{M}} = -\frac{1}{\left|\ln\bar{M}\right|} \simeq -0.0415\,.$$

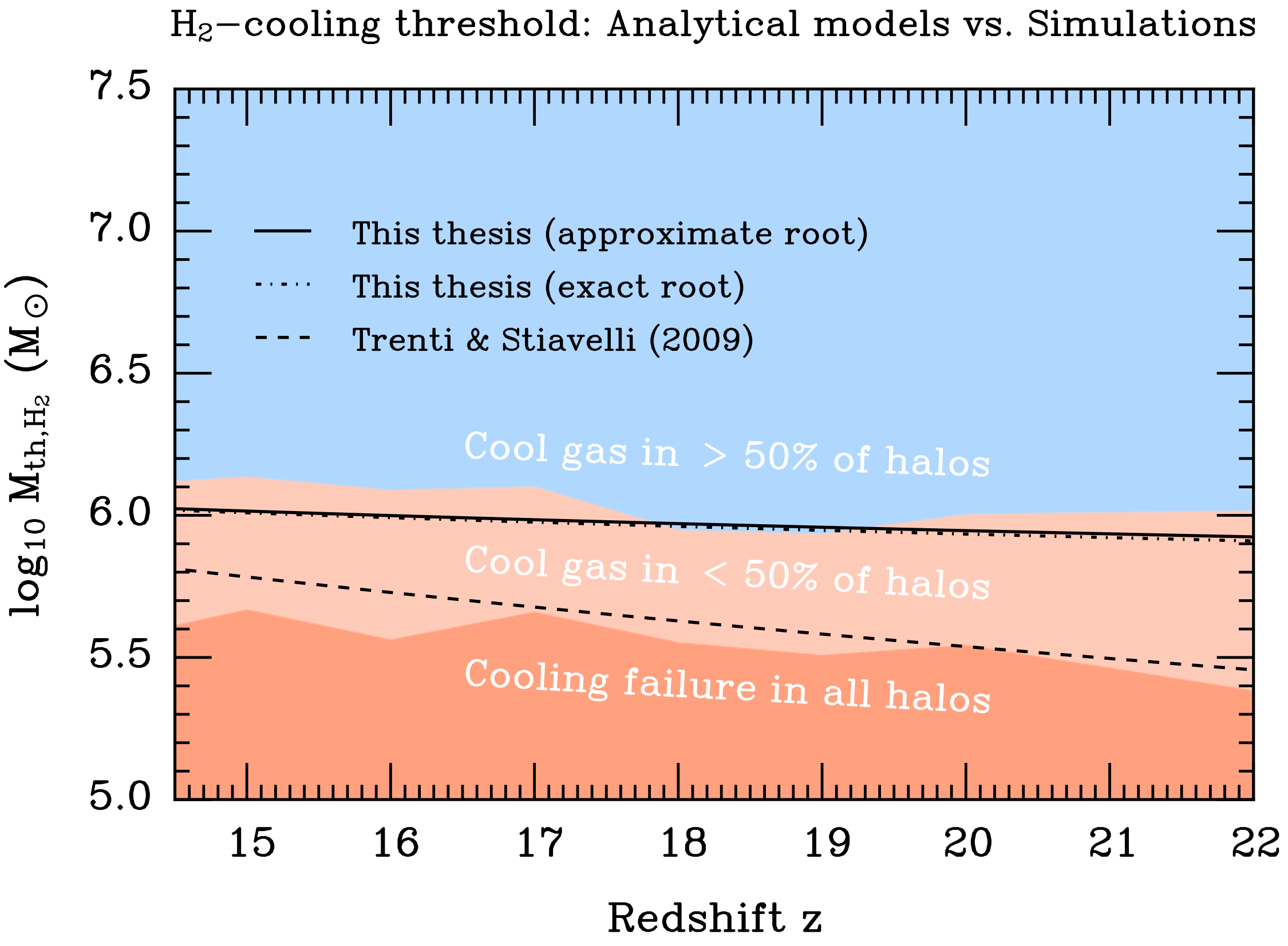

**Figure 7.** A plot of the $H_2$-cooling threshold in minihalos in the absence of radiative feedback. The solid line show the result of numerically finding the root to Eq. (43), while the dash-dotted line shows the approximate root in Eq. (45). Also shown is the analytical model of Trenti & Stiavelli (2009). The colored background shows data from the cosmological hydrodynamical simulations of Schauer *et al.* (2020). As can be seen, unlike the atomic-cooling threshold, the $H_2$-cooling threshold is not particularly sharp (due to the cooling function in Eq. 17 not being extremely sensitive to temperature, along with the actual scatter in central densities of minihalos evident from Figure 6). However, the model of this thesis is seen to better reproduce the halo mass above which > 50% of minihalos contain cold and dense gas than the model of Trenti & Stiavelli (2009). **Credit:** Dr. Anna T. P. Schauer of UT Austin for kindly providing the data from the simulations in Schauer *et al.* (2020).

**Halo mass threshold for efficient $H_2$-cooling in the absence of radiative feedback**

$$M_{\mathrm{th,H_2}} \simeq 1.4\times10^6\,\frac{e^{0.015\,[(1+z)/10]^{0.2707}}}{\ln^{0.2703}\!\left(1+1.8\,[(1+z)/10]^{-2.25}\right)}\left(\frac{1+z}{10}\right)^{-1.094}\,\mathrm{M}_\odot\,. \tag{45}$$

This approximate root to Eq. (43) is accurate to within 4% (6%) for $5<z<25$ ($5<z<35$), and will therefore be adopted in Anaxagoras. In Figure 7 I have plotted the approximate root in Eq. (45) as well as the exact root to Eq. (43),[19] comparing it to the expression derived by Trenti & Stiavelli (2009). The colored background is data derived from the cosmological hydrodynamical simulations of Schauer *et al.*

[19]Found numerically using `scipy.optimize.bisect`.

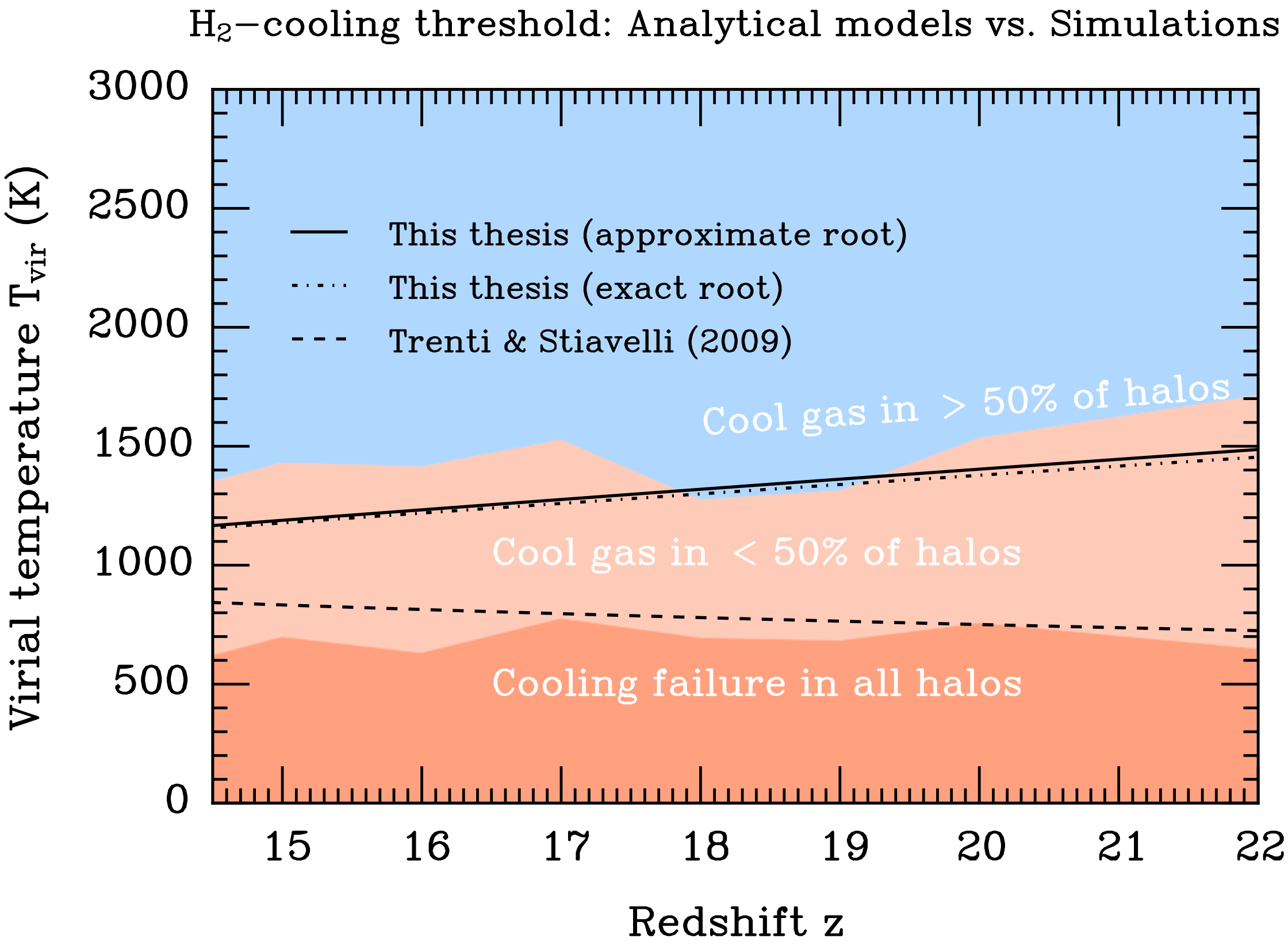

**Figure 8.** Same as Figure 7, but for the corresponding virial temperature (i.e. the threshold virial temperature for efficient $H_2$-cooling in the absence of radiative feedback). **Credit:** Dr. Anna T. P. Schauer of UT Austin for kindly providing the data from the simulations in Schauer *et al.* (2020).

(2020), kindly provided to this student by Dr. Anna T. P. Schauer (Department of Astronomy at UT Austin). In Figure 8 an analogous plot is shown for the threshold virial temperature $T_{\rm vir}$ rather than halo mass, evaluated using Eq. (6). Schauer *et al.* (2019, 2020) studied the halo mass threshold below which no halos contained cool gas, or less than 50% contained cool gas. These authors found that in the absence of radiative feedback, the $H_2$-cooling threshold below which $< 50\%$ of halos contained cool and dense gas has a very weak redshift dependence, as can be seen in Figure 7 (see Yoshida *et al.*, 2003; Hummel *et al.*, 2012, for similar results). We see that Eq. (45) reproduces this threshold fairly accurately compared to the model of Trenti & Stiavelli (2009) — both the magnitude and the weak redshift dependence. This improvement is not surprising for the reasons given earlier in this section: Trenti & Stiavelli (2009) adopted an incorrect gas density for the minihalos, along with results for the maximum $H_2$ abundance from Tegmark *et al.* (1997) that used the wrong recombination coefficient (Case-A rather than Case-B).

Our derivation of the $H_2$-cooling threshold in Eq. (45) did not consider the effect of radiative feedback, which can suppress the $H_2$-cooling rate and hence increase the $H_2$-cooling threshold above Eq. (45). Two processes in particular have been discussed in the literature:

1. *Lyman-Werner (LW) feedback:* Photons in the LW band ($11.2 - 13.6$ eV) can photodissociate $H_2$ (see e.g. pp. 346-347 in Draine, 2011), and therefore suppress the $H_2$-cooling rate in minihalos.

2. *Photodetachment of $H^-$:* Photons with energies $\gtrsim 0.754$ eV can photodetach $H^-$ (Miyake *et al.*, 2010; McLaughlin *et al.*, 2017), hence reducing the rate of $H_2$ formation. With fewer $H_2$ molecules around, the $H_2$-cooling rate in minihalos is suppressed.

The bulk of the relevant literature on cooling in minihalos over the years have focused on LW feedback (e.g. Haiman *et al.*, 1997, 2000; Machacek *et al.*, 2001; Yoshida *et al.*, 2003; Wise & Abel, 2007b; O'Shea & Norman, 2008; Trenti & Stiavelli, 2009; Safranek-Shrader *et al.*, 2012; Pawlik *et al.*, 2013; Visbal *et al.*, 2014b; Latif & Khochfar, 2019; Kulkarni *et al.*, 2020; Schauer *et al.*, 2020; Skinner & Wise, 2020). Early studies found that LW feedback could be important, leading to the wide-spread belief that it could raise the cooling threshold globally to the atomic-cooling limit $T_{\rm vir} \simeq 10^4$ K well before the IGM is reionized (e.g. Haiman *et al.*, 1997; Trenti & Stiavelli, 2009; Pawlik *et al.*, 2013) even for modest LW intensities of $J_{\rm LW,21} \sim 0.01 - 0.1$, where $J_{\rm LW,21} \equiv J_{\rm LW}/10^{-21}$ erg cm$^{-2}$ s$^{-1}$ Hz$^{-1}$ sr$^{-1}$ is the normalized LW intensity at $h\nu = 12.4$ eV (i.e. the middle of the LW band). However, more recent studies have found that LW feedback has been greatly exaggerated due to a neglect of self-shielding by the $H_2$ in minihalos against external LW radiation (e.g. Latif & Khochfar, 2019; Kulkarni *et al.*, 2020; Schauer *et al.*, 2020; Skinner & Wise, 2020). Because of this, Cen (2017) has suggested that photodetachment of $H^-$ could be more important. Modelling the effect of LW feedback with self-shielding cannot be done analytically in the manner we did for the case without radiative feedback,[20] and so we have to turn to simulations. Recently Latif & Khochfar (2019) studied the effect of LW and $H^-$ photodetachment feedback from Pop II galaxies on cooling in minihalos, including self-shielding by $H_2$. Modelling the spectra of Pop II galaxies as thermal with temperature $2 \times 10^4$ K these authors found that the cooling threshold can be fitted by

**HALO MASS THRESHOLD FOR EFFICIENT $H_2$-COOLING IN THE PRESENCE OF RADIATIVE FEEDBACK**

$$M_{\rm th,LW} = \exp\left\{19.38 - 4.3\exp(-0.16\,\ln J_{21})\right\}\ {\rm M}_\odot\,, \tag{46}$$

where $J_{21} \equiv J_\nu(h\nu = 13.6\ {\rm eV})/10^{-21}$ erg cm$^{-2}$ s$^{-1}$ Hz$^{-1}$ sr$^{-1}$ (not to be confused with $J_{\rm LW,21}$), and the fit is valid for the intensities considered by these authors, namely $0.1 \leqslant J_{21} \leqslant 1000$. As shown in Appendix 4, the normalized intensity $J_{21}$ a distance $r$ from a galaxy of star formation rate $\dot{M}_{\star,\rm gal}$ is approximately given by

$$\begin{aligned} J_{21} &\simeq 4900\left(\frac{\dot{M}_{\star,\rm gal}}{1\ {\rm M}_\odot\ {\rm yr}^{-1}}\right)\left(\frac{r}{1\ {\rm kpc}}\right)^{-2} \\ &\simeq 400\,M_{10}^{1.58}\left(\frac{1+z}{10}\right)^{2.20}\left(\frac{r}{1\ {\rm kpc}}\right)^{-2}, \end{aligned} \tag{47}$$

[20] This is because we cannot estimate the $H_2$ fraction in the minihalo analytically due to its dependence on self-shielding, which in turn depends on the $H_2$ fraction. Only if self-shielding is ignored (which is unrealistic) can the effect of LW feedback be studied analytically.

where we have used the average star formation rate of galaxies in host halos of mass $M$ derived from the state-of-the-art FIRE-2 simulations of galaxies during the Epoch of Reionization (EoR) on the second line (Ma *et al.*, 2018). We will use this result later to model the effect of LW feedback from galaxies on the cooling threshold in minihalos.

### 3.1.2. Lyman-$\alpha$ cooling in atomic-cooling halos

Even in the absence of $H_2$, efficient cooling by atomic processes becomes possible at sufficiently high temperatures. In particular, collisional excitation of the Lyman-$\alpha$ (Ly$\alpha$) line by electrons give a large cooling rate above $T > 8000$ K, independent of the gas metallicity or the $H_2$ abundance. We therefore expect that in the absence of $H_2$ there is a cooling threshold of $T_{\rm vir} \simeq 10^4$ K, and this is indeed found in simulations (e.g. Greif *et al.*, 2008; Safranek-Shrader *et al.*, 2014b; Fernandez *et al.*, 2014; Kimm *et al.*, 2016; Latif *et al.*, 2020). However, since we will see that the disk accretion rate has a fairly strong dependence on virial temperature, it is prudent to derive the cooling threshold in more detail. The atomic-cooling rate for temperatures $T \ll 10^5$ K (dominated by the Ly$\alpha$ line) is given by (e.g. Black, 1981; Cen, 1992; Katz *et al.*, 1996)

$$\Lambda_{\rm Ly\alpha}(T,n) \simeq 7.5\times10^{-19}\, e^{-11.8348/T_4} n_{\rm e} n_{\rm HI}\ {\rm erg\ cm^{-3}\ s^{-1}}\,. \tag{48}$$

At the relevant temperatures and gas densities, the ionization state of the gas is set by collisional ionization. For simplicity we only consider ionization of hydrogen gas, so that $n_{\rm e} = n_{\rm HII}$. Then we get $n_{\rm e}n_{\rm HI} = x(1-x)n_{\rm H}^2 = 0.857\,x(1-x)n^2$, where we have again used $n_{\rm H}/n = 0.926$ for our primordial gas, with the ionization fraction in collisional ionization equilibrium given by

$$\begin{aligned} x &= 1 - \frac{k_1}{k_1 + k_{\rm collion}} \\ &\simeq 1 - \frac{2.54\times10^{-13}\,T_4^{-0.8163-0.0208\ \ln T_4}}{2.54\times10^{-13}\,T_4^{-0.8163-0.0208\ \ln T_4} + 5.85\times10^{-9}\,T_4^{1/2}e^{-15.7809/T_4}}\,, \end{aligned} \tag{49}$$

where we have used the collisional ionization rate from Katz *et al.* (1996) (along with $T \ll 10^5$ K), and the Case-B recombination coefficient from Draine (2011) (p. 139, Eq. 14.6), valid for 30 K $< T < 3\times10^4$ K (see Appendix 1 for details). The cooling time-scale in Eq. (9), with $n_{\rm core} \simeq 3.5\,[(1+z)/10]^3$ cm$^{-3}$ (Eq. 11), is then

$$t_{\rm cool} \simeq 0.36\,T_{\rm vir,4}\,\frac{e^{11.8348/T_{\rm vir,4}}}{x(1-x)}\left(\frac{1+z}{10}\right)^{-3}\ {\rm yrs}\,. \tag{50}$$

For efficient cooling to be possible, this should be smaller than the free-fall time-scale $t_{\rm ff} \simeq 1.8\times10^7\,[(1+z)/10]^{-3/2}$ yrs (Eq. 25). This can be solved numerically for the minimum virial temperature for efficient cooling (i.e. the atomic-cooling threshold) at a given redshift $z$ (the only quantity the threshold depends on). The resulting atomic-cooling threshold ($T_{\rm vir,Ly\alpha}$) can be fitted to an accuracy of 1.3% over the range $0 < z < 100$ by the following expression:

$$T_{\rm vir,Ly\alpha}(z) \simeq 9946\left(\frac{1+z}{10}\right)^{-0.052}\ {\rm K}\,. \tag{51}$$

The accuracy of this fit can be seen in Figure 25 in Appendix . The above fit yields $T_{\rm vir,Ly\alpha} \simeq 1.01\times10^4$ K at $z = 6$ and $T_{\rm vir,Ly\alpha} \simeq 0.95\times10^4$ K at $z = 25$, so we have indeed found that the atomic-cooling threshold is fairly insensitive to the assumed redshift, but there is a dependence there non-the-less. The corresponding threshold halo mass for efficient atomic-cooling is:

**Halo mass threshold for efficient atomic-cooling**

$$M_{\rm th,Ly\alpha} \simeq 5.07 \times 10^7 \left(\frac{1+z}{10}\right)^{-1.578} \ {\rm M}_\odot\,. \tag{52}$$

#### 3.1.3. Summary of the cooling model

In summary, we assume that efficient cooling is only possible in halos with mass $M > M_{\rm th}$, where

**General halo mass threshold for efficient cooling**

$$M_{\rm th} = \min[\max(M_{\rm th,H_2},\, M_{\rm th,LW}),\, M_{\rm th,Ly\alpha}]\,, \tag{53}$$

with $M_{\rm th,H_2}$ given by Eq. (45), $M_{\rm th,LW}$ by Eq. (46), and $M_{\rm th,Ly\alpha}$ by Eq. (52).

## 3.2. A self-similar model of gas accretion in low-mass halos

Our eventual estimate of the star formation rate and final stellar mass in low-mass halos will depend crucially on the central gas accretion rate in the halo, given that this will determine the gas mass reservoir available for star formation. A very rough order-of-magnitude estimate of the gas accretion rate can be obtained as follows (see e.g. Begelman *et al.*, 2006; Lodato & Natarajan, 2006; Choi *et al.*, 2013; Inayoshi *et al.*, 2018). If the halo contains a gas mass $M_{\rm gas}$ that collapse on a free-fall time $t_{\rm ff} \sim R_{\rm vir}/v_{\rm vir}$, we get an estimated gas accretion rate of:

$$\dot{M}_{\rm gas} \sim \frac{\text{Gas mass reservoir}}{\text{Free-fall time-scale}} \sim \frac{M_{\rm gas}}{R_{\rm vir}/v_{\rm vir}}\,. \tag{54}$$

Prior to a halo reaching the cooling threshold (Eq. 53), we expect that $M_{\rm gas} \simeq f_{\rm B} M$ since the baryons follow the dark matter accretion rate (at least above the Jeans scale). Using this and $M/R_{\rm vir} = v_{\rm vir}^2/G$, we then arrive at

$$\dot{M}_{\rm gas} \sim \frac{f_{\rm B} v_{\rm vir}^3}{G}\,. \tag{55}$$

This is the scaling with halo properties we expect for a pressureless collapse, but we have not determined how this will be modified by collapsing gas having a non-zero temperature, or the numerical prefactor to the scaling in Eq. (55). Thus, in the pursuit of better accuracy we can try to develop a more detailed model of gas accretion. As we will now show, a new exact self-similar solution of accretion of isothermal gas, similar in spirit to the solution found by Shu (1977),[21] can be found if one approximates the halo as an isothermal sphere (i.e. as having a flat rotation curve). We start from the fluid equation assuming spherical symmetry:

$$\begin{aligned} \frac{\partial M_{\rm gas}}{\partial t} + v\frac{\partial M_{\rm gas}}{\partial r} &= 0 \\ \frac{\partial M_{\rm gas}}{\partial r} &= 4\pi r^2 \rho_{\rm gas} \\ \frac{\partial v}{\partial t} + v\frac{\partial v}{\partial r} &= -\frac{c_{\rm s,h}^2}{\rho_{\rm gas}}\frac{\partial \rho_{\rm gas}}{\partial r} - \frac{GM_{\rm gas}}{r^2} - \frac{GM_{\rm DM}}{r^2}\,, \end{aligned} \tag{56}$$

where $c_{\rm s,h} = (k_{\rm B}T_{\rm h}/\mu_{\rm h} m_{\rm H})^{1/2}$ is the isothermal sound speed ($T_{\rm h}$ and $\mu_{\rm h}$ being the temperature and mean molecular weight of circumgalactic gas in the halo, respectively), $\rho_{\rm gas}(r,t)$ is the gas density, $v(r,t)$ the fluid velocity, $M_{\rm gas}(<r,t)$ the total gas mass enclosed within a radius $r$, and $M_{\rm DM}(<r,t)$ the total enclosed DM mass within a radius $r$. We can now make the self-similarity Ansatz for the fluid variables (Shu, 1977):

$$\rho_{\rm gas}(r,t) = \frac{\bar{\rho}(x)}{4\pi G t^2}, \qquad M_{\rm gas}(<r,t) = \frac{c_{\rm s,h}^3 t}{G}\bar{m}(x), \qquad v(r,t) = c_{\rm s,h}\bar{v}(x)\,, \tag{57}$$

where $x \equiv r/c_{\rm s,h}t$. The central gas accretion rate will therefore be given by

$$\dot{M}_{\rm gas} = \frac{c_{\rm s,h}^3}{G}\bar{m}(x=0)\,. \tag{58}$$

[21]Shu (1977) studied the accretion of self-gravitating isothermal gas and found that the central accretion rate is given by $\dot{M}_{\rm gas} = 0.975\,c_{\rm s}^3/G$. This estimate cannot be applied in our case however since it does not take into account the gravitational influence of the DM halo.

As shown in Appendix 5, with the self-similarity Ansatz in Eq. (57) the fluid equations takes the form:

$$\frac{\mathrm{d}\bar{m}}{\mathrm{d}x}(x-\bar{v}) = \bar{m} \tag{59}$$

$$\frac{\mathrm{d}\bar{m}}{\mathrm{d}x} = x^2\bar{\rho} \tag{60}$$

$$\frac{\mathrm{d}\bar{v}}{\mathrm{d}x}\left\{(x-\bar{v})^2-1\right\} = \beta\frac{(x-\bar{v})}{x}+\frac{\bar{m}}{x^2}(x-\bar{v})\,, \tag{61}$$

where $\beta \equiv (1-f_\mathrm{B})(v_\mathrm{vir}/c_\mathrm{s,h})^2-2$ is a constant parameter (in the absence of a DM halo we have $\beta=-2$), where I have used $GM_\mathrm{DM}(<r)/r^2=(1-f_\mathrm{B})v_\mathrm{vir}^2/r$ for the assumed isothermal structure of the DM halo.[22] For future reference, $\beta$ evaluates to (assuming $\mu_\mathrm{h}=1.23$):

$$\beta = 2.29\,\frac{T_\mathrm{vir}}{T_\mathrm{h}}-2\,. \tag{62}$$

We see that Eqs. (59) and (60) can be combined to give $\bar{m}=(x-\bar{v})x^2\bar{\rho}$, so that we are free to focus on only the two variables $\bar{m}$ and $\bar{v}$. Boundary conditions can be imposed at $x\to\infty$ for consistency with the assumed initial density profile at large radii. Cosmological hydrodynamical simulations and simple physical arguments indicate that, prior to efficient cooling, the baryons settle in a cored density profile (see e.g. Yoshida *et al.*, 2003; Wise & Abel, 2007a; Visbal *et al.*, 2014a; Inayoshi *et al.*, 2015; Chon *et al.*, 2016). The constant-density core has a radius $R_\mathrm{core}\simeq 0.1\,R_\mathrm{vir}$, beyond which the density falls like $\rho_\mathrm{gas}(r)\propto r^{-2}$ (Yoshida *et al.*, 2003; Visbal *et al.*, 2014a). If the central gas density is $\rho_\mathrm{core}$, and we define a core radius $R_\mathrm{core}$ via $\rho_\mathrm{gas}(r=R_\mathrm{core})\equiv\rho_\mathrm{core}/2$ following Chon *et al.* (2016), we can approximate the initial gas density profile as

$$\rho_\mathrm{gas}(r)\simeq\frac{\rho_\mathrm{core}}{1+(r/R_\mathrm{core})^2}\,. \tag{63}$$

This model reproduces the correct asymptotic behaviour at small and large radii. The enclosed gas mass at large radii is then

$$\begin{aligned} M_\mathrm{gas}(<r) &= \int_0^r \mathrm{d}r'\,4\pi r'^2\rho_\mathrm{gas}(r') \\ \text{(for } r\gg R_\mathrm{core}) &\simeq 4\pi R_\mathrm{core}^2\rho_\mathrm{core}r\,. \end{aligned} \tag{64}$$

We therefore find:

$$\bar{m}(x\to\infty)=\frac{4\pi G R_\mathrm{core}^2\rho_\mathrm{core}}{c_\mathrm{s,h}^2}\,x\,, \tag{65}$$

or using $R_\mathrm{core}\simeq 0.1\,R_\mathrm{vir}$,

$$\bar{m}(x\to\infty)\simeq\frac{4\pi G R_\mathrm{vir}^2\rho_\mathrm{core}}{100c_\mathrm{s,h}^2}\,x\equiv m_\infty x\,. \tag{66}$$

As discussed in Section 3.1, the core density $\rho_\mathrm{core}$ is different for minihalos and more massive halos, and so the value of $m_\infty$ will differ in both cases. We will therefore consider both mass regimes separately. Before doing so we should also determine the boundary condition for $\bar{v}$. Imposing the boundary condition $\bar{v}\to 0$ as $x\to\infty$ yields

$$\begin{aligned} \frac{\mathrm{d}\bar{v}}{\mathrm{d}x} &\simeq \frac{\beta}{x^2}+\frac{\bar{m}(x\to\infty)}{x^3} \\ &\simeq \frac{\beta+m_\infty}{x^2} \qquad \text{as } x\to\infty\,. \end{aligned} \tag{67}$$

[22]This follows since $M_\mathrm{DM}(<r)=(1-f_\mathrm{B})Mr/R_\mathrm{vir}$ for an isothermal sphere.

Upon integration we find

$$\bar{v}(x \to \infty) = -\frac{\beta + m_\infty}{x}\,, \tag{68}$$

where the integration constant has been set to zero to satisfy our boundary condition for $\bar{v}$. In Appendix 5 we find an approximate solution using the boundary conditions in Eqs. (66) and (68). The resulting approximate expression for the gas accretion rate, denoted $\dot{M}_{\rm gas,approx}$, is shown to be

$$\frac{\dot{M}_{\rm gas,approx}}{v_{\rm vir}^3/G} = m_\infty(\beta + m_\infty)^{1/2}\left(\frac{1 - f_{\rm B}}{\beta + 2}\right)^{3/2}\,. \tag{69}$$

This result will be extremely useful in understanding the scaling of the exact numerical solution, and finding a fit to the numerical results that is more accurate than Eq. (69). Let us now consider the two halo mass regimes separately, starting with the high-mass regime.

### 3.2.1. Halos with masses $M > 2.92 \times 10^6\,[(1+z)/10]^{3/2}$ $M_\odot$

For halos with masses $M > 2.92 \times 10^6\,[(1+z)/10]^{3/2}$ $M_\odot$ the core density is given by $\rho_{\rm core} = n_{\rm core,HM}\mu_{\rm h} m_{\rm H}$, with $n_{\rm core,HM}$ given by Eq. (11). Assuming $\mu_{\rm h} \simeq 1.23$ (appropriate for neutral gas of primordial composition), this yields $\rho_{\rm core} \simeq 7.2 \times 10^{-24}\,[(1+z)/10]^3$ g cm$^{-3} \simeq 15\,\rho_{\rm vir}$, and so

$$\bar{m}(x \to \infty) \simeq \frac{15}{100}\frac{4\pi G R_{\rm vir}^2 \rho_{\rm vir}}{c_{\rm s,h}^2}\,x\,. \tag{70}$$

Using $G R_{\rm vir}^2 \rho_{\rm vir} = 3 v_{\rm vir}^2/4\pi$ then gives us

$$\begin{aligned} \bar{m}(x \to \infty) &\simeq \frac{45}{100}\frac{v_{\rm vir}^2}{c_{\rm s,h}^2}\,x \\ &= \frac{45}{100\,(1 - f_{\rm B})}\,(\beta + 2)\,x \\ &\simeq 0.53\,(\beta + 2)\,x\,. \end{aligned} \tag{71}$$

Thus, since $m_\infty \equiv \bar{m}(x \to \infty)/x$ and $\bar{v}(x \to \infty)$ is given by Eq. (68), we find:

$$m_\infty \simeq 0.53\,(\beta + 2)\,, \qquad \bar{v}(x \to \infty) \simeq \frac{1.53\,\beta + 1.06}{x}\,. \tag{72}$$

Furthermore, from Eq. (69) the approximate expression for the gas accretion rate is given by

$$\frac{\dot{M}_{\rm gas,approx}}{v_{\rm vir}^3/G} = 0.410\left\{\frac{\beta}{\beta + 2} + 0.53\right\}^{1/2}\,. \tag{73}$$

Some justification for the validity of this approximate result can be found in Appendix 5. We also solve the fluid equations numerically in `Julia`[23] with the boundary conditions in Eq. (72) for different values of $\beta$. In Figure 9 we plot the ratio $\dot{M}_{\rm gas}/\dot{M}_{\rm gas,approx}$ between the exact numerical result for the gas accretion rate $\dot{M}_{\rm gas}$ and the approximate result in Eq. (73).

It seen that $\dot{M}_{\rm gas}/\dot{M}_{\rm gas,approx}$, which only depend on $\beta$, is well-fitted by

$$\frac{\dot{M}_{\rm gas}}{\dot{M}_{\rm gas,approx}} = 0.8453 + 1.7\,\exp[-6\,\log_{10}^{0.52}(\beta + 1.693)]\,. \tag{74}$$

In other words, we find numerically that the central gas accretion rate is given by:

[23] `Julia` is a relatively new programming language developed at MIT that is quickly gaining popularity. An impressive feature of `Julia` is that it combines the simple syntax of a language like `Python` with the numerical speed of languages like `C` or `Fortran`. For more information about `Julia`, see this link.

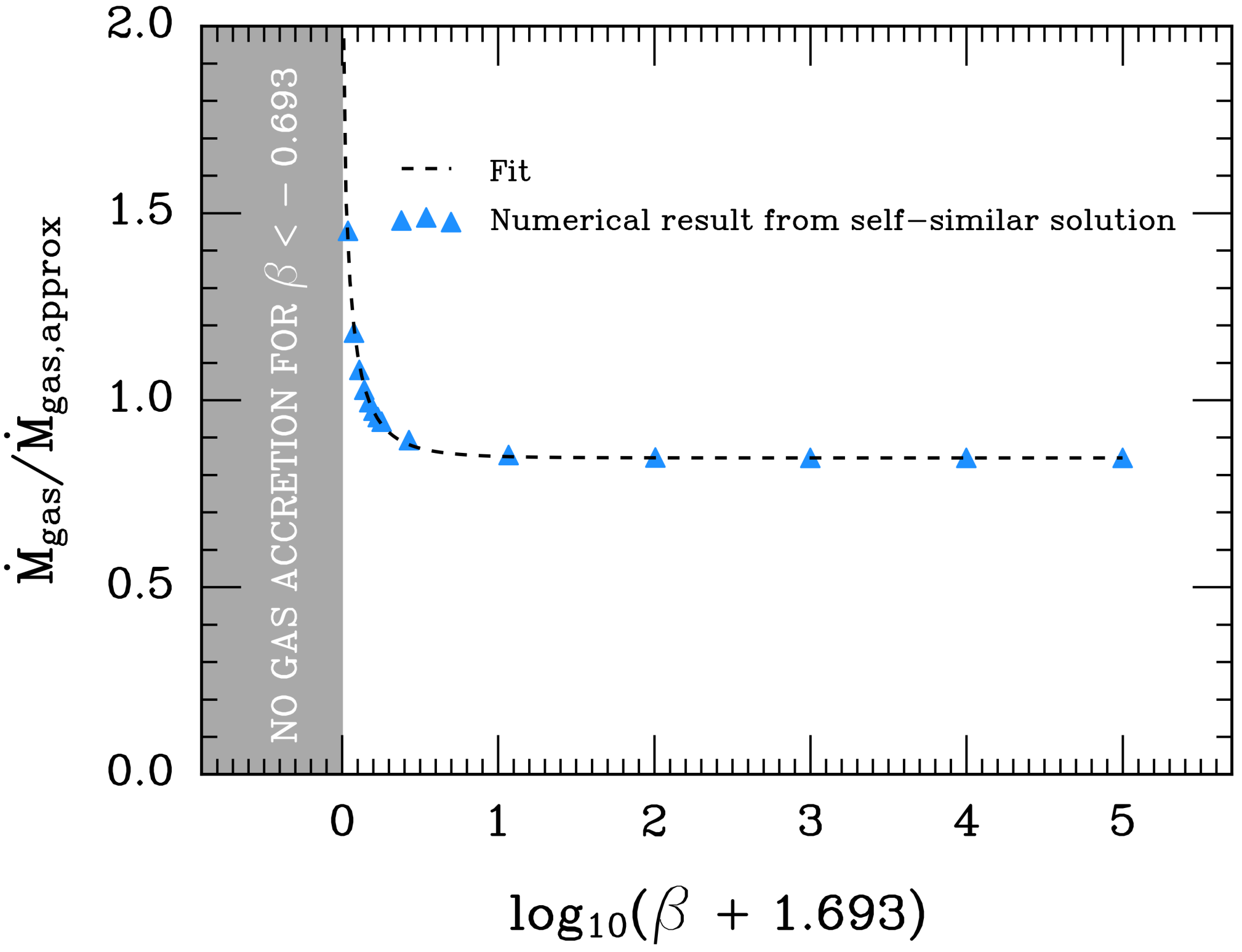

**Figure 9.** A comparison between the numerical ($\dot{M}_{\rm gas}$) and approximate analytical ($\dot{M}_{\rm gas,approx}$) solution for the central gas accretion rate as a function of $\beta = 2.29\,(T_{\rm vir}/T_{\rm h}) - 2$. A fit is also shown for the ratio of the two result (Eq. 74). In the gray region ($\beta < -0.693$) the gas accretion rate is zero since the gas temperature is sufficiently high that the pressure of the gas can resist gravitational collapse.

**Predicted central gas accretion rate in halos with masses $M > 2.92 \times 10^6\,[(1+z)/10]^{3/2}$ M$_\odot$**

$$\dot{M}_{\rm gas} = 0.347\,\frac{v_{\rm vir}^3}{G}\,\left\{1 + 2.01\,\exp[-6\,\log_{10}^{0.52}(\beta + 1.693)]\right\}\left\{\frac{\beta}{\beta+2} + 0.53\right\}^{1/2}, \quad (75)$$

$$\beta \equiv (1 - f_{\rm B})(v_{\rm vir}/c_{\rm s,h})^2 - 2\,.$$

In halos reaching the atomic-cooling threshold $T_{\rm vir} = 10^4$ K with the gas undergoing isothermal cooling at $T_{\rm h} = 8000$ K we find $\beta = 0.8625$ (see Eq. 62), and so a predicted central gas accretion rate $\dot{M}_{\rm gas} = 0.331\,v_{\rm vir}^3/G = 0.194$ M$_\odot$ yr$^{-1}$. This is in good agreement with the results of cosmological hydrodynamical simulations. For example, using high-resolution simulations, Latif & Volonteri (2015) followed gas collapse in three atomic-cooling halos, consdering a range of Lyman-Werner intensities for each halo. As can be seen in their figures, $10-100$ pc from the center of the halos the mass accretion rate was typically found to be $0.1$ – few $\times\,0.1$ M$_\odot$ yr$^{-1}$ (with higher accretion rates for higher Lyman-Werner intensities). Similarly, Visbal *et al.* (2017) used simulations to determine the infall time-scale $t_{\rm inf} \equiv r/v(r)$ as a function of enclosed gas mass in three halos. For an enclosed gas mass of $10^6$ M$_\odot$ they find infall time-scales of $\simeq 8 - 10$ Myrs for a virial temperature of $T_{\rm vir} \simeq 10^4$ K at collapse,[24] giving a rough estimated accretion rate of $\simeq 0.1$ M$_\odot$ yr$^{-1}$. Finally, from the simulation of Kimm *et al.* (2016) where rapid gas accretion at the center of two atomic-cooling halos led to the formation of GC-like objects with stellar masses $\simeq 6\times10^5$ M$_\odot$ within less than $\simeq 10$ Myrs, we get a lower bound on the gas accretion rate of $\simeq 0.06$ M$_\odot$ yr$^{-1}$. Taking into account the accreted central gas (and not just the mass of the stars), and the fact that most of the stars form within *less* than 10 Myrs would obviously raise the estimated central gas accretion rate from their simulation. In summary, we find that the self-similar model of gas collapse developed in this section is in good agreement with simulations.

### 3.2.2. Halos with masses $M \leqslant 2.92 \times 10^6\,[(1+z)/10]^{3/2}$ M$_\odot$

Next we consider the low-mass limit $M \leqslant 2.92\times10^6\,[(1+z)/10]^{3/2}$ M$_\odot$ relevant for minihalos. In this case the the core density $\rho_{\rm core} = n_{\rm core,LM}\mu_{\rm h} m_{\rm H}$ is given by (see Eq. 12)

$$\begin{aligned}\rho_{\rm core} &\simeq \bar{\rho}_{\rm B}\left(\frac{T_{\rm vir}}{T_{\rm IGM}}\right)^{3/2}\\ &= \frac{f_{\rm B}\bar{\rho}_{\rm vir}}{\Delta_{\rm vir}}\left(\frac{T_{\rm vir}}{T_{\rm IGM}}\right)^{3/2}.\end{aligned} \tag{76}$$

This yields

$$\bar{m}(x\to\infty) \simeq \frac{4\pi f_{\rm B} G\bar{\rho}_{\rm vir} R_{\rm vir}^2}{100\Delta_{\rm vir} c_{\rm s,h}^2}\left(\frac{T_{\rm vir}}{T_{\rm IGM}}\right)^{3/2} x\,. \tag{77}$$

Using $G\rho_{\rm vir}R_{\rm vir}^2 = 3v_{\rm vir}^2/4\pi$ then gives us

$$\begin{aligned}\bar{m}(x\to\infty) &\simeq \frac{3f_{\rm B}}{100\Delta_{\rm vir}}\left(\frac{v_{\rm vir}}{c_{\rm s,h}}\right)^2\left(\frac{T_{\rm vir}}{T_{\rm IGM}}\right)^{3/2} x\\ &= 2.65\times10^{-5}\left(\frac{v_{\rm vir}}{c_{\rm s,h}}\right)^2\left(\frac{T_{\rm vir}}{T_{\rm IGM}}\right)^{3/2} x\\ &= 3.14\times10^{-5}\,(\beta+2)\left(\frac{T_{\rm vir}}{T_{\rm IGM}}\right)^{3/2} x\\ &\equiv 3.14\times10^{-5}\,(\beta+2)\zeta\, x\,,\end{aligned} \tag{78}$$

and so $m_\infty = 3.14\times10^{-5}\,(\beta+2)\zeta$ in this case, where $\zeta \equiv (T_{\rm vir}/T_{\rm IGM})^{3/2}$. The corresponding boundary condition for $\bar{v}$ in Eq. (68) then becomes:

$$\bar{v}(x\to\infty) \simeq -\frac{\beta + 3.14\times10^{-5}\,(\beta+2)\zeta}{x}\,, \tag{79}$$

and the approximate solution for the gas accretion rate in Eq. (69) is given by:

$$\frac{\dot{M}_{\rm gas,approx}}{v_{\rm vir}^3/G} = 2.43\times10^{-5}\,\zeta\left\{\frac{\beta}{\beta+2} + 3.14\times10^{-5}\,\zeta\right\}^{1/2}. \tag{80}$$

[24] More specifically, from their Figure 2 we see that gas collapse occur when $T_{\rm vir} \simeq 10^4$ K for a background ionizing intensity $F_{\rm bg}/F_0 = 0.01 - 0.1$. In their Figures 7 and 8 one can then read off the corresponding infall time-scales.

Gas cooling in minihalos is driven primarily by the presence of $H_2$, the cooling function of which (c.f. Eq. 20) is not as steep as in the case of Ly$\alpha$-cooling in atomic-cooling halos. Because of this, the assumption of isothermal gas in our model is not strictly correct (although the self-similar solution should become exact in the pressureless limit $\beta \to \infty$). As a rough approximation, however, we can adopt a mass-weighted temperature of $\langle T_{\rm h} \rangle \simeq 0.5\, T_{\rm vir}$, derived from the the temperature evolution of collapsing gas in minihalos (see e.g. Figure 7 in Yoshida *et al.*, 2003).[25] Using Eq. (62) this yields $\beta = 2.58$, and an approximate gas accretion rate of:

$$\frac{\dot{M}_{\rm gas,approx}}{v_{\rm vir}^3/G} = 2.43 \times 10^{-5}\, \zeta\, (0.563 + 3.14 \times 10^{-5} \zeta)^{1/2} \,. \tag{81}$$

In Figure 10 we compare Eq. (81) with the "exact" (i.e. numerical) result for the gas accretion rate as a function of $\zeta \equiv (T_{\rm vir}/T_{\rm IGM})^{3/2}$ (with $\beta = 2.56$ held fixed). As seen in the figure, the ratio between the numerical result and the approximate solution can be well-fitted by $\dot{M}_{\rm gas}/\dot{M}_{\rm gas,approx} = 0.8239\, \exp[0.00281\, \log_{10}^3(\zeta/3$ over the range $32 < \zeta < 1.65 \times 10^4$ of interest to us — the lower limit corresponding to the cosmological Jeans mass (see e.g. Barkana & Loeb, 2001),[26] and the upper limit to the halo mass $M = 2.92 \times 10^6\, [(1+z)/10]^{3/2}$ $M_\odot$ above which the self-similar solution of the previous sub-section becomes relevant.

Thus, for halos with masses $M \leqslant 2.92 \times 10^6\, [(1+z)/10]^{3/2}$ $M_\odot$ we take the central gas accretion rate to be given by

**Predicted central gas accretion rate in halos with masses $M \leqslant 2.92 \times 10^6\, [(1+z)/10]^{3/2}$ $M_\odot$**

$$\dot{M}_{\rm gas} = 2.00 \times 10^{-5}\, \frac{v_{\rm vir}^3}{G}\, e^{0.00281\, \log_{10}^3(\zeta/32)} \zeta\, (0.563 + 3.14 \times 10^{-5}\, \zeta)^{1/2} \,, \tag{82}$$

$$\zeta \equiv \left(\frac{T_{\rm vir}}{T_{\rm IGM}}\right)^{3/2} = 5650\, M_6 \left(\frac{1+z}{10}\right)^{-3/2} \,,$$

where $T_{\rm IGM} \simeq 2.29\, [(1+z)/10]^2$ K (see Eq. 13) and Eq. (6) for $T_{\rm vir}$ have been used to evaluate $\zeta$ as a function of halo mass and redshift.

[25] This can be derived as follows. From Figure 7 in Yoshida *et al.* (2003) we see that the temperature drops — almost as a perfect power law — from $T \simeq T_{\rm vir} \simeq 2000$ K to $T \simeq 150$ K as the hydrogen number density goes from $n_{\rm H} = 1$ cm$^{-3}$ to $n_{\rm H} = 10^3$ cm$^{-3}$. This yields $T(n_{\rm H})/T_{\rm vir} \simeq (n_{\rm H}/n_{\rm vir})^{-0.434}$, where $n_{\rm vir}$ is the gas number density near the virial radius. We expect the infalling gas to have the density profile $n_{\rm H} \sim r^{-2}$, and so $T(\bar{r})/T_{\rm vir} \simeq \bar{r}^{0.868}$ where $\bar{r} \equiv r/R_{\rm vir}$. The mass-weighted mean temperature is then

$$\langle T_{\rm h} \rangle = T_{\rm vir} \frac{\int_0^1 {\rm d}\bar{r}\, \bar{r}^2 n_{\rm H}(\bar{r}) \bar{r}^{0.868}}{\int_0^1 {\rm d}\bar{r}\, \bar{r}^2 n_{\rm H}(\bar{r})} = T_{\rm vir} \frac{\int_0^1 {\rm d}\bar{r}\, \bar{r}^{0.868}}{\int_0^1 {\rm d}\bar{r}} = 0.536\, T_{\rm vir} \,.$$

Given the crudeness of this estimate along with the fact that more massive minihalos could probably cool to lower temperatures relative to the virial temperature we can simply adopt $\langle T_{\rm h} \rangle \simeq 0.5\, T_{\rm vir}$, as stated.

[26] The cosmological Jeans mass is given by (see e.g. Barkana & Loeb, 2001) $M_{\rm J} \simeq 5.8 \times 10^3\, [(1+z)/10]^{3/2}$ $M_\odot$. This corresponds to a virial temperature $T_{\rm vir} \simeq 23$ K $[(1+z)/10]^2$, and so a minimum value $\zeta \simeq 32$, independent of redshift. For the upper limit, we have $M < 2.92 \times 10^6\, [(1+z)/10]^{3/2}$ $M_\odot$, and so $T_{\rm vir} < 1480\, [(1+z)/10]^2$ K, giving us $\zeta < 1.65 \times 10^4$.

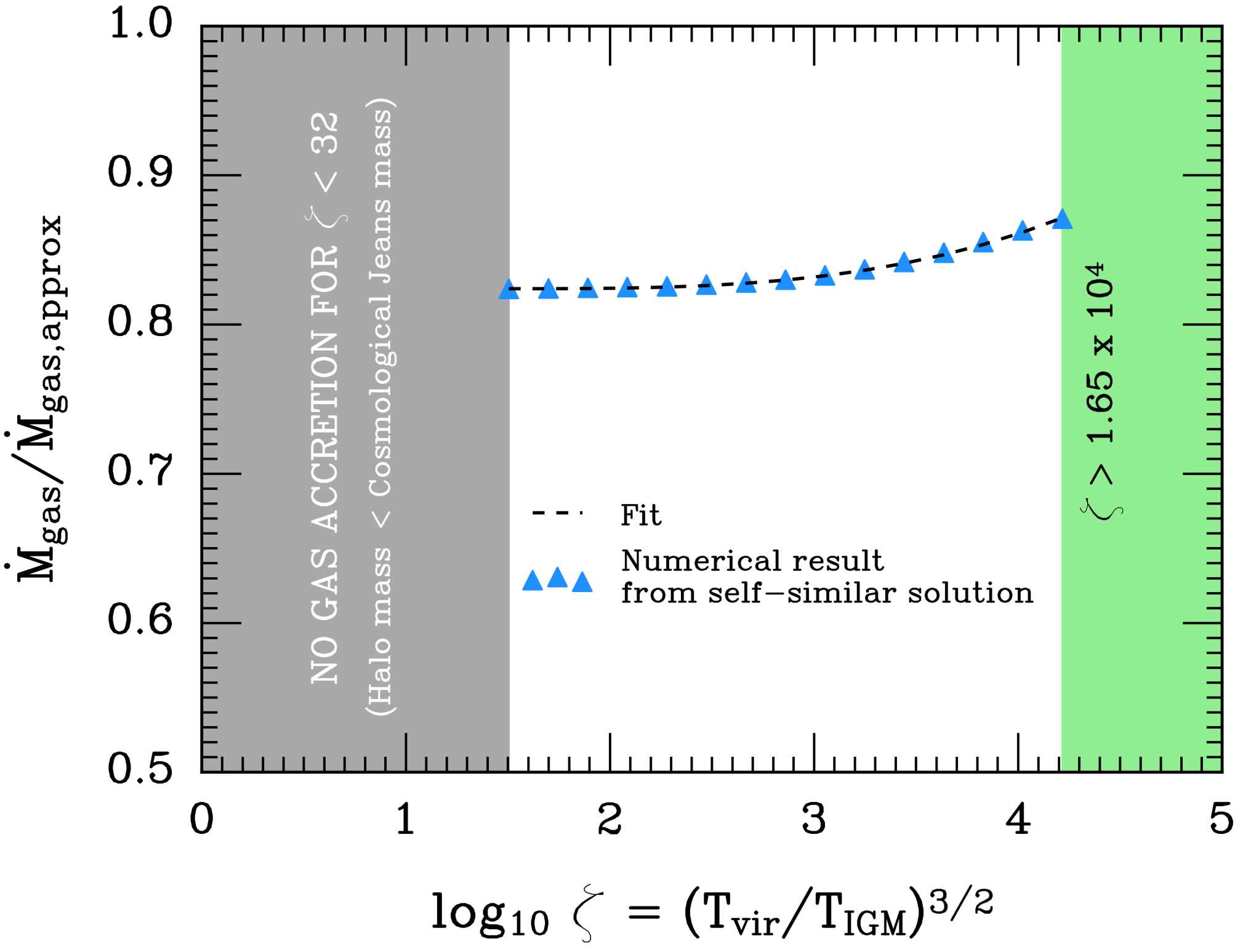

**Figure 10.** Similar plot to Figure 9, but as a function of $\zeta = 5650\, M_6[(1+z)/10]^{-3/2}$ in halos with masses $M \leqslant 2.92 \times 10^6\, [(1+z)/10]^{3/2}$ $M_\odot$ (corresponding to $\zeta \leqslant 1.65 \times 10^4$). In the gray region the halo mass is below the cosmological Jeans mass, and so no gas accretion onto the halo (and hence star formation) is possible in the first place.

### 3.2.3. Summary of results for the gas accretion rate

To summarize, we have found that the central gas accretion rate $\dot{M}_{\rm gas}$ in a halo of mass $M$ following efficient gas cooling and subsequent collapse can be written as:

$$\dot{M}_{\rm gas} = \frac{v_{\rm vir}^3}{G}\mathscr{G}_{\rm gas} = 0.585\, T_{\rm vir,4}^{3/2}\,\mathscr{G}_{\rm gas}\ {\rm M_\odot\ yr^{-1}}\,,$$

$$\mathscr{G}_{\rm gas} \equiv \begin{cases} 2.00\times 10^{-5}\, e^{0.00281\,\log_{10}^3(\zeta/32)}\zeta\,(0.563+3.14\times10^{-5}\,\zeta)^{1/2} & \text{for } M_6 \leqslant 2.92\,[(1+z)/10]^{3/2}\ {\rm M_\odot} \\ 0.347\,\left\{1+2.01\,e^{-6\,\log_{10}^{0.52}(\beta+1.693)}\right\}\,\{\beta/(\beta+2)+0.53\}^{1/2} & \text{for } M_6 > 2.92\,[(1+z)/10]^{3/2}\ {\rm M_\odot} \end{cases}, \quad (83)$$

where $\beta = 2.29\,T_{\rm vir}/T_{\rm h} - 2$ and $\zeta = 5650\,M_6[(1+z)/10]^{-3/2}$. For reference, in the case of Ly$\alpha$-cooling at $T_{\rm h} = 8000$ K in halos with $T_{\rm vir} = 10^4$ K we have $\mathscr{G}_{\rm gas} = 0.331$. We can use this and Eq. (7) for $M$ to evaluate the fraction $\mathcal{F} = \dot{M}_{\rm gas}t/f_{\rm B}M$ of baryons in the halo that reside in the disk a time $t$ after disk formation:

$$\mathcal{F} = 0.0241\,\mathscr{G}_{\rm gas,0.331}t_{\rm Myrs}\left(\frac{1+z}{10}\right)^{3/2}, \quad (84)$$

where $t_{\rm Myrs} \equiv t/1$ Myr. Apart from possibly $\mathscr{G}_{\rm gas,0.331} \equiv \mathscr{G}_{\rm gas}/0.331$, at fixed redshift this is independent of halo mass. Thus, the large central gas accretion rate implied by Eq. (83) can at most be sustained for a time $t_{\rm max} = 41.5\,\mathscr{G}_{\rm gas,0.331}^{-1}[(1+z)/10]^{-3/2}$ Myrs (i.e. until $\mathcal{F} = 1$), after which we expect the gas accretion to either be zero (e.g. as a result of efficient internal stellar feedback pushing away surrounding gas in the IGM), or be dictated by the halo mass growth rate. In this thesis we are mostly interested in short starbursts lasting a time $< t_{\rm max}$, so we will adopt the gas accretion rate in Eq. (83).

# 4. Disk formation and properties

## 4.1. Gaseous disks at Cosmic Dawn

Galactic disks are ubiquitous in the Universe mainly because matter perturbations acquire angular momentum from tidal torques up until halo formation (e.g. Peebles, 1969; Fall & Efstathiou, 1980; Heavens & Peacock, 1988; Mo *et al.*, 1998). Below we derive the expected properties of gaseous disks at Cosmic Dawn, wherein the first stars, star clusters, and galaxies are expected for form.

## 4.2. Size of the disk

In this section we will derive the size of the gaseous disk that is expected to form and grow in a halo. We begin this section by discussing the distribution of angular momentum in halos, which will be crucial in predicting the scatter in properties of disks, and hence the resulting scatter in properties of early galaxies and star clusters.

### 4.2.1. The spin parameter

The total angular momentum $\mathcal{J}_{\rm B}$ of the baryons within a halo of mass $M$ can be parametrized in terms of the dimensionless (Bullock) spin parameter $\lambda$ for the gas (Bullock *et al.*, 2001a):

$$\mathcal{J}_{\rm B} = \lambda\sqrt{2} f_{\rm B} M v_{\rm vir} R_{\rm vir}\,, \tag{85}$$

Cosmological simulations (e.g. Zjupa & Springel, 2017) as well as analytical calculations based on tidal torque theory (e.g. Peebles, 1969; Heavens & Peacock, 1988; Steinmetz & Bartelmann, 1995) predicts that $\lambda$ should be nearly independent of halo mass and redshift, have a median value of $\bar{\lambda} \sim$ few $\times\, 0.01$, and an approximately log-normal distribution:

$$p(\lambda)\,{\rm d}\lambda = \frac{1}{\sqrt{2\pi}\sigma_{\ln\lambda}} \exp\left\{-\frac{\ln^2(\lambda/\bar{\lambda})}{2\sigma^2_{\ln\lambda}}\right\} \frac{{\rm d}\lambda}{\lambda}\,. \tag{86}$$

Several authors have investigated the distribution of $\lambda$ in low-mass halos at high redshifts using cosmological simulations:

1. Jang-Condell & Hernquist (2001) studied the distribution of $\lambda$ for halo masses $4\times10^5$ M$_\odot$ $< M <$ $4\times10^8$ M$_\odot$ at redshift $z = 10$. When they limited their sample to those 389 halos in their simulation containing at least 50 particles (corresponding to $M = 2.3\times10^6$ M$_\odot$ for their DM particle mass of $4.6\times10^4$ M$_\odot$) they found the best-fit parameters $\bar{\lambda} = 0.033$ and $\sigma_{\ln\lambda} = 0.52$.

2. At $z = 10$ Knebe & Power (2008) found that the spin parameter varied (weakly) with halo mass, finding $\bar{\lambda} = 0.042$ and $\sigma_{\ln\lambda} = 0.54$ for halo masses around $M = 5.95\times10^6 h^{-1}$ M$_\odot$ and $\bar{\lambda} = 0.040$ and $\sigma_{\ln\lambda} = 0.54$ near $M = 8.19\times10^7 h^{-1}$ M$_\odot$.

3. Hirano *et al.* (2014) studied minihalos with masses $M \sim 10^5 - 10^6$ M$_\odot$ at redshifts $z \sim 10 - 35$, and found $\bar{\lambda} = 0.0498$ and $\sigma_{\ln\lambda} = 0.750$ for the gas (for the DM they found $\bar{\lambda} = 0.0495$ and $\sigma_{\ln\lambda} = 0.545$).

4. Sasaki *et al.* (2014) also studied minihalos with masses $10^4 h^{-1}$ M$_\odot$ $< M <$ $2\times10^6 h^{-1}$ M$_\odot$ at redshifts $z \geqslant 15$, all of them containing at least 1000 particles. For the whole halo population they find $\bar{\lambda} = 0.0262$ and $\sigma_{\ln\lambda} = 0.495$, with no evidence of any dependence on halo mass or redshift in both parameters. For $M = 2\times10^6 h^{-1}$ M$_\odot$ in particular (closer to the halo masses of interest to us) they find $\bar{\lambda} \simeq 0.029$ at $z = 15$ (see their Figure 3). The value for $\bar{\lambda}$ found by these authors is significantly lower than the one found by Hirano *et al.* (2014), and they attribute this to selection effects and a different halo definition employed — Hirano *et al.* (2014) focused only on $\sim 100$ minihalos in their simulations where gas can cool efficiently.

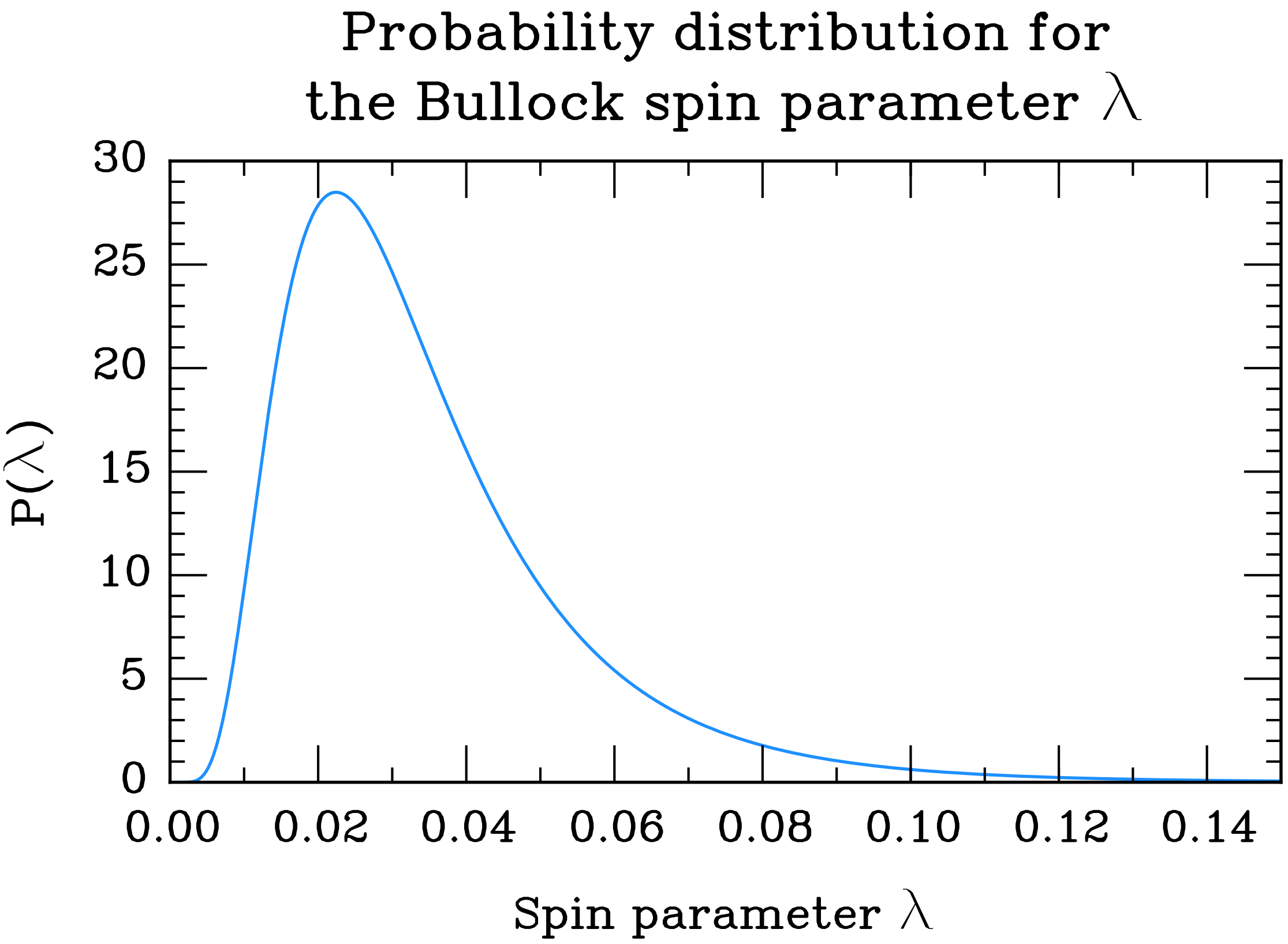

**Figure 11.** The log-normal distribution — see Eq. (86) — for the spin parameter $\lambda$ used in this thesis. As discussed in the main text, the adopted parameters are $\bar{\lambda} = 0.03$ and $\sigma_{\ln\lambda} = 0.54$.

5. Angel *et al.* (2016) found, similar to Knebe & Power (2008), a weak halo mass dependency of $\bar{\lambda} = (0.025 \pm 0.009)(M/10^{10}h^{-1}\ \mathrm{M}_\odot)^{-0.023\pm0.016}$ for $10^8 h^{-1}\ \mathrm{M}_\odot < M < 4 \times 10^{11} h^{-1}\ \mathrm{M}_\odot$ (they do not provide a log-normal fit). At $M = 10^8 h^{-1}\ \mathrm{M}_\odot$ this yields a median value of $\bar{\lambda} \simeq 0.028$, somewhat smaller than the value found by Knebe & Power (2008) and closer to the result of Jang-Condell & Hernquist (2001).

6. Druschke *et al.* (2018) used a large sample of well-resolved ($\gtrsim 500$ particles of both DM and gas) halos with $6.5 \times 10^4\ \mathrm{M}_\odot < M \lesssim 10^7\ \mathrm{M}_\odot$, finding an excellent log-normal fit with $\bar{\lambda} = 0.0253$ to the data at $z = 14$, with no evidence of significant redshift evolution over $14 \leqslant z \leqslant 24$. Unfortunately the authors do not provide their fitted value for $\sigma_{\ln\lambda}$.

As is evident from the literature, there is some scatter in the exact fitted parameters, some of it possibly due to different simulation resolutions and halo definitions. **In Anaxagoras where the focus is on halo masses $M \sim 10^6 - 10^8\ \mathrm{M}_\odot$ we can adopt $\bar{\lambda} = 0.03$ and $\sigma_{\ln\lambda} = 0.54$ as a reasonable middle ground.** A plot of the resulting distribution is shown in Figure 11. From the log-normal distribution we find that $\simeq 68\%$ of halos should have spin parameters in the range $0.0175 < \lambda < 0.0515$ (i.e. the $1\sigma$ interval $e^{-\sigma_{\ln\lambda}} < \lambda/\bar{\lambda} < e^{\sigma_{\ln\lambda}}$).

### 4.2.2. Estimated disk size

We are now ready to derive the predicted size of the gaseous disk that would form at the center of the halo, assuming angular momentum conservation. This approach was made famous by Mo *et al.* (1998) for massive disk galaxies (e.g. the Milky Way disk), but has later also been applied to the study of disks forming in low-mass halos at Cosmic Dawn (e.g. Oh & Haiman, 2002; Lodato & Natarajan, 2006; Devecchi & Volonteri, 2009; Inayoshi *et al.*, 2018). To proceed we need to assume a surface density profile $\Sigma_{\rm disk}(R)$ for the disk. In the literature mainly two surface density profiles have been considered in analytical works — an exponential profile with $\Sigma_{\rm disk} \propto e^{-R/R_{\rm disk}}$ (e.g. Oh & Haiman, 2002; Begelman *et al.*, 2006; Inayoshi *et al.*, 2018), or a Mestel disk Mestel (1963) with $\Sigma_{\rm disk} \propto R^{-1}$ for $R \leqslant R_{\rm disk}$ and $\Sigma_{\rm disk}(R > R_{\rm disk}) = 0$ (e.g. Begelman *et al.*, 2006; Lodato & Natarajan, 2006; Devecchi & Volonteri, 2009; Devecchi *et al.*, 2010). In this work we will adopt the Mestel disk for a number of reasons:

1. The radial distribution of angular momentum in a given halo found by Bullock *et al.* (2001a) yields a theoretically expected Mestel-like profile of the form $\Sigma_{\rm disk} \propto R^{-1}$ in the central region, with a more steeper dependence in the outer region (at most $\Sigma_{\rm disk} \propto R^{-3}$) (Bullock *et al.*, 2001a). Perhaps more interestingly than these theoretical arguments, Choi *et al.* (2013) used an idealized high-resolution simulation of gas collapse in an atomic-cooling halo, with an initial angular momentum distribution based on the results of Bullock *et al.* (2001a). These authors found that a steadily growing disk formed at the center of the halo with $\Sigma_{\rm disk} \propto R^{-1.3}$, and a fairly sharp outer boundary. Furthermore, the density profile was seen to be very close to isothermal with $\rho_{\rm disk} \propto R^{-2}$, which is precisely what one expects for an isothermal Mestel disk, as we shall see.

2. The Mestel disk makes the modelling analytically tractable. In particular, in a Mestel disk numerous quantities are either independent of radius (e.g. the Toomre parameter or the rotational velocity) or have a simple dependence on radius (a power law), which simplify the calculations dramatically. In contrast, an exponential disk has a complicated rotation curve, and radial dependences (e.g. the Toomre parameter is radially dependent).

Thus, we proceed under the assumption that a Mestel disk form in the halo following gas cooling and collapse. To be clear, the surface density of the Mestel disk, along with its total mass $M_{\rm disk}$,[27] are as follows:

$$\begin{aligned} \Sigma_{\rm disk}(R) &= \begin{cases} \Sigma_{\rm disk,0} R_{\rm disk}/R & \text{for } R \leqslant R_{\rm disk} \\ 0 & \text{for } R > R_{\rm disk} \end{cases} \\ M_{\rm disk} &= 2\pi \Sigma_{\rm disk,0} R_{\rm disk}^2 \,. \end{aligned} \tag{87}$$

We now have two unknown parameters, the disk radius $R_{\rm disk}$ and the normalization of the surface density, $\Sigma_{\rm disk,0}$ (which is equal to the disk surface density at $R = R_{\rm disk}$). Since we only have one possible known quantity here (the disk mass $M_{\rm disk}$, which can be obtained from the accretion rate) we need an additional equation. This comes from the assumed angular momentum conservation. If the disk contains a fraction

[27] The mass enclosed within a radius $R$ in the disk is given by

$$M_{\rm disk}(< R) = 2\pi \int_0^R {\rm d}R' \, R' \Sigma_{\rm disk}(R') \,.$$

Thus, the total disk mass is simply

$$M_{\rm disk} = 2\pi \int_0^{R_{\rm disk}} {\rm d}R' \, R' \Sigma_{\rm disk}(R') \,.$$

$\mathcal{F} \leqslant 1$ of the baryons in the halo (i.e. $M_{\rm disk} = \mathcal{F} f_{\rm B} M$), the angular momentum in the disk, assuming a constant rotation curve $v_{\rm rot}$, becomes[28]

$$\mathcal{J}_{\rm disk} = \frac{1}{2}\mathcal{F} f_{\rm B} M v_{\rm rot} R_{\rm disk} \,. \tag{88}$$

If the angular momentum of the collapsing baryons that ultimately form the disk is $j_{\rm disk}\mathcal{J}_{\rm B}$ with $j_{\rm disk} \leqslant 1$, angular momentum conservation yields $\mathcal{J}_{\rm disk} = j_{\rm disk}\mathcal{J}_{\rm B}$. Upon using Eqs. (85) and (88) for $\mathcal{J}_{\rm B}$ and $\mathcal{J}_{\rm disk}$ respectively, we find the following expression for the disk radius:

$$R_{\rm disk} = \sqrt{8}\lambda \left(\frac{j_{\rm disk}}{\mathcal{F}}\right)\left(\frac{v_{\rm vir}}{v_{\rm rot}}\right) R_{\rm vir} \,. \tag{89}$$

To proceed, we note that $j_{\rm disk}/\mathcal{F}$ is simply the normalized specific angular momentum of the disk baryons (i.e. $j_{\rm disk}/\mathcal{F} = 1$ for $M_{\rm disk} = f_{\rm B} M$). Since we assume angular momentum conservation, this inherits the specific angular momentum distribution of all the baryons in the enclosed mass $\mathcal{F} f_{\rm B} M$ prior to cooling. As noted earlier, Bullock *et al.* (2001a) studied precisely this distribution, and found that $j_{\rm disk}/\mathcal{F} = \mathcal{F}^s$ with $s = 1.0 \pm 0.3$, and also provide theoretical arguments based on tidal torque theory and minor mergers for $s \sim 1$. Furthermore, O'Shea & Norman (2007, 2008) used high-resolution cosmological simulations of minihalos at Cosmic Dawn and found that $j_{\rm disk}/\mathcal{F} = \mathcal{F}^s$ with $s$ close to unity (see also e.g. Figure 4 in Abel *et al.*, 2002), a result later used by Meece *et al.* (2014) for the initial conditions in their idealized simulations. Given the reasonable consistency with simulations and theoretical arguments we adopt $s = 1$, which has the advantage of simplifying future calculations. Thus, the disk radius becomes

$$R_{\rm disk} = \sqrt{8}\mathcal{F}\lambda \left(\frac{v_{\rm vir}}{v_{\rm rot}}\right) R_{\rm vir} \,. \tag{90}$$

Next we need to evaluate $v_{\rm vir}/v_{\rm rot}$. The contribution to the rotation curve from the Mestel disk is simply $v_{\rm disk} = (GM_{\rm disk}/R_{\rm disk})^{1/2}$ (see e.g. p. 100 in Binney & Tremaine, 2008). Approximating the DM halo as isothermal (i.e. having a constant rotation curve equal to $v_{\rm vir}$) and using $v_{\rm rot} = (v_{\rm disk}^2 + v_{\rm vir}^2)^{1/2}$ then yields[29]

$$v_{\rm rot} = v_{\rm disk}\left(1 + \frac{R_{\rm disk}}{\mathcal{F} f_{\rm B} R_{\rm vir}}\right)^{1/2} \,. \tag{91}$$

Inserting this into Eq. (90), introducing the definition $x \equiv R_{\rm disk}/R_{\rm vir}$, and using $v_{\rm vir}/v_{\rm disk} = (x/\mathcal{F} f_{\rm B})^{1/2}$ yields

$$x\left(1 + \frac{x}{\mathcal{F} f_{\rm B}}\right)^{1/2} = \sqrt{8}\mathcal{F}\lambda \left(\frac{x}{\mathcal{F} f_{\rm B}}\right)^{1/2} \,, \tag{92}$$

[28]To see this, we have

$$\mathcal{J}_{\rm disk} = \int {\rm d}M_{\rm disk}\, v_{\rm rot} R = 2\pi \int_0^{R_{\rm disk}} {\rm d}R\, R^2 v_{\rm rot}\Sigma_{\rm disk}(R) \,.$$

Using Eq. (87) along with $v_{\rm rot} = $ const. yields

$$\begin{aligned} \mathcal{J}_{\rm disk} &= 2\pi R_{\rm disk}^3 \Sigma_{\rm disk,0} v_{\rm rot} \int_0^1 {\rm d}x\, x \\ &= \frac{1}{2}(2\pi R_{\rm disk}^2 \Sigma_{\rm disk,0}) v_{\rm rot} R_{\rm disk} \\ &= \frac{1}{2} M_{\rm disk} v_{\rm rot} R_{\rm disk} \,, \end{aligned}$$

which is precisely the stated result if $M_{\rm disk} = \mathcal{F} f_{\rm B} M$.

[29]To see this, note that $v_{\rm vir}/v_{\rm disk} = (MR_{\rm disk}/M_{\rm disk}R_{\rm vir})^{1/2} = (R_{\rm disk}/\mathcal{F} f_{\rm B} R_{\rm vir})^{1/2}$, where the last step uses $M_{\rm disk} = \mathcal{F} f_{\rm B} M$.

or:

$$x^2 + \mathcal{F} f_\mathrm{B} x - 8\lambda^2 \mathcal{F}^2 = 0 \,. \tag{93}$$

Solving this quadratic equation for $x \equiv R_\mathrm{disk}/R_\mathrm{vir}$ gives us the following (physically acceptable, i.e. positive) solution:

$$\begin{aligned} \frac{R_\mathrm{disk}}{R_\mathrm{vir}} &= -\frac{\mathcal{F} f_\mathrm{B}}{2} + \left\{ \left( \frac{\mathcal{F} f_\mathrm{B}}{2} \right)^2 + 8\lambda^2 \mathcal{F}^2 \right\}^{1/2} \\ &= -\frac{\mathcal{F} f_\mathrm{B}}{2} + \frac{\mathcal{F} f_\mathrm{B}}{2} \left\{ 1 + \frac{32\lambda^2}{f_\mathrm{B}^2} \right\}^{1/2} \,. \end{aligned} \tag{94}$$

Two limits of interest exist in this expression, corresponding to whether the baryonic disk is gravitationally dominant ($32\lambda^2/f_\mathrm{B}^2 \lesssim 1$) or not ($32\lambda^2/f_\mathrm{B}^2 \gtrsim 1$):

$$\frac{R_\mathrm{disk}}{R_\mathrm{vir}} \simeq \begin{cases} 8\mathcal{F}\lambda^2/f_\mathrm{B} & \text{Gravitationally dominant baryonic disk } (32\lambda^2/f_\mathrm{B}^2 \lesssim 1) \\ \sqrt{8}\mathcal{F}\lambda & \text{Gravitationally dominant DM halo } (32\lambda^2/f_\mathrm{B}^2 \gtrsim 1) \end{cases} \,. \tag{95}$$

To clearly see how the quantity $32\lambda^2/f_\mathrm{B}^2$ connects to which component (baryons or DM) dominate gravitationally, note that the condition $32\lambda^2/f_\mathrm{B}^2 \gtrsim 1$ is equivalent to demanding that the rotation curve in the disk is $v_\mathrm{rot} \simeq v_\mathrm{vir}$ in Eq. (90), as expected for a gravitationally dominant (isothermal) halo. For our chosen median value for the spin parameter ($\bar{\lambda} = 0.03$) we find $32\bar{\lambda}^2/f_\mathrm{B}^2 = 1.17$, and so it is convenient to write the disk radius in Eq. (94) in terms of the disk-dominated limit as follows:

### Estimated disk radius in a halo

$$\begin{aligned} R_\mathrm{disk} &= \frac{8\mathcal{F}}{f_\mathrm{B}} f_\mathrm{disk} \lambda^2 R_\mathrm{vir} \\ f_\mathrm{disk}(\lambda) &\equiv \frac{f_\mathrm{B}^2}{16\lambda^2} \left[ \left\{ 1 + \frac{32\lambda^2}{f_\mathrm{B}^2} \right\}^{1/2} - 1 \right] \,. \end{aligned} \tag{96}$$

This is the expression used for the disk radius in Anaxagoras. In Figure 12 the dimensionless quantity $f_\mathrm{disk}$ as a function of the spin parameter $\lambda$ is plotted, showing that it is typically of order unity in most halos (e.g. 95% of halos have $0.525 < f_\mathrm{disk} < 1$).[30] For the median spin parameter $\bar{\lambda} = 0.03$ we find

[30] From Eq. (86), the probability $P(\lambda < \lambda_0) = \int_0^{\lambda_0} \mathrm{d}\lambda \, p(\lambda)$ of a halo having a spin parameter below $\lambda$ is

$$P(\lambda < \lambda_0) = \int_{-\infty}^{x \equiv \ln(\lambda_0/\bar{\lambda})} \mathrm{d}y \, \frac{1}{\sqrt{2\pi}\sigma_{\ln\lambda}} \, e^{-y^2/2\sigma_{\ln\lambda}^2} = \frac{1}{2} \left\{ 1 + \mathrm{erf}\left( \frac{x}{\sqrt{2}\sigma_{\ln\lambda}} \right) \right\} \,.$$

This can be inverted to yield $\lambda_0$ as a function of the probability $P(\lambda < \lambda_0)$:

$$\lambda_0 = \bar{\lambda} \, e^{\sqrt{2}\sigma_{\ln\lambda} \, \mathrm{erf}^{-1}(2P-1)} \,,$$

where $\mathrm{erf}^{-1}$ is the inverse error function.

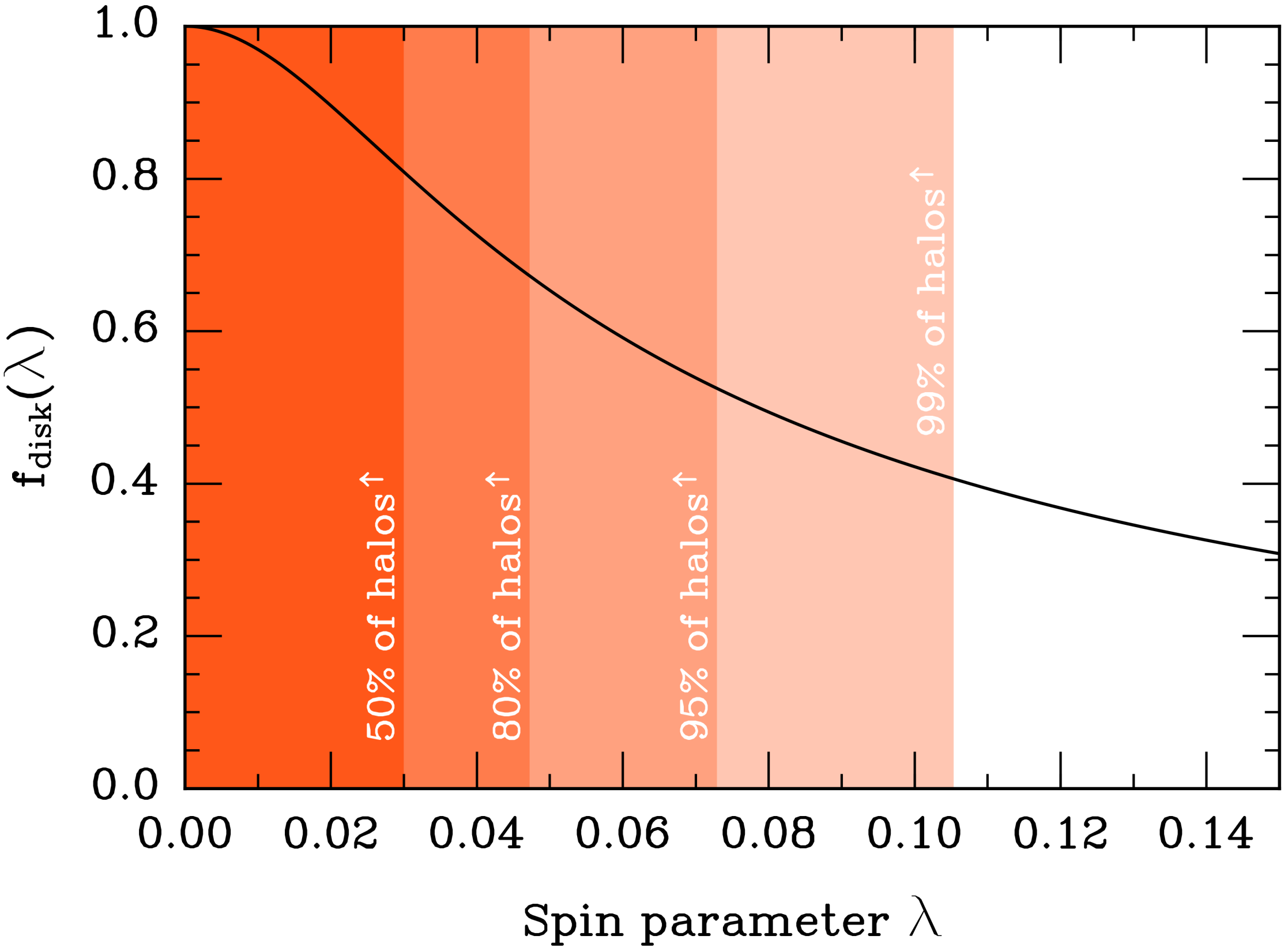

**Figure 12.** A plot of $f_{\rm disk}(\lambda)$ (solid line) along with colored regions showing the probability of having a smaller spin parameter $\lambda$, calculated from the distribution in Eq. (86). It turns out that 80% of halos have $0.672 < f_{\rm disk} < 1$, and 95% of halos fall in the range $0.525 < f_{\rm disk} < 1$. Thus, $f_{\rm disk}$ is of order unity in most halos.

$f_{\rm disk}(\bar{\lambda}) \simeq 0.81$, which we will use as a reference value when evaluating some of the derived expressions numerically.

We observe that $R_{\rm disk}$ scales linearly with $\mathcal{F} \equiv M_{\rm disk}/f_{\rm B}M \leq 1$, the fraction of baryons within the halo that resides in the disk. Shortly after efficient gas cooling we have seen in Section 3.2 that $M_{\rm disk} = \dot{M}_{\rm gas}t$, where $\dot{M}_{\rm gas}$ is the (constant) gas accretion rate. Thus, we deduce that the disk size should grow linearly with time during this stage. The steady growth of the gaseous disk is supported at least qualitatively by the idealized high-resolution simulations of gas collapse in atomic-cooling halos Choi *et al.* (2013), who assume a similar specific angular momentum distribution in the halo. It is also seen in cosmological high-resolution simulations of atomic-cooling halos (e.g. Patrick *et al.*, 2020).

#### 4.2.3. Numerical estimates and comparison with simulations

What is the expected disk size from Eq. (96), and how does it compare with simulations? To get some numerical estimates we can use Eq. (7) for $R_{\rm vir}$,

$$R_{\rm disk} = 44.6 \text{ pc } \mathcal{F} f_{\rm disk,0.81} \lambda_{0.03}^2 T_{\rm vir,4}^{1/2} \left(\frac{1+z}{10}\right)^{-3/2}, \tag{97}$$

where $f_{\rm disk,0.81} \equiv f_{\rm disk}/0.81$, and $\lambda_{0.03} \equiv \lambda/0.03$. Thus, most disks forming following gas collapse at the atomic-cooling threshold ($T_{\rm vir,4} = 1$) will at most have radii of few $\times$ 10 pc. We expect them to be smaller still during the early stages of gas accretion, since $\mathcal{F} \ll 1$ then. Using Eq. (84) for $\mathcal{F}$ we find

$$R_{\rm disk} = 1.07 \text{ pc } \mathscr{G}_{\rm gas,0.331} t_{\rm Myrs} f_{\rm disk,0.81} \lambda_{0.03}^2 T_{\rm vir,4}^{1/2} \,. \tag{98}$$

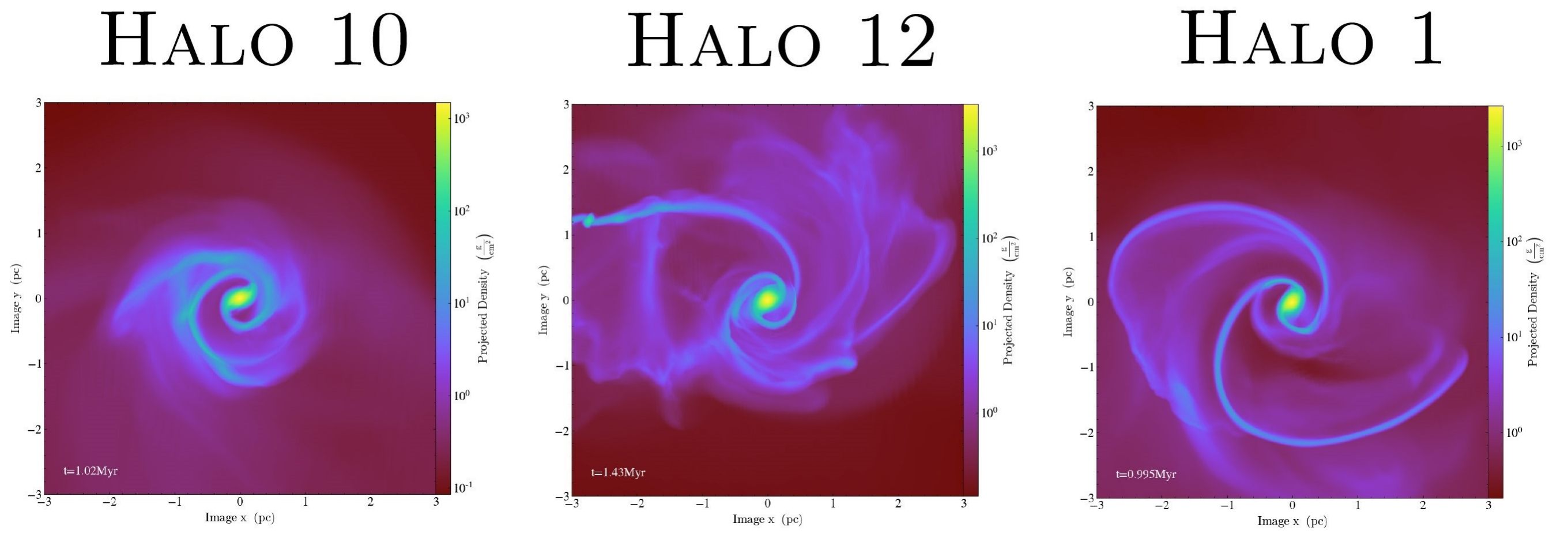

**Figure 13.** Snapshots of the surface density (in g cm$^{-2}$) in three of the atomic-cooling halos simulated by Patrick *et al.* (2020). Each snapshot has a side length of 6 pc. The disk sizes appear to be consistent with our predicted $1\sigma$ range ($0.45$ pc $\lesssim R_{\rm disk} \lesssim 2.8$ pc for $t_{\rm Myrs} \simeq 1$), with the caveat that it is hard to make out a sharp disk boundary in the simulations. **Credit:** The halo snapshots are taken from Figure 3 in Patrick *et al.* (2020).

Thus, atomic-cooling halos with $T_{\rm vir,4} \simeq 1$ are expected to host disks with radii $0.45 \text{ pc}\, t_{\rm Myrs} \lesssim R_{\rm disk} \lesssim 2.8 \text{ pc}\, t_{\rm Myrs}$ ($1\sigma$ interval). This analytical prediction is in fairly good agreement with the gaseous disks found at the center of atomic-cooling halos in high-resolution simulations (e.g. Safranek-Shrader *et al.*, 2014b; Kimm *et al.*, 2016; Latif *et al.*, 2020; Patrick *et al.*, 2020). For a simple comparison between our model and simulations, consider the results of Patrick *et al.* (2020) who followed gas collapse in 8 atomic-cooling halos at Cosmic Dawn, all of which formed gaseous disks (see also Latif *et al.*, 2020). The authors show snapshots of the central region in five of these halos at different times, three of which fall in the range $0.930 \leqslant T_{\rm vir,4} \leqslant 1.42$ (their halos 1, 10, and 12) as calculated from their halo masses and collapse redshifts.[31] In Figure 13 we see snapshots at times $t_{\rm Myrs} \sim 1$ in these halos, seeing disks with radii in rough agreement with the predicted $1\sigma$ interval.[32]

## 4.3. Disk structure: Density, temperature, and turbulence

Having found the size of the disk, we are now ready to derive its density, which will be needed to derive the free-fall time-scale — a crucial ingredient for the predicted star formation rate. We start with the

[31] The two other halos, 20 and 2, have only single snapshots, making it harder to ascertain their evolution. Furthermore, halo 2 has a rather high virial temperature of $T_{\rm vir,4} = 2.17$, while the single snapshot of the central region in halo 20 shows a binary disk, making it hard to compare with the predictions of our model.

[32] Patrick *et al.* (2020) also provide the *total* (i.e. DM + baryon) spin parameter $\lambda_{\rm tot}$ for their halos which could give a more accurate comparison between our modelling and their simulation results. However, while $\lambda_{\rm tot}$ and $\lambda$ are expected to have almost exactly the same distribution, simulations show that in a given halo there is only a weak correlation between $\lambda_{\rm tot}$ and $\lambda$ (Druschke *et al.*, 2018), preventing us from giving a rigorous comparison.

surface density $\Sigma_{\rm disk}(R)$ for $R < R_{\rm disk}$:

$$\Sigma_{\rm disk}(R) = \frac{M_{\rm disk}}{2\pi R_{\rm disk} R} \equiv \frac{\Sigma_{\rm disk,0}}{\bar{R}}, \tag{99}$$

where $\Sigma_{\rm disk,0} \equiv M_{\rm disk}/2\pi R_{\rm disk}^2$ is the outer disk surface density, and $\bar{R} \equiv R/R_{\rm disk}$ is a convenient dimensionless radial variable. Using $M_{\rm disk} = \mathcal{F} f_{\rm B} M$ and $R_{\rm disk} = 8\mathcal{F} f_{\rm disk}\lambda^2 R_{\rm vir}/f_{\rm B}$ (see Eq. 96) we find that

$$\Sigma_{\rm disk,0} = \frac{f_{\rm B}^3 M}{128\pi \mathcal{F} f_{\rm disk}^2 \lambda^4 R_{\rm vir}^2} = \frac{422\ {\rm M_\odot\ pc^{-2}}}{\mathcal{F} f_{\rm disk}^2 \lambda_{0.03}^4} T_{\rm vir,4}^{1/2} \left(\frac{1+z}{10}\right)^{3/2}, \tag{100}$$

where we have used Eq. (7) for $M$ and $R_{\rm vir}$ to get the numerical result on the second line. Another convenient way of evaluating $\Sigma_{\rm disk,0}$ can be obtained by using $\mathcal{F} = \dot{M}_{\rm gas} t/f_{\rm B} M$ and $\dot{M}_{\rm gas} = 0.331\, \mathscr{G}_{\rm gas,0.331}\, v_{\rm vir}^3/G$ (see Eq. 83), giving us

$$\Sigma_{\rm disk,0} = \frac{1}{128\pi \times 0.331 f_{\rm disk}^2 \lambda^4} \frac{f_{\rm B}^4 v_{\rm vir}}{\mathscr{G}_{\rm gas,0.331} G t} = \frac{1.73 \times 10^4\ {\rm M_\odot\ pc^{-2}}}{f_{\rm disk}^2 \lambda_{0.03}^4 \mathscr{G}_{\rm gas,0.331} t_{\rm Myrs}} T_{\rm vir,4}^{1/2}, \tag{101}$$

an expression that will be useful since we shall see later that $\Sigma_{\rm disk,0}$ is a key parameter controlling the efficiency of momentum feedback in the disk. Finally, for some numerical estimates it will also be convenient to have $\Sigma_{\rm disk}$ as a function of the physical radius $R$. Using $\Sigma_{\rm disk} = M_{\rm disk}/2\pi R_{\rm disk} R$ with $M_{\rm disk} = 8.02 \times 10^6\, \mathcal{F} T_{\rm vir,4}^{3/2}[(1+z)/10]^{-3/2}\ {\rm M_\odot}$ and (97) for $R_{\rm disk}$, we find:

$$\Sigma_{\rm disk}(R) = \frac{2.32 \times 10^4\ {\rm M_\odot\ pc^{-2}}}{f_{\rm disk}\lambda_{0.03}^2} T_{\rm vir,4} \left(\frac{R}{1\ {\rm pc}}\right)^{-1}. \tag{102}$$

Thus, at a fixed radius $R$ from the center of the disk, halos with higher virial temperatures have higher disk surface densities (for equal spin parameters at least). To get the volume (rather than surface) density of the disk we need to consider its vertical structure. For an isothermal self-gravitating disk we can follow earlier work on gaseous disks in atomic-cooling halos (Oh & Haiman, 2002; Inayoshi *et al.*, 2018) and apply the Spitzer "isothermal sheet" model (Spitzer, 1942). In this case the density $\rho_{\rm disk}$ (at some given $R$) at a height $z$ above (or below) the midplane of the disk is given by

$$\rho_{\rm disk}(z) = \rho_{\rm mid}\, {\rm sech}^2\left(\frac{z}{2z_0}\right), \tag{103}$$

where ${\rm sech}x \equiv 2/(e^x + e^{-x})$, and the scale-height $z_0$ depends on the gas velocity dispersion $\sigma_{\rm gas}$ and the midplane density $\rho_{\rm mid}$:

$$z_0 = \frac{\sigma_{\rm gas}}{(8\pi G \rho_{\rm mid})^{1/2}}. \tag{104}$$

By definition this must be related to the disk surface density via:

$$\begin{aligned}\Sigma_{\rm disk} &= \int_{-\infty}^{\infty} {\rm d}z\, \rho_{\rm disk}(z) \\ &= \rho_{\rm mid} z_0 \int_{-\infty}^{\infty} {\rm d}x\, {\rm sech}^2\left(\frac{x}{2}\right) \\ &= 4\rho_{\rm mid} z_0\,.\end{aligned} \tag{105}$$

Using our expression for $z_0$, this yields the following relation between $\rho_{\rm mid}$ and $\Sigma_{\rm disk}$:

$$\rho_{\rm mid} = \frac{\pi G \Sigma_{\rm disk}^2}{2\sigma_{\rm gas}^2}\,. \tag{106}$$

Furthermore, upon plugging this into Eq. (104) we also find an expression for $z_0$ in terms of $\Sigma_{\rm disk}$:

$$z_0 = \frac{\sigma_{\rm gas}^2}{2\pi G \Sigma_{\rm disk}}\,. \tag{107}$$

To estimate both $\rho_{\rm mid}$ and $z_0$ we clearly need to know the gas velocity dispersion $\sigma_{\rm gas}$. Both thermal and turbulent motion of the gas in the disk contribute to $\sigma_{\rm gas}$, and so we have

$$\begin{aligned}\sigma_{\rm gas}^2 &= c_{\rm s}^2 + \sigma_{\rm t}^2 \\ &\equiv c_{\rm s}^2\left(1 + \frac{1}{3}\mathcal{M}^2\right)\,,\end{aligned} \tag{108}$$

where $c_{\rm s} = (k_{\rm B}T/\mu m_{\rm H})^{1/2} = 8.20 \times 10^5\, T_4^{1/2} \mu_{1.23}^{-1}$ cm s$^{-1}$ is the sound speed of the gas in the disk ($T$ being the temperature of the disk, and $\mu$ the mean molecular weight), $\sigma_{\rm t}$ the (1D) turbulent velocity dispersion, and $\mathcal{M} = \sqrt{3}\sigma_{\rm t}/c_{\rm s}$ the (3D) turbulent Mach number. For the temperature of the gaseous disk we will adopt a simplified scheme:

$$T = \begin{cases} 8000\ {\rm K} & \text{if } Z/Z_\odot < 10^{-4} \text{ and in the absence of H}_2 \\ 200\ {\rm K} & \text{if } Z/Z_\odot < 10^{-4} \text{ and with H}_2 \text{ present} \\ \max[10\ {\rm K},\, T_{\rm CMB}(z)] & \text{if } Z/Z_\odot > 10^{-4} \text{ and with H}_2 \text{ present} \end{cases}\,. \tag{109}$$

This is motivated as follows. In the absence of sufficient metals ($Z/Z_\odot < 10^{-4}$) and $H_2$, the most important cooling process is Ly$\alpha$-cooling, which becomes exponentially inefficient below $T \simeq 8000$ K due to the absence of a sufficient number of energetic electrons (especially in collisional ionization equilibrium) to excite neutral hydrogen (e.g. Oh & Haiman, 2002; Spaans & Silk, 2006; Choi *et al.*, 2013; Kimm *et al.*, 2016; Latif *et al.*, 2020; Patrick *et al.*, 2020). When we allow the formation of $H_2$ in the gas but have $Z/Z_\odot < 10^{-4}$, cooling is possible down to $T \simeq 200$ K in both minihalos and atomic-cooling halos (e.g. Oh & Haiman, 2002; Safranek-Shrader *et al.*, 2012, 2014b; Schauer *et al.*, 2017). Finally, when metals are also present in sufficient quantities, metal cooling (e.g. by the C II 157.74 $\mu$m fine-structure line) can cool the gas down to either $T \simeq 10$ K or the CMB temperature $T_{\rm CMB}(z) = 27.26\,[(1+z)/10]$ K depending on which is the largest (e.g. Krumholz, 2012; Safranek-Shrader *et al.*, 2014b). A more detailed model of the gas thermal state is left for a future extension of Anaxagoras, but we note that the exact details are not very important here because the star formation efficiency will mostly depend on the surface density (which is temperature-independent) rather than the volume density. Next we discuss the turbulence in the disk. Prior to star formation and efficient stellar feedback the turbulence of disks in low-mass halos are expected to be driven by disk accretion (e.g. Greif *et al.*, 2008; Safranek-Shrader *et al.*, 2012; Choi *et al.*, 2013). To make a rough order-of-magnitude estimate for $\mathcal{M}$ in the case of accretion-driven turbulence, we can loosely follow the analytical model of accretion-driven turbulence in disks by Elmegreen & Burkert (2010). As the disk is accreting gas from the surrounding halo, turbulent energy is injected into the disk at a rate $\dot{E}_{\rm acc} \sim \frac{1}{2}\epsilon_{\rm t}\dot{M}_{\rm disk}v_{\rm vir}^2$, where it has been assumed that a fraction $\epsilon_{\rm t}$ of the energy of accretion is driving turbulence (Klessen & Hennebelle, 2010; Elmegreen & Burkert, 2010), and that the gas falls into the disk with velocity $\sim v_{\rm vir}$. In the absence of driving, turbulence is expected to decay in a (vertical) disk crossing time $t_{\rm cross} \sim 2z_0/\sigma_{\rm t} = \sigma_{\rm gas}^2/\pi G\Sigma_{\rm disk}\sigma_{\rm t}$ (e.g. Elmegreen & Burkert, 2010; Faucher-Giguère *et al.*, 2013). The turbulent energy decay rate is therefore $\dot{E}_{\rm decay} \sim \frac{3}{2}M_{\rm disk}\sigma_{\rm t}^2/t_{\rm cross}$, or

$$\dot{E}_{\rm decay} \sim \frac{3}{2}M_{\rm disk}\pi G\Sigma_{\rm disk}\frac{\sigma_{\rm t}^3}{\sigma_{\rm gas}^2}\,. \tag{110}$$

In terms of the Mach number this becomes:

$$\dot{E}_{\rm decay} \sim \frac{1}{2}M_{\rm disk}\pi G\Sigma_{\rm disk}\frac{\mathcal{M}^3 c_{\rm s}}{(1+\mathcal{M}^2/3)}\,. \tag{111}$$

Setting this equal to the turbulent energy injection rate $\dot{E}_{\rm in} \sim \frac{1}{2}\epsilon_{\rm t}\dot{M}_{\rm disk}v_{\rm vir}^2$, using $M_{\rm disk} = \dot{M}_{\rm disk}t$, and using $\Sigma_{\rm disk} \sim 2E_{\rm disk,0}$ (i.e. evaluating the decay time-scale at the half-mass radius $R = \frac{1}{2}R_{\rm disk}$ of the disk) yields an equilibrium Mach number determined by:

$$\frac{\mathcal{M}^3}{1+\mathcal{M}^2/3} \sim \frac{2\epsilon_{\rm t}v_{\rm vir}^2}{\pi G\Sigma_{\rm disk,0}c_{\rm s}t}\,. \tag{112}$$

where the right-hand side is independent of time since $\Sigma_{\rm disk,0} \propto t^{-1}$ (see Eq. 101). When the right-hand side is small we get $\mathcal{M} \sim (2\epsilon_{\rm t}v_{\rm vir}^2/\pi G\Sigma_{\rm disk,0}c_{\rm s}t)^{1/3}$, whereas when the right-hand side is large we find $\mathcal{M} \sim 2\epsilon_{\rm t}v_{\rm vir}^2/3\pi G\Sigma_{\rm disk,0}c_{\rm s}t$. The following fit to the exact root of Eq. (112) is accurate to within 6.7% for all values of $2\epsilon_{\rm t}v_{\rm vir}^2/\pi G\Sigma_{\rm disk,0}c_{\rm s}t$, and contains the correct asymptotic limits just mentioned:

$$\mathcal{M} \sim \left\{\left(\frac{2\epsilon_{\rm t}v_{\rm vir}^2}{\pi G\Sigma_{\rm disk,0}c_{\rm s}t}\right)^{2/3} + \left(\frac{2\epsilon_{\rm t}v_{\rm vir}^2}{3\pi G\Sigma_{\rm disk,0}c_{\rm s}t}\right)^{2}\right\}^{1/2}\,. \tag{113}$$

To get some numerical estimates we can use

$$\frac{2\epsilon_{\rm t}v_{\rm vir}^2}{\pi G\Sigma_{\rm disk,0}c_{\rm s}t} = \frac{0.0186}{f_{\rm disk}^2\lambda_{0.03}^4\mathscr{G}_{\rm gas,0.331}}\left(\frac{\epsilon_{\rm t}}{0.1}\right)\left(\frac{T_{\rm vir}}{T}\right)^{1/2}\,. \tag{114}$$

If we use this in Eq. (113) we can get the following estimates (using $\epsilon_{\rm t} = 0.1$ and Eq. 96 for $f_{\rm disk}$):

- For halos at the atomic-cooling threshold ($T_{\rm vir} = 10^4$ K) where the absence of both metals and $H_2$ leads to a disk with $T = 8000$ K we predict $0.18 < \mathcal{M} < 0.60$ for the 68% spin parameter range $0.0175 < \lambda < 0.0515$ (with lower spin halos having larger $\mathcal{M}$).

- For halos at the atomic-cooling threshold ($T_{\rm vir} = 10^4$ K) with negligible metallicity and wherein $H_2$-cooling leads to a disk with $T = 200$ K we predict $0.33 < \mathcal{M} < 1.2$ for the 68% spin parameter range.

- For halos at the atomic-cooling threshold ($T_{\rm vir} = 10^4$ K) with enough metals present to push the temperature down to $T = T_{\rm CMB} = 27$ K (at $z = 9$) we predict $0.46 < \mathcal{M} < 2.0$ for the 68% spin parameter range.

And for a typical minihalo of mass $M \sim 8 \times 10^5$ M$_\odot$ at redshift $z = 20$ containing primordial gas, we find $0.07 < \mathcal{M} < 0.23$ for $0.0175 < \lambda < 0.0515$. Thus, the gas is not expected to be highly supersonic, consistent with simulations of both atomic-cooling halos and minihalos (see e.g. Greif *et al.*, 2011; Safranek-Shrader *et al.*, 2012; Choi *et al.*, 2013; Wise *et al.*, 2019). Using this we find the following numerical estimates for the midplane gas number density $n_{\rm mid} = \rho_{\rm mid}/\mu m_{\rm H}$ and the scale-height $z_0$ (assuming $\mu_{1.23} = 1$):

$$
\begin{aligned}
n_{\rm mid} &= \frac{1.78 \times 10^6 \text{ cm}^{-3}}{f_{\rm disk}^2 \lambda_{0.03}^4 T_4 (1 + \mathcal{M}^2/3)} T_{\rm vir,4}^2 \left(\frac{R}{1 \text{ pc}}\right)^{-2} \\
z_0 &= 0.107 \; f_{\rm disk} \lambda_{0.03}^2 \left(\frac{T_4}{T_{\rm vir,4}}\right) \left(1 + \frac{1}{3}\mathcal{M}^2\right) \left(\frac{R}{1 \text{ pc}}\right) \text{ pc} \,.
\end{aligned} \tag{115}
$$

As a quick sanity check we can plug in values for a typical (i.e. $\lambda_{0.03} = 1$, $f_{\rm disk} = 0.81$) atomic-cooling halo that is free of both $H_2$ and metals ($T_{\rm vir,4} = 1$, $T_4 = 0.8$) and find $n_{\rm mid} \sim 3 \times 10^6 \, (R/1 \text{ pc})^{-2} \text{ cm}^{-3}$, which is in nice agreement with the gaseous disks found in the simulations of Kimm *et al.* (2016), as can be seen in panel (e) of Figure 5. Having discussed both the assumed disk temperature and the calculation of the Mach number, we can now proceed to calculate the free-fall time-scale in the disk midplane, $t_{\rm ff} = (3\pi/32 G \rho_{\rm mid})^{1/2}$, a quantity that will be crucial for the predicted star formation rate in the disk later. Using Eq. (106) for $\rho_{\rm mid}$ yields $t_{\rm ff} = \sqrt{3}\sigma_{\rm gas}/4G\Sigma_{\rm disk}$ or

$$
\begin{aligned}
t_{\rm ff}(\bar{R}) &= \frac{\sqrt{3} c_{\rm s}}{4 G \Sigma_{\rm disk,0}} \left(1 + \frac{1}{3}\mathcal{M}^2\right)^{1/2} \bar{R} \\
&= 4.67 \times 10^4 \left(\frac{T}{T_{\rm vir}}\right)^{1/2} f_{\rm disk}^2 \lambda_{0.03}^4 \mathscr{G}_{\rm gas,0.331} t_{\rm Myrs} \left(1 + \frac{1}{3}\mathcal{M}^2\right)^{1/2} \bar{R} \text{ yrs} \,.
\end{aligned} \tag{116}
$$

Thus, even in a disk with $T = 8000$ K forming in an atomic-cooling halo with $T_{\rm vir} = 10^4$ K we typically have $t_{\rm ff} \lesssim 10^5$ yrs, also in agreement with the simulations of Kimm *et al.* (2016).

## 4.4. Toomre instability and fragmentation

A prerequisite for star formation is that the gaseous disk can fragment into smaller collapsing gas clouds. Linear perturbations with wavenumber $k$ (e.g. in the surface density) in a gaseous disk grow or oscillate as $\sim e^{-i\omega t}$, where

$$
\omega^2 = k^2 \sigma_{\rm gas}^2 - 2\pi G \, |k| \, \Sigma_{\rm disk} + \kappa^2 \,, \tag{117}
$$

and $\kappa \equiv \sqrt{2 v_{\rm disk}(\partial_R v_{\rm disk} + v_{\rm disk}/R)/R} = \sqrt{2} v_{\rm disk}/R$ is the so-called epicycle frequency, and the second equality follows for a flat rotation curve, as assumed in our case (see e.g. Binney & Tremaine, 2008; Mo *et al.*, 2010, for derivations of this result). Gravitational collapse will commence most rapidly when $\omega^2$ is minimized, and as long as $\omega^2 < 0$ (so that $i\omega > 0$). We find that $\omega^2$ is minimized for a scale $k_{\rm Toomre} = \pi G \Sigma_{\rm disk}/\sigma_{\rm gas}^2$, and $\omega^2(k = k_{\rm Toomre}) < 0$ if the **Toomre condition** is satisfied:

$$
Q_{\rm Toomre} \equiv \frac{\kappa \sigma_{\rm gas}}{\pi G \Sigma_{\rm disk}} < 1 \,. \tag{118}
$$

Only if $Q_{\rm Toomre} < 1$ will the disk be susceptible to fragmentation and able to form stars. A particularly convenient property of the Mestel disk assumed here is that $Q_{\rm Toomre}$ is independent of both time and radial position, since at fixed radius $R$ we have $\Sigma_{\rm disk} \propto R^{-1}$ and $\kappa \propto R^{-1}$, and both quantities being time-independent.

# 5. Stellar feedback, star formation, and derived properties

## 5.1. Momentum feedback and the maximum star formation efficiency

Before we estimate the star formation rate in the disk we will need to know the effects of momentum feedback from radiation and stellar winds, which, as we shall see, will put an upper limit on the star formation efficiency even before expanding H II regions or supernova remnants can eject the gas from the disk. Let us first consider radiative injection of momentum into gas. A photon of energy $E_\nu = h\nu$ has a momentum $p_\nu = E_\nu/c$. If the radiation flux vector is $\boldsymbol{F}_\nu$ (in erg cm$^{-2}$ s$^{-1}$ Hz$^{-1}$),[33] then this carries a momentum flux $\boldsymbol{F}_\nu/c$. The momentum injection rate per unit mass — i.e. the acceleration in a given volume element $\mathrm{d}^3\boldsymbol{x}$ — then (e.g. Rybicki & Lightman, 1986; Kimm *et al.*, 2017; Hopkins *et al.*, 2020)

$$\frac{\partial \boldsymbol{v}}{\partial t} = \frac{1}{c}\int_0^\infty \mathrm{d}\nu\, \boldsymbol{F}_\nu \kappa_\nu \,, \tag{119}$$

where $\kappa_\nu$ is the mass absorption coefficient (i.e. the cross-section per unit mass, or opacity, in cm$^2$ g$^{-1}$). We can write this in terms of the luminosity as follows. The magnitude of the radiative momentum injection rate into a gas cloud of volume $\mathcal{V}$ and density $\rho_{\mathrm{gas}}$ in a direction $\hat{\boldsymbol{\ell}}$ is then

$$\begin{aligned} \dot{P}_{\mathrm{rad}} &= \int \mathrm{d}^3\boldsymbol{x}\, \rho_{\mathrm{gas}} \frac{\partial \boldsymbol{v}}{\partial t} \cdot \hat{\boldsymbol{\ell}} \\ &= \frac{1}{c}\int_0^\infty \mathrm{d}\nu \int_{\mathcal{V}} \mathrm{d}^3\boldsymbol{x}\, \kappa_\nu \rho_{\mathrm{gas}} \boldsymbol{F}_\nu \cdot \hat{\boldsymbol{\ell}} \,. \end{aligned} \tag{120}$$

Let us consider the flux from a point source (e.g. a star, or far from a stellar population) with $\hat{\boldsymbol{\ell}}$ pointing in the same direction as $\boldsymbol{F}_\nu$. We then have

$$\boldsymbol{F}_\nu \cdot \hat{\boldsymbol{\ell}} = \frac{L_\nu e^{-\tau_\nu(\boldsymbol{x})}}{4\pi r^2} \,, \tag{121}$$

where $L_\nu$ is the specific luminosity, $r$ the distance to the point source, and $\tau_\nu$ the optical depth. Given the spherical symmetry, we have $\mathrm{d}^3\boldsymbol{x} = 4\pi r^2\,\mathrm{d}r$ so that the momentum injection rate into a gas cloud of radius $R$ is:

$$\dot{P}_{\mathrm{rad}} = \frac{1}{c}\int_0^\infty \mathrm{d}\nu\, L_\nu \int_0^R \mathrm{d}r\, \kappa_\nu \rho_{\mathrm{gas}} e^{-\tau'_\nu(\boldsymbol{x})} \,, \tag{122}$$

or, since $\mathrm{d}r\,\kappa_\nu\rho_{\mathrm{gas}} = \mathrm{d}\tau'_\nu$:[34]

$$\begin{aligned} \dot{P}_{\mathrm{rad}} &= \frac{1}{c}\int_0^\infty \mathrm{d}\nu\, L_\nu \int_0^{\tau_\nu} \mathrm{d}\tau'_\nu\, e^{-\tau'_\nu} \\ &= \frac{1}{c}\int_0^\infty \mathrm{d}\nu\, L_\nu (1 - e^{-\tau_\nu}) \,. \end{aligned} \tag{123}$$

For convenience we can divide the spectrum into several frequency bands. The discrete frequency (or wavelength) band division that I will adopt for Pop II stars is based on the work of Hopkins *et al.* (2018a) (see also Hopkins *et al.*, 2020) for the state-of-the-art FIRE-2 galaxy formation simulations. These authors divided the spectrum into five discrete bands. Excluding the mid/far-IR band ($\lambda > 3 \times 10^4$ Å) for which there is negligible stellar emission, the remaining four bands of interest and their main stellar contributors are as follows (Hopkins *et al.*, 2020):

[33] The radiation flux vector is defined as $\boldsymbol{F}_\nu = \int \mathrm{d}\Omega\, I_\nu \hat{\boldsymbol{n}}$, where $I_\nu$ is the specific intensity, and $\hat{\boldsymbol{n}}$ the unit vector pointing in the same direction as the ray (see e.g. p. 15 in Rybicki & Lightman, 1986).

[34] A similar derivation with the same result can be found in Appendix A.1. of Raskutti *et al.* (2016).

1. Hydrogen ionizing radiation (ion) ($\lambda < 912$ Å): Dominated by young, short-lived, massive stars.

2. Far-UV radiation (FUV) (912 Å $< \lambda <$ 1550 Å): Dominated by young stars.

3. Near-UV (NUV) (1550 Å $< \lambda <$ 3600 Å): Dominated by young stars.

4. Optical/near-IR (opt) (3600 Å $< \lambda < 3 \times 10^4$ Å): Dominated by long-lived stars.

On top of direct stellar emission in these bands, it has also been realized recently that radiation pressure from Ly$\alpha$ photons produced by recombining hydrogen atoms in a photoionized region can be a significant, if not dominant, source of feedback (see e.g. Abe & Yajima, 2018; Kimm *et al.*, 2018; Tomaselli & Ferrara, 2021). If we let $\Psi_{\rm band}$ be the IMF-averaged luminosity per unit stellar mass formed (i.e. the light-to-mass ratio) in a given band, the resulting IMF-averaged momentum injection rate per stellar mass $\left\langle \dot{P}/m_\star \right\rangle$ into the gas from both radiation and stellar winds is then

$$\begin{aligned} \left\langle \frac{\dot{P}}{m_\star} \right\rangle &= \left(\frac{\Psi_{\rm ion}}{c}\right)(1-e^{-\bar{\tau}_{\rm ion}}) \\ &+ (1+\tau_{\rm IR}^{\rm eff}) \sum_{\rm band\,=\,\{FUV,NUV,opt\}} \left(\frac{\Psi_{\rm band}}{c}\right)(1-e^{-\bar{\tau}_{\rm band}}) \\ &+ M_{\rm F}\left(\frac{\Psi_{\rm Ly\alpha}}{c}\right) + \left\langle \frac{\dot{P}}{m_\star} \right\rangle_{\rm wind} . \end{aligned} \tag{124}$$

The terms in this equation can be summarized as follows:

1. The first term on the right is due to radiation pressure from absorbed hydrogen ionizing photons (e.g. Haehnelt, 1995; Kimm *et al.*, 2016, 2017; Hopkins *et al.*, 2018a; Tomaselli & Ferrara, 2021), with $\bar{\tau}_{\rm ion}$ being a suitably band-averaged optical depth due to photoionization of neutral hydrogen.

2. The second term(s) on the right include direct stellar emission from the rest of the bands, again with suitably band-averaged optical depths (in this case due to dust). Unlike the case of the ionizing band we have also included a factor $(1+\tau_{\rm IR}^{\rm eff})$ where $\tau_{\rm IR}^{\rm eff}$ is the effective far-infrared (FIR) optical depth from dust. This factor arises because when photons from stars are absorbed by dust, they will be re-radiated in the FIR and scattered multiple times if $\tau_{\rm IR}^{\rm eff}$ is large, in effect boosting the momentum injected into the ISM (see e.g. discussion in Murray *et al.*, 2010; Hopkins *et al.*, 2011; Faucher-Giguère *et al.*, 2013; Rahner *et al.*, 2017; Zhang & Davis, 2017; Costa *et al.*, 2018; Kruijssen *et al.*, 2019a; Hopkins *et al.*, 2020). A rigorous derivation of the factor $1+\tau_{\rm IR}^{\rm eff}$ can be found in Appendix 6. The factor $(1+\tau_{\rm IR}^{\rm eff})$ is not present for the ionizing band because ionizing photons are efficiently absorbed by neutral hydrogen rather than dust. Recent observational evidence from H II regions in the Milky Way suggest that multiple scattering by dust can be an important, or even dominant, stellar feedback process (Olivier *et al.*, 2021).

3. The third term on the right includes the radiation pressure from Ly$\alpha$ photons produced by recombining hydrogen atoms. Due to the large Ly$\alpha$ optical depth in H I regions, these photons are expected to be scattered multiple times, and hence impart more momentum than expected from a single scattering. This effect is captured by the so-called "force multiplier" factor $M_{\rm F} \geqslant 1$ (Abe & Yajima, 2018; Kimm *et al.*, 2018; Tomaselli & Ferrara, 2021).[35]

[35] Here I am following the notation of Kimm *et al.* (2018) and Tomaselli & Ferrara (2021). Abe & Yajima (2018) uses the notation $f_{\rm boost}$ instead of $M_{\rm F}$, but these are the same quantities.

4. Finally, the last term on the right is the momentum injection rate into the gas from stellar winds. This is primarily driven by radiation pressure on metal lines in the photosphere of massive stars. Thus, the mass loss rate — and hence the momentum injected from winds — is dependent on metallicity.

Let us consider the four contributions in more detail below for both Pop II and Pop III stars.

### 5.1.1. Radiation pressure from absorbed stellar emission and FIR scattering by dust

We start by considering the first two terms in Eq. (124). To evaluate these terms we need to know the light-to-mass ratios in each band ($\Psi_{\rm band}$), the band-averaged optical depths ($\bar{\tau}_{\rm band}$), and the effective FIR optical depth due to dust ($\tau_{\rm IR}^{\rm eff}$). In the case of Pop II stars we can use the results of Hopkins *et al.* (2018a) who calculated $\Psi_{\rm band}$ and $\bar{\tau}_{\rm band}$ for the mentioned wavelength bands above using `STARBURST99` (Leitherer *et al.*, 1999), assuming a Kroupa IMF (Kroupa, 2001, 2002).[36] The result is summarized in Table 2, where we have time-averaged[37] the light-to-mass ratios over the first 3.5 Myrs following an instantaneous stellar burst. This time-averaging is needed to ensure that Anaxagoras remain analytically tractable. The ionizing, NUV, and optical bands have constant or near-constant light-to-mass ratios, so for these bands this averaging has little or no effect.[38] Only the FUV band has a significant time-dependence, rising from $\Psi_{\rm FUV}(t=0) = 271\,{\rm L}_\odot\,{\rm M}_\odot^{-1}$ to $\Psi_{\rm FUV}(t=3.4\ {\rm Myrs}) = 572\,{\rm L}_\odot\,{\rm M}_\odot^{-1}$, which can be compared to the adopted time-averaged value of $\bar{\Psi}_{\rm FUV} = 367\,{\rm L}_\odot\,{\rm M}_\odot^{-1}$. Neglecting this time-dependence is unlikely to have a big effect on any predictions of Anaxagoras because the other bands have constant (or near-constant) light-to-mass ratios, and it is only their sum that is of interest to us. To calculate the band-averaged optical depths we also adopt the dust opacities from Hopkins *et al.* (2018a) for the non-ionizing bands. For the ionizing band we have assumed a band-averaged photoionization cross-section $\bar{\sigma}_{\rm ion} = 3.3\times10^{-18}\ {\rm cm}^2$ to estimate the opacity to radiation from Pop II stars (Rosdahl *et al.*, 2013; Agertz *et al.*, 2020).[39] The resulting opacities are listed in Table 2. The band-averaged optical depth would then be given by $\bar{\tau}_{\rm band} = \kappa_{\rm band}\Sigma_{\rm disk}$, which would be a function of the radius $R$ from the center of the disk. This would render the model analytically intractable because we would end up with integrals that could not be solved analytically. As a simplified procedure we instead calculate $\bar{\tau}_{\rm band}$ via the mass-weighted average of $e^{-\kappa_{\rm band}\Sigma_{\rm disk}}$, since it is this exponential that controls the fraction of the the momentum from radiation that is actually absorbed by the gas. Thus,

[36] More specifically, they considered stars with masses $0.1\,{\rm M}_\odot < m_\star < 100\,{\rm M}_\odot$, for which the Kroupa IMF adopted in `STARBURST99` is given by

$$\frac{{\rm d}N}{{\rm d}m_\star} \propto \begin{cases} m_\star^{-1.3} & \text{for } 0.1 < m_\star/{\rm M}_\odot < 0.5 \\ m_\star^{-2.3} & \text{for } 0.5 \leqslant m_\star/{\rm M}_\odot < 100 \end{cases} .$$

[37] That is, the time-averaged light-to-mass ratio $\bar{\Psi}_{\rm band}$ for a given band is given by $\bar{\Psi}_{\rm band} \equiv (1/\Delta t)\int_0^{\Delta t} {\rm d}t\,\Psi_{\rm band}(t)$ with $\Delta t = 3.5$ Myrs in this case.

[38] The ionizing band has a constant value during the first 3.5 Myrs, and so time-averaging yields the same value. In Hopkins *et al.* (2018a) the approximation for the NUV is time-dependent, and has an unphysical negative dip. I reached out to Prof. Philip Hopkins in March of 2021 regarding this issue. He explained that this issue is due to a simplification in Hopkins *et al.* (2018a) of how the NUV band is actually calculated in their simulations. Instead he suggested that I could replace the NUV band with a constant value of $\Psi_{\rm NUV} = 200\,{\rm L}_\odot\,{\rm M}_\odot^{-1}$ for $t < 3.5$ Myrs to get results approximately consistent with their direct NUV fits.

[39] This is smaller than the value at the Lyman limit (912 Å, $h\nu = 13.6$ eV) — $\sigma_{\rm ion}(h\nu = 13.6\ {\rm eV}) \simeq 6\times10^{-18}\ {\rm cm}^2$ — since the photoionization cross-section drops approximately as $\nu^{-3}$. Thus, photons from stars with $h\nu > 13.6$ eV encounter smaller cross-sections. To get a value for the opacity in the ionization band we use $\kappa_{\rm ion}\rho_{\rm gas} = \bar{\sigma}_{\rm ion}n_{\rm H}$ so that $\kappa_{\rm ion} = \bar{\sigma}_{\rm ion}n_{\rm H}/\rho_{\rm gas}$. Since $n_{\rm H} = X\rho_{\rm gas}/m_{\rm H}$ we get $\kappa_{\rm ion} = \bar{\sigma}_{\rm ion}X/m_{\rm H}$, which is used to evaluate the opacity.

Table 2. Radiative momentum feedback during the first $\sim 3.5$ Myrs following an instantaneous Population II stellar burst

| Wavelength band | Light-to-mass ratio† $\bar{\Psi}_{\rm band}$ ($\mathrm{L_\odot\,M_\odot^{-1}}$) | Momentum injection rate $\bar{\Psi}_{\rm band}/c$ (cm s$^{-2}$) | Opacity $\kappa_{\rm band}$ (g cm$^{-2}$) |
|---|---|---|---|
| Hydrogen ionizing (ion) | $\bar{\Psi}_{\rm ion} = 500$ | $\bar{\Psi}_{\rm ion}/c = 3.27 \times 10^{-8}$ | $\kappa_{\rm ion} = 1.5 \times 10^{6}$ |
| Far-UV (FUV) | $\bar{\Psi}_{\rm FUV} = 367$ | $\bar{\Psi}_{\rm FUV}/c = 2.40 \times 10^{-8}$ | $\kappa_{\rm FUV} = 2000\,(Z/Z_\odot)$ |
| Near-UV (NUV) | $\bar{\Psi}_{\rm NUV} = 200$ | $\bar{\Psi}_{\rm NUV}/c = 1.31 \times 10^{-8}$ | $\kappa_{\rm NUV} = 1800\,(Z/Z_\odot)$ |
| Optical/near-IR (opt) | $\bar{\Psi}_{\rm opt} = 103$ | $\bar{\Psi}_{\rm opt}/c = 6.73 \times 10^{-9}$ | $\kappa_{\rm opt} = 180\,(Z/Z_\odot)$ |

† $\bar{\Psi}_{\rm band} \equiv (1/3.5\ \text{Myrs}) \int_0^{3.5\ \text{Myrs}} \mathrm{d}t\, \Psi_{\rm band}(t)$ is the time-averaged light-to-mass ratio.

since $\Sigma_{\rm disk} = \Sigma_{\rm disk,0}\bar{R}^{-1}$:[40]

$$
\begin{aligned}
e^{-\bar{\tau}_{\rm band}} &= \int_0^1 \mathrm{d}\bar{R}\, e^{-\kappa_{\rm band}\Sigma_{\rm disk,0}\bar{R}^{-1}} \\
&\simeq e^{-\kappa_{\rm band}\Sigma_{\rm disk,0} - 0.883\,(\kappa_{\rm band}\Sigma_{\rm disk,0})^{0.48}}\,,
\end{aligned}
\tag{125}
$$

where the second line is a fit to the integral, accurate to within 7% for $0 < \kappa_{\rm band}\Sigma_{\rm disk,0} < 5$. At larger values of $\kappa_{\rm band}\Sigma_{\rm disk,0}$ both the fit and the exact expression for $e^{-\bar{\tau}_{\rm band}}$ are very close to (and approach) zero, and inaccuracies in the fit is of no importance to us since the only quantity determining the radiation pressure on gas is $1 - e^{-\bar{\tau}_{\rm band}}$, which is practically unity for $\kappa_{\rm band}\Sigma_{\rm disk,0} > 5$.

Next we need to evaluate effective FIR optical depth due to dust, $\tau_{\rm IR}^{\rm eff}$. Since the gas distribution in a turbulent medium is not isotropic, the effective FIR optical depth $\tau_{\rm IR}^{\rm eff}$ will in general not be equal to the naïve estimate $\tau_{\rm IR} = \kappa_{\rm R}\Sigma_{\rm gas}$, where $\kappa_{\rm R}$ is the Rosseland mean opacity.[41] In particular, IR photons will be

[40]The mass-weighted average value is

$$
e^{-\bar{\tau}_{\rm band}} = \frac{2\pi R_{\rm disk}^2 \int_0^1 \mathrm{d}\bar{R}\,\bar{R}\Sigma_{\rm disk}\, e^{-\kappa_{\rm band}\Sigma_{\rm disk,0}\bar{R}^{-1}}}{2\pi R_{\rm disk}^2 \int_0^1 \mathrm{d}\bar{R}\,\bar{R}\Sigma_{\rm disk}}\,.
$$

Using $\Sigma_{\rm disk} = \Sigma_{\rm disk,0}\bar{R}^{-1}$, evaluating the bottom integral, and cancelling $2\pi R_{\rm disk}^2\Sigma_{\rm disk,0}$, yields the integral in Eq. (125).

[41]The Rosseland mean absorption coefficient $\kappa_{\rm R}$ is defined via (see e.g. p. 41 in Rybicki & Lightman, 1986)

$$
\kappa_{\rm R} \equiv \frac{\int_0^\infty \mathrm{d}\nu\,(\partial B_\nu/\partial T)}{\int_0^\infty \mathrm{d}\nu\,\kappa_\nu^{-1}(\partial B_\nu/\partial T)}\,.
$$

This mean absorption coefficient is the relevant to use (instead of e.g. the Planck mean, $\kappa_{\rm P}$) because it arises naturally if the IR optical depth is large and IR photons diffuse, as can be understood using the following argument by Krumholz *et al.*

able to escape easier through lines of sight with lower column densities (for a nice analytical discussion, see Appendix B in Hopkins *et al.*, 2011). The effective optical depth in detailed radiation hydrodynamic simulations is found to be $\tau_{\rm IR}^{\rm eff} = \eta_{\rm IR}\tau_{\rm IR}$ with $\eta_{\rm IR} \sim 0.5 - 0.9$ (Zhang & Davis, 2017), or $\eta_{\rm IR} \simeq 1$ (Costa *et al.*, 2018).[42] I will adopt $\eta_{\rm IR} = 0.9$ as a middle ground between the results of (Zhang & Davis, 2017) and (Costa *et al.*, 2018). This magnitude of $\kappa_{\rm R}$ is dependent on the both the gas metallicity $Z$ and the grain temperature $T_{\rm gr}$, but for 150 K $\lesssim T_{\rm gr} \lesssim$ 1500 K and Solar metallicity it has a roughly constant value of $\kappa_{\rm R} \simeq 5$ cm$^2$ g$^{-1}$ (Semenov *et al.*, 2003).[43] Thus, to avoid having to model the grain temperature we will follow Kimm *et al.* (2017) and simply adopt $\kappa_{\rm R} \simeq 5\,(Z/Z_\odot)$ cm$^2$ g$^{-1}$, where the simple linear scaling with gas metallicity is commonly assumed in models of dust and cooling in the early Universe (e.g. Safranek-Shrader *et al.*, 2014a; Kimm *et al.*, 2017). We have $\tau_{\rm IR}^{\rm eff} > 1$ for $\Sigma_{\rm gas} > 3.5 \times 10^4\,[(Z/Z_\odot)/0.03]^{-1}$ M$_\odot$ pc$^{-2}$, which exceeds the typical surface densities encountered in metal-poor disks (see Eq. 101). Thus, we do not expect multiple IR scattering from dust to be a dominant feedback process in the possible formation of metal-poor GCs and UFDs with $Z/Z_\odot \lesssim 0.03$, but we include it for physical realism and completeness.

So far we have only discussed radiation pressure from Pop II stars in metal-enriched (albeit typically metal-poor) gas. Pop III stars form in either metal-free or extremely metal-poor gas. The resulting absence or near-absence of dust in the star-forming gas means that there can be no significant direct radiation pressure in non-ionizing bands, and we therefore only consider the ionizing band in this case. As discussed in Section 3, while Pop III stars are thought to have been more massive than today, their IMF is still uncertain and under a great deal of study using state-of-the-art high-resolution simulations. For simplicity we will follow Ricotti *et al.* (2016) and consider a Dirac delta IMF (i.e. every Pop III star in a halo is born with the same mass $m_{\star,\rm PopIII}$),[44] but with two different possible stellar masses:

- $m_{\star,\rm PopIII} = 25$ M$_\odot$: We take this as our fiducial mass for Pop III stars, based on the results of the most recent sophisticated simulations addressing the Pop III IMF by Latif *et al.* (2022), the recent analytical calculation of the characteristic stellar mass as a function of metallicity by Sharda & Krumholz (2022), and the recent observational evidence for ultra metal-poor stars being enriched by Pop III hypernovae with progenitor masses $m_{\star,\rm PopIII} = 20 - 30$ M$_\odot$ (Placco *et al.*, 2021; Skúladóttir *et al.*, 2021; Yong *et al.*, 2021). Pop III stars of this mass are expected to explode as a hypernovae, releasing an energy $E_{\rm SN} = 13 \times 10^{51}$ erg (Wise *et al.*, 2012; Kimm *et al.*, 2017).

---

(2007). In this optically thick limit, the Rosseland approximation (e.g. pp. 39-42 in Rybicki & Lightman, 1986) gives a flux of $\boldsymbol{F}_\nu \propto \kappa_\nu^{-1}(\partial B_\nu/\partial T)\boldsymbol{\nabla}T$ for a temperature gradient $\boldsymbol{\nabla}T$. The acceleration in Eq. (119) due to absorbed radiation scales like $\propto \int {\rm d}\nu\,\kappa_\nu \boldsymbol{F}_\nu \equiv \bar{\kappa}\boldsymbol{F}$, and so depend on the flux-averaged absorption coefficient $\bar{\kappa} \equiv \int_0^\infty {\rm d}\nu\,\kappa_\nu \boldsymbol{F}_\nu / \int_0^\infty {\rm d}\nu\,\boldsymbol{F}_\nu$. Using the Rosseland approximation then yields

$$\bar{\kappa} = \frac{\int_0^\infty {\rm d}\nu\,\kappa_\nu \times \kappa_\nu^{-1}(\partial B_\nu/\partial T)\boldsymbol{\nabla}T}{\int_0^\infty {\rm d}\nu\,\kappa_\nu^{-1}(\partial B_\nu/\partial T)\boldsymbol{\nabla}T} = \frac{\int_0^\infty {\rm d}\nu\,(\partial B_\nu/\partial T)}{\int_0^\infty {\rm d}\nu\,\kappa_\nu^{-1}(\partial B_\nu/\partial T)} = \kappa_{\rm R}\,.$$

This explains why authors adopt this mean mass absorption coefficient without explanation (e.g. Murray *et al.*, 2010; Rahner *et al.*, 2017; Kruijssen *et al.*, 2019a).

[42] Using radiation-hydrodynamic simulations, Costa *et al.* (2018) found that $\eta_{\rm IR} \simeq 1$ when outflow velocities are negligible. When outflow velocities $v_{\rm outflow}$ are large, multiple scattering becomes inefficient (i.e. $\eta_{\rm IR} \to 0$) since the outflow expand faster than the scattered photons, effectively leaving them behind before they can impart radiation pressure corresponding to $\eta_{\rm IR} = 1$. Costa *et al.* (2018) find that this occurs when $\tau_{\rm IR} \gtrsim 10^4\,(v_{\rm outflow}/10$ km s$^{-1})^{-1}$, which far exceeds the typical IR optical depths encountered in gaseous disks at Cosmic Dawn for all relevant outflow velocities. As discussed in Costa *et al.* (2018), this effect could instead be important in AGN-driven outflows.

[43] For $T_{\rm gr} \lesssim$ 150 K Semenov *et al.* (2003) found that the Rosseland mean opacity for the porous 5-layered spheres model of grains can be fitted by $\kappa_{\rm R} \propto T_{\rm gr}^{3/2}$ (see e.g. footnote 34 in Hopkins *et al.*, 2018a), which is a slightly different scaling compared to the relation $\kappa_{\rm R} \propto T_{\rm gr}^2$ one obtains for non-porous grains, stemming from the absorption efficiency $Q_{\rm abs} \propto \lambda^{-2}$ at long wavelengths (see e.g. Draine, 2011). At high grain temperatures ($T_{\rm gr} \gtrsim$ 1500 K) grains sublimate, and the results from Semenov *et al.* (2003) in this temperature regime can be fitted by $\kappa_{\rm R} \propto T_{\rm gr}^{-24}$, qualitatively similar to the dependence adopted by Dopcke *et al.* (2011), but steeper. In the intermediate temperature regime the opacity is well described by $\kappa_{\rm R} \simeq 5\,(Z/Z_\odot)$ cm$^2$ g$^{-1}$ (Kimm *et al.*, 2017; Hopkins *et al.*, 2018a).

[44] Choosing a Dirac delta IMF also allow us to circumvent the problem of the possibile poor sampling of the Pop III IMF in low-mass halos, and hence faulty calculations of stellar feedback that are based on the IMF being well-sampled.

- $m_{\star,\mathrm{PopIII}} = 140\ \mathrm{M}_\odot$: We also consider the possibility of more massive Pop III stars, since these have also been found to form in earlier simulations of the first stars (Hirano *et al.*, 2014, 2015; Hosokawa *et al.*, 2016).[45] Compared to the fiducial case of $m_{\star,\mathrm{PopIII}} = 25\ \mathrm{M}_\odot$, Pop III stars of this mass are even more luminous and more efficient at producing ionizing photons. At the end of their life they explode as pair-instability SNe with energy $E_{\mathrm{SN}} = 6.3 \times 10^{51}$ erg (Wise *et al.*, 2012; Kimm *et al.*, 2017).

For a given Pop III star mass $m_{\star,\mathrm{PopIII}}$ we estimate the lifetime-averaged momentum injection rate $\bar{\Psi}_{\mathrm{ion}}/c$ using the lifetime-averaged ionizing photon emission rates from Schaerer (2002) and the corresponding (ZAMS) average LyC photon energy from Mas-Ribas *et al.* (2016).[46] To estimate the band-averaged opacity in the ionizing band we calculate the flux-weighted average photoionization cross-sections assuming a black-body spectrum with an effective temperature taken from Mas-Ribas *et al.* (2016).[47] The resulting values of $\bar{\Psi}_{\mathrm{ion}}/c$ and $\kappa_{\mathrm{ion}}$ are shown in Table 3. By comparison with Table 2 it is evident that Pop III stars exert significantly larger direct radiation pressure on gas than Pop II stars — a factor 2.7 and 18 greater for $m_{\star,\mathrm{PopIII}} = 25\ \mathrm{M}_\odot$ and $140\ \mathrm{M}_\odot$ respectively (assuming $\bar{\tau}_{\mathrm{band}} \gg 1$ in all bands). Thus, we expect that radiative feedback from Pop III stars will be much more efficient in preventing the formation of massive Pop III star clusters in low-mass halos.

[45] As explained by Latif *et al.* (2022), the difference between their simulations and those of Hirano *et al.* (2015) and Hosokawa *et al.* (2016) is that Latif *et al.* (2022) for the first time considered radiative feedback from multiple stars forming in the disk, rather than just a single star.

[46] The momentum injection rate from a Pop III star of mass $m_{\star,\mathrm{PopIII}}$ is then given by $\bar{\Psi}_{\mathrm{ion}}/c = \dot{\bar{N}}_{\mathrm{ion}} \bar{E}_{\mathrm{LyC}}/m_{\star,\mathrm{PopIII}}c$, where $\dot{\bar{N}}_{\mathrm{ion}}$ is the lifetime-averaged photon emission rate, and $\bar{E}_{\mathrm{LyC}}$ the average LyC photon energy. The fact that we use the ZAMS value for $\bar{E}_{\mathrm{LyC}}$ is an approximation, but it is good enough for our purposes since it mainly depends on the effective temperature of the star, which in turn is approximately constant over the main sequence lifetime (close to $10^5$ K for massive Pop III stars). Mas-Ribas *et al.* (2016) do not provide a value of $\bar{E}_{\mathrm{LyC}}$ for $m_{\star,\mathrm{PopIII}} = 140\ \mathrm{M}_\odot$, so a linear interpolation has been used between $m_{\star,\mathrm{PopIII}} = 120\ \mathrm{M}_\odot$ and $m_{\star,\mathrm{PopIII}} = 200\ \mathrm{M}_\odot$ (two masses for which values of $\bar{E}_{\mathrm{LyC}}$ are provided). This approximation is fine since $\bar{E}_{\mathrm{LyC}}$ only varies weakly over this mass range. The result is $\bar{E}_{\mathrm{LyC}} = 28.4$ eV for $m_{\star,\mathrm{PopIII}} = 25\ \mathrm{M}_\odot$ and $\bar{E}_{\mathrm{LyC}} = 34.9$ eV for $m_{\star,\mathrm{PopIII}} = 140\ \mathrm{M}_\odot$.

[47] The flux-weighted average cross-section $\bar{\sigma}_{\mathrm{ion}} = \int_{\nu_{\mathrm{LyC}}}^{\infty} \mathrm{d}\nu\, \boldsymbol{F}_\nu \sigma_{\mathrm{ion}} / \int_{\nu_{\mathrm{LyC}}}^{\infty} \mathrm{d}\nu\, \boldsymbol{F}_\nu$ is used since the acceleration induced in the gas from absorbed ionizing photons is $\propto \int_0^\infty \mathrm{d}\nu\, \boldsymbol{F}_\nu \sigma_{\mathrm{ion}}(\nu)$ (from Eq. 119). Following Pawlik *et al.* (2013) we approximate the spectrum from Pop III stars as a black-body spectrum, using the ZAMS values for the effective temperature provided by Mas-Ribas *et al.* (2016). This yields $T_{\mathrm{eff}} = 7.08 \times 10^4$ K for $m_{\star,\mathrm{PopIII}} = 25\ \mathrm{M}_\odot$ and $T_{\mathrm{eff}} = 9.67 \times 10^4$ K for $m_{\star,\mathrm{PopIII}} = 140\ \mathrm{M}_\odot$, with the latter value found by linear interpolation between $m_{\star,\mathrm{PopIII}} = 120\ \mathrm{M}_\odot$ ($T_{\mathrm{eff}} = 9.57 \times 10^4$ K) and $m_{\star,\mathrm{PopIII}} = 200\ \mathrm{M}_\odot$ ($T_{\mathrm{eff}} = 9.98 \times 10^4$ K). Using Eq. (13.1) from Draine (2011) for the photoionization cross-section $\sigma_{\mathrm{ion}}(\nu)$, the resulting band-averaged (i.e. flux-weighted) photoionization cross-sections are $\bar{\sigma}_{\mathrm{ion}} = 1.73 \times 10^{-18}\ \mathrm{cm}^2$ for $m_{\star,\mathrm{PopIII}} = 25\ \mathrm{M}_\odot$ and $\bar{\sigma}_{\mathrm{ion}} = 1.15 \times 10^{-18}\ \mathrm{cm}^2$ for $m_{\star,\mathrm{PopIII}} = 140\ \mathrm{M}_\odot$.

TABLE 3. POP III RADIATIVE MOMENTUM FEEDBACK

| Stellar mass $m_{\star,\mathrm{PopIII}}$ ($\mathrm{M}_\odot$) | Lifetime[†] $t_{\star,\mathrm{PopIII}}$ (Myrs) | Momentum injection rate $\bar{\Psi}_{\mathrm{ion}}/c$ (cm s$^{-2}$) | Opacity $\kappa_{\mathrm{ion}}$ (g cm$^{-2}$) |
|---|---|---|---|
| 25 | 6.07 | $2.10 \times 10^{-7}$ | $7.8 \times 10^5$ |
| 140 | 2.44 | $1.35 \times 10^{-6}$ | $5.2 \times 10^5$ |

[†] Calculated from the fits provided by Schaerer (2002).

#### 5.1.2. Radiation pressure from scattered Ly$\alpha$ photons

*"[The] impact of Lya radiation pressure cannot be anymore neglected as simulations and models struggle for a more complete physical description of the ISM and its dynamics. This is even more urgent in early galaxies models, where the lower dust content strongly enhances the role of Lya radiation pressure."* — TOMASELLI & FERRARA (2021)

*"We find that the Lya radiation feedback significantly suppresses the star formation when the metallicity of the cloud is low. (...) Thus we suggest that the Lya radiation feedback can be a key role in the formation of GCs."* — ABE & YAJIMA (2018)

*"Our experiments demonstrate that Lya pressure may be a critical mechanism by which potential globular cluster candidates with low metallicities are disrupted. (...) In this regard, future simulations aiming to understand the detailed formation histories of metal-poor GCs may need to include Lya feedback along with other radiation feedback processes."* — KIMM *et al.* (2018)

Next we consider the momentum feedback from Ly$\alpha$ photons. Only in the last few years have researchers noted the importance of this feedback process in low-metallicity environments, with the bulk of the work concluding that it cannot be neglected, and may even be more important than direct radiation pressure from stars (Kimm *et al.*, 2018; Tomaselli & Ferrara, 2021). Unfortunately, implementing this feedback process in numerical simulations is extremely computationally demanding, and at the time of this writing only Kimm *et al.* (2018) have done so in high-resolution simulations of an individual metal-poor dwarf galaxy. This is further motivation for the development of a physically realistic model of starbursts like ANAXAGORAS. Feedback from Ly$\alpha$ photons work as follows. Stars emit ionizing photons that will be photoionize at least some of the surrounding hydrogen gas. As hydrogen recombines, a fraction $f_{\mathrm{Ly}\alpha} \simeq 2/3$ of the times this leads to the emission of a Ly$\alpha$ photon. The optical depth $\tau_0$ at the center of the Ly$\alpha$ line is usually huge, and so the trapping time $t_{\mathrm{trap}}$ of Ly$\alpha$ photons will exceed the time $t_{\mathrm{light}} \equiv R/c$ it would otherwise take a photon to escape the cloud (i.e. $t_{\mathrm{trap}}/t_{\mathrm{light}} > 1$). Because of this, the energy density $u_{\mathrm{Ly}\alpha}$ of Ly$\alpha$ photons in the gas cloud is increased by a factor $\sim t_{\mathrm{trap}}/t_{\mathrm{light}}$, and the resulting acceleration due to radiation pressure ($\propto \boldsymbol{\nabla} P_{\mathrm{Ly}\alpha} \propto \boldsymbol{\nabla} u_{\mathrm{Ly}\alpha}$) is increased by the same factor $\sim t_{\mathrm{trap}}/t_{\mathrm{light}}$ (e.g. Haehnelt, 1995; Abe & Yajima, 2018; Kimm *et al.*, 2018; Tomaselli & Ferrara, 2021). Thus, we expect that the force multiplier $M_{\mathrm{F}}$ in Eq. (124) is $M_{\mathrm{F}} \sim t_{\mathrm{trap}}/t_{\mathrm{light}}$. This result is shown using a slightly more quantitative, yet approximate, manner in Appendix 7. The trapping time $t_{\mathrm{trap}}$ for Ly$\alpha$ photons is determined by the time it takes for them to diffuse in frequency space (due to many absorptions and re-emissions) far enough into the wings of the Voigt profile that the optical depth becomes low enough for the Ly$\alpha$ photons to escape. This is given by $t_{\rm trap}/t_{\rm light} \sim (a_{\rm V}\tau_0)^{1/3}$, where $a_{\rm V} = 4.7\times10^{-4}\,T_4^{-1/2}$ is the Voigt parameter for the Ly$\alpha$ line (e.g. Spaans & Silk, 2006; Dijkstra, 2017). Thus, we expect that in the absence of dust that can destroy Ly$\alpha$ photons (see Appendix 7 for a more detailed discussion):

$$M_{\rm F}^{\rm nodust} \sim (a_{\rm V}\tau_0)^{1/3}\,. \tag{126}$$

This expectation has been confirmed by the detailed analytical radiative transfer solutions found by Lao & Smith (2020) and Tomaselli & Ferrara (2021) employing the Eddington approximation. For a central point source of Ly$\alpha$ radiation in a uniform spherical gas cloud they find $M_{\rm F}^{\rm nodust} = 3.51\,(a_{\rm V}\tau_0)^{1/3}$, or

$$M_{\rm F}^{\rm nodust}(\tau_0) = 127\,T_4^{-1/6}\left(\frac{\tau_0}{10^8}\right)^{1/3}\,. \tag{127}$$

Dust in the gas can absorb Ly$\alpha$ and prevent further scattering. Thus, for $\tau_0 \to \infty$ there is an upper limit to $M_{\rm F}$ set by the dust content of the gas (Kimm *et al.*, 2018; Tomaselli & Ferrara, 2021). Tomaselli & Ferrara (2021) find that the upper limit is reached approximately when $\tau_0 > \tau_0^* = 3.58\times10^6\,T_4^{-1/4}(Z/Z_\odot)^{-3/4}$, where I have assumed that the dust-to-gas mass ratio is proportional to the gas metallicity (as done elsewhere in this thesis). The resulting value of $M_{\rm F}$ is then given by

$$\begin{aligned} M_{\rm F}(\tau_0) &= \min[M_{\rm F}^{\rm nodust}(\tau_0),\, M_{\rm F}^{\rm nodust}(\tau_0^*)] \\ &= \min\left[127\,T_4^{-1/6}\left(\frac{\tau_0}{10^8}\right)^{1/3},\, 41.9\,T_4^{-1/4}\left(\frac{Z}{Z_\odot}\right)^{-1/4}\right]\,. \end{aligned} \tag{128}$$

To implement this in Anaxagoras we need to evaluate the optical depth at line center, $\tau_0$, which depends on the neutral hydrogen column density $N_{\rm HI}$ (e.g. Kimm *et al.*, 2018):

$$\begin{aligned} \tau_0 &= 5.88\times10^{-14}\,T_4^{-1/2}N_{\rm HI} \\ &\simeq 5.53\times10^6\,T_4^{-1/2}\left(\frac{\Sigma_{\rm disk}}{1~{\rm M_\odot~pc^{-2}}}\right)\,, \end{aligned} \tag{129}$$

where on the second line I have expressed the column density in terms of the disk surface density.[48] This clearly has a radial dependence which again would render the model analytically intractable. Because of this we calculate a mass-weighted average value of $M_{\rm F}^{\rm nodust}(\tau_0)$. With $\Sigma_{\rm disk} = \Sigma_{\rm disk,0}\,\bar{R}^{-1}$ we find[49]

$$M_{\rm F}(\Sigma_{\rm disk,0}) = \min\left[72.6\,T_4^{-1/3}\left(\frac{\Sigma_{\rm disk,0}}{1~{\rm M_\odot~pc^{-2}}}\right)^{1/3},\, 41.9\,T_4^{-1/4}\left(\frac{Z}{Z_\odot}\right)^{-1/4}\right]\,. \tag{130}$$

This is the magnitude of the force multiplier that we will adopt in Anaxagoras. Next we need to evaluate the Ly$\alpha$ light-to-mass ratio $\Psi_{\rm Ly\alpha}$ to get the resulting momentum injection rate (see Eq. 124). Assuming

[48]The line center optical depth is $\tau_0 = \sigma_0 N_{\rm HI}$ with $\sigma_0 = 5.88\times10^{-14}\,T_4^{-1/2}$ cm$^2$. In the early stages of star formation prior to the expansion of any H II region that can destroy the disk, we simply have $N_{\rm HI} \simeq X\Sigma_{\rm disk}/m_{\rm H}$, giving the quoted result.

[49]The mass-weighted average of $M_{\rm F}^{\rm nodust}(\tau_0)$ is

$$M_{\rm F}^{\rm nodust}(\Sigma_{\rm disk,0}) = 48.4\,T_4^{-1/3}\left(\frac{\Sigma_{\rm disk,0}}{1~{\rm M_\odot~pc^{-2}}}\right)^{1/3}\frac{\int_0^1 {\rm d}\bar{R}\,\bar{R}\Sigma_{\rm disk}\,\bar{R}^{-1/3}}{\int_0^1 {\rm d}\bar{R}\,\bar{R}\Sigma_{\rm disk}}\,.$$

Since $\bar{R}\Sigma_{\rm disk} = {\rm const.}$, this yields $\int_0^1 {\rm d}\bar{R}\,\bar{R}^{-1/3}/\int_0^1 {\rm d}\bar{R} = 3/2$, and the quoted result.

Table 4. Ly$\alpha$ radiative feedback: Values for $(\Psi_{\mathrm{Ly}\alpha}/c)/(1-e^{-\bar{\tau}_{\mathrm{ION}}})$

| Pop II $(\Psi_{\mathrm{Ly}\alpha}/c)/(1-e^{-\bar{\tau}_{\mathrm{ion}}})$ (cm s$^{-2}$) | Pop III ($m_{\star,\mathrm{PopIII}} = 25$ M$_\odot$) $(\Psi_{\mathrm{Ly}\alpha}/c)/(1-e^{-\bar{\tau}_{\mathrm{ion}}})$ (cm s$^{-2}$) | Pop III ($m_{\star,\mathrm{PopIII}} = 140$ M$_\odot$) $(\Psi_{\mathrm{Ly}\alpha}/c)/(1-e^{-\bar{\tau}_{\mathrm{ion}}})$ (cm s$^{-2}$) |
|---|---|---|
| $9.13 \times 10^{-9}$ | $5.02 \times 10^{-8}$ | $2.62 \times 10^{-7}$ |

that a fraction $f_{\mathrm{Ly}\alpha} \simeq 2/3$ of recombinations lead to Ly$\alpha$ emission, this is given by (e.g. Abe & Yajima, 2018; Tomaselli & Ferrara, 2021)

$$\Psi_{\mathrm{Ly}\alpha} = \frac{2}{3}\dot{Q}_{\mathrm{ion}}(1-e^{-\bar{\tau}_{\mathrm{ion}}})E_{\mathrm{Ly}\alpha}\,, \tag{131}$$

where $\dot{Q}_{\mathrm{ion}}$ (s$^{-1}$ M$_\odot^{-1}$) is the IMF-averaged ionizing photon emission rate per stellar mass formed, $1-e^{-\bar{\tau}_{\mathrm{ion}}}$ is the fraction of ionizing photons that are absorbed (where $\bar{\tau}_{\mathrm{ion}}$ is the band-averaged optical depth in the ionizing band), and $E_{\mathrm{Ly}\alpha} = 10.2$ eV is the energy of a Ly$\alpha$ photon. For Pop II stars we adopt $\dot{Q}_{\mathrm{ion}} = 5 \times 10^{46}\,\mathrm{s}^{-1}$ M$_\odot^{-1}$, which is an appropriate average value for the first $\sim 3.5$ Myrs following an instantaneous stellar burst with a Kroupa IMF and metallicity $Z = 0.05\,Z_\odot$ (e.g. Ma *et al.*, 2016; Kimm *et al.*, 2017; Rosdahl *et al.*, 2018). While many GCs and practically all UFDs have lower metallicities than this, the effect of metallicity on $\dot{Q}_{\mathrm{ion}}$ is small (see e.g. Stanway & Eldridge, 2019) and can be neglected for our purposes.[50] In the case of Pop III stars with the Dirac delta function IMF that we have adopted, we get $\dot{Q}_{\mathrm{ion}} = \dot{\bar{N}}_{\mathrm{ion}}/m_{\star,\mathrm{PopIII}}$ (see Table 3 for values of $\dot{\bar{N}}_{\mathrm{ion}}$). The resulting values of $\Psi_{\mathrm{Ly}\alpha}/c$ can be found in Table 4.

### 5.1.3. Momentum feedback from stellar winds

Finally we need to consider the momentum injection rate $\left\langle \dot{P}/m_\star \right\rangle_{\mathrm{wind}}$ due to stellar winds (or, more generally, stellar mass loss). For Pop II stars with a Kroupa IMF, Hopkins *et al.* (2018a) give the following fits to results from `STARBURST99` for the IMF-averaged wind velocity $v_{\mathrm{w}}$ and mass loss rate $\dot{m}_{\mathrm{w}}$ following an instantaneous burst of mass $m_\star$:

$$\begin{aligned} \dot{m}_{\mathrm{w}}(t) &= f_{\mathrm{w}}(t)\, m_\star \;\mathrm{Gyr}^{-1} \\ v_{\mathrm{w}}(t) &= \left\{2\psi(t) \times 10^{12}\ \mathrm{erg\ g}^{-1}\right\}^{1/2}, \end{aligned} \tag{132}$$

[50]For example, in Figure 3 of Stanway & Eldridge (2019) we see that for an upper IMF slope of 2.35, lowering the metallicity from $Z = 0.05\,Z_\odot$ ($Z = 10^{-3}$) to $Z = 5 \times 10^{-4}\,Z_\odot$ ($Z = 10^{-5}$) would only raise $\dot{Q}_{\mathrm{ion}}$ by a factor $\sim 0.1$ dex, i.e. $\sim 26\%$.

where

$$
\begin{aligned}
f_\mathrm{w}(t) \;&=\; 4.763\left(0.01+\frac{Z}{Z_\odot}\right)\times\begin{cases}1 & \text{for } t_\mathrm{Myrs}<1\\ t_\mathrm{Myrs}^{1.45+0.8\,\ln(Z/Z_\odot)} & \text{for } 1<t_\mathrm{Myrs}<3.5\end{cases}\\
\psi(t) \;&=\; \frac{5.94\times10^4}{1+(t_\mathrm{Myrs}/2.5)^{1.4}+(t_\mathrm{Myrs}/10)^{5.0}}+4.83 \qquad \text{for } t_\mathrm{Myrs}<100\,.
\end{aligned}
\tag{133}
$$

The momentum injection rate at time $t$ following an instantaneous burst is then

$$
\left\langle\frac{\dot{P}}{m_\star}\right\rangle_\mathrm{wind}=\frac{\dot{m}_\mathrm{w}(t)v_\mathrm{w}(t)}{m_\star}\,,
$$

which is independent of $m_\star$ as expected. The metallicity dependence of the mass loss rate $\dot{m}_\mathrm{w}$ stems from the fact that winds in massive stars are primarily driven by metal lines. As in the case of radiative momentum feedback, we can time-average over $0<t_\mathrm{Myrs}<3.5$ to keep Anaxagoras analytically tractable. We note that almost the exact same procedure was used by Agertz *et al.* (2013), who also time-averaged results from `STARBURST99`, but over 6.5 Myrs. The only differences here is that we are averaging over 3.5 Myrs and adopting a more accurate fit to the metallicity dependence. Using Eqs. (132) and (133) we find that the time-averaged momentum injection rate due to stellar winds can be well-fitted by:

$$
\left\langle\frac{\dot{P}}{m_\star}\right\rangle_\mathrm{wind}=1.6\times10^{-8}\,e^{8\times10^{-4}\,\log_{10}^{4.3}[(Z/Z_\odot)/10^{-6}]}\left(0.01+\frac{Z}{Z_\odot}\right)\ \mathrm{cm\ s^{-2}}\,. \tag{134}
$$

This fit is accurate to within $\simeq 7.4\%$ for $10^{-6}<Z/Z_\odot<1$ (although the calculations from `STARBURST99`, on which the these fits are based, only extend down to $Z/Z_\odot=0.02$). We will adopt Eq. (134) for the wind momentum injection rate from Pop II stars in Anaxagoras. For Solar metallicity ($Z=Z_\odot$) we find a wind momentum injection rate of $9.5\times10^{-8}$ cm s$^{-2}$, which slightly exceeds the radiative momentum injection rate from *all* bands in Table 2. However, in a more metal-poor environment with $Z/Z_\odot=0.01$, fairly typical of a metal-poor GC, we only get $4.4\times10^{-10}$ cm s$^{-2}$ — significantly smaller than the contribution from *any* of the bands in Table 2. Thus, at low metallicities we do not expect stellar feedback from winds to be important compared to radiation pressure. In the case of Pop III stars we can therefore safely neglect mass loss from winds altogether, since line or continuum-driven winds are expected to yield negligible mass-loss rates in this case (e.g. Krtička & Kubát, 2006).

### 5.1.4. Upper limit on star formation efficiency from momentum feedback

We have now discussed momentum feedback from radiation pressure and stellar winds. Momentum can also be injected into the gas from SNe, but this only kicks in after after several Myrs (3.4 Myrs in the case of Pop II stars), typically well after observed Giant Molecular Clouds are disrupted by feedback (e.g. Kruijssen *et al.*, 2019a), and will therefore be considered separately in Section . For future convenience we can write the total momentum injection rate from radiation pressure and stellar winds in Eq. (124) as follows:

$$
\left\langle\frac{\dot{P}}{m_\star}\right\rangle=(1+\eta_\mathrm{IR}\kappa_\mathrm{R}\Sigma_\mathrm{gas})\left\langle\frac{\dot{P}_\mathrm{rad}}{m_\star}\right\rangle_\text{no ion}+\left\langle\frac{\dot{P}}{m_\star}\right\rangle_\mathrm{rest}\,, \tag{135}
$$

where I have used $\tau_{\rm IR}^{\rm eff} = \eta_{\rm IR}\kappa_{\rm R}\Sigma_{\rm gas}$, and where

$$\begin{aligned} \left\langle \frac{\dot{P}_{\rm rad}}{m_\star} \right\rangle_{\rm no\ ion} &\equiv \sum_{\rm band\,=\,\{FUV,NUV,opt\}} \left( \frac{\bar{\Psi}_{\rm band}}{c} \right) (1 - e^{-\bar{\tau}_{\rm band}}) \\ \left\langle \frac{\dot{P}}{m_\star} \right\rangle_{\rm rest} &\equiv \left( \frac{\bar{\Psi}_{\rm ion}}{c} \right) (1 - e^{-\bar{\tau}_{\rm ion}}) + M_{\rm F}(\Sigma_{\rm disk,0}) \left( \frac{\Psi_{\rm Ly\alpha}}{c} \right) + \left\langle \frac{\dot{P}}{m_\star} \right\rangle_{\rm wind} . \end{aligned} \tag{136}$$

The direct relevance of $\left\langle \dot{P}/m_\star \right\rangle$ for the efficiency of star formation in giant molecular clouds is by now well understood (e.g. Fall *et al.*, 2010; Raskutti *et al.*, 2016; Grudić *et al.*, 2018; Li *et al.*, 2019; Mok *et al.*, 2021). To see how we can follow the simple force balance derivation of Grudić *et al.* (2018). Consider a region of radius $R$ in a gas disk with total surface mass density $\Sigma_{\rm disk}$, which can be divided into gas ($\Sigma_{\rm gas}$) and stars ($\Sigma_\star$) such that $\Sigma_{\rm disk} = \Sigma_{\rm gas} + \Sigma_\star$. The force due to momentum injection feedback from the stellar mass $\Sigma_\star \pi R^2$ in this region will be given by

$$\dot{P} = \left\{ (1 + \eta_{\rm IR}\kappa_{\rm R}\Sigma_{\rm gas}) \left\langle \frac{\dot{P}_{\rm rad}}{m_\star} \right\rangle_{\rm no\ ion} + \left\langle \frac{\dot{P}}{m_\star} \right\rangle_{\rm rest} \right\} \Sigma_\star \pi R^2 \, . \tag{137}$$

We expect star formation to be possible up until $\dot{P}$ balances the gravitational force holding the gas mass $\Sigma_{\rm gas}\pi R^2$ together, $\Sigma_{\rm gas}\pi R^2 g$, where $g = 2\pi G \Sigma_{\rm disk}$ is the gravitational acceleration in the disk. Thus, star formation can *at most* continue until we have the following equality:

$$\left\{ (1 + \eta_{\rm IR}\kappa_{\rm R}\Sigma_{\rm gas}) \left\langle \frac{\dot{P}_{\rm rad}}{m_\star} \right\rangle_{\rm no\ ion} + \left\langle \frac{\dot{P}}{m_\star} \right\rangle_{\rm rest} \right\} \Sigma_\star = 2\pi G \Sigma_{\rm disk}\Sigma_{\rm gas} \, . \tag{138}$$

Since $\Sigma_{\rm gas} = \Sigma_{\rm disk} - \Sigma_\star$, after a little algebra we get the following quadratic equation for the maximum integrated star formation efficiency $\epsilon_{\star,\rm max} \equiv \Sigma_\star/\Sigma_{\rm disk}$:

$$\begin{aligned} 0 &= \langle \tau_{\rm IR} \rangle \, \epsilon_{\star,\rm max}^2 - \left( 1 + \frac{\Sigma_{\rm crit}}{\Sigma_{\rm disk}} + \langle \tau_{\rm IR} \rangle \right) \epsilon_{\star,\rm max} + 1 \\ \Sigma_{\rm crit} &\equiv \frac{1}{2\pi G} \left\{ \left\langle \frac{\dot{P}_{\rm rad}}{m_\star} \right\rangle_{\rm no\ ion} + \left\langle \frac{\dot{P}}{m_\star} \right\rangle_{\rm rest} \right\} \\ \langle \tau_{\rm IR} \rangle &\equiv \frac{\eta_{\rm IR}\kappa_{\rm R}}{2\pi G} \left\langle \frac{\dot{P}_{\rm rad}}{m_\star} \right\rangle_{\rm no\ ion} . \end{aligned} \tag{139}$$

This is a quadratic equation that can be solved exactly for $\epsilon_{\star,\rm max}$, but the solution is complicated and would render Anaxagoras analytically intractable. Instead we can use an approximation, which can be quickly deduced from Eq. (139) as follows. Consider first the limit of no IR feedback from dust, i.e. $\langle \tau_{\rm IR} \rangle = 0$. In this case $\langle \tau_{\rm IR} \rangle \, \epsilon_{\star,\rm max}^2 = 0$ and we find

$$\epsilon_{\star,\rm max} = \frac{1}{1 + \Sigma_{\rm crit}/\Sigma_{\rm disk}} \qquad \text{(Negligible IR feedback)} \, , \tag{140}$$

in agreement with the results of Grudić *et al.* (2018) who ignored IR feedback. In the opposite limit where IR feedback is extremely important, $\langle \tau_{\rm IR} \rangle \gg 1 + \Sigma_{\rm crit}/\Sigma_{\rm disk} > 1$ we expect that $\epsilon_{\star,\rm max} \ll 1$ and so we can neglect $\langle \tau_{\rm IR} \rangle \, \epsilon_{\star,\rm max}^2$ again to get

$$\epsilon_{\star,\rm max} \simeq \frac{1}{1 + \Sigma_{\rm crit}/\Sigma_{\rm disk} + \langle \tau_{\rm IR} \rangle} \qquad \text{(Dominant IR feedback)} \, , \tag{141}$$

which indeed is smaller than unity under our assumption $\langle\tau_{\rm IR}\rangle \gg 1+\Sigma_{\rm crit}/\Sigma_{\rm disk} > 1$. The two asymptotic limits above suggest that we could simply adopt Eq. (141) as our expression for $\epsilon_{\star,\rm max}$ since it also correctly reproduce the limit of no IR feedback (Eq. 140) for $\langle\tau_{\rm IR}\rangle \to 0$. To summarize, we take the maximum star formation efficiency $\epsilon_{\star,\rm max}$ to be given by:

**Maximum star formation efficiency in a disk of surface density $\Sigma_{\rm disk}$**

$$
\begin{aligned}
\epsilon_{\star,\rm max} &= \frac{1}{1+\Sigma_{\rm crit}/\Sigma_{\rm disk}+\langle\tau_{\rm IR}\rangle} \\
\Sigma_{\rm crit} &\equiv \frac{1}{2\pi G}\left\{\left\langle\frac{\dot{P}_{\rm rad}}{m_\star}\right\rangle_{\rm no\ ion}+\left\langle\frac{\dot{P}}{m_\star}\right\rangle_{\rm rest}\right\} \\
\langle\tau_{\rm IR}\rangle &\equiv \frac{\eta_{\rm IR}\kappa_{\rm R}}{2\pi G}\left\langle\frac{\dot{P}_{\rm rad}}{m_\star}\right\rangle_{\rm no\ ion},
\end{aligned}
\tag{142}
$$

where $\left\langle\dot{P}_{\rm rad}/m_\star\right\rangle_{\rm no\ ion}$ and $\left\langle\dot{P}/m_\star\right\rangle_{\rm rest}$ are given in Eq. (136).

## 5.2. Star formation rate

Having discussed momentum feedback from stars, we are now in a position to derive the total stellar mass formed by some time $t$. As long as the gaseous disk is Toomre unstable, the disk will be gravitationally unstable, and we expect that gas will convert into stars on a free-fall timescale $t_{\rm ff}$ with some efficiency $\epsilon_{\rm ff} \lesssim 1$ (the value of which will be discussed below). This theoretical expectation is borne out by observations (e.g. Krumholz *et al.*, 2012), and is implemented in state of the art high-resolution simulations of galaxy formation (e.g. Kimm *et al.*, 2016; Ricotti *et al.*, 2016; Kimm *et al.*, 2017; Hopkins *et al.*, 2018a).[51] This includes the high-resolution simulations of GC and UFD formation in low-mass halos at Cosmic Dawn by Kimm *et al.* (2016) and Ricotti *et al.* (2016) that are highly relevant for the work in this thesis. Schematically, we can write our star formation law as follows:

$$
\dot{\Sigma}_\star = \epsilon_{\rm ff}\times\frac{\text{Gas surface density reservoir available for star formation}}{t_{\rm ff}}, \tag{143}
$$

where $\Sigma_\star$ is the stellar surface density of the disk at some fixed radius, so that $\dot{\Sigma}_\star$ is the star formation surface density. As discussed earlier in Section 5.1.4, only a maximum fraction $\epsilon_{\star,\rm max}$ of the total (initial) disk surface density $\Sigma_{\rm disk}$ can be converted into stars due to momentum feedback. Thus, the gas surface density reservoir available for star formation is $\epsilon_{\star,\rm max}\Sigma_{\rm disk}-\Sigma_\star$, and so we get the following star formation rate:

$$
\dot{\Sigma}_\star = \epsilon_{\rm ff}\frac{(\epsilon_{\star,\rm max}\Sigma_{\rm disk}-\Sigma_\star)}{t_{\rm ff}}. \tag{144}
$$

Assuming constant values for $\epsilon_{\rm ff}$ and $t_{\rm ff}$ at fixed disk radius, the above differential equation can be separated and integrated to yield

$$
\Sigma_\star(R,t) = \epsilon_{\star,\rm max}\Sigma_{\rm disk}\left\{1-e^{-\epsilon_{\rm ff}(t-t_{\rm min})/t_{\rm ff}}\right\}, \tag{145}
$$

[51] In fact, this recipe for star formation was adopted in the very early galaxy formation simulations by Katz (1992) and Katz *et al.* (1996).

where $t_{\rm min}$ is the initial time at which gas exist at the radius $R$ of interest, or in other words, the time when $R_{\rm disk} = R$. The key takeaway here is that at early times ($t - t_{\rm min} \ll t_{\rm ff}/\epsilon_{\rm ff}$) the stellar surface density grows linearly with time like $\Sigma_\star \simeq \epsilon_{\rm ff}\epsilon_{\star,\rm max}\Sigma_{\rm disk}(t - t_{\rm min})/t_{\rm ff}$, and at sufficiently late times ($t - t_{\rm min} \gg t_{\rm ff}/\epsilon_{\rm ff}$) saturates to $\Sigma_\star \simeq \epsilon_{\star,\rm max}\Sigma_{\rm disk}$, as expected from momentum feedback. The exact solution in Eq. (145) is unlikely to be an exact solution in reality, because of the somewhat idealized setup (e.g. constant $\epsilon_{\rm ff}$, $t_{\rm ff}$, and $\epsilon_{\star,\rm max}$). So within the modelling uncertainty we are free to choose almost any function of time with the same asymptotic limits as Eq. (145), and not too different for intermediate times (i.e. $t - t_{\rm min} \sim t_{\rm ff}/\epsilon_{\rm ff}$). We will choose the following approximation instead of Eq. (145):

$$\Sigma_\star(R,t) = \epsilon_{\star,\rm max}\Sigma_{\rm disk}\,\frac{\epsilon_{\rm ff}(t - t_{\rm min})/t_{\rm ff}}{1 + \epsilon_{\rm ff}(t - t_{\rm min})/t_{\rm ff}}\,. \tag{146}$$

This expression has the exact same asymptotic limits as Eq. (145). The advantage of Eq. (146) over Eq. (145) is that it will allow us to write simpler analytical expression for the total stellar mass formed within a radius $R$ in the disk.

## 5.3. Total stellar mass formed as a function of time

We are now ready to derive the stellar mass formed in the disk by some time $t$ after disk formation. The total stellar mass within a dimensionless radius $\bar{R} \equiv R/R_{\rm disk}$ at time $t$ is given by

$$M_\star(<\bar{R},t) = 2\pi R_{\rm disk}^2 \int_0^{\bar{R}} {\rm d}\bar{R}'\,\bar{R}'\Sigma_\star(\bar{R}',t)\,. \tag{147}$$

Using Eq. (146) this yields (suppressing the explicit dependence on $\bar{R}$ of variables to reduce clutter):

$$M_\star(<\bar{R},t) = 2\pi R_{\rm disk}^2 \int_0^{\bar{R}} {\rm d}\bar{R}'\,\bar{R}'\epsilon_{\star,\rm max}\Sigma_{\rm disk}\,\frac{\epsilon_{\rm ff}(t - t_{\rm min})/t_{\rm ff}}{1 + \epsilon_{\rm ff}(t - t_{\rm min})/t_{\rm ff}}\,. \tag{148}$$

Using Eq. (142) for $\epsilon_{\star,\rm max}$ yields

$$M_\star(<\bar{R},t) = 2\pi R_{\rm disk}^2 \int_0^{\bar{R}} {\rm d}\bar{R}'\,\bar{R}'\frac{\Sigma_{\rm disk}^2}{\Sigma_{\rm disk}(1 + \langle\tau_{\rm IR}\rangle) + \Sigma_{\rm crit}}\,\frac{\epsilon_{\rm ff}(t - t_{\rm min})/t_{\rm ff}}{1 + \epsilon_{\rm ff}(t - t_{\rm min})/t_{\rm ff}}\,. \tag{149}$$

To proceed we can use $\Sigma_{\rm disk} = \Sigma_{\rm disk,0}\bar{R}^{-1}$ and $2\pi R_{\rm disk}^2\Sigma_{\rm disk,0} = M_{\rm disk}$ (see Eq. 87), giving us (after a little algebra)

$$M_\star(<\bar{R},t) = \left(\frac{\Sigma_{\rm disk,0}}{\Sigma_{\rm crit}}\right) M_{\rm disk} \int_0^{\bar{R}} {\rm d}\bar{R}'\,\frac{1}{\Sigma_{\rm disk,0}(1 + \langle\tau_{\rm IR}\rangle)/\Sigma_{\rm crit} + \bar{R}'}\,\frac{\epsilon_{\rm ff}(t - t_{\rm min})/t_{\rm ff}}{1 + \epsilon_{\rm ff}(t - t_{\rm min})/t_{\rm ff}}\,. \tag{150}$$

Next we note that $t_{\rm ff} \propto \bar{R} \propto R$ (see Eq. 116), and furthermore from Eq. (98) we know that $R_{\rm disk} \propto \mathcal{F} \propto t$, so that $t_{\rm min} \propto R$. Thus, $\epsilon_{\rm ff}t_{\rm min}/t_{\rm ff} \equiv \bar{t}_{\rm min}$ is independent of $\bar{R}$, whereas $\epsilon_{\rm ff}t/t_{\rm ff} \propto \bar{R}^{-1}$. We can therefore also define $\epsilon_{\rm ff}t/t_{\rm ff} \equiv \bar{t}\bar{R}^{-1}$, where $\bar{t}$ is independent of $\bar{R}$ (equivalently we have $\bar{t} = \epsilon_{\rm ff}t/t_{\rm ff}(\bar{R} = 1)$, where $t_{\rm ff}(\bar{R} = 1)$ is the free-fall time-scale at $R = R_{\rm disk}$). Furthermore, we note that since $t = t_{\rm min}$ at $R = R_{\rm disk}$, it follows that $\bar{t}_{\rm min} = \bar{t}$. Thus, in terms of $\bar{t}$ we get

$$M_\star(<\bar{R},t) = \left(\frac{\Sigma_{\rm disk,0}}{\Sigma_{\rm crit}}\right) M_{\rm disk}\bar{t} \int_0^{\bar{R}} {\rm d}\bar{R}'\,\frac{1}{\Sigma_{\rm disk,0}(1 + \langle\tau_{\rm IR}\rangle)/\Sigma_{\rm crit} + \bar{R}'}\,\frac{1 - \bar{R}'}{\bar{t} + (1 - \bar{t})\bar{R}'}\,. \tag{151}$$

This yields[52]

$$M_\star(<\bar{R},t) = \left(\frac{\Sigma_{\rm disk,0}}{\Sigma_{\rm crit}}\right) M_{\rm disk}\bar{t}\,\frac{\ln\left|(\bar{t}^{-1}-1)\bar{R}+1\right| + (\bar{\Sigma}+1)(\bar{t}-1)\ln(1+\bar{R}/\bar{\Sigma})}{(\bar{t}-1)\{(\bar{\Sigma}+1)\bar{t}-\bar{\Sigma}\}}$$

$$= \left(\frac{\Sigma_{\rm disk,0}}{\Sigma_{\rm crit}}\right) M_{\rm disk}\left\{\frac{\bar{t}\,\ln\left|(\bar{t}^{-1}-1)\bar{R}+1\right|}{(\bar{t}-1)\{(\bar{\Sigma}+1)\bar{t}-\bar{\Sigma}\}} + \frac{(\bar{\Sigma}+1)\bar{t}\,\ln(1+\bar{R}/\bar{\Sigma})}{\{(\bar{\Sigma}+1)\bar{t}-\bar{\Sigma}\}}\right\}, \tag{152}$$

where $\bar{\Sigma} \equiv \Sigma_{\rm disk,0}(1+\langle\tau_{\rm IR}\rangle)/\Sigma_{\rm crit}$. From this expression we can deduce that the total stellar $M_{\star,\rm tot}$ formed within the entire disk (i.e. $\bar{R}=1$) by time $t$ is given by (rewriting the pre-factor in terms of $\bar{\Sigma}$):

$$M_{\star,\rm tot}(t) = \frac{\bar{\Sigma}M_{\rm disk}}{(1+\eta_{\rm IR}\kappa_{\rm R}\Sigma_{\rm crit,rad})}\left\{\frac{(\bar{\Sigma}+1)\bar{t}\,\ln(1+1/\bar{\Sigma})}{\{(\bar{\Sigma}+1)\bar{t}-\bar{\Sigma}\}} - \frac{\bar{t}\,\ln\bar{t}}{(\bar{t}-1)\{(\bar{\Sigma}+1)\bar{t}-\bar{\Sigma}\}}\right\}$$

$$\bar{\Sigma} \equiv \Sigma_{\rm disk,0}(1+\langle\tau_{\rm IR}\rangle)/\Sigma_{\rm crit} \tag{153}$$

$$\bar{t} \equiv \epsilon_{\rm ff}t/t_{\rm ff}(\bar{R}=1)\,.$$

We can evaluate the parameter $\bar{t}$ using Eq. (116) for $t_{\rm ff}$ and Eq. (101) for $\Sigma_{\rm disk,0}$ to get:

$$\bar{t} = \frac{4G\Sigma_{\rm disk,0}\epsilon_{\rm ff}t}{\sqrt{3}c_{\rm s}\left(1+\mathcal{M}^2/3\right)^{1/2}}$$

$$= \frac{\epsilon_{\rm ff}f_{\rm B}^4 v_{\rm vir}}{32\pi\sqrt{3}c_{\rm s}f_{\rm disk}^2\lambda^4\mathcal{G}_{\rm gas}\left(1+\mathcal{M}^2/3\right)^{1/2}}\,, \tag{154}$$

which is independent of the time $t$ since disk formation ($\bar{t}$ could instead be seen as a parameter characterizing the star formation time-scale in the disk). As we expect, given enough time for star formation — or short enough free-fall time in the disk — (i.e. $\bar{t} \gg 1$) and inefficient momentum feedback (i.e. $\bar{\Sigma} \gg 1$), the stellar mass formed becomes[53] $M_{\star,\rm tot} \simeq \bar{\Sigma}M_{\rm disk} \times \bar{\Sigma}^{-1}/(1+\langle\tau_{\rm IR}\rangle) = M_{\rm disk}/(1+\langle\tau_{\rm IR}\rangle)$. In other words, if momentum feedback is negligible and star formation can proceed sufficiently long, a fraction $1/(1+\langle\tau_{\rm IR}\rangle)$ of the disk mass $M_{\rm disk}$ will be converted into stars. Ignoring FIR feedback from dust (i.e. $\langle\tau_{\rm IR}\rangle \to 0$) this would mean that the whole disk would be converted into stars (as expected in these limits). The additional suppression due to the inclusion of FIR feedback from dust — even for an extremely dense disk ($\bar{\Sigma} \to \infty$) — reflects its unique property of scaling as rapidly with gas surface density as the the self-gravity of the disk (as pointed out by Murray *et al.*, 2010).

Another interesting property of Eq. (153) is the insensitivity of the predicted stellar mass to variations in the per-free-fall star formation efficiency $\epsilon_{\rm ff}$, which only enters through $\bar{t}$ ($M_{\rm disk}$ scales linearly with time due to gas accretion from the IGM and/or the outer halo, but it does not depend on $\epsilon_{\rm ff}$). In particular, as long as the free-fall time is sufficiently short in the disk we will have $\bar{t} \gg 1$, and $M_{\star,\rm tot}(t)$ becomes almost independent of $\bar{t}$, and hence $\epsilon_{\rm ff}$ — as gas is accreted onto the disk it is converted into

[52]Here we have used the following integral (for some constant $\mathcal{A}$):

$$\int {\rm d}\bar{R}'\,\frac{1}{\mathcal{A}+\bar{R}'}\,\frac{1-\bar{R}'}{\bar{t}+(1-\bar{t})\bar{R}'} = \frac{\ln\left|(\bar{t}-1)\bar{R}-\bar{t}\right| + (\mathcal{A}+1)(\bar{t}-1)\ln(\mathcal{A}+\bar{R})}{(\bar{t}-1)\{(\mathcal{A}+1)\bar{t}-\mathcal{A}\}} + {\rm const.}.$$

[53]To see this, we can use $\ln(1+1/\bar{\Sigma}) \simeq \bar{\Sigma}^{-1}$ for large $\bar{\Sigma}$. Furthermore, $(\bar{\Sigma}+1)\bar{t}/\{(\bar{\Sigma}+1)\bar{t}-\bar{\Sigma}\} \simeq 1$ for large $\bar{\Sigma}$ and $\bar{t}$. The second term in the brackets in Eq. (153) scales as $\ln\bar{t}/\bar{\Sigma}\bar{t}$ which becomes negligible in the same limits.

stars almost immediately, and hence the stellar mass formed by time $t$ is only sensitive to momentum feedback and the mass of the disk, not $\epsilon_{\rm ff}$. This conclusion is also supported by the simulations of Ricotti *et al.* (2016) who found that varying $\epsilon_{\rm ff}$ ($\epsilon_\star$ in their notation) in the range $0.01 \leqslant \epsilon_{\rm ff} \leqslant 1$ led to negligible qualitative differences in the stellar masses of compact GC-like objects that formed in low-mass halos at Cosmic Dawn. Similarly, using high-resolution simulations of dwarf galaxies, Hislop *et al.* (2021) recently found that varying $\epsilon_{\rm ff}$ in the range $0.002 \leqslant \epsilon_{\rm ff} \leqslant 0.5$ did not change the predicted star formation rate nor the outflow rates from the galaxy. This is also in line with analytical models of star formation in galaxies, showing that the galactic star formation rate is regulated by momentum feedback, and is largely independent of $\epsilon_{\rm ff}$ (Faucher-Giguère *et al.*, 2013). Similarly, using high-resolution simulations of GMCs that include both radiative and SN feedback, Grudić *et al.* (2018, 2019) found that $\epsilon_{\rm ff} \sim 1$ in the absence of stellar feedback, with momentum feedback controlling the total stellar mass produced. Motivated by this, we choose $\epsilon_{\rm ff} = 1$ as a fiducial value, as done e.g. in the FIRE-2 simulations (Hopkins *et al.*, 2018a). An interesting future development would be to use a turbulence-dependent $\epsilon_{\rm ff}$ as implemented in the recent high-resolution galaxy formation simulations by Kimm *et al.* (2017) that focused on star formation in low-mass halos (including minihalos).[54]

## 5.4. Destruction of the star-forming disk

### 5.4.1. Photoionization feedback: Expanding H II regions

Here we will discuss photoionization feedback, specifically its thermal heating effects. As massive stars form, their ionizing radiation will produce H II regions within which the gas is photoheated to $T_{\rm HII} \sim 10^4$ K. Usually the ambient temperature $T$ of the star-forming gas will be $T \ll T_{\rm HII}$, so that the H II region will be overpressurized and hence expand outwards. If the H II region continues to expand until the whole star-forming gas cloud is encompassed, the gas cloud could be disrupted, which would put an end to to the starburst. In fact, recent observations support the notion that expanding H II regions are one of the key processes that explain why Giant Molecular Clouds (GMCs) in the local Universe are destroyed after only $\sim 1-2$ Myrs, well before the first SN explosion (Kruijssen *et al.*, 2019a; Chevance *et al.*, 2020). Here we will estimate the time-scale for photoionization feedback to potentially destroy the disk. We will assume a simplified setup where stars form near the midplane of the disk, producing a vertically expanding H II region of height $z_{\rm HII}$, as sketched in Figure 14. Ignoring the external pressure (a complication to be discussed later) the equation of motion for $z_{\rm HII}$ is given by (see Spitzer, 1978; Raga *et al.*, 2012, for derivations)

$$\dot{z}_{\rm HII} \simeq c_{\rm s,HII} \left(\frac{\rho_{\rm HII}}{\rho_{\rm mid}}\right)^{1/2}, \tag{155}$$

where $c_{\rm s,HII}$ is the sound speed within the H II region, $\rho_{\rm HII}$ is the gas density in the H II region, and $\rho_{\rm mid}$ (the midplane density) is approximately the density of gas in the H I regions. Consider an arbitrary area $\mathcal{A}$ in the disk. The ionization rate from the stars in this area will be $\dot{N}_{\rm ion} = \dot{Q}_{\rm ion}\Sigma_\star\mathcal{A}$, where $\dot{Q}_{\rm ion}$ (in $\rm s^{-1}\ M_\odot^{-1}$) is the hydrogen ionizing photon rate per Solar mass of stars formed. For Pop II stars we adopt $\dot{Q}_{\rm ion} = 5\times10^{46}\,\rm s^{-1}\ M_\odot^{-1}$, which is an appropriate average value for the first $\sim 3.5$ Myrs following an instantaneous stellar burst with a Kroupa IMF and metallicity $Z = 0.05\,Z_\odot$ (e.g. Ma *et al.*, 2016; **?**; Rosdahl *et al.*, 2018). While many GCs and practically all UFDs have lower metallicities than this, the effect of metallicity on $\dot{Q}_{\rm ion}$ is small (see e.g. Stanway & Eldridge, 2019) and can be neglected for our purposes.[55] In the case of Pop III stars we can estimate $\dot{Q}_{\rm ion}$ as $\dot{Q}_{\rm ion} = \dot{N}_{\rm ion}/m_{\star,\rm PopIII}$ using the values of $\dot{N}_{\rm ion}$ from Schaerer (2002). The resulting values of $\dot{Q}_{\rm ion}$ are summarized in Table 5. The recombination rate in the H II region of vertical thickness $2z_{\rm HII}$ is simply $\dot{N}_{\rm rec} = k_1 n_{\rm e} n_{\rm p} \times 2z_{\rm HII}\mathcal{A}$, where $k_1$ is the case-B recombination coefficient as usual. In photoionization equilibrium ($\dot{N}_{\rm ion} = \dot{N}_{\rm rec}$) we then find $n_{\rm e}n_{\rm p} = \dot{Q}_{\rm ion}\Sigma_\star/2z_{\rm HII}k_1$. For a primordial H II region wherein helium is singly ionized we have $n_{\rm e}n_{\rm p} = f_Y\rho_{\rm HII}^2/m_{\rm H}^2$ where $f_Y = X(X+Y/4) = 0.614$ is a numerical factor.[56] Thus, the density in the H II region is given by:

$$\rho_{\rm HII} = \left(\frac{\dot{Q}_{\rm ion}\Sigma_\star m_{\rm H}^2}{2f_Y z_{\rm HII}k_1}\right)^{1/2} . \tag{156}$$

[54] It has been suggested that the density distribution of supersonic isothermal turbulence could naturally give rise to values as low as $\epsilon_{\rm ff} \sim 0.01$, depending on the Mach number of the gas and how gravitationally unstable the gas cloud is (captured by the virial parameter $\alpha_{\rm vir} \equiv 2E_{\rm kin}/\left|E_{\rm grav}\right|$ of the gas) (see e.g. Federrath & Klessen, 2012; Krumholz, 2015, and discussion and references therein). While the simulations by Grudić *et al.* (2018, 2019) indicate that turbulence is unimportant, this is likely due to the decay of turbulence in these simulations. Recently Pokhrel *et al.* (2021) found observational evidence for $\epsilon_{\rm ff} \ll 1$ even in regions containing no massive stars, and so presumably negligible or no feedback from radiation pressure, photoionization, or SNe. These authors suggest feedback from low-mass stars, e.g. protostellar outflows, could maintain turbulence in the gas cloud and lead to $\epsilon_{\rm ff} \ll 1$ even in the absence of feedback from massive stars.

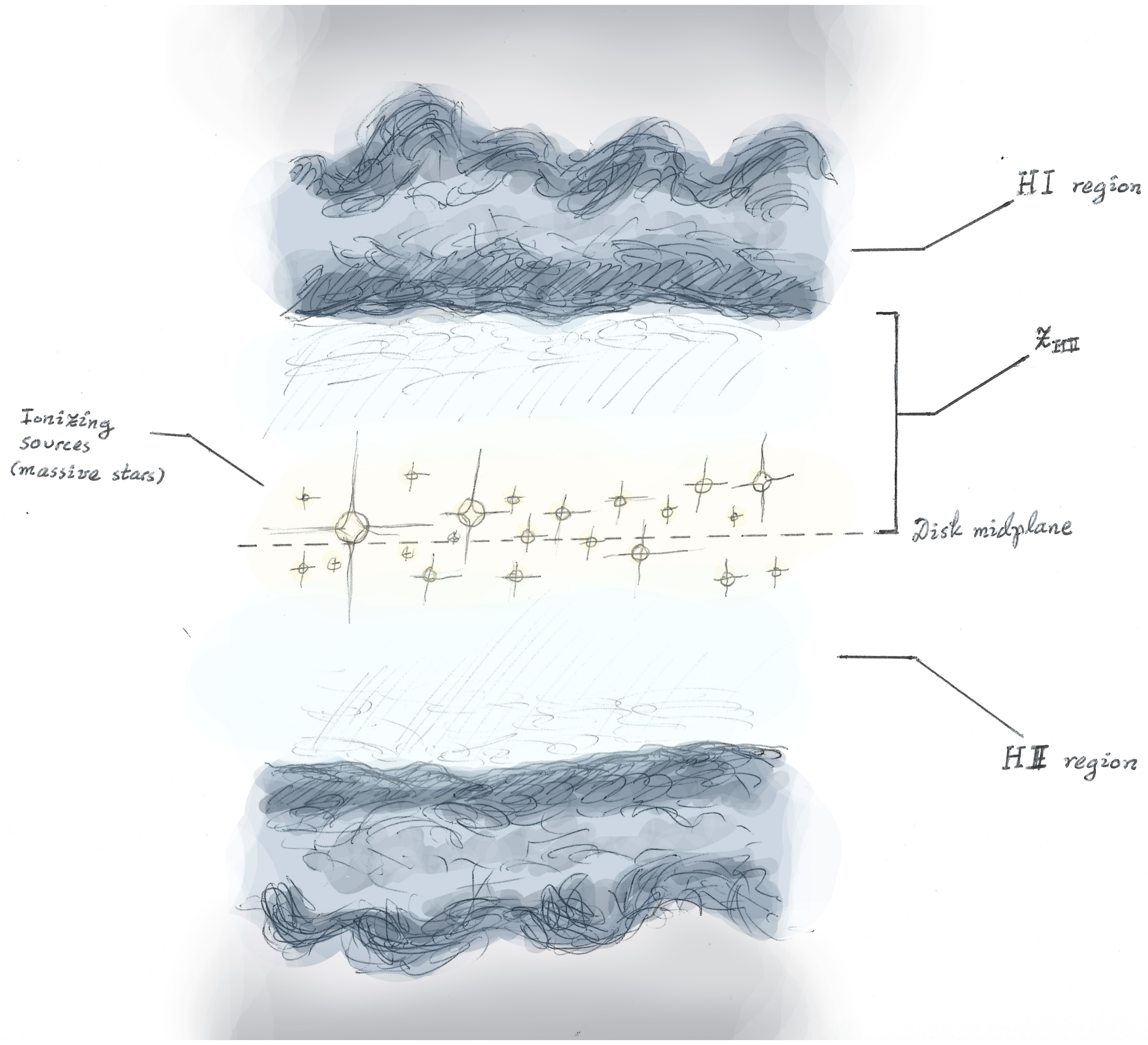

**Figure 14.** The setup for the calculation of the expanding H II region in the disk. The sketch show a vertical cross-section of the disk, with young massive stars near the midplane producing a vertically expanding H II region of height $z_{\rm HII}$ over the midplane.

[55]For example, in Figure 3 of Stanway & Eldridge (2019) we see that for an upper IMF slope of 2.35, lowering the metallicity from $Z = 0.05\,Z_\odot$ ($Z = 10^{-3}$) to $Z = 5\times10^{-4}\,Z_\odot$ ($Z = 10^{-5}$) would only raise $\dot{Q}_{\rm ion}$ by a factor $\sim 0.1$ dex, i.e. $\sim 26\%$.

[56]The electron number density in a gas containing only ionized hydrogen and singly ionized helium is $n_{\rm e} = n_{\rm HII} + n_{\rm HeII} = n_{\rm H} + n_{\rm He}$. Furthermore, $n_{\rm p} = n_{\rm HII} = n_{\rm H}$. Since $n_{\rm H} = X\rho/m_{\rm H}$ and $n_{\rm He} = Y\rho/m_{\rm H}$ this yields $n_{\rm e}n_{\rm p} = (X+Y/4)X\rho^2/m_{\rm H}^2$ as stated.

Table 5. Photoionization feedback: Values for $\dot{Q}_{\rm ION}$

| Pop II | Pop III ($m_{\star,\rm PopIII} = 25$ M$_\odot$) | Pop III ($m_{\star,\rm PopIII} = 140$ M$_\odot$) |
|---|---|---|
| $\dot{Q}_{\rm ion}$ | $\dot{Q}_{\rm ion}$ | $\dot{Q}_{\rm ion}$ |
| (s$^{-1}$ M$_\odot^{-1}$) | (s$^{-1}$ M$_\odot^{-1}$) | (s$^{-1}$ M$_\odot^{-1}$) |
| $5 \times 10^{46}$ | $2.75 \times 10^{47}$ | $1.44 \times 10^{48}$ |

Using this and Eq. (146) for $\Sigma_\star$ in Eq. (155) yields

$$\frac{{\rm d}z_{\rm HII}}{{\rm d}t'} \simeq c_{\rm s,HII} \left( \frac{\dot{Q}_{\rm ion} \epsilon_{\star,\rm max} \Sigma_{\rm disk} m_{\rm H}^2}{2 f_Y k_1 \rho_{\rm mid}^2} \right)^{1/4} \frac{1}{z_{\rm HII}^{1/4}} \left\{ \frac{\epsilon_{\rm ff} t'/t_{\rm ff}}{1 + \epsilon_{\rm ff} t'/t_{\rm ff}} \right\}^{1/4} , \tag{157}$$

where $t' \equiv t - t_{\rm min}$. This is the differential equation governing $z_{\rm HII}$ at some radius $R < R_{\rm disk}$. As shown in Appendix 8, this can be solved approximately, with a resulting fairly messy looking final result. We consider the disk to be destroyed by photoionization feedback at time $t_{\rm HII,destr}$ when the time needed for the H II region to break out of the disk at $R = R_{\rm disk}$ becomes comparable to the disk accretion time-scale (i.e. the time since the formation of the disk). The result is:

$$t_{{\rm HII,destr},T=0} \simeq \frac{t_{\rm arbitrary}}{(2^{4/5}-1)^2} \left\{ \frac{4\epsilon_{\rm ff}}{5 c_{\rm s,HII} t_{\rm ff}(t_{\rm arbitrary})} \right\}^2 \left\{ \frac{f_Y k_1 \sigma_{\rm gas}^6}{2\pi^3 \dot{Q}_{\rm ion} \epsilon_{\star,\rm max}(t_{{\rm HII,destr},T=0}) m_{\rm H}^2 G^3 \Sigma_{\rm disk,0}^2(t_{\rm arbitrary})} \right\}^{1/2} , \tag{158}$$

where $t_{\rm arbitrary}$ is a completely arbitrary time-scale, which can be set to 1 Myrs without loss of generality, and the subscript $T = 0$ emphasize that this was calculated in the absence of external pressure (i.e. assuming a temperature $T = 0$ in the H I region). Eq. (158) cannot be solved analytically due to the dependence on $\epsilon_{\star,\rm max}(t_{{\rm HII,destr},T=0}) \equiv \epsilon_{\star,\rm max}[R = R_{\rm disk}(t_{{\rm HII,destr},T=0})]$ on the right-hand side. In the Anaxagoras code we estimate $t_{{\rm HII,destr},T=0}$ using 5 iterations,[57] starting with the initial "guess" $\epsilon_{\star,\rm max}(t_{{\rm HII,destr},T=0}) = 1$. The actual disk destruction time-scale $t_{\rm HII,destr}$ due to photoionization feedback will in general not be equal to $t_{{\rm HII,destr},T=0}$ because we have neglected external pressure and the gravitational effects of the disk. Indeed, in the simulations of Kimm *et al.* (2016) H II feedback was ineffective in terminating star formation because the ambient gas in the disk had a very high temperature ($T \simeq 8000$ K) since the disk gas did not contain $H_2$, nor metals prior to the first SNe. And in the simulations of star formation in GMCs by Grudić *et al.* (2018) photoionization feedback had practically no effect on the predicted stellar masses because the escape velocity of their GMCs exceeded the sound speed of the H II regions (see also Dale *et al.*, 2012). An expanding H II region will come to a halt if the pressure within the H II region, $\rho_{\rm HII} c_{\rm s,HII}^2$, becomes equal to the ambient pressure in the H I region, $\rho_{\rm mid} c_{\rm s}^2$. Thus, the photoionization feedback will only be effective in expelling the gas if $\rho_{\rm HII} c_{\rm s,HII}^2 > \rho_{\rm mid} c_{\rm s}^2$. Furthermore, in order for the H II region to break out of the disk we demand that $c_{\rm s,HII} > v_{\rm disk}/\sqrt{3}$, which ensures that the thermal energy of the gas exceed its

[57] In experiments the result was found to converge quickly (which is not surprising given the insensitive dependence on $\epsilon_{\star,\rm max}$), and so no more iterations are needed.

binding energy.[58] In summary, photoionization can destroy the disk and put an end to the starburst in a time $t_{\rm HII,destr}$ given by:

**DISK DESTRUCTION TIME-SCALE DUE TO H II FEEDBACK**

$$t_{\rm HII,destr} = \begin{cases} t_{{\rm HII,destr},T=0} & \text{if } \rho_{\rm HII} c_{\rm s,HII}^2 > \rho_{\rm mid} c_{\rm s}^2 \text{ and } c_{\rm s,HII} > v_{\rm disk}/\sqrt{3} \\ \infty & \text{else} \end{cases} . \tag{159}$$

In this thesis we do not calculate the equilibrium temperature $T_{\rm HII}$ of the H II region explicitly. Instead we simply assume it to be $2 \times 10^4$ K in the metal-poor H II regions around Pop II stars (e.g. Shapiro *et al.*, 2004; Kimm *et al.*, 2016), and $3 \times 10^4$ K for metal-free gas in H II regions around Pop III stars (e.g. Kitayama *et al.*, 2004; Abel *et al.*, 2007; Hirano *et al.*, 2014). This temperature is used in the calculation of the case-B recombination coefficient $k_1$ using Eq. (197) from Draine (2011). The sound speed in the H II region is $c_{\rm s,HII} = (k_{\rm B} T_{\rm HII}/\mu_{\rm HII} m_{\rm H})^{1/2}$, where $\mu_{\rm HII} = 0.614$ is the mean molecular weight in the H II region.[59]

### 5.4.2. Feedback from supernovae

Even if expanding H II regions fail to destroy the disk, the disk in a low-mass halo is unlikely to survive multiple SN consecutive explosions which inject energy and momentum into the gas. Here we will discuss and derive the time-scale of disk destruction (which could terminate both Pop II and Pop III star formation) or metal-enrichment by SNe (which could terminate further Pop III star formation). Consider a single SN exploding with energy $E_{\rm SN}$. During the early stages of the SN remnant (SNR), energy is conserved and the expansion of the SNR shock radius $R_{\rm s}$ is governed by the well-known Sedov-Taylor solution (here we

[58] This can be derived as follows. If H II regions cover all the disk gas, the energy of the gas in the disk is given by

$$E_{\rm gas} = \frac{3}{2} M_{\rm gas} c_{\rm s,HII}^2 - \frac{1}{2} M_{\rm gas} v_{\rm disk}^2 ,$$

where $-\frac{1}{2} M_{\rm gas} v_{\rm disk}^2$ is the sum of kinetic and gravitational binding energy of the gas (using the virial theorem). For the gas to become unbound we need $E_{\rm gas} > 0$, and so $c_{\rm s,HII} > v_{\rm disk}/\sqrt{3}$.

[59] For primoridal gas wherein hydrogen is fully ionized and helium singly ionized, the mean particle mass is $\bar{m} = (n_{\rm e} m_{\rm e} + n_{\rm HII} m_{\rm H} + n_{\rm HeII} m_{\rm He})/n \simeq (n_{\rm HII} m_{\rm H} + n_{\rm HeII} 4 m_{\rm H})/n$, so that

$$\mu \equiv \frac{\bar{m}}{m_{\rm H}} = \frac{n_{\rm HII} + 4 n_{\rm HeII}}{n_{\rm e} + n_{\rm HII} + n_{\rm HeII}} . \tag{160}$$

Since $n_{\rm HII} = X \rho_{\rm B}/m_{\rm H}$, $n_{\rm HeII} = Y \rho_{\rm B}/4 m_{\rm H}$, and $n_{\rm e} = n_{\rm HII} + n_{\rm HeII}$, we get

$$\mu = \frac{X + Y}{2(X + Y/4)} . \tag{161}$$

Since $X + Y = 1$ for the primordial gas of interest, we get $\mu = [2(X + Y/4)]^{-1} = 0.614$. This expression is consistent with the more general expression provided by Black & Bodenheimer (1975).

assume that the disk has a density $\rho_{\rm mid}$):

$$R_{\rm s}(t) = \left(\frac{\xi_{\rm ST} E_{\rm SN} t^2}{\rho_{\rm mid}}\right)^{1/5} \qquad \text{(Energy-conserving Sedov-Taylor phase)}$$

$$= 0.323\, E_{\rm SN,51}^{1/5} n_{\rm mid}^{-1/5} t_{\rm yr}^{2/5}\ {\rm pc}$$

where $\xi_{\rm ST} = 2.02597$ (e.g. Ostriker & McKee, 1988), $E_{\rm SN,51} \equiv E_{\rm SN}/10^{51}$ erg, $n_{\rm mid}$ is the ambient gas number density in cm$^{-3}$, and $t_{\rm yr} \equiv t/1$ yr. Eventually the shock-heated gas in the SNR reaches a temperature low enough that cooling becomes efficient, and the assumption of energy conservation breaks down. This occurs at a radius $R_{\rm cool} \simeq 28.4\, E_{\rm SN,51}^{2/7} n_{\rm mid}^{-3/7} f(Z)$ pc, where

$$\begin{aligned} R_{\rm cool} &\simeq 28.4\, E_{\rm SN,51}^{2/7} n_{\rm mid}^{-3/7} f_Z\ {\rm pc}\,, \\ f_Z &= \begin{cases} 2 & \text{for } Z/Z_\odot < 0.01 \\ (Z/Z_\odot)^{-0.14} & \text{for } Z/Z_\odot \geqslant 0.01 \end{cases}, \end{aligned} \tag{162}$$

where the absence of a metallicity dependence for $Z/Z_\odot < 0.01$ stems from the fact that in this regime atomic cooling dominates (Kimm & Cen, 2014; Hopkins *et al.*, 2018b). We see that $R_{\rm s}(t_{\rm cool}) = R_{\rm cool}$ at a time

$$t_{\rm cool} \simeq 7.25 \times 10^4\, E_{\rm SN,51}^{3/14} n_{\rm mid}^{-4/7} f_Z^{5/2}\ {\rm yrs}\,, \tag{163}$$

which has the nearly the same scaling and magnitude as other approximate expressions found in the literature (e.g. Dekel & Silk, 1986; Draine, 2011). For times $t > t_{\rm cool}$ the SNR shock has lost significant amount of thermal energy, and is now driven by momentum rather than energy conservation (the so-called snowplow phase), for which the shock expand like $R_{\rm s} \propto t^{1/4}$.[60] Thus, we can approximate the evolution of the SNR shock as follows:

$$R_{\rm s}(t) \simeq \begin{cases} R_{\rm cool}\,(t/t_{\rm cool})^{2/5} & \text{for } t < t_{\rm cool} \\ R_{\rm cool}\,(t/t_{\rm cool})^{1/4} & \text{for } t \geqslant t_{\rm cool} \end{cases}. \tag{164}$$

Next we want to compute the total volume occupied by all SNR in order to estimate the time it takes for these SNR to encompass the whole disk (after which the disk could be disrupted if enough energy and momentum is injected at that point). Let $\mathrm{d}n_{\rm SN}(t)$ be the number of SNe going off in the time interval $[t,\, t + \mathrm{d}t]$ per unit volume in the disk. Then the fraction of volume occupied by SNR in the disk is given by

$$f_{\rm SN}(t) \simeq \int_0^{n_{\rm SN}(t)} \mathrm{d}n_{\rm SN}(t_{\rm form})\, \frac{4\pi}{3} R_{\rm s}^3(t - t_{\rm form})\,. \tag{165}$$

Let $\mathscr{R}_{\rm SN}$ be the rate of SNe per unit stellar mass per year. Then the rate $\mathrm{d}\dot{n}_{\rm SN}(t)$ of SNe at time $t$ from an infinitesimal stellar surface mass density $\mathrm{d}\Sigma_\star(t_{\rm form})$ that formed at time $t_{\rm form}$ is $\mathrm{d}\dot{n}_{\rm SN}(t) = \mathscr{R}_{\rm SN}(t - t_{\rm form})\, \mathrm{d}\Sigma_\star(t_{\rm form})/2z_0$, where $2z_0$ is the disk thickness. The total SN rate per unit volume is therefore (using $\mathrm{d}\Sigma_\star(t_{\rm form}) = \dot{\Sigma}_\star\, \mathrm{d}t_{\rm form}$)

$$\dot{n}_{\rm SN}(t) = \frac{1}{2z_0} \int_0^t \mathrm{d}t_{\rm form}\, \mathscr{R}_{\rm SN}(t - t_{\rm form}) \dot{\Sigma}_\star(t_{\rm form})\,, \tag{166}$$

which can then be integrated again to find $n_{\rm SN}(t)$. The form of $\mathscr{R}_{\rm SN}$ will differ depending on whether we are considering Pop II stars that have a Kroupa IMF, or Pop III stars with the assumed Dirac delta IMF in this thesis. Let us therefore consider both cases separately below.

[60] During the snowplow phase momentum is conserved, so $M(< R_{\rm s})\dot{R}_{\rm s}$ is constant. Since $M(< R_{\rm s}) \propto R_{\rm s}^3$ this means that $R_{\rm s}^3 \dot{R}_{\rm s}$ is constant, which has the solution $R_{\rm s} \propto t^{1/4}$.

### Pop II stars

In the case of Pop II stars with a Kroupa IMF, $\mathscr{R}_{\rm SN}$ for type II SNe and times $t < 10.37$ Myrs is approximately given by (e.g. Hopkins *et al.*, 2018a)[61]

$$\mathscr{R}_{\rm SN}(t) = \begin{cases} 0 & \text{for } t < t_{\rm SN} \\ \\ \bar{\mathscr{R}}_{\rm SN} & \text{for } t > t_{\rm SN} \end{cases} , \tag{167}$$

where $\bar{\mathscr{R}}_{\rm SN} = 5.408 \times 10^{-10}\,{\rm SNe\ yr^{-1}\ M_\odot^{-1}}$, and $t_{\rm SN} = 3.401$ Myrs is the time of the first SN in this case ($t_{\rm SN}$ will take on other values for Pop III stars). The integral in Eq. (166) therefore only gets contribution from $t - t_{\rm form} > t_{\rm SN}$, and so

$$\begin{aligned} \dot{n}_{\rm SN}(t) &= \frac{1}{2z_0}\bar{\mathscr{R}}_{\rm SN}\int_0^{t-t_{\rm SN}} {\rm d}t_{\rm form}\,\dot{\Sigma}_\star(t_{\rm form}) \\ &= \frac{1}{2z_0}\bar{\mathscr{R}}_{\rm SN}\Sigma_\star(t-t_{\rm SN})\,. \end{aligned} \tag{168}$$

Thus, ${\rm d}n_{\rm SN}(t_{\rm form}) = \bar{\mathscr{R}}_{\rm SN}\Sigma_\star(t_{\rm form}-t_{\rm SN})\,{\rm d}t_{\rm form}/2z_0$ and so the resulting fraction of volume occupied by SNR in the disk (Eq. 165) becomes:

$$f_{\rm SN}(t) \simeq \frac{2\pi}{3z_0}\bar{\mathscr{R}}_{\rm SN}\int_0^t {\rm d}t_{\rm form}\,\Sigma_\star(t_{\rm form}-t_{\rm SN})\,R_{\rm s}^3(t-t_{\rm form})\,. \tag{169}$$

Since $\Sigma_\star(t) \propto [\epsilon_{\rm ff}(t-t_{\rm min})/t_{\rm ff}]/\{1+[\epsilon_{\rm ff}(t-t_{\rm min})/t_{\rm ff}]\}$ (Eq. 146), the integrand is only non-zero for $t_{\rm form}-t_{\rm SN} > t_{\rm min}$. Furthermore, since we expect the SNe to go off for times $\epsilon_{\rm ff}t_{\rm SN}/t_{\rm ff} \gtrsim 1$, we can make the approximation $\Sigma_\star(t_{\rm form}-t_{\rm SN}) \simeq \epsilon_{\star,\rm max}\Sigma_{\rm disk}$. Taken together, this yields

$$f_{\rm SN}(t) \simeq \frac{2\pi}{3z_0}\bar{\mathscr{R}}_{\rm SN}\epsilon_{\star,\rm max}\Sigma_{\rm disk}\int_{t_{\rm SN}+t_{\rm min}}^t {\rm d}t_{\rm form}\,R_{\rm s}^3(t-t_{\rm form})\,. \tag{170}$$

As shown in more detail in Appendix 9, this can be evaluated using Eq. (164) for $R_{\rm s}$ to to get

$$f_{\rm SN}(t) \simeq \frac{2\pi}{3z_0}\bar{\mathscr{R}}_{\rm SN}\epsilon_{\star,\rm max}\Sigma_{\rm disk}R_{\rm cool}^3 t_{\rm cool}\left\{-\frac{9}{77}+\frac{4}{7}\left[\frac{t-(t_{\rm SN}+t_{\rm min})}{t_{\rm cool}}\right]^{7/4}\right\}. \tag{171}$$

Thus, in the vicinity of the at the radial position $\bar{R}$ of interest in the disk, SN remnants cover all of the interstellar medium ($f_{\rm SN} = 1$) at a time $t_{\rm SN,destr}$ when

$$t_{\rm SN,destr} \simeq t_{\rm SN} + t_{\rm min} + \left[\frac{7}{4}\left\{\frac{3z_0}{2\pi\bar{\mathscr{R}}_{\rm SN}\epsilon_{\star,\rm max}\Sigma_{\rm disk}R_{\rm cool}^3 t_{\rm cool}} + \frac{9}{77}\right\}\right]^{4/7} t_{\rm cool}\,. \tag{172}$$

To evaluate this we need to choose a value of $t_{\rm min}$, which is equivalent to choosing a value of $R$. A reasonable choice would be the half-mass radius of the disk at time $t_{\rm SN}$ when the first SN explode. This yields

[61] Type 1a SNe are ignored because they occur first after $t > 37.53$ Myrs (Hopkins *et al.*, 2018a), which exceeds the length of the short starbursts considered in this thesis. Furthermore, for times $10.37 < t_{\rm Myr} < 37.53$ Hopkins *et al.* (2018a) give a type II SNe rate of $\mathscr{R}_{\rm SN} = 2.516 \times 10^{-10}\,{\rm SNe\ yr^{-1}\ M_\odot^{-1}}$. But since we are mainly considering bursts lasting shorter than this, we take their value of $5.408 \times 10^{-10}\,{\rm SNe\ yr^{-1}\ M_\odot^{-1}}$ for $t_{\rm Myr} > t_{\rm SN}$. In any case, the two non-zero rates are equal to within a factor of $\sim 2$, and so even if the burst lasts longer than 10.37 Myrs, we do not expect the predicted stellar mass to be effected too much (especially within the scatter due to halo properties). For reference, the starbursts that led to the two GCs in the simulations by Kimm *et al.* (2016) lasted for $\sim 5$ Myrs or so.

$t_{\rm min} = t_{\rm SN}/2$ since $t_{\rm min} \propto R$. Furthermore, at this radius $R = R_{\rm disk}(t_{\rm SN})/2$ we also have $\Sigma_{\rm disk} = 2\Sigma_{\rm disk,0}(t_{\rm SN})$, and $z_0 = \sigma^2_{\rm gas}/2\pi G\Sigma_{\rm disk} = \sigma^2_{\rm gas}/4\pi G\Sigma_{\rm disk,0}(t_{\rm SN})$ (using Eq. 107). Similarly, the gas number density that enter $t_{\rm cool}$ and $R_{\rm cool}$ is evaluated at a density $n_{\rm mid} = n_{\rm mid}[R = R_{\rm disk}(t_{\rm SN})/2]$, and $\epsilon_{\star,\rm max}$ is evaluated for a surface density $\Sigma_{\rm disk} = 2\Sigma_{\rm disk,0}(t_{\rm SN})$. Thus, we assume that Pop II SNe put an end to star formation after a time $t_{\rm SN,destr}$ given by

$$t_{\rm SN,destr} \simeq \frac{3}{2}t_{\rm SN} + \left[\frac{7}{4}\left\{\frac{3\sigma^2_{\rm gas}}{16\pi^2 G\bar{\mathscr{R}}_{\rm SN}\epsilon_{\star,\rm max}\Sigma^2_{\rm disk,0}(t_{\rm SN})R^3_{\rm cool}t_{\rm cool}} + \frac{9}{77}\right\}\right]^{4/7} t_{\rm cool}\,. \tag{173}$$

In most cases the disk density is so high and $t_{\rm cool}$ so small that $t_{\rm SN,destr} \simeq 3t_{\rm SN}/2$, independent of the disk properties.

#### Pop III stars

Regardless of whether SNe can destroy the disk, they will enrich the gaseous disk with metals, which in turn is likely to put an end to further Pop III star formation very quickly (see e.g. Xu *et al.*, 2016). Thus, in a sense Pop III star formation is more sensitive to SN feedback. For simplicity we therefore assume that a Pop III star formation phase can last at most a time $t_{\rm SN,destr} = t_{\star,\rm PopIII}$, where $t_{\star,\rm PopIII}$ is the lifetime of the Pop III stars (see Table 3). A more detailed approach would consider both the expansion of SNR and the turbulent mixing of metals in the disk, but is expected to give a time-scale $t_{\rm SN,destr} \sim t_{\star,\rm PopIII}$ in any case, similar to the case of Pop II starbursts considered earlier.

## 5.5. Total stellar mass formed

With the feedback time-scales due to photoionization ($t_{\rm HII,destr}$) and SNe ($t_{\rm SN,destr}$) determined, the total stellar mass formed in a starburst is simply given by

$$M_{\star,\rm tot} = \begin{cases} M_{\star,\rm tot}(t_{\rm burst}) & \text{if } Q_{\rm Toomre} < 1 \\ 0 & \text{else} \end{cases} \quad \text{where } t_{\rm burst} = \min(t_{\rm HII,destr},\, t_{\rm SN,destr})\,, \tag{174}$$

where we have taken into account the fact that star formation cannot take place unless the disk is Toomre unstable (see Section 4.4), i.e. $Q_{\rm Toomre} < 1$. In other words, the fastest of the two mechanisms will determine the starburst time-scale $t_{\rm burst}$.

## 5.6. Half-mass radii of GC and UFD candidates

### 5.6.1. Cluster half-mass radius prior to gas expulsion

Prior to gas evacuation of the disk we expect the stars to quickly re-arrange themselves in a spheroidal or cluster-like configuration (Kimm *et al.*, 2016; Ricotti *et al.*, 2016). The initial half-mass radius $R_{\rm h,i}$ of this star cluster can be estimated as follows. The total energy $E_{\star,\rm tot}$ of all stars is given by

$$E = -\frac{1}{2}M_{\star,\rm tot}v^2_{\rm disk}\,. \tag{175}$$

To get this expression I have simply used the virial theorem, which states that the kinetic ($K$) and potential ($W$) energies are related via $2K + W = 0$, so that $E = -K$, or equivalently $E = W/2$. The potential energy of the cluster can be defined in terms of the so-called (initial) gravitational radius $R_{\rm g,i}$:

$$W = -\frac{GM_{\star,\rm tot}M_{\rm disk}}{R_{\rm g,i}}\,. \tag{176}$$

Since $E = W/2$ we find $R_{\rm g,i} = -GM_{\star,\rm tot}^2/2E$, or upon using Eq. (175):

$$\begin{aligned} R_{\rm g,i} &= \frac{GM_{\rm disk}}{v_{\rm disk}^2} \\ &= R_{\rm disk}\,. \end{aligned} \tag{177}$$

Thus, the initial cluster gravitational radius is simply equal to the disk radius. The gravitational radius $R_{\rm g,i}$ can now be related to the half-mass radius of the cluster, $R_{\rm cl}$, by assuming a density profile for the stars. A reasonable assumption for the stellar density profile is the King sphere. The King sphere is a steady state solution to the Fokker-Planck equation for the velocity distribution of the stars, with the result being a "lowered" Gaussian distribution, since the high-velocity tail of a full Gaussian distribution would exceed the escape velocity of the cluster (King, 1965, 2008). Observations of the local young star-forming Orion Nebula Cluster also show fairly good agreement with a King distribution, even prior to gas expulsion (Hillenbrand & Hartmann, 1998). The stellar density profile of a King sphere is qualitatively similar to an isothermal sphere with a constant-density core of radius $R_0$,[62] but with a finite radius $R_{\rm t}$ (known as the "tidal radius"), and so finite mass. The only free parameter in this case is the so-called concentration (not to be confused with the concentration $c$ of NFW DM halos):

$$c_{\rm King} \equiv \log_{10}(R_{\rm t}/R_0)\,. \tag{178}$$

A given value of $c_{\rm King}$ will relate the gravitational radius $R_{\rm g,i}$ to the (initial) cluster half-mass radius $R_{\rm cl,i}$ (see e.g. figure 4.10 on p. 310 in Binney & Tremaine, 2008). The assumed initial concentration of the cluster is fairly uncertain. We can follow the modelling of star clusters formed in atomic-cooling halos at Cosmic Dawn by Ramirez-Ruiz *et al.* (2015), who chose a ratio of core to half-mass radii of $R_0/R_{\rm cl,i} = 0.1$, which is equivalent to $c_{\rm King} \simeq 2$, for which $R_{\rm cl,i} \simeq 0.45\,R_{\rm g,i}$. The same concentration was also used to successfully model the observed Orion Nebula Cluster by Hillenbrand & Hartmann (1998). Fortunately, our result for $R_{\rm cl,i}$ is fairly insensitive to variations in $c_{\rm King}$ — one has $0.40 \lesssim R_{\rm cl,i}/R_{\rm g,i} \lesssim 0.52$ for concentrations $0.5 \lesssim c_{\rm King} \lesssim 3.4$ (see also p. 12 in Spitzer, 1987, where it is pointed out that that this holds for most realistic distributions). Using Eq. (177), we then find an initial cluster half-mass radius of

$$R_{\rm cl,i} = 0.45\,R_{\rm disk}\,. \tag{179}$$

### 5.6.2. Final cluster configuration and mass after gas expulsion

Eventually stellar feedback is expected to expel the gas within the star cluster on a time-scale short compared to the dynamical time-scale (or crossing time-scale). Hills (1980) first analyzed the case where a fraction $f_\star$ of a gas cloud is converted into stars, with the rest expelled by feedback on a time-scale short compared to the dynamical time-scale (for a nice overview see also e.g. Chapter 11.2 in Krumholz, 2015). Using the virial theorem, Hills (1980) found that the stars will remain gravitationally bound (perhaps forming a star cluster) if $f_\star > 1/2$, while for $f_\star < 1/2$ the stars will not remain bound by their own gravity after gas expulsion. This analytic result is also qualitatively borne out by simulations, but not in detail. For example, Li *et al.* (2019) recently used hydrodynamical simulations of star formation in GMCs, including momentum feedback from stars to study this issue. These authors found that the fraction of stars bound in clusters saturated at close to unity if more than 50% of the mass of the GMC is

[62] The core radius is also known as the King radius, and is defined in terms of the central stellar density $\rho_0$ and velocity dispersion $\sigma$ via (see e.g. p. 305 in Binney & Tremaine, 2008)

$$R_0 \equiv \left(\frac{9\sigma^2}{4\pi G\rho_0}\right)^{1/2}.$$

converted into stars, in reasonable agreement with Hills (1980). However, Li *et al.* (2019) also found that a smaller fraction of stars can still remain bound for star formation efficiencies $< 50\%$ since stars in natal clusters have a distribution of velocities, with only a fraction of the stars having enough energy to escape. Similar results have been found in earlier more idealized (but controlled) analytical calculations and N-body simulations of star clusters in which gas is expelled, with the exact bound fraction of stars after gas expulsion depending on the assumed initial distribution of gas and stars (e.g. Lada *et al.*, 1984; Adams, 2000; Geyer & Burkert, 2001; Baumgardt & Kroupa, 2007; Pfalzner & Kaczmarek, 2013a; Farias *et al.*, 2015, 2018). As gas and stars with sufficient energy escape their natal cluster, the cluster comprising of the bound stars are expected to expand (as expected from the virial argument by Hills, 1980). Baumgardt & Kroupa (2007) used N-body simulations to study the ratio between the final and initial half-mass radius of the cluster (i.e. $R_{\rm cl,f}/R_{\rm cl,i}$) and the fraction $f_{\rm bound}$ of bound stars as a function of $f_\star$, and their result is plotted in Figure 15. Also shown are fits to their results that we will adopt in this thesis, namely:

$$\begin{aligned} \frac{R_{\rm cl,f}}{R_{\rm cl,i}} &= e^{4.4\,(f_\star-1)^2} \\ f_{\rm bound} &= \begin{cases} 0.01\,e^{-25\,(f_\star-0.33)^2} + \left\{1 - e^{-25\,(f_\star-0.33)^2}\right\}^{1/2} & \text{if } f_\star \geqslant 0.33 \\ 0 & \text{if } f_\star < 0.33 \end{cases} . \end{aligned} \tag{180}$$

Thus, in Anaxagoras no bound cluster can be formed if $f_\star \equiv M_{\star,\rm tot}/M_{\rm disk} < 0.33$. While I have tailored these fits to the data from Baumgardt & Kroupa (2007) for application in this thesis, other studies are in good agreement with their results for equal-mass stars, regardless of the assumed initial density profile of the stars, as can be seen in Figure 15 (Lada *et al.*, 1984; Geyer & Burkert, 2001; Pfalzner & Kaczmarek, 2013b). In summary, we take the final stellar mass $M_{\star,\rm cl}$ of the cluster and its half-mass radius $R_{\rm cl,f}$ to be given by

**Predicted stellar mass ($M_{\star,\rm CL}$) and half-mass radius ($R_{\rm CL,F}$) of bound star clusters (potential GC candidates)**

$$M_{\star,\rm cl} = M_{\star,\rm tot} \times \begin{cases} 0.01\,e^{-25\,(f_\star-0.33)^2} + \left\{1 - e^{-25\,(f_\star-0.33)^2}\right\}^{1/2} & \text{if } f_\star \geqslant 0.33 \\ 0 & \text{if } f_\star < 0.33 \end{cases} \tag{181}$$

$$R_{\rm cl,f} = 0.45\,e^{4.4\,(f_\star-1)^2}\,R_{\rm disk}\,. \tag{182}$$

### 5.6.3. The fate of the unbound stars: Ultra-Faint Dwarf candidates?

A fraction $1 - f_{\rm bound}$ of the total stellar mass $M_{\star,\rm tot}$ formed will not remain bound in their natal cluster at the center of the halo. In the simulations of Ricotti *et al.* (2016) it was found that these stars expand outward into the DM halo, potentially coming to a halt if the enclosed mass (mainly from the DM) could keep the stars gravitationally bound again — these dSph, DM-dominated luminous objects are potential UFD candidates if they have low enough luminosities (and assuming we are dealing with Pop II stars). Ricotti *et al.* (2016) also note that stars with too high velocities can also escape the DM halo altogether, eventually ending up the IGM. We can make a rough estimate of the fraction $f_{\mathrm{bound,halo}}$ of stars escaping their natal cluster that will remain bound in the DM halo as follows. The RMS velocity in the natal cluster, prior to gas expulsion, is simply equal to the disk velocity $v_{\mathrm{disk}}$. Immediately after gas expulsion the escape velocity from the cluster will be $\sqrt{2f_\star}v_{\mathrm{disk}}$. If we assume a distribution $p_{\mathrm{vel}}(v_{\star,\mathrm{i}})$ for the initial velocities $v_{\star,\mathrm{i}}$ of stars in the natal cluster it follows that the fraction of stars $f_{\mathrm{bound,DM}}$ that will leave their natal cluster but remain bound in the DM halo is

$$f_{\mathrm{bound,halo}} = (1 - f_{\mathrm{bound}})\,\frac{\int_{\sqrt{2f_\star}v_{\mathrm{disk}}}^{v_{\star,\mathrm{i,esc}}} \mathrm{d}v_{\star,\mathrm{i}}\, p_{\mathrm{vel}}(v_{\star,\mathrm{i}})}{\int_{\sqrt{2f_\star}v_{\mathrm{disk}}}^{\infty} \mathrm{d}v_{\star,\mathrm{i}}\, p_{\mathrm{vel}}(v_{\star,\mathrm{i}})}\,, \tag{183}$$

where $v_{\star,\mathrm{i,esc}}$ is the initial velocity of a star in the cluster needed to escape both the cluster and the assumed NFW DM halo. This depends on both the escape velocity $v_{\mathrm{NFW,esc}}$ of the DM halo (see Appendix 11 for a derivation), and the star formation efficiency $f_\star$ due to the gravitational influence of the star cluster:[63]

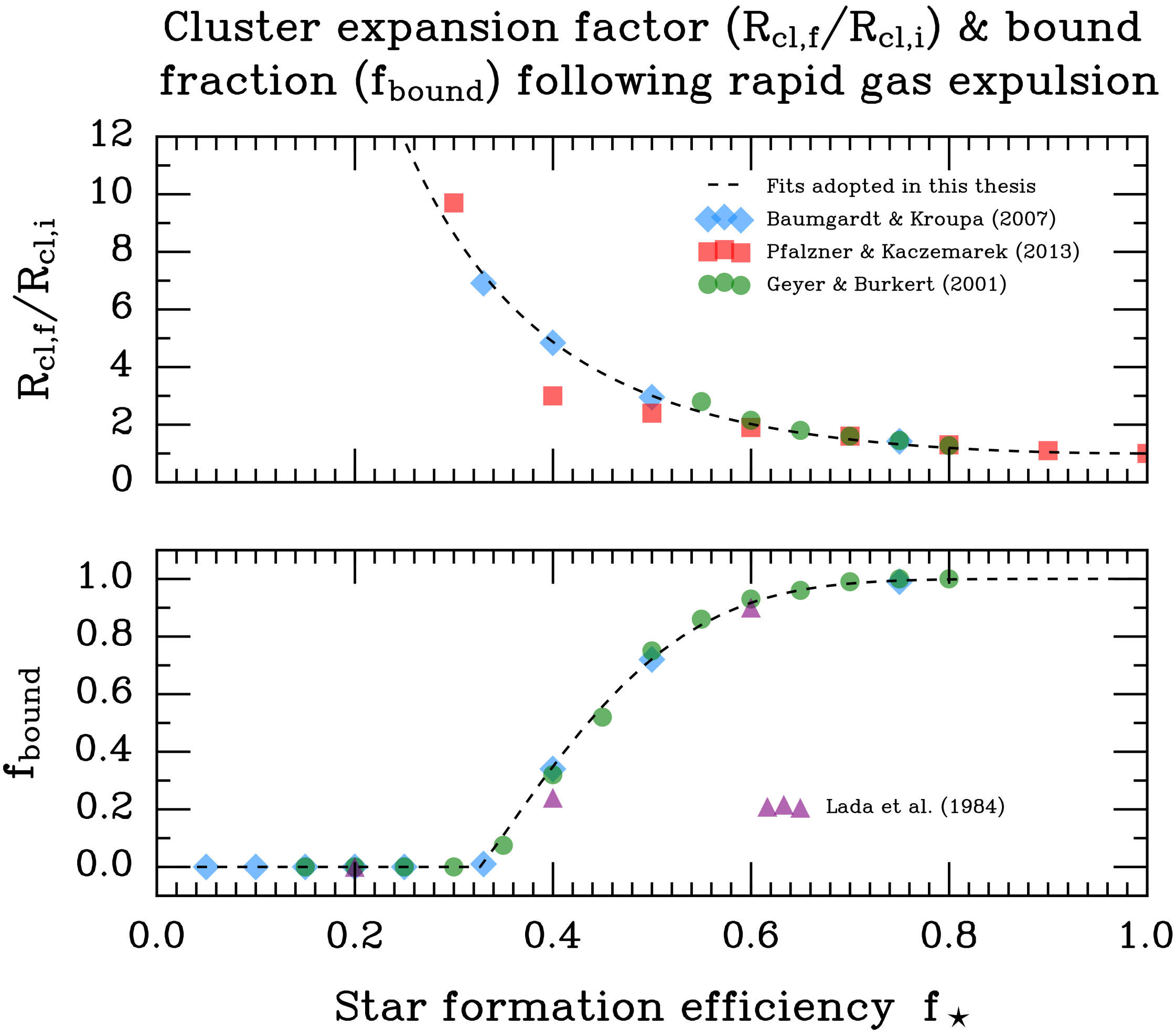

**Figure 15.** *Upper panel:* The expansion factor $R_{\mathrm{cl,f}}/R_{\mathrm{cl,i}}$ of star clusters following rapid gas expulsion as a function of the star formation efficiency $f_\star$. The symbols denote results from simulations, and the dashed line the fit adopted in this thesis. *Lower panel:* The fraction $f_{\mathrm{bound}}$ of stars that remain gravitationally bound following rapid gas expulsion as a function $f_\star$. Here too the symbols denote simulation results and the dashed line the adopted fit.

$$
\begin{aligned}
v_{\star,\mathrm{i,esc}} &= \sqrt{v^2_{\mathrm{NFW,esc}} + 2 f_\star v^2_{\mathrm{disk}}} \\
v_{\mathrm{NFW,esc}} &= \sqrt{2} v_{\mathrm{vir}} \left\{ \frac{c}{f(c)} \right\}^{1/2} \left\{ \frac{\ln(1 + 4.5\, R_{\mathrm{disk}}/R_{\mathrm{s}})}{4.5\, R_{\mathrm{disk}}/R_{\mathrm{s}}} - \frac{\ln(1+c)}{c} \right\}^{1/2} \\
f(c) &\equiv \ln(1+c) - \frac{c}{1+c} ,
\end{aligned}
\tag{184}
$$

where we have assumed that the escaping stars have velocity $v_{\star,\mathrm{i,esc}}$ at the outer (i.e. tidal) radius $R_{\mathrm{t}} = 4.5\, R_{\mathrm{disk}}$ of the natal cluster.[64] To determine the halo mass and redshift-dependent halo concentration $c = c(M, z)$ we use the following fit by Correa *et al.* (2015b), valid for $z > 4$ and all halo masses $M$:[65]

$$
\begin{aligned}
c(M, z) &= 10^\alpha \left( \frac{M}{1\ \mathrm{M}_\odot} \right)^\beta \\
\alpha &= 1.3081 - 0.1078\,(1+z) + 0.00398\,(1+z)^2 \\
\beta &= 0.0223 - 0.0944\,(1+z)^{-0.3907} .
\end{aligned}
\tag{185}
$$

The fit itself is physically motivated by the predicted mass accretion history of halos in extended Press-Schechter theory (Neistein *et al.*, 2006; Correa *et al.*, 2015a; Salcido *et al.*, 2018). Stars with $v_{\star,\mathrm{i}} > v_{\star,\mathrm{i,esc}}$ will escape the halo into the IGM, which therefore sets the upper limit of integration in Eq. (183). If we assume a nearly Maxwellian distribution of stellar velocities — in line with the theoretical expectations of a King distribution, and simulations (Farias *et al.*, 2015, 2018; Li *et al.*, 2019) — we find

$$
f_{\mathrm{bound,halo}} = (1 - f_{\mathrm{bound}}) \frac{(2/\pi)^{1/2} \left( x_{\mathrm{min}}\, e^{-x^2_{\mathrm{min}}/2} - x_{\mathrm{max}}\, e^{-x^2_{\mathrm{max}}/2} \right) - \mathrm{erf}\left( x_{\mathrm{min}}/\sqrt{2} \right) + \mathrm{erf}\left( x_{\mathrm{max}}/\sqrt{2} \right)}{(2/\pi)^{1/2} x_{\mathrm{min}}\, e^{-x^2_{\mathrm{min}}/2} - \mathrm{erf}\left( x_{\mathrm{min}}/\sqrt{2} \right) + 1} , \tag{186}
$$

where $x_{\mathrm{min}} = \sqrt{6 f_\star}$, $x_{\mathrm{max}} = \sqrt{3} v_{\star,\mathrm{i,esc}}/v_{\mathrm{disk}}$, and erf is the error function. The stellar mass of the putative dSph that remain bound by the DM halo is then $M_{\star,\mathrm{dSph}} = f_{\mathrm{bound,halo}} M_{\star,\mathrm{tot}}$. Next we proceed to estimate the half-mass radius $R_{\mathrm{dSph}}$ of this dSph. An ejected star with velocity $v_\star = \sqrt{v^2_{\star,\mathrm{i}} - 2 f_\star v^2_{\mathrm{disk}}}$ far from the cluster (but still in the central region of the halo) will reach a new final maximum radius $r_{\mathrm{max}}$ that can be found by integrating the equation of motion in an NFW DM halo, as is done in Appendix 11 (see Shull, 2014, for a calculation of the maximum radius reached in an NFW halo), and using an accurate fitting

[63] The dependence on $f_\star$ comes from the gravitational effect of the cluster itself, which slows down the ejected star, and hence increases the needed initial velocity to escape the DM halo. The relationship can be derived as follows. The energy of a star with velocity $v_{\star,\mathrm{i}}$ and mass $m_\star$ immediately following instantaneous gas expulsion is $E = \frac{1}{2} m_\star v^2_{\star,\mathrm{i}} - G M_{\star,\mathrm{tot}} m_\star / R_{\mathrm{g,i}}$. Using $G M_{\star,\mathrm{tot}} m_\star / R_{\mathrm{g,i}} = f_\star G M_{\mathrm{disk}} m_\star / R_{\mathrm{g,i}} = f_\star v^2_{\mathrm{disk}}$ (Eq. 177), this yields $E/m_\star = \frac{1}{2} v^2_{\star,\mathrm{i}} - f_\star v^2_{\mathrm{disk}}$. We want the final velocity of the star (i.e. far from the natal cluster) to be $(2E/m_\star)^{1/2} = v_{\mathrm{NFW,esc}}$, thus giving the cited result:

$$
v_{\star,\mathrm{i,esc}} = \sqrt{v^2_{\mathrm{NFW,esc}} + 2 f_\star v^2_{\mathrm{disk}}} .
$$

[64] This tidal radius follows from Eq. 179 and our chosen value of $c_{\mathrm{King}} = 2$.

[65] Correa *et al.* (2015b) give fits to the halo concentration for different assumed cosmological parameters (WMAP1, WMAP3, WMAP5, WMAP9, and Planck). We adopt their fit that assume cosmological parameters ($\Omega_{\mathrm{M}} = 0.317$, $\Omega_\Lambda = 0.683$, $h = 0.67$, $\sigma_8 = 0.834$, $n_{\mathrm{s}} = 0.962$) consistent with Planck, and most similar to the parameters chosen in this thesis.

function for $r_{\rm max}(v_\star)$, also discussed in the Appendix:

$$r_{\rm max} = \frac{2\bar{v}_\star^2 R_{\rm vir}}{c(1-1.012\,\bar{v}_\star^2+0.01815\,\bar{v}_\star^4)^{1.325}}\,, \tag{187}$$

where

$$\bar{v}_\star^2 \equiv \frac{v_\star^2}{2v_{\rm vir}^2}\frac{f(c)}{c}+1-\frac{\ln(1+4.5\,R_{\rm disk}/R_{\rm s})}{4.5\,R_{\rm disk}/R_{\rm s}}\,. \tag{188}$$

If we use the average value of $v_{\star,\rm i}$ to get a typical value of $v_\star = \sqrt{v_{\star,\rm i}^2 - 2f_\star v_{\rm disk}^2}$, and assume that the half-mass radius $R_{\rm dSph}$ of the dSph is $\sim 1/2$ of the maximum radius reached by stars, we find:

$$\begin{aligned} R_{\rm dSph} &= \frac{\bar{v}_{\star,\rm typical}^2 R_{\rm vir}}{c(1-1.012\,\bar{v}_{\star,\rm typical}^2+0.01815\,\bar{v}_{\star,\rm typical}^4)^{1.325}} \\ \bar{v}_{\star,\rm typical}^2 &\equiv \frac{v_{\star,\rm typical}^2}{2v_{\rm vir}^2}\frac{f(c)}{c}+1-\frac{\ln(1+4.5\,R_{\rm disk}/R_{\rm s})}{4.5\,R_{\rm disk}/R_{\rm s}}\,, \end{aligned} \tag{189}$$

where $v_{\star,\rm typical} = \sqrt{\langle v_{\star,\rm i}\rangle^2 - 2f_\star v_{\rm disk}^2}$, and $\langle v_{\star,\rm i}\rangle$ is the average value of $v_{\star,\rm i}$ among stars that escape the cluster but remain bound in the DM halo:[66]

$$\begin{aligned} \langle v_{\star,\rm i}\rangle &= \frac{\int_{\sqrt{2f_\star}v_{\rm disk}}^{v_{\star,\rm i,esc}} {\rm d}v_{\star,\rm i}\, v_{\star,\rm i} p_{\rm vel}(v_{\star,\rm i})}{\int_{\sqrt{2f_\star}v_{\rm disk}}^{v_{\star,\rm i,esc}} {\rm d}v_{\star,\rm i}\, p_{\rm vel}(v_{\star,\rm i})} \\ &= \frac{v_{\rm disk}}{\sqrt{3}}\,\frac{(2/\pi)^{1/2}\left\{(x_{\rm min}^2+2)\,e^{-x_{\rm min}^2/2}-(x_{\rm max}^2+2)\;e^{-x_{\rm max}^2/2}\right\}}{(2/\pi)^{1/2}\left(x_{\rm min}\,e^{-x_{\rm min}^2/2}-x_{\rm max}\,e^{-x_{\rm max}^2/2}\right)-{\rm erf}\left(x_{\rm min}/\sqrt{2}\right)+{\rm erf}\left(x_{\rm max}/\sqrt{2}\right)}\,, \end{aligned} \tag{191}$$

where again $x_{\rm min} = \sqrt{6f_\star}$ and $x_{\rm max} = \sqrt{3}v_{\star,\rm i,esc}/v_{\rm disk}$. In summary, in Anaxagoras the stellar masses and half-mass radii of faint dSph galaxies (potential UFD candidates) are estimated using

[66] For the Maxwellian velocity distribution the average value of $x = \sqrt{3}v_{\star,\rm i}/v_{\rm disk}$ of escaping stars ($v_{\star,\rm i} > \sqrt{2f_\star}v_{\rm disk}$) that remain bound in the halo ($v_{\star,\rm i} < v_{\star,\rm i,esc}$) is given by

$$\langle x\rangle = \frac{(2/\pi)^{1/2}\int_{x_{\rm min}}^{x_{\rm max}}{\rm d}x\,x^3 e^{-x^2/2}}{(2/\pi)^{1/2}\int_{x_{\rm min}}^{x_{\rm max}}{\rm d}x\,x^2 e^{-x^2/2}}\,, \tag{190}$$

where $x_{\rm max} = \sqrt{3}v_{\star,\rm i,esc}/v_{\rm disk}$ and $x_{\rm min} = \sqrt{6f_\star}$. We then have $\langle v_{\star,\rm i}\rangle = \langle x\rangle\, v_{\rm disk}/\sqrt{3}$.

**Predicted stellar mass ($M_{\rm dSph}$) and half-mass radius ($R_{\rm dSph}$) of dwarf spheroidal galaxies (potential UFD candidates)**

$$
\begin{aligned}
M_{\rm dSph} &= f_{\rm bound,halo} M_{\star,\rm tot} \\
R_{\rm dSph} &= \frac{\bar{v}^2_{\star,\rm typical} R_{\rm vir}}{c(1 - 1.012\,\bar{v}^2_{\star,\rm typical} + 0.01815\,\bar{v}^4_{\star,\rm typical})^{1.325}}\,, \\
\bar{v}^2_{\star,\rm typical} &\equiv \frac{\left\{\langle v_{\star,\rm i}\rangle^2 - 2 f_\star v^2_{\rm disk}\right\}}{2 v^2_{\rm vir}} \frac{f(c)}{c} + 1 - \frac{\ln(1 + 4.5\,R_{\rm disk}/R_{\rm s})}{4.5\,R_{\rm disk}/R_{\rm s}}\,,
\end{aligned}
\tag{192}
$$

with $f_{\rm bound,halo}$ and $\langle v_{\star,\rm i}\rangle$ given by Eqs. (186) and (191), respectively, and the halo concentration estimated using Eq. (185).

# Part IV. Applications & Results

In this section we apply Anaxagoras to study starbursts in two main scenarios:

- *Pop II star formation in low-mass halos accreted by the Milky Way:* This is the main focus in this thesis and is most amenable to observational constraints at the present. Here we will see whether the properties of observed UFDs can be reproduced, and whether GCs can form in these halos.
- *Pop III star formation in minihalos at Cosmic Dawn:* Next we consider Pop III star formation in minihalos with masses $M \sim 10^6$ $M_\odot$at Cosmic Dawn. The efficiency of star formation is investigated for the two different assumed masses of Pop III stars (25 $M_\odot$ and 140 $M_\odot$).

Below we start with the case of Pop II star formation.

# 1. Population II star formation in low-mass halos accreted by the Milky Way

## 1.1. The setup

To study Pop II star formation in low-mass halos accreted by the Milky Way I have made use of the Parkinson *et al.* (2008) halo merger algorithm, programmed into `Julia`, to estimate the number of star-forming low-mass halos a Milky Way-like host halo would accrete by $z = 0$, and their redshifts and masses when they start to form Pop II stars. For the Milky Way (MW) I have assumed a present-day ($z = 0$) halo mass of $M_{\rm MW} = 1.1 \times 10^{12}$ M$_\odot$, consistent with the recent estimate of Rodriguez Wimberly *et al.* (2021) who compared *Gaia* data of satellite galaxies with the predictions of cosmological simulations to constrain the MW halo mass to fall in the range $1 - 1.2 \times 10^{12}$ M$_\odot$. It also happens to be the estimate derived by Harris *et al.* (2017) using the GC mass-halo mass relation. The halo merger code was run with a mass resolution of $7 \times 10^5$ M$_\odot$, which is sufficient to resolve halos down to the $H_2$-cooling threshold.[67] For computational reasons only first-order subhalos were counted (i.e. subhalos of subhalos were not considered).[68] The resulting cumulative unevolved subhalo mass function is plotted in Figure 16, showing good agreement with cosmological simulations. The mass history of each subhalo is then evolved backwards in time using the same halo merger code, given the redshift the subhalo is accreted onto the MW progenitor. At the (time-dependent) position of each subhalo, a time-dependent Lyman-Werner (LW) intensity $J_{21}$ is calculated by summing the contribution from a uniform cosmological background ($J_{\rm bg,21}$) and the MW progenitor galaxy ($J_{\rm gal,21}$). The contribution from Pop III stars is expected to be sub-dominant and is therefore neglected. The uniform LW background is estimated using the cosmic star formation rate density (SFRD) for $6 < z < 12$ derived from the FIRE-2 simulations of Pop II galaxies during the EoR (Ma *et al.*, 2018), which is consistent with current observational constraints. As shown in Appendix 4, the resulting LW background is well approximated by the following expression:

$$J_{\rm bg,21}(z) = 10\, e^{-(z-6.0)/3.3}\,. \tag{193}$$

Similar LW backgrounds have been found and applied by other authors (e.g. Dijkstra *et al.*, 2008; Ahn *et al.*, 2009; Pawlik *et al.*, 2013). The LW intensity $J_{\rm gal,21}$ from the MW progenitor is estimated as follows.

Using fits from Ma *et al.* (2018) that relate the star formation rate of halos with mass $> 3 \times 10^7$ M$_\odot$ during the EoR to their halo mass and redshift, the resulting LW intensity a distance $r$ from the MW progenitor (of mass $M_{\rm MW}(z)$ at redshift $z$) is estimated to be

$$J_{\rm gal,21}(z, r) \simeq 400\, M_{\rm MW,10}^{1.58} \left(\frac{1+z}{10}\right)^{2.20} \left(\frac{r}{1~{\rm kpc}}\right)^{-2}, \tag{194}$$

where $M_{\rm MW,10} \equiv M_{\rm MW}(z)/10^{10}$ M$_\odot$ (see Appendix 4 for a derivation). The distance $r(z)$ between a halo and the MW progenitor at redshift $z$ can be estimated using the spherical collapse model in Eq. (5). In particular, if the halo is accreted at redshift $z_{\rm acc}$ onto the MW progenitor of mass $M_{\rm MW}(z_{\rm acc})$, it is expected to be at a distance

$$\begin{aligned} r(z > z_{\rm acc}) &\simeq \frac{1}{2} R_{\rm vir}[M_{\rm MW}(z_{\rm acc}),\, z_{\rm acc}] \left\{12 \sin\left(\frac{\pi t}{t_{\rm vir}}\right)\right\}^{2/3} \left\{1 - 0.237 \sin^{3/5}\left(\frac{\pi t}{t_{\rm vir}}\right)\right\} \\ \frac{t}{t_{\rm vir}} &= 0.909 \left(\frac{1+z_{\rm acc}}{1+z}\right)^{3/2}, \end{aligned} \tag{195}$$

[67]Benson *et al.* (2013) showed that the Parkinson *et al.* (2008) was able to successfully reproduce the results of the Aquarius simulations (Springel *et al.*, 2008) that had a resolution of $10^6$ M$_\odot$, and so they concluded that the algorithm is applicable to the study of very low-mass halos.

[68]It took 27.51 minutes to run the halo merger code on my PC in this case. Taking into account all subhalos of subhalos would probably multiply this figure several fold.

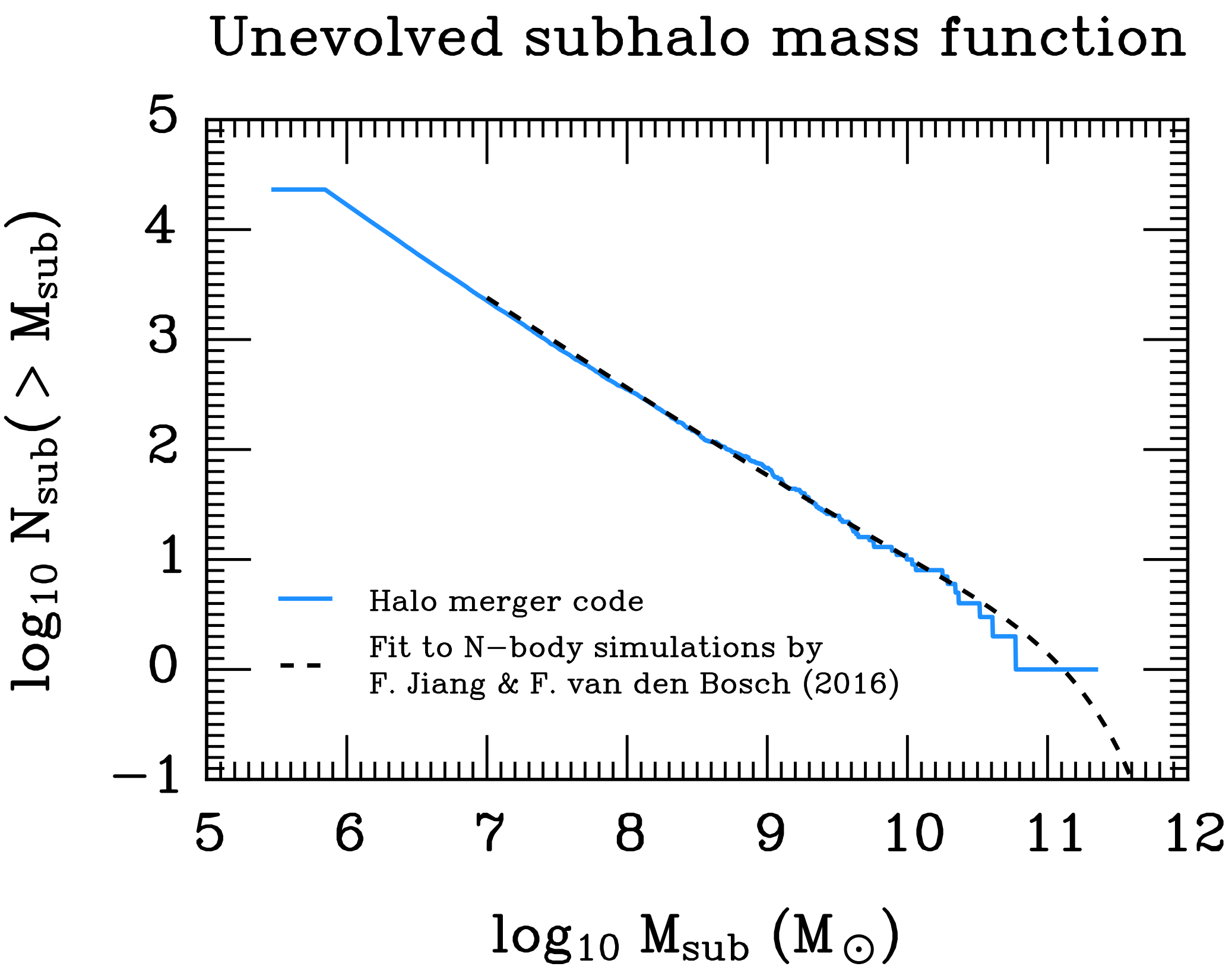

**Figure 16.** A comparison between the predicted cumulative, first-order unevolved subhalo mass function (SHMF) for the MW-like halo ($M_{\rm MW} = 1.1 \times 10^{12}$ M$_\odot$) and the fit to cosmological simulations by Jiang & van den Bosch (2016) (integrating their Eq. 14 down to a subhalo mass $M_{\rm sub} = 10^7$ M$_\odot$, beyond which some numerical problems in the integration was encountered). A good agreement between the halo merger code and cosmological simulations is clearly seen. The predicted cumulative SHMF plateaus at $7 \times 10^5$ M$_\odot$ since this was the adopted mass resolution.

where the factor 0.909 relates the virialization (or "collapse") time when $r = 0$ to the actual accretion time, defined by $r(z_{\rm acc}) \equiv R_{\rm vir}[M_{\rm MW}(z_{\rm acc}), z_{\rm acc}]$, at which point the halo has become a subhalo of the MW progenitor. The time-dependent LW intensity ($J_{21} = J_{\rm gal,21} + J_{\rm bg,21}$) can then be estimated for each low-mass halo the MW progenitor will accrete. The resulting cooling threshold in Eq. (53) then yields the minimum halo mass for star formation in that halo. The first stars in a low-mass halo is likely to be Pop III stars, at least some of which would die in energetic SNe that expel all the gas from the halo. Only after a "recovery time" $t_{\rm recov}$ will the halo be able to accrete newly enriched gas and form Pop II stars. Estimating the recovery time-scale analytically from first principle is an interesting future problem, but is well beyond the scope of this thesis. Instead we seek guidance from numerical simulations. Several studies have found that $t_{\rm recov}$ has a strong dependence on the strength of feedback from both radiation and SNe (see e.g. Greif *et al.*, 2010; Ritter *et al.*, 2012; Jeon *et al.*, 2014; Latif & Schleicher, 2020; Abe *et al.*, 2021). Massive Pop III stars in minihalos that explode as energetic pair-instability SNe typically have $t_{\rm recov} \sim 30 - 300$ Myrs (Greif *et al.*, 2010; Latif & Schleicher, 2020; Abe *et al.*, 2021), whereas less massive Pop III stars with weaker radiative feedback that explode as normal core-collapse SNe typically have $t_{\rm recov} \sim$ few $- 30$ Myrs (Ritter *et al.*, 2012; Jeon *et al.*, 2014; Abe *et al.*, 2021). As discussed in Section 5.1.1, the fiducial Pop III stellar mass in this thesis is 25 $M_\odot$, with some recent observational evidence of such Pop III stars exploding as hypernovae with energies comparable to that of pair-instability SNe. While the radiative feedback from such stars is weaker than the more massive stars that would end their lives in pair-instability SNe, Anaxagoras typically predicts that more than one such star is likely to form per minihalo. The recovery time-scale is therefore likely to be long, and we adopt a constant value $t_{\rm recov} = 100$ Myrs, as in the "slow recovery" case recently considered by Magg *et al.* (2021) in their modelling of the 21-cm signal from Cosmic Dawn. Future modelling should attempt to estimate $t_{\rm recov}$ from first principles, and its dependence on both Pop III stellar feedback and the minihalo environment wherein the Pop III stars are born. We only consider starbursts in halos that form stars at redshifts $z > 6$ (i.e. prior to reionization) and that have peak virial velocities $v_{\rm vir} < 21$ km s$^{-1}$. This is motivated by the self-similar solution in Section 3.2, which predict that halos with peak virial velocities $v_{\rm vir} \gtrsim 21$ km s$^{-1}$ can still accrete gas and form stars after reionization heats the IGM to $\simeq 2 \times 10^4$ K, and so the luminous objects produced by $z = 0$ in halos with $v_{\rm vir} \gtrsim 21$ km s$^{-1}$ are unlikely to be UFD or GC candidates that form in short bursts at $z > 6$.[69] This is also consistent with recent cosmological simulations that also find that halos with $v_{\rm vir} \lesssim 20 - 30$ km s$^{-1}$ are quenched of gas following reionization (e.g. Zhu *et al.*, 2016; Fitts *et al.*, 2017; Graus *et al.*, 2019). For the halos with peak virial velocities $v_{\rm vir} < 21$ km s$^{-1}$ that can form stars before $z > 6$ we draw random spin parameters from the distribution in Eq. (86) and then use Anaxagoras to predict the properties of the galaxies and/or star clusters produced within them, assuming a metallicity $Z/Z_\odot = 0.01$.[70]

## 1.2. Results

We find a total of 292 subhalos meeting our requirements for Pop II starbursts at $z > 6$. The halo masses $M(z_{\rm burst})$ and redshifts $z_{\rm burst}$ at the moment of the starbursts are plotted in Figure 17. Except for one early outlier, all starbursts take place at redshifts $6 < z < 15$, and in halo masses $4\times10^6$ $M_\odot < M < 5\times10^7$ $M_\odot$, centered around $M \sim 10^7$ $M_\odot$. Only one of the halos form its first Pop II stars above the atomic-cooling threshold, with most of the rest forming their stars at or slightly above the $H_2$-cooling threshold (taking LW feedback into account), indicative of the effectiveness of self-shielding of $H_2$ against the LW background. This result is inconsistent with the assumption the assumption of Kimm *et al.* (2016) that LW feedback and/or long gas recovery would suppress Pop II star formation until the atomic-cooling threshold is reached. However, our result is consistent with simulations that explicitly model LW feedback (with self-shielding) that find the first Pop II starbursts to take place when halos reach masses of $\sim 10^6$ $M_\odot$ − few $\times 10^7$ $M_\odot$ (e.g. Ricotti *et al.*, 2016; Kimm *et al.*, 2017; Latif & Schleicher, 2020; Abe *et al.*, 2021). For example, in the recent simulations of Abe *et al.* (2021), the first Pop II stars in the halos they name M8run, M9run, and M9runH form when they reach halo masses of $\sim 2\times10^7$ $M_\odot$, $\sim 2\times10^7$ $M_\odot$, and $\sim 1\times10^7$ $M_\odot$ respectively (as seen by comparing their Figures 6 and 7), in good agreement with the distribution we find.

The predicted stellar masses and half-mass radii of the luminous objects produced in these halos are plotted in Figure 18. Bound and compact ($\sim 0.1 - 5$ pc) GC-like star clusters form in 288 out of the 292 halos, with stellar masses in the range $10 - 10^6$ $M_\odot$. The less massive star clusters form in gaseous disks that are disrupted by H II feedback before the first SN go off, and so do not have enough time to attain large masses. The more massive star clusters ($\sim 4 \times 10^4 - 10^6$ $M_\odot$) on the other hand form in starbursts lasting 3.4 Myrs $< t_{\rm burst} \lesssim 5$ Myrs, and so are self-enriched by SNe. Massive star clusters with $\gtrsim 10^5$ M$_\odot$ are more likely to survive tidal disruption and two-body effects to the present-day (e.g. Baumgardt & Makino, 2003; Gnedin *et al.*, 2014; Katz & Ricotti, 2014), and in our run a total of 40 (65) GC candidates form with stellar masses $\geqslant 10^5$ M$_\odot$ ($\geqslant 5 \times 10^4$ M$_\odot$), which could explain at least a significant fraction of the observed old and metal-poor GCs. *We conclude that GCs can form at the center of DM halos at high redshifts, and the best candidates for such GCs around the Milky Way should have initial (present-day) masses* $< 10^6$ $M_\odot$ *(*$\lesssim 6 \times 10^5$ $M_\odot$*) and form in starbursts lasting* $> 3.4$ *Myrs.* This renders the non-detection of DM in the two GCs NGC 2419 and MGC1 (Conroy *et al.*, 2011; Ibata *et al.*, 2013) unsurprising since they have present-day stellar masses of $\simeq 10^6$ M$_\odot$ (Conroy *et al.*, 2011). Several candidates for GCs formed in DM halos have been found more recently (Bianchini *et al.*, 2019; Boldrini & Vitral, 2021; Carlberg & Grillmair, 2021). These candidate GCs, and the potential evidence for them having formed in DM halos, are listed in Table 6. Interestingly, all of them have present-day stellar masses $\lesssim 6 \times 10^5$ M$_\odot$, estimated initial stellar masses $< 10^6$ M$_\odot$ (Wirth *et al.*, 2021), and estimated starburst time-scales 3.4 Myrs $\leqslant t_{\rm burst} \leqslant 5$ Myrs on the basis of their internal iron spreads (Bailin, 2019; Wirth *et al.*, 2021) — all consistent with the predictions of Anaxagoras.

[69] This can be seen from Eq. (75), where the gas accretion is zero for $\beta < -0.693$. Since $\beta \equiv (1 - f_{\rm B})(v_{\rm vir}/c_{\rm s,h})^2 - 2$ this yields $v_{\rm vir} > 1.25\,c_{\rm s,h}$ as a requirement for gas accretion to be possible. After reionization the IGM is photoheated to $T_{\rm h} \simeq 2 \times 10^4$ K, with the gas being ionized and having a mean molecular weight $\mu_{\rm h} \simeq 0.6$, implying that halos with $v_{\rm vir} \lesssim 21$ km s$^{-1}$ are quenched of gas by reionization. With no inflow of new gas star formation ceases, and by $z = 0$ the galaxy and/or star cluster would only contain old stars that formed prior to reionization.

[70] Since we do not model the enrichment of the gas by Pop III stars or nearby galaxies, we need to assume a metallicity. The choice $Z/Z_\odot = 0.01$ is similar to that of typical UFDs and old GCs. Our results are not sensitive to this choice, since the dominant feedback processes in metal-poor gas are insensitive to the gas metallicity (e.g. $Z/Z_\odot = 0.001$ do not make much difference).

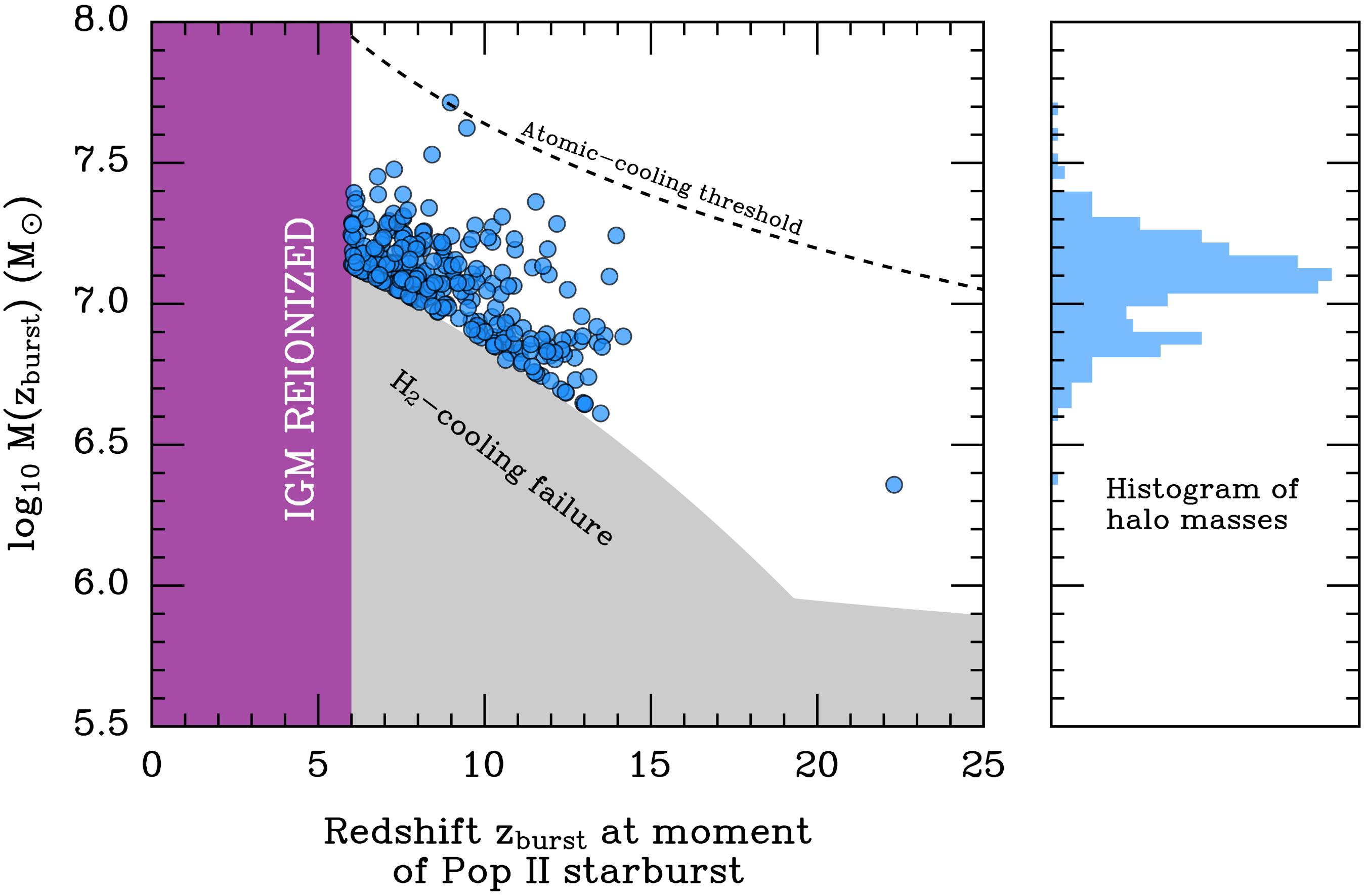

**Figure 17.** The halo masses and redshifts at the point of star formation in our run. The gray area denote the region where $H_2$-cooling becomes inefficient, and atomic-cooling becomes efficient above the dashed line. To the right the distribution of the halo masses is shown in a histogram for clarity.

More detailed studies will have to confirm that these GCs indeed reside in DM halos, which is hard

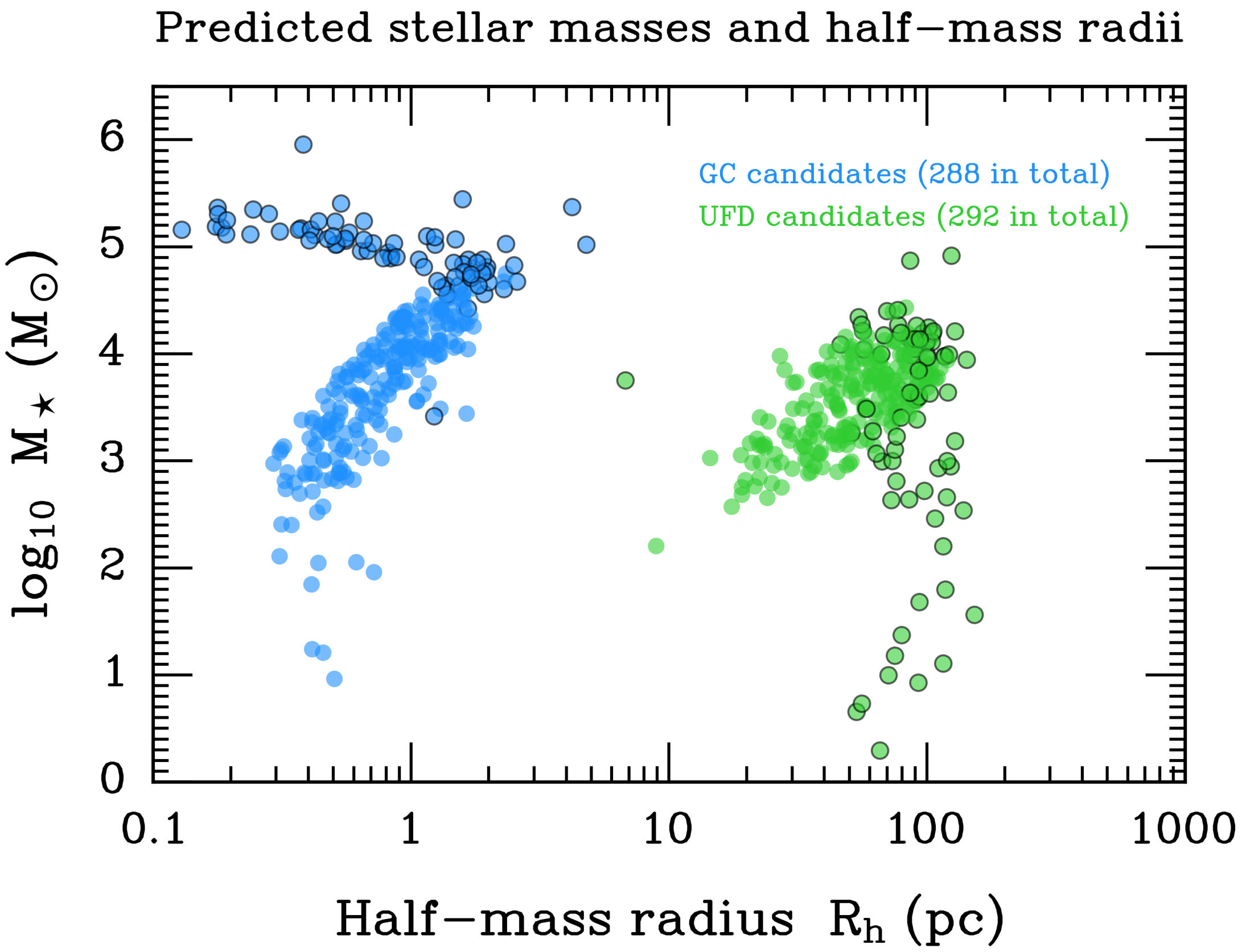

**Figure 18.** The predicted stellar masses and half-mass radii of the dSph galaxies (green symbols) and gravitationally bound star clusters (blue symbols). The symbol representing a given object has an edge if the starburst that produced it lasted longer than 3.4 Myrs (the time of the first SN explosion).

to do since the DM mass interior to the half-mass radius is expected to be completely subdominant and hard to detect. For example, for a $10^5$ M$_\odot$ GC with a half-mass radius of 3 pc born at $z = 10$ in a $M = 10^7$ M$_\odot$ NFW halo, the DM contribute only $\sim 3\%$ of the total mass inside the half-mass radius. Recent modelling by Vitral & Boldrini (2021) of GCs in minihalos indicate that DM only start to dominate the density around $\gtrsim 60$ pc from the center, corresponding to many half-light radii beyond which there are few stars to detect and use to infer the presence of DM. In all halos, some non-zero fraction of stars $0 < 1 - f_{\rm bound} \leqslant 1$ do not remain bound following gas expulsion and expand to greater radii. In this case extended dSph galaxies form, with stellar masses $M_{\rm dSph} \sim$ few $- 10^5$ M$_\odot$, with most of them having half-mass radii $R_{\rm dSph} \sim 20 - 200$ pc, in good agreement with the simulations of Ricotti *et al.* (2016). On the face of it, these are very similar to observed UFD galaxies. For a more clear comparison to observations, we also plot the V-band luminosities instead of the stellar masses in Figure 19 by adopting a V-band mass-to-light ratio of $M/L_{\rm V} = 2.59$ M$_\odot$ L$_\odot^{-1}$ for a 13.04 Gyr old $Z = 0.0004$ stellar population with a Kroupa IMF. This estimate was derived using the *Yggdrasil* spectral synthesis model (Zackrisson *et al.*, 2011), the V-band passband in the range 4700 Å $< \lambda <$ 7000 Å from Bessell (1990), and the Solar V-band luminosity of $1\,{\rm L}_\odot = 4.64 \times 10^{32}$ erg s$^{-1}$ from Binney & Merrifield (1998) (a subscript V is not used for convenience of notation here).[71] We have increased the half-mass radii of the GCs by a factor $1/(1-0.4) = 5/3$ in this

[71]The galaxy research group at the local Department of Astronomy was also consulted for comments and potential confirmation of this result. Dr. Angela Adamo found $M/L_{\rm V} = 2.64$ M$_\odot$ L$_\odot^{-1}$ assuming the same IMF and the F555W passband, which is close (but not exactly equal) to the classic V-band passband from Bessell (1990) according to Dr. Christopher Usher. In contrast, Dr. Christopher Usher found $M/L_{\rm V} = 2.15$ M$_\odot$ L$_\odot^{-1}$ for the Bessell (1990) passband, but this was derived using the FSPS model (Conroy *et al.*, 2009), which has other model ingredients (including the IMF

## TABLE 6. KNOWN CANDIDATE MILKY WAY GCs THAT COULD HAVE FORMED IN DM HALOS

| Globular Cluster† | Potential evidence of a DM† | Present-day mass⋆ ($M_\odot$) | Initial mass‡ ($M_\odot$) | Starburst time-scale‡ (Myrs) |
|---|---|---|---|---|
| NGC 6397 | No tidal tails & rising velocity dispersion | $9.66 \pm 0.13 \times 10^4$ | $2.62 \times 10^5$ | 3.5 |
| NGC 6752 | Rising velocity dispersion at large radii | $2.76 \pm 0.04 \times 10^5$ | $4.81 \times 10^5$ | 3.6 |
| NGC 362 | Flat velocity dispersion at large radii | $2.84 \pm 0.04 \times 10^5$ | $8.69 \times 10^5$ | 5.0 |
| NGC 5272 (*M 3*) | Flat velocity dispersion at large radii | $4.06 \pm 0.17 \times 10^5$ | $5.95 \times 10^5$ | 4.8 |
| NGC 4590 (*M 68*) | Flat velocity dispersion at large radii | $1.22 \pm 0.09 \times 10^5$ | $2.06 \times 10^5$ | 3.5 |
| NGC 7078 (*M 15*) | Flat velocity dispersion at large radii | $6.33 \pm 0.07 \times 10^5$ | $8.50 \times 10^5$ | 3.5 |
| NGC 1904 (*M 79*) | Flat velocity dispersion at large radii | $1.39 \pm 0.11 \times 10^5$ | $6.46 \times 10^5$ | 3.6 |
| NGC 5024 (*M 53*) | Flat velocity dispersion at large radii | $4.55 \pm 0.32 \times 10^5$ | $6.47 \times 10^5$ | 3.6 |
| NGC 6205 | Rising velocity dispersion at large radii | $5.45 \pm 0.21 \times 10^5$ | $9.58 \times 10^5$ | 4.6 |
| NGC 3201 | Flat velocity dispersion at large radii | $1.60 \pm 0.03 \times 10^5$ | $2.41 \times 10^5$ | 3.8 |

†: The case of NGC 6397 was analyzed by Boldrini & Vitral (2021), NGC 3201 by Bianchini *et al.* (2019), and the rest by Carlberg & Grillmair (2021). The GCs with the strongest evidence of DM have rising velocity dispersions at large radii.

⋆: Present-day masses are taken from the database of Prof. Holger Baumgardt, available here.

‡: Estimated initial masses and starburst time-scales come from Wirth *et al.* (2021). The initial masses were estimated by these authors on the basis of their present-day masses and the expected mass-loss rate from Baumgardt & Makino (2003)

plot to take into account adiabatic expansion due to the loss of $\sim 40\%$ of the cluster mass over 13 Gyrs from stellar evolution (Hills, 1980; Prieto & Gnedin, 2008; Shin *et al.*, 2013).

Also shown are data of confirmed and candidate Milky Way UFDs as compiled by Simon (2019) (purple stars), and data for 158 Milky Way GCs (white diamonds) derived by Baumgardt (2017); Baumgardt & Hilker (2018); Baumgardt *et al.* (2020).[72] V-band magnitudes of observed GCs and UFDs (or UFD candidates) were converted to Solar V-band luminosities using the V-band magnitude of the Sun, $M_{\odot,V} = 4.83$ (Binney & Merrifield, 1998). The properties of the UFD candidates predicted by ANAXAGORAS are in very good agreement with observations when it comes to their luminosities, half-mass radii, and the scatter in these quantities. On the other hand, most of the GC candidates predicted by ANAXAGORAS — including those with masses $> 10^5$ $M_\odot$ that are most likely to survive to the present — are smaller than observed GCs. However, this result is not unexpected: We have not modelled tidal disruption and two-body effects in these clusters upon their accretion onto the Milky Way. GCs that are born very compact are expected to expand until their half-mass radii are controlled by the local tidal field (Gieles & Baumgardt, 2008; Shin *et al.*, 2013), which would push the GC candidates predicted by ANAXAGORAS closer to observed GCs.

possibly). Thus, we adopt $M/L_{\rm V} = 2.59$ $M_\odot$ $L_\odot^{-1}$, and note that the uncertainties involved in this figure are relatively small compared to the scatter in predicted luminosities. To convert a stellar mass $M_\star$ (of a star cluster or galaxy) to a V-band luminosity $L_{\rm V}$ after 13 Gyrs of stellar evolution we then use $L_{\rm V} = f_{\rm remain} M_\star/(M/L_{\rm V})$, where $f_{\rm remain} = 0.676$ is the fraction of the stellar mass remaining after 13.04 Gyrs of stellar evolution (Zackrisson *et al.*, 2011).

[72] The structural parameters can be found online here.

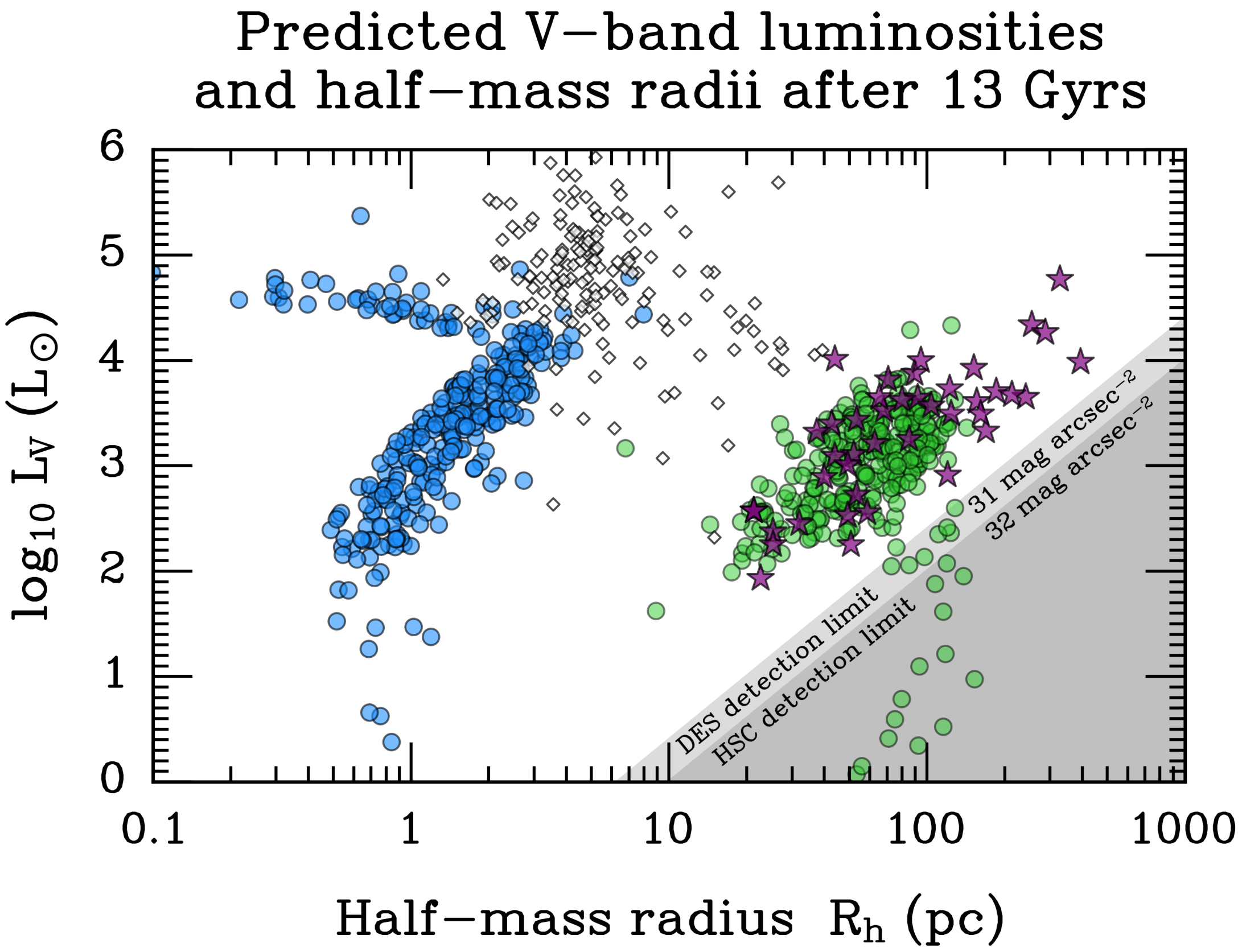

**Figure 19.** The predicted V-band luminosities and half-mass radii of the GC (blue) and UFD (green) candidates formed in the run after 13 Gyrs of stellar evolution (but not taking into account evolution in a tidal field). For comparison we also show 158 Milky Way GCs (white diamonds) and confirmed and candidate UFDs (purple stars) — see the main text for references. The gray areas represent surface brightness limits below which a given object would be too faint to be detectable. The light and dark gray areas represent the detection limits $\sim 31$ mag arcsec$^{-2}$ and $\sim 32$ mag arcsec$^{-2}$ of the Dark Energy Survey (DES) and the Hyper Suprime-Cam Strategic Survey Program (HSC), respectively, taken from Nadler *et al.* (2020). Some of the predicted UFD candidates fall into the gray area and would not be detectable by current telescopes.

The UFD candidates predicted by Anaxagoras are expected to be DM dominant due to the low stellar masses and the large half-mass radii. In Figure 20 we compare the V-band mass-to-light ratio $M/L_{\rm V}$ of predicted and observed UFDs. The predicted mass-to-light ratios for the Anaxagoras UFD candidates are in good agreement with observations, although better observational constraints are needed since the uncertainties in the derived mass-to-light ratios are large. It is also possible that we have underestimated the scatter in $M/L_{\rm V}$ since we have not taken into account the scatter in the NFW halo concentration, but rather used the median value (Eq. 185) for a given halo mass and redshift to compute the enclosed DM mass within the half-mass radius.

The formation redshifts and corresponding ages of the predicted UFD (with stellar masses $M_{\rm dSph} >$ 100 M$_\odot$) and GC candidates (with initial stellar masses $M_{\star,\rm cl} > 10^5$ M$_\odot$) are listed in Table 7. For the fiducial cosmological parameters adopted in this thesis we find that the GC and UFD candidates have ages $13.27^{+0.21}_{-0.39}$ Gyrs and $13.18^{+0.29}_{-0.31}$ Gyrs respectively (median and 95% interval). Accurate absolute ages of observed GCs and UFDs are hard to determine due to systematic errors of $\sim 1$ Gyr associated with

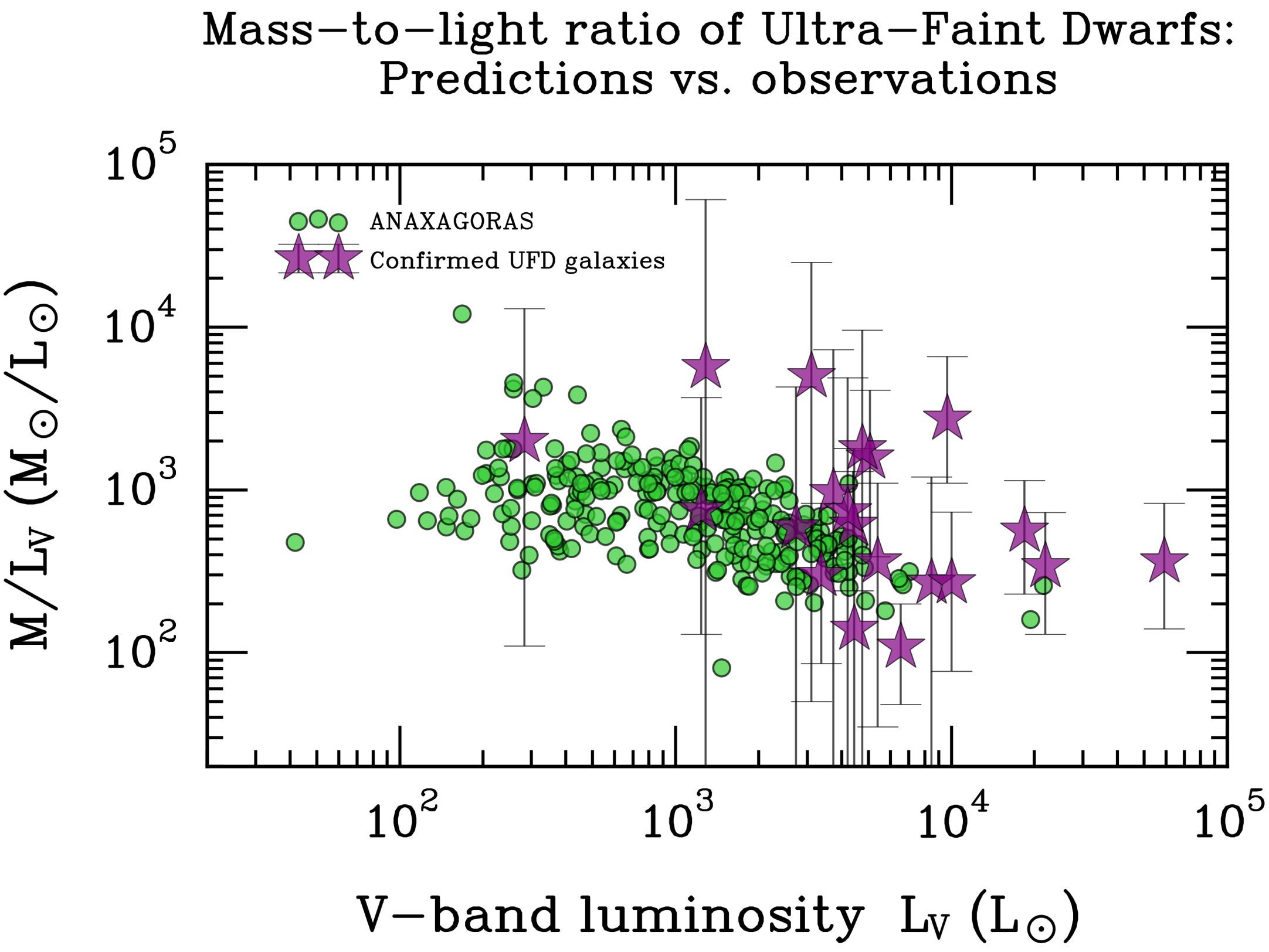

**Figure 20.** The V-band mass-to-light ratios $M/L_{\rm V}$ of predicted UFD candidates by Anaxagoras and observed confirmed UFDs (with observational data taken from the compilation of Simon, 2019). I have only plotted the UFD candidates from Anaxagoras above the surface brightness limit 31 mag arcsec$^{-2}$, roughly the detection limit of DES (see Figure 19) — candidates much fainter than this would not be detectable and should therefore not be compared with observations. Reasonable agreement is seen between observations and predictions from Anaxagoras to within observation uncertainties (the error bars correspond to approximately the 95% confidence interval).

determining distances, reddening, and stellar evolution modelling (see e.g. discussion in O'Malley *et al.*, 2017; Forbes *et al.*, 2018b; Boylan-Kolchin & Weisz, 2021). Nevertheless, the oldest and most metal-poor GCs have estimated ages centered around $12.5 - 13.5$ Gyrs (e.g. VandenBerg *et al.*, 2013; O'Malley *et al.*, 2017; Camargo & Minniti, 2019; Kruijssen *et al.*, 2019b; Jimenez *et al.*, 2019), with similar findings for observed UFD galaxies (Brown *et al.*, 2014), consistent with the predictions of our run. Wagner-Kaiser *et al.* (2017) found that six old GCs in the Large Magellanic Cloud (LMC) had the same mean age as the oldest Milky Way GCs to within $0.2 \pm 0.4$ Gyrs. This is consistent with our model prediction as well since the LW background would be very similar around the LMC, so that GCs would form in subhalos around the LMC progenitor at similar redshifts. At the time of this writing there is significant tension between the Hubble constant determined indirectly from early-time CMB observations ($h = 0.6736 \pm 0.0054$) in the context of $\Lambda$CDM, and directly from late-time observations ($h = 0.732 \pm 0.013$) (for an excellent recent review, see Di Valentino *et al.*, 2021). This has significant implications for the age of the Universe, and consequently the inferred formation redshifts of observed ancient GCs and UFDs (Boylan-Kolchin & Weisz, 2021). Because of this, we also list the predicted ages of objects if the formation redshifts remain the same as in our run,[73] but using the cosmological parameters $h = 0.710$ and $\Omega_{\rm M} = 0.302$, consistent with the

[73]The formation redshifts are expected to remain almost exactly the same since they mainly depend on the linear growth factor, and hence $\Omega_{\rm M}$ and $\Omega_\Lambda$, which remain almost the same in the Early Dark Energy scenario as in the fiducial Planck

Table 7. Formation redshifts and ages of objects predicted by Anaxagoras

| Class of objects | Formation redshifts (median & 95% interval) | Ages assuming fiducial cosmological parameters (median & 95% interval) | Ages assuming $h = 0.710$ and $\Omega_{\rm M} = 0.302$ (median & 95% interval) |
|---|---|---|---|
| GC candidates ($M_{\star,\rm cl} > 10^5$ $\rm M_\odot$) | $9.22^{+4.20}_{-3.14}$ | $13.27^{+0.21}_{-0.39}$ Gyrs | $12.74^{+0.20}_{-0.38}$ Gyrs |
| UFD candidates ($M_{\rm dSph} > 100$ $\rm M_\odot$) | $8.27^{+4.87}_{-2.22}$ | $13.18^{+0.29}_{-0.31}$ Gyrs | $12.66^{+0.28}_{-0.30}$ Gyrs |

Early Dark Energy (EDE) solution to the Hubble tension (Boylan-Kolchin & Weisz, 2021). In this case the predicted ages of GCs and UFDs are shifted down by $\sim 0.5$ Gyrs compared to the fiducial Planck cosmology. If it is determined that any ΛCDM-like cosmology (this includes EDE scenarios) *definitely* predict the formation of GCs in low-mass halos at Cosmic Dawn, and if the systematic uncertainties in the ages of observed ancient GCs could be reduced significantly, then this opens up the interesting possibility of constraining the Hubble constant by matching observed and predicted ages of GCs. Lastly, we note that the predicted number of UFD galaxies with (present-day) V-band magnitudes $M_{\rm V} < 0$ by our run (277) is similar to the inferred number of Milky Way galaxies with $M_{\rm V} < 0$: $220 \pm 50$ (68% confidence interval) (Nadler *et al.*, 2020). However, a rigorous comparison between predictions and observations in this regard would require a detailed model of tidal disruption of satellites around the Milky Way, as well as taking into account second and higher-order subhalos (i.e. satellites of accreted subhalos which would contribute to the number of satellites around the Milky Way).

cosmology.

# 2. Population III star formation in minihalos at Cosmic Dawn

Next we apply Anaxagoras to study Pop III star formation in isolated, chemically pristine minihalos in the Universe. We use the uniform cosmological LW background from Eq. (193) to compute the cooling threshold at redshifts $6 < z < 30$ in redshift increments $\Delta z \simeq 1$, and draw 200 random spin parameters for each redshift in this range. The halo masses (from the cooling threshold), redshifts, and the randomly drawn spin parameters are then fed into Anaxagoras to predict the resulting distribution of total Pop III stellar mass (and the number of Pop III stars) produced in minihalos. The result is plotted in Figure 21 for both the fiducial case of 25 $M_\odot$ stars (upper panel), as well as 140 $M_\odot$ stars (lower panel). In the fiducial case, the median number of Pop III stars formed per minihalo is $\sim 5-6$ for $z > 20$, with a 68% (95%) interval of roughly $\sim 2-15$ ($\sim 1-40$). This is in fairly good agreement with the recent sophisticated simulations of Latif *et al.* (2022) that resolved the individual formation of Pop III stars, and included their stellar feedback (including radiation pressure and photoionization feedback). Unlike the simplified setup in Anaxagoras, these authors did make any assumptions regarding the Pop III IMF, but rather *predicted* it from first principles, and find $\sim 6-23$ Pop III stars per minihalo at redshifts $20 \leqslant z \leqslant 28$. While the characteristic stellar mass in their simulation is close to that of our fiducial run (indeed, their simulation result was part of the motivation for this choice of stellar mass), they also find Pop III stars of lower masses, which could boost the number of stars per minihalo relative to our prediction. The total Pop III stellar mass produced in their minihalos, $\sim 30-160$ $M_\odot$ is also consistent with our run.

The increase in the stellar mass produced per minihalo at redshifts $z < 20$ in our run is due to the increase in the cooling threshold (as can be seen in Figure 17), which delays Pop III star formation until halos become more massive and have larger central gas accretion rates. A small fraction ($\lesssim 2\%$ or so) of rare chemically pristine halos at $z < 15$ could produce Pop III star clusters of hundreds of stars in this case, corresponding to stellar masses of $\sim$ several $\times 10^3 - 10^4$ $M_\odot$. Pop III stellar masses of this magnitude were observed in several halos in the simulations of Kimm *et al.* (2017) too, which assumed a fixed Pop III IMF but still included many important feedback processes (including radiation pressure and photoioniztaion feedback). In contrast to our fiducial run with 25 $M_\odot$ Pop III stars, the case of 140 $M_\odot$ stars lead to the formation of only a single Pop III star per minihalo (of mass $\sim 7 \times 10^5 - 9 \times 10^5$ $M_\odot$) at redshifts $z > 20$. This is due to both the strong radiative feedback from a single such massive Pop III star, and the larger gas reservoir needed to form many of them. Our finding of only a single 140 $M_\odot$ Pop III star per minihalo is consistent with the simulations of Susa *et al.* (2014) who found that in most cases, Pop III stars with masses $> 140$ $M_\odot$ are born alone. At later times ($z < 20$), it becomes possible to form $\sim 1-15$ Pop III stars (95% interval) of mass 140 $M_\odot$ in more massive chemically pristine halos.

While Pop III stars and star clusters in low-mass halos are expected to be very faint and short-lived, it is conceivable that gravitational lensing could magnify them enough to be detectable (e.g. Vikaeus *et al.*, 2021; Welch *et al.*, 2022). A key quantity to know in this case is the comoving number density of the objects in question, since a greater number density entail a greater chance of finding objects with high enough magnification. An upper bound to the comoving number density of Pop III stars $n_{\rm PopIII}$ in minihalos is plotted in Figure 22. The comoving number density was estimated using the following expression:[74]

$$n_{\rm PopIII} = \frac{\mathrm{d}}{\mathrm{d}t} n_{\rm ST}(z, M > M_{\rm th}) \times t_{\rm lifetime}\,, \tag{196}$$

where $n_{\rm ST}(z, M > M_{\rm th})$ is the comoving number density of halos with mass $M > M_{\rm th}$ at redshift $z$, and

[74]This expression can be derived from the differential equation governing the time evolution of $n_{\rm PopIII}$:

$$\frac{\mathrm{d}n_{\rm PopIII}}{\mathrm{d}t} = \frac{\mathrm{d}}{\mathrm{d}t} n_{\rm ST}(z, M > M_{\rm th}) - \frac{n_{\rm PopIII}}{t_{\rm lifetime}}\,.$$

If $n_{\rm ST}(z, M > M_{\rm th})$ varies on a time-scale longer than $t_{\rm lifetime} \lesssim \mathcal{O}(10\ \mathrm{Myrs})$, a steady state will be reached where the right-hand side is almost equal to zero, giving the stated result.

$t_{\rm lifetime}$ is the total lifetime of the Pop III star or star cluster (i.e. the time it remains luminous and detectable). To estimate the comoving number density of halos the Sheth-Tormen halo mass function (Sheth & Tormen, 2002) was adopted, which is known to be reproduce the results of cosmological simulations at high redshifts and low halo masses fairly well (e.g. Reed *et al.*, 2007; Lukić *et al.*, 2007; Sasaki *et al.*, 2014). We take $t_{\rm lifetime}$ to fall in the range 2.44 Myrs $< t_{\rm lifetime} <$ 12.14 Myrs, with the lower bound representing the lifetime of a single 140 $M_{\odot}$ Pop III star, and the upper bound being twice the lifetime of a 25 $M_{\odot}$Pop III star (as expected for a starburst lasting 6.07 Myrs, so that if several 25 $M_{\odot}$Pop III stars form, the last star to form in the burst would die $\simeq$ 12 Myrs after the first star is born). The above estimate is an upper bound since, like Stiavelli & Trenti (2010), we do not take into account external enrichment of halos,[75] which could reduce the comoving number density by at least $\sim$ few $\times$ 10% (Stiavelli & Trenti, 2010). It is seen in Figure 22 that $n_{\rm PopIII} \sim 0.1 - 10$ cMpc$^{-3}$ for $z \sim 6 - 30$, which will be used to predict observational detection rates by Dr. Erik Zackrisson at Uppsala University (E. Zackrisson, private communication, Jan. 2022).

[75]Internal enrichment can be ignored because the first stars form when the halo crosses the cooling threshold, so no internal enrichment could have taken place prior to this point.

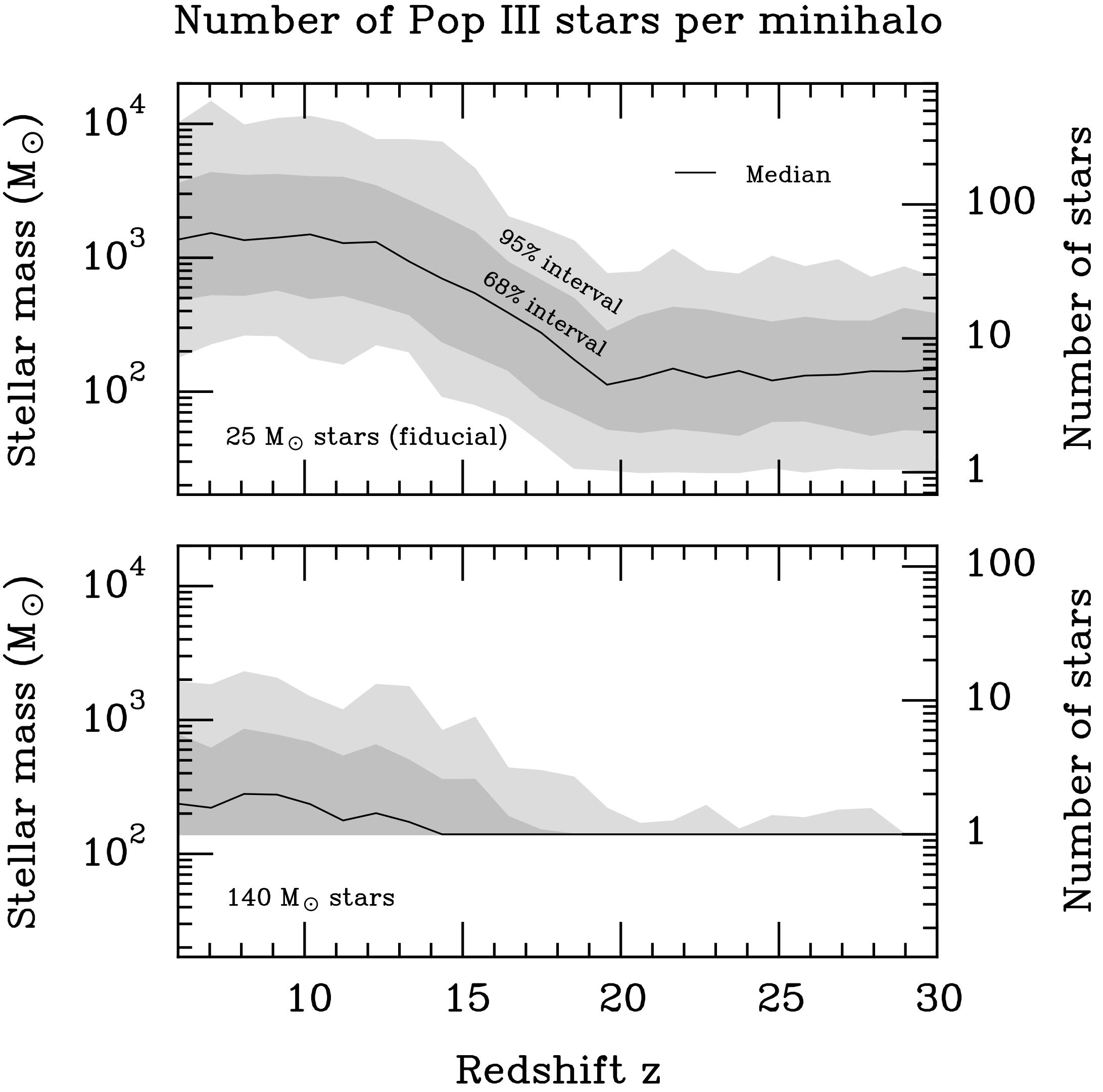

**Figure 21.** The total stellar mass of Pop III stars (and the corresponding number of Pop III stars) produced in minihalos at different redshifts. *Upper panel*: The fiducial case of 25 $M_\odot$ Pop III stars. The black line shows the median stellar mass (left $y$-axis) and the corresponding number of stars (right $y$-axis). The dark and light gray regions show the 68% and 95% intervals, respectively. *Lower panel*: Same as the upper panel but for 140 $M_\odot$ Pop III stars.

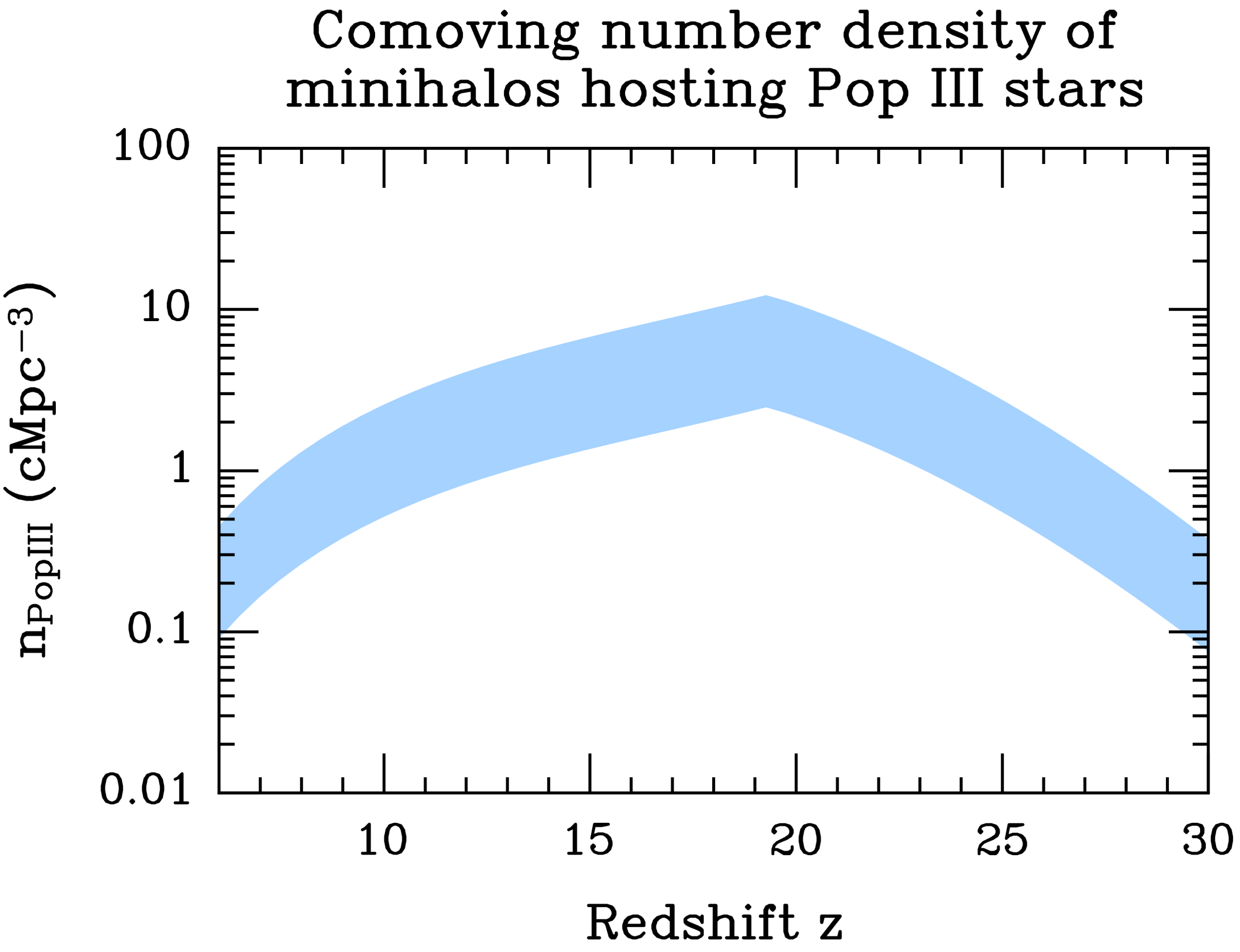

**Figure 22.** The predicted comoving number density $n_{\rm PopIII}$ of minihalos hosting luminous (i.e. detectable) Pop III stars. The blue band correspond to the range of possible lifetimes of either individual Pop III stars or star clusters (2.44 Myrs $< t_{\rm lifetime} <$ 12.14 Myrs). As discussed in the text, external metal enrichment has not been taken into account in this calculation, so this is an upper bound on $n_{\rm PopIII}$.

# Part V. Summary and discussion

In this thesis project I have developed Anaxagoras, a detailed *ab initio* analytical model of short starbursts in low-mass halos in the early Universe. Using this framework, we can make fairly robust predictions for the properties of the first star clusters and faint galaxies in the context of ΛCDM. An advantage of Anaxagoras over previous numerical simulations and models is that it is not limited by computational time, resolution, or the neglect of important stellar feedback processes. The model include new detailed derivations of the cooling threshold and central gas accretion rate in low-mass halos, a comprehensive model of stellar feedback to calculate the stellar mass formed, and a detailed calculation of the half-mass radii of the final object(s). In this thesis we applied Anaxagoras to two cases of interest:

1. *Pop II star formation at high redshifts in low-mass halos that would later accrete onto the Milky Way halo and become satellites.* We found that Anaxagoras together with a halo merger code predict the existence of a population of faint dwarf galaxies with luminosities ($L_V <$ few$\times 10^4$ $L_\odot$), half-mass radii ($\sim 10 - 200$ pc), mass-to-light ratios ($M/L_V \sim 100 -$ few $\times\, 10^3$ $M_\odot/L_\odot$), and ages ($13.18^{+0.29}_{-0.31}$ Gyrs) in good agreement with observed Ultra-Faint Dwarfs. Furthermore, in some minihalos the model predict the formation of compact star clusters with properties similar to that of old globular clusters (initial half-mass radii $\sim 0.1 - 5$ pc, initial stellar masses $10^5 - 10^6$ $M_\odot$, and ages $13.27^{+0.21}_{-0.39}$ Gyrs). Since Anaxagoras include all known relevant stellar feedback processes, this lend strong support for the idea that ΛCDM predict the formation of GCs in low-mass halos, first put forth by Peebles (1984). Intriguingly, the properties of these predicted GC-like objects are consistent with the GCs recently identified as possibly residing in dark matter halos (Bianchini *et al.*, 2019; Boldrini & Vitral, 2021; Carlberg & Grillmair, 2021).

2. *Pop III star formation in minihalos.* The flexibility of Anaxagoras allowed us to briefly study the formation of the first stars in minihalos. Depending on the characteristic masses of Pop III stars, it was found that typically only $\sim 1 - 30$ stars form per minihalo at redshifts $z > 15$. At lower redshifts the number of Pop III stars per minihalo could increase to $\lesssim$ few $\times$ 100 if the stars are not overly massive (and hence their feedback not overly strong), since Lyman-Werner feedback limit star formation to higher-mass halos with larger gas reservoirs. However, due to efficient $H_2$ self-shielding (neglected in previous work), we find that Lyman-Werner feedback is not strong enough to allow the formation of massive Pop III star clusters/galaxies in atomic-cooling halos.

While Anaxagoras is fairly comprehensive as it is, and the predictions fairly robust to small model changes, several future additions could be made to make the model even more realistic and make additional predictions. These include:

- *A more detailed calculation of the temperature of gas in the disk:* Instead of assuming a fixed (though well-motivated) disk temperature, the gas temperature could in principle be estimated from the equilibrium of cooling and heating processes in the disk. This is unlikely to have a big impact on predicted properties of galaxies and star clusters since these predictions mainly rely on the disk surface density (which is independent of gas temperature) rather than volume density. However, it is important for the broader goal of making Anaxagoras a completely *ab initio* model.

- *Estimating the metallicity distribution of stars:* Instead of assuming a fixed gas and star metallicity, an improvement would be to model the diffusion of metals and enrichment of the disk gas from supernovae. This could be used to derive the metallicity distribution of stars, and hence offer an additional way of comparing predictions to observations. The pre-enrichment of gas could also be modelled by considering starbursts in progenitor halos, which would remove the need to make any assumptions regarding the initial gas metallicity.

- *A more detailed treatment of supernova feedback:* Supernova feedback is treated quite crudely in Anaxagoras, especially for Pop III star formation. Improvements would include a more detailed consideration of whether the gas is indeed expelled from the halo after supernova remnants cover the disk, and if not, how much longer the starburst would proceed. Furthermore, to model the transition from Pop III to Pop II star formaton, the spread and diffusion of metals following the first supernovae would be important to model.
- *The gas recovery time-scale and multiple starbursts:* We have focused on single starbursts that, together with reionization feedback, subsequently shut down further star formation. However, if the feedback is not sufficiently strong (e.g. due to few supernova explosions), the gas could re-accrete onto the halo after a "recovery time-scale" (which would need to be estimated) and form stars in a new burst. One would then have to apply Anaxagoras to predict the outcome of this burst, and the characteristics of the objects formed out of multiple bursts of star formation.

A follow-up project should attempt to address this, building upon the version of Anaxagoras presented in this thesis. It is hoped that this will help in solidifying our understanding of the first stars and galaxies, as well as deriving predictions to test our current cosmological framework.

# Part VI.
# Acknowledgments

First I would like to thank Garrelt Mellema for supervising a project of this magnitude, always being there when I needed help, and for inviting me to scientific seminars and the journal club (though because of part-time work I wish I had more time for it), always making me feel welcome at, and a part of, the Oskar Klein Centre. I have also received many helpful comments and answers to questions by several other researchers. In particular, I would like to thank:

- Erik Zackrisson (Uppsala University) for many discussions concerning the formation of Pop III star clusters in low-mass halos, and the possibility of detecting them. *Yggdrasil* has also been very helpful in this project for deriving observational predictions (Zackrisson *et al.*, 2011).
- Anna T. P. Schauer (The University of Texas at Austin) for kindly providing data from cosmological simulations she and her co-workers have ran to study the cooling threshold in minihalos, as well as discussing their results. This data was key in verifying the analytical results derived in this thesis.
- Philip F. Hopkins (Caltech) for answering questions regarding the light-to-mass ratios (and hence radiation pressure) in different radiation bands from the FIRE-2 simulations.
- Angela Adamo and Chris Usher (both at Stockholm University) for checking my calculation of the V-band light-to-mass ratio, and other discussions related to globular clusters.
- Taysun Kimm (Yonsei University) for answering questions I had, well before the start of this thesis, regarding their seminal paper on globular cluster formation in atomic-cooling halos (Kimm *et al.*, 2016). It is this work that sparked my interest in the formation of star clusters in low-mass dark matter halos at Cosmic Dawn, and the intricate and interesting stellar feedback loops at play.

# Part VII. Appendix

## 1. The recombination coefficient $k_1$

As discussed in Section 3.1.1, the recombination rate in the gaseous constant-density core of a minihalo will determine how efficiently molecular hydrogen can form, and hence whether it can cool efficiently or not. It was also noted that the relevant recombination coefficient to use is the Case-B recombination coefficient, since the core is optically thick at the Lyman limit. Draine (2011) (p. 139, Eq. 14.6) gives the following fit for the Case-B recombination coefficient, valid for 30 K $< T < 3 \times 10^4$ K:

$$k_1 \simeq 2.54 \times 10^{-13}\, T_4^{-0.8163-0.0208 \ln T_4}\ \mathrm{cm}^3\ \mathrm{s}^{-1}\,, \tag{197}$$

where $T_4 \equiv T/10^4$ K. Padmanabhan (2000) (p. 323) arrives at the following approximate formula for the Case-B recombination coefficient using quantum mechanical calculations for the photoionization cross-section and then using detailed balance (along with an approximate evaluation of the sum over energy levels $n$ in the hydrogen atom):

$$k_1 \simeq 5.2 \times 10^{-14}\, x_2^2 \left\{ x_2 e^{x_2} E_1(x_2) \left( \frac{1}{2} + \frac{1}{x_2} \right) + \gamma + \ln x_2 \right\}\ \mathrm{cm}^3\ \mathrm{s}^{-1}\,, \tag{198}$$

where $x_2 = 39500$ K$/T$, $\gamma \simeq 0.5772$ is the Euler–Mascheroni constant, and $E_1(x) \equiv \int_x^\infty \mathrm{d}y\, e^{-y}/y$ is the exponential integral. In Figure 23 we compare these fits to the fit adopted in this thesis, and the fit adopted by Tegmark *et al.* (1997) (see Table 1). Also shown is a fit to the Case-A recombination coefficient by Draine (2011) (p. 139, Eq. 14.5). It is clearly seen that the fit adopted by Tegmark *et al.* (1997) overestimates the relevant (i.e. Case-B) recombination coefficient, whereas the fit to the Case-B recombination coefficient adopted in this thesis shows good agreement with Draine (2011) and Padmanabhan (2000). Tegmark *et al.* (1997) used a rate coefficient by Hutchins (1976), who in turn had derived it from fitting the Case-A recombination coefficient by Spitzer (1956) (pp. 91-92). As can be seen in Figure 23, the fit used by Tegmark *et al.* (1997) is much closer to the Case-A recombination coefficient from Draine (2011).

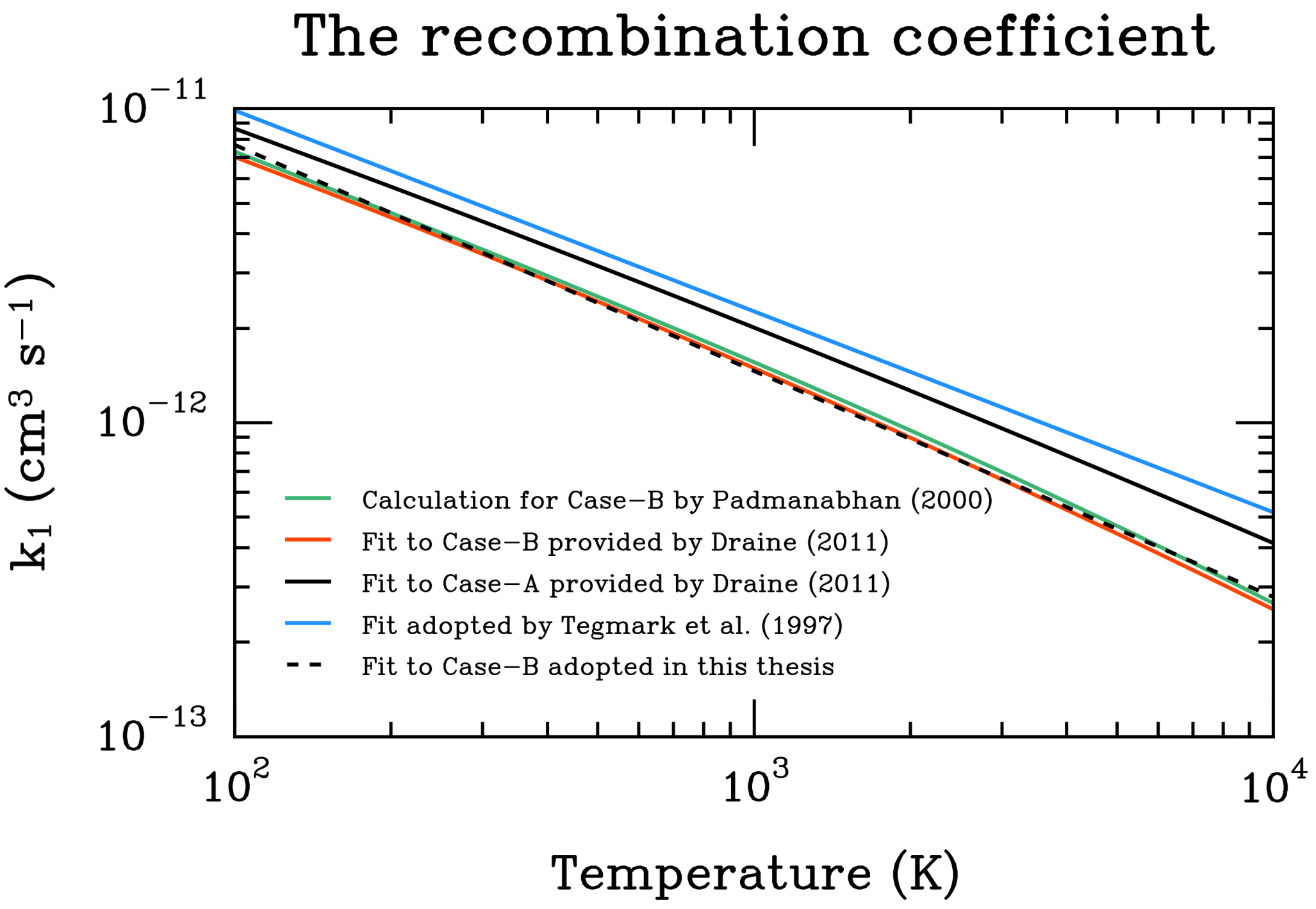

**Figure 23.** A comparison of fits to either the Case-A or Case-B recombination coefficients. For applications to the dense gas in minihalos prior to efficient cooling one should adopt the Case-B recombination coefficient, for which our power law fit in Table 1 is an excellent approximation. In contrast, the fit used by Tegmark *et al.* (1997) is seen to overestimate the recombination rate significantly, probably because they use a fit that derives from the Case-A recombination coefficient given by Spitzer (1956). It is indeed seen that the fit used by Tegmark *et al.* (1997) is closer in value to the newer fit to the Case-A recombination coefficient given by Draine (2011).

# 2. The RMS linear overdensity $\sigma(M)$ in ΛCDM: A convenient approximation

The RMS linear overdensity $\sigma(M) = \sqrt{\langle \delta_M^2 \rangle}$ on a mass-scale $M = 4\pi R^3 \bar{\rho}_{\mathrm{M},0}/3$ (corresponding to a comoving radius $R$) is given by (see e.g. Peebles, 1993; Peacock, 1999; Barkana & Loeb, 2001; Tegmark *et al.*, 2006)

$$\sigma^2(M) \equiv \int_0^\infty \frac{\mathrm{d}k}{2\pi^2} k^2 P(k) \left|\tilde{W}(kR)\right|^2 , \tag{199}$$

where $P(k) \equiv \left\langle |\delta_{\boldsymbol{k}}|^2 \right\rangle$ is the power spectrum of linear density fluctuations, and $\tilde{W}(kR)$ is the Fourier transform of the window function $W(\boldsymbol{x})$. Usually one considers spherical top-hat density fluctuations, for which

$$W(\boldsymbol{x}) = \begin{cases} 1 & \text{if } |\boldsymbol{x}| < R \\ 0 & \text{else} \end{cases} . \tag{200}$$

The corresponding Fourier transformed window function is $\tilde{W}(kR) = 3[\sin(kR) - kR\cos(kR)]/(kR)^3$. Qualitatively we see that

$$\left|\tilde{W}(kR)\right|^2 \simeq \begin{cases} 1 & \text{if } k \lesssim R^{-1} \\ \\ 0 & \text{if } k \gtrsim R^{-1} \end{cases} . \tag{201}$$

Thus, (for reasonable power spectra that do not vary faster than $\left|\tilde{W}(kR)\right|^2$) the RMS linear overdensity is roughly given by

$$\sigma^2(M) \simeq \int_0^{R^{-1}} \frac{\mathrm{d}k}{2\pi^2}\, k^2 P(k)\,. \tag{202}$$

The power spectrum can be written as $P(k) = P_{\rm prim}(k)T^2(k)$, where $P_{\rm prim}(k)$ is the primordial power spectrum (e.g. due to inflation), and the so-called transfer function $T(k)$ encapsulates the scale-dependent subsequent evolution of density perturbations. Quantum fluctuations in the inflaton scalar field driving inflation is expected to give $P_{\rm prim}(k) \propto k^{n_{\rm s}}$, with $n_{\rm s}$ (the spectral index) very close to but sligthly smaller than unity (e.g. Bardeen *et al.*, 1983; Guth & Pi, 1985; Peebles, 1993; Tegmark, 2005). This is borne out by recent CMB observations, which find $n_{\rm s} = 0.965 \pm 0.004$ (Planck Collaboration *et al.*, 2020). Density fluctuations evolve differently depending on if the scale of the perturbation is smaller or larger than the horizon (sub-horizon or super-horizon scales, respectively). Furthermore, the time-dependence of density fluctuations is different during the radiation-dominated epoch compared to the matter-dominated epoch. This selects the horizon scale at matter-radiation equality as a key size scale. The approximate form of $T(k)$ in the context of CDM can be derived as follows (see e.g. Peacock, 1999; Padmanabhan, 2002, 2006):

1. First consider a dark matter density fluctuation on a comoving scale $k \gg 1/R_{\rm eq}$, where $R_{\rm eq} \equiv R_{\rm H}(a_{\rm eq})$ is the comoving horizon radius at matter-radiation equality. Shortly after inflation (at some scale factor $a_{\rm infl}$) any relevant scale $k$ will be a super-horizon scale since the nearly constant horizon radius during inflation $\simeq cH_{\rm infl}^{-1}$ (where $c$ is the speed of light) is microscopically small because $H_{\rm infl}$, the Hubble parameter during inflation, is extremely large. Density fluctuations on super-horizon scales grow as $\delta \propto 1/\bar{\rho}a^2$ where $\bar{\rho}$ is the mean density of the Universe (e.g. Peebles, 1993; Padmanabhan, 2002). During radiation-domination we have $\bar{\rho} \propto a^{-4}$ and so $\delta \propto a^2$. After inflation, but before matter domination, the horizon grows in size to eventually become larger than $k^{-1}$ at some scale factor $a_{\rm enter}$. During this time the density fluctuation grow only logarithmically, $\delta \propto \ln a$. Finally, after matter-radiation equality the density fluctuation grow as $\delta \propto a$ before dark energy starts to dominate. Summarizing all of this, at some scale factor $a > a_{\rm eq}$ we roughly have

$$\begin{aligned} \delta_{\boldsymbol{k}}(a) &\simeq \delta_{\boldsymbol{k},{\rm prim}} \left(\frac{a_{\rm enter}}{a_{\rm infl}}\right)^2 \ln\left(\frac{a_{\rm eq}}{a_{\rm enter}}\right) \frac{a}{a_{\rm eq}} \\ &= \delta_{\boldsymbol{k},{\rm prim}} \left(\frac{a_{\rm eq}}{a_{\rm infl}}\right)^2 \left(\frac{a_{\rm eq}}{a_{\rm enter}}\right)^{-2} \ln\left(\frac{a_{\rm eq}}{a_{\rm enter}}\right) \frac{a}{a_{\rm eq}} \qquad \text{for } k \gg 1/R_{\rm eq}\,. \end{aligned} \tag{203}$$

   where $\delta_{\boldsymbol{k},{\rm prim}}$ is the primordial amplitude of the density fluctuation on the scale $k$, the RMS value of which is given by $P_{\rm prim}^{1/2}(k)$. For the radiation-dominated era we have $k \sim 1/R_{\rm H}(a_{\rm enter}) = H_0\sqrt{\Omega_{\rm M}}a_{\rm eq}^{1/2}/ca_{\rm enter}$,[76] so that $a_{\rm enter} \sim H_0\sqrt{\Omega_{\rm M}}a_{\rm eq}^{1/2}/ck$ and $a_{\rm eq}/a_{\rm enter} \sim cka_{\rm eq}^{1/2}/H_0\sqrt{\Omega_{\rm M}}$. Further-

[76] This is calculated from

$$R_{\rm H}(a) = c\int_0^{t(a)} \frac{\mathrm{d}t'}{a(t')} = c\int_0^a \frac{\mathrm{d}a'}{H(a')a'^2}\,.$$

During radiation domination we have $H(a) \simeq H_0\sqrt{\Omega_{\rm R}}a^{-2}$, and so $R_{\rm H}(a \ll a_{\rm eq}) \simeq ca_{\rm enter}/H_0\sqrt{\Omega_{\rm R}}$. Since $a_{\rm eq} = \Omega_{\rm R}/\Omega_{\rm M}$ where $\Omega_{\rm R}$ is the radiation density parameter, this yields $R_{\rm H}(a \ll a_{\rm eq}) \simeq ca_{\rm enter}/H_0\sqrt{\Omega_{\rm M}}a_{\rm eq}^{1/2}$.

more, $a_{\rm eq}^{1/2} = R_{\rm eq} H_0 \sqrt{\Omega_{\rm M}}/2(\sqrt{2}-1)c$ so that $a_{\rm eq}/a_{\rm enter} \sim kR_{\rm eq}$.[77] Thus, we expect that the RMS value of $\delta_{\boldsymbol{k}}$ during matter-domination is given by

$$\delta_{\boldsymbol{k}}(a) \simeq \mathcal{C} P_{\rm prim}^{1/2}(k) \, \frac{\ln[\eta k R_{\rm eq}]}{[\eta k R_{\rm eq}]^2} \, \frac{a}{a_{\rm eq}} \qquad \text{for } k \gg 1/R_{\rm eq} \, , \tag{204}$$

for some constants $\mathcal{C}$ and $\eta$.

2. Next consider a dark matter density fluctuation on a comoving scale $k \ll 1/R_{\rm H}(z_{\rm eq})$. In this case the perturbation (unless it is larger than the current horizon) is expected to enter the horizon during the matter-dominated era. As before, we have $\delta \propto 1/\bar{\rho} a^2 \propto a^2$ on super-horizon scales during the radiation-dominated era, and then $\delta \propto a$ during the matter-dominated era, regardless of whether the fluctuation has entered the horizon or not. Thus, in this case we have

$$\delta_{\boldsymbol{k}}(a) \simeq \delta_{\boldsymbol{k},\rm prim} \left( \frac{a_{\rm eq}}{a_{\rm infl}} \right)^2 \frac{a}{a_{\rm eq}} \qquad \text{for } k \ll 1/R_{\rm H}(z_{\rm eq}) \, . \tag{205}$$

This yields a RMS value of

$$\delta_{\boldsymbol{k}}(a) \simeq \mathcal{C} P_{\rm prim}^{1/2}(k) \, \frac{a}{a_{\rm eq}} \qquad \text{for } k \ll 1/R_{\rm H}(z_{\rm eq}) \, . \tag{206}$$

Using the above results, and the definition $P(k) = P_{\rm prim}(k) T^2(k)$, we arrive at the following transfer function for cold dark matter density fluctuations:

$$T(k) \propto \begin{cases} [\eta k R_{\rm eq}]^{-2} \ln[\eta k R_{\rm eq}] & \text{for } k \gg 1/R_{\rm eq} \\ \\ 1 & \text{for } k \ll 1/R_{\rm eq} \end{cases} \, . \tag{207}$$

The mass $M_{\rm eq} = 4\pi R_{\rm eq}^3 \bar{\rho}_{\rm M,0}/3$ within the horizon radius at matter-radiation equality $R_{\rm eq} = 2(\sqrt{2} - 1) a_{\rm eq}^{1/2} c/H_0\sqrt{\Omega_{\rm M}}$ is

$$M_{\rm eq} = 2.4 \times 10^{17} \left( \frac{\Omega_{\rm M} h^2}{0.14} \right)^{-1/2} \left( \frac{1+z_{\rm eq}}{3400} \right)^{-3/2} \, \mathrm{M}_\odot \, , \tag{208}$$

where I have normalized to $a_{\rm eq}^{-1} = 1 + z_{\rm eq} \simeq 3400$, the matter-radiation equality redshift found by Planck (Planck Collaboration *et al.*, 2020). Since galaxies have masses $\ll M_{\rm eq}$ (i.e. $k \gg 1/R_{\rm eq}$), let us consider the behaviour of $\sigma^2(M)$ in this regime.

[77] The horizon radius at matter-radiation equality is given by

$$R_{\rm eq} = c \int_0^{a_{\rm eq}} \frac{{\rm d}a}{H(a) a^2} = \frac{c}{H_0} \int_0^{a_{\rm eq}} \frac{{\rm d}a}{\sqrt{\Omega_{\rm M} a + \Omega_{\rm R}}} \, ,$$

where I have used $H(a) \simeq H_0 \sqrt{\Omega_{\rm M} a^{-3} + \Omega_{\rm R} a^{-4}}$. Evaluating the integral yields

$$R_{\rm eq} = \frac{c}{H_0} \left\{ \frac{2\sqrt{\Omega_{\rm M} a_{\rm eq} + \Omega_{\rm R}}}{\Omega_{\rm M}} - \frac{2\sqrt{\Omega_{\rm R}}}{\Omega_{\rm M}} \right\} \, .$$

Using $a_{\rm eq} = \Omega_{\rm R}/\Omega_{\rm M}$ then simplify this to $R_{\rm eq} = 2(c/H_0)(\sqrt{2}-1)\sqrt{\Omega_{\rm R}}/\Omega_{\rm M} = 2(c/H_0)(\sqrt{2}-1)(a_{\rm eq}/\Omega_{\rm M})^{1/2}$.

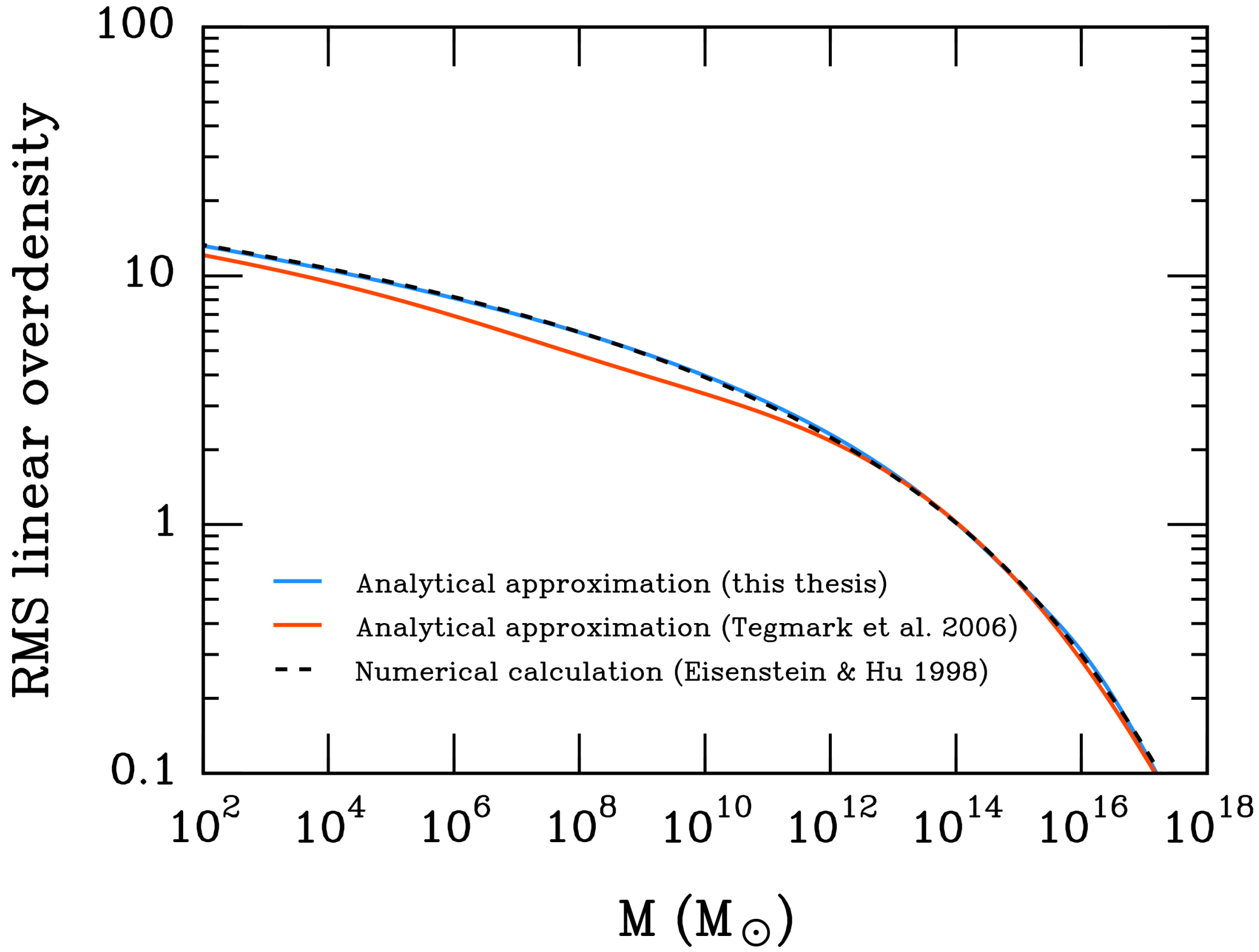

**Figure 24.** A comparison between the RMS linear overdensity $\sigma(M)$ computed numerically from the CDM transfer function in Eisenstein & Hu (1998) (with $n_{\rm s} = 0.965$, $\sigma_8 = 0.811$) and the analytical approximations provided in Eq. (211) and by Tegmark *et al.* (2006). It is seen that Eq. (211) reproduces the numerical result with greater accuracy than the approximation of Tegmark *et al.* (2006).

Using $n_{\rm s} \simeq 1$ and the transfer function in Eq. (207) and integrating Eq. (202) up to some radius $R$ yields

$$\begin{aligned} \sigma^2(M \ll M_{\rm eq}) &\propto \text{const.} + \int^{\kappa} {\rm d}\kappa' \, \kappa'^3 \times \frac{\ln^2(\kappa')}{\kappa'^4} \\ &= \text{const.} + \int^{\ln\kappa} {\rm d}\ln\kappa' \, \ln^2(\kappa') \\ &= \text{const.} + \ln^3(\kappa) \, , \end{aligned} \tag{209}$$

where $\kappa \equiv \eta R_{\rm eq}/R = (\eta^3 M_{\rm eq}/M)^{1/3}$, and so for $M \ll M_{\rm eq}$ we expect that the RMS linear overdensity in $\Lambda$CDM scale as

$$\sigma^2(M \ll M_{\rm eq}) \propto \text{const.} - \ln^3(M/\eta^3 M_{\rm eq}) \, . \tag{210}$$

Motivated by this asymptotic scaling, after some experimentation the following approximation for $\sigma(M)$

(at $z = 0$) was found to reproduce the numerical calculation of $\sigma(M)$ fairly well:

$$\begin{aligned} \sigma(M) &\simeq \mathcal{N}\left\{\frac{36}{1+3\bar{M}} - \ln^3\left(\frac{\bar{M}}{1+\bar{M}}\right)\right\}^{1/2} \\ \mathcal{N} &= 0.0845\,\sigma_8 \\ \bar{M} &\equiv \frac{8M}{M_{\rm eq}}\,. \end{aligned} \tag{211}$$

In Figure 24, Eq. (211) is compared with a direct numerical integration of Eq. (199) using the CDM transfer function from Eisenstein & Hu (1998) with the fiducial value $n_{\rm s} = 0.965$, showing good agreement between the two over a vast range of mass scales, including the range of halo masses of interest to us ($\sim 10^5 - 10^8\ {\rm M}_\odot$). Also shown is the analytical approximation provided by Tegmark *et al.* (2006), which is seen to be significantly less accurate than the approximation in Eq. (211). The advantage of the approximation in Eq. (211) is that it allow us to derive both analytical results (e.g. for the halo growth rate), as well as speed up numerical calculations that rely on $\sigma(M)$ (e.g. in the halo merger code).

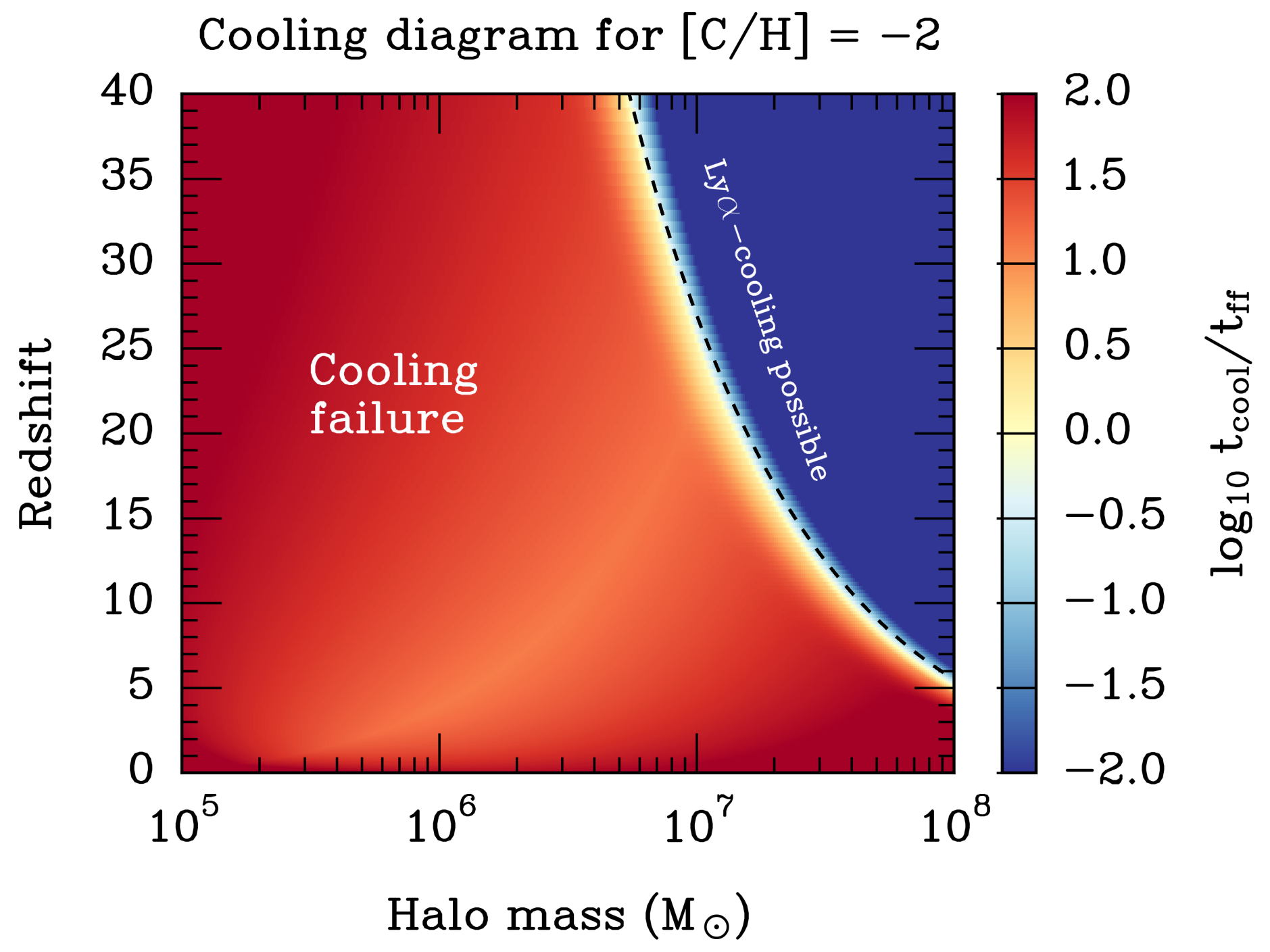

**Figure 25.** A cooling diagram showing the ratio $t_{\rm cool}/t_{\rm ff}$ as a function of redshift and halo mass, assuming only (atomic) metal and Lyman-$\alpha$ cooling. Efficient cooling (and hence star formation) is expected when $t_{\rm cool}/t_{\rm ff} < 1$. The black dashed line show the atomic-cooling threshold from Eq. (52), which correctly reproduce the dividing line $t_{\rm cool}/t_{\rm ff} = 1$ for Lyman-$\alpha$ cooling. It is seen that metal cooling is inefficient ($t_{\rm cool}/t_{\rm ff} > 1$) in halos below the atomic-cooling threshold.

## 3. Metal-cooling halos?

In the main text we derived the cooling threshold halo mass assuming only $H_2$ and Lyman-$\alpha$ cooling. In this Appendix we justify the neglect of metal cooling. Both the density and metallicity in primordial halos is so low that we can assume that the metals not locked up in dust grains is in the form of atoms rather than molecules like CO, OH, and $H_2O$ (see e.g. the discussion in Glover & Jappsen, 2007). The cooling rate due to atomic fine-structure metal lines is approximately given by (Hopkins *et al.*, 2022):

$$\Lambda_{Z,\rm atomic}(T,n) = 10^{-27} z_{\rm C} \left\{ \left(0.47\, T^{0.15} + 4890\, xT^{-0.5}\right) e^{-91.211/T} + 0.0208\, e^{-23.6/T} \right\} n_{\rm H}^2 \ {\rm erg\ cm^{-3}\ s^{-1}}\,, \quad (212)$$

where the dominant term proportional to $e^{-91.211/T}$ is due to cooling via the [C II]-158 $\mu$m line (corresponding to a photon energy $\Delta E/k_{\rm B} = 91.211$ K), the sub-dominant term proportional to $e^{-23.6/T}$ comes from [C I]-609 $\mu$m line cooling (with $\Delta E/k_{\rm B} = 23.6$ K) (see e.g. Table 8 in Hollenbach & McKee, 1989), and $z_{\rm C} \equiv 10^{[{\rm C/H}]}$. For a gas of (nearly) primordial composition, the hydrogen number density is $n_{\rm H} \simeq 0.926\, n$. For the ionization fraction $x$ we take it to be the maximum of the residual primordial value ($2.33 \times 10^{-4}$) and the value in collisional ionization equilibrium (Eq. 49). We take the total cooling rate to be $\Lambda(T,n) = \Lambda_{Z,\rm atomic}(T,n) + \Lambda_{\rm Ly\alpha}(T,n)$, where $\Lambda_{\rm Ly\alpha}(T,n)$ is the Lyman-$\alpha$ cooling rate (Eq. 48), and we neglect $H_2$-cooling for the purpose of isolating the importance of metal cooling in minihalos. The cooling time-scale is computed using $t_{\rm cool} = \frac{3}{2} n_{\rm core} k T_{\rm vir}/\Lambda(T_{\rm vir}, n_{\rm core})$ (see Eq. 9), where $n_{\rm core}$ is the central gas number density (Eq. 15) and $T_{\rm vir}$ the virial temperature of the halo (Eq. 6). We then plot the ratio $t_{\rm cool}/t_{\rm ff}$ as a function of halo mass and redshift, assuming a carbon abundance $[{\rm C/H}] = -2$ (as implicitly assumed in our Anaxagoras runs). The result is shown in Figure 25, wherein we see that metal cooling is too inefficient (i.e. $t_{\rm cool}/t_{\rm ff} > 1$) to lead to gas collapse and star formation in minihalos below the atomic-cooling threshold (Eq. 52). This justify our neglect of metal-cooling.

# 4. Lyman-Werner flux from high-redshift galaxies

## 4.1. Lyman-Werner intensity from a single high-redshift galaxy

Here we want to estimate the Lyman-Werner (LW) intensity at a given distance $r$ from a galaxy with specific luminosity $L_\nu$. This will be needed to assess whether nearby galaxies can suppress $H_2$-cooling in minihalos. We begin by noting that that since the flux is related to the intensity via $F_\nu = 4\pi J_\nu$, we have

$$J_{\mathrm{gal},\nu} = \frac{L_{\mathrm{gal},\nu}}{16\pi^2 r^2}\,. \tag{213}$$

Following Latif & Khochfar (2019) we can model the spectrum in the LW band (11.2 eV $< h\nu <$ 13.6 eV) as a black-body with temperature $T_\star = 2\times10^4$ K. If the total luminosity in the LW band is $L_{\mathrm{LW}}$ we then have

$$L_{\mathrm{gal},\nu} = L_{\mathrm{LW}}\,\frac{B_\nu(T_\star)}{\int_{(11.2\ \mathrm{eV})/h}^{(13.6\ \mathrm{eV})/h}\mathrm{d}\nu\, B_\nu(T_\star)}\,, \tag{214}$$

where $B_\nu$ is the Planck function. At the Lyman limit ($h\nu = 13.6$ eV) we then find

$$L_{\mathrm{gal},\nu}(h\nu = 13.6\ \mathrm{eV}) = 4.32\times10^{18}\left(\frac{L_{\mathrm{LW}}}{1\,\mathrm{L}_\odot}\right)\ \mathrm{erg\ s^{-1}\ Hz^{-1}}\,. \tag{215}$$

Upon using this in Eq. (213) with the definition $J_{\mathrm{gal},21} \equiv J_{\mathrm{gal},\nu}(h\nu = 13.6\ \mathrm{eV})/10^{-21}\ \mathrm{erg\ cm^{-2}\ s^{-1}\ Hz^{-1}\ sr^{-1}}$ we find

$$J_{\mathrm{gal},21} = 2.87\times10^{-6}\left(\frac{L_{\mathrm{LW}}}{1\,\mathrm{L}_\odot}\right)\left(\frac{r}{1\ \mathrm{kpc}}\right)^{-2}\,. \tag{216}$$

Consider a galaxy with star formation rate $\dot{M}_{\star,\mathrm{gal}}$. The LW luminosity at time $t$ is given by

$$L_{\mathrm{LW}}(t) = \int_0^t \mathrm{d}t_{\mathrm{form}}\,\Psi_{\mathrm{LW}}(t - t_{\mathrm{form}})\dot{M}_{\star,\mathrm{gal}}(t_{\mathrm{form}})\,, \tag{217}$$

where $\Psi_{\mathrm{LW}}(\Delta t)$ is the LW light-to-mass ratio a time $\Delta t$ after an instantaneous burst. We are in essence summing over the formation times $t_{\mathrm{form}}$ of stars in the galaxy here. If $\Psi_{\mathrm{LW}}(t - t_{\mathrm{form}})$ varies on a time-scale significantly shorter than $\dot{M}_{\star,\mathrm{gal}}$, this simplifies to

$$L_{\mathrm{LW}}(t) \simeq \dot{M}_{\star,\mathrm{gal}}(t)\int_0^t \mathrm{d}t_{\mathrm{form}}\,\Psi_{\mathrm{LW}}(t - t_{\mathrm{form}})$$

Or, with $\bar{t} = t - t_{\mathrm{form}}$, $\mathrm{d}\bar{t} = -\mathrm{d}t_{\mathrm{form}}$,

$$L_{\mathrm{LW}}(t) \simeq -\dot{M}_{\star,\mathrm{gal}}(t)\int_t^0 \mathrm{d}\bar{t}\,\Psi_{\mathrm{LW}}(\bar{t})$$

Which gives us:

$$\begin{aligned} L_{\mathrm{LW}}(t) &\simeq -\dot{M}_{\star,\mathrm{gal}}(t)\int_t^0 \mathrm{d}\bar{t}\,\Psi_{\mathrm{LW}}(\bar{t}) \\ &= \dot{M}_{\star,\mathrm{gal}}(t)\int_0^t \mathrm{d}\bar{t}\,\Psi_{\mathrm{LW}}(\bar{t})\,. \end{aligned} \tag{218}$$

The light-to-mass ratio in the LW band is approximately given by (Hopkins *et al.*, 2020)

$$\Psi_{\mathrm{LW}}(\bar{t}) = \begin{cases} 109\ (1+\bar{t}_{3.4}^2)\,\mathrm{L}_\odot\ \mathrm{M}_\odot^{-1} & \text{for } \bar{t}_{3.4} \le 1 \\ 243\ \bar{t}_{3.4}^{-1.6}e^{-\bar{t}_{3.4}/118}\,\mathrm{L}_\odot\ \mathrm{M}_\odot^{-1} & \text{for } \bar{t}_{3.4} > 1 \end{cases}\,, \tag{219}$$

where $\bar{t}_{3.4} \equiv \bar{t}/3.4$ Myrs. Thus, for $t \gg 3.4$ Myrs we can evaluate the integral to find:

$$\begin{aligned} L_{\rm LW}(t) &\simeq \dot{M}_{\star,\rm gal}(t) \times 3.4 \text{ Myrs} \int_0^\infty {\rm d}\bar{t}_{3.4}\, \Psi_{\rm LW}(\bar{t}) \\ &= 1.71 \times 10^9 \left(\frac{\dot{M}_{\star,\rm gal}}{1\,{\rm M}_\odot \text{ yr}^{-1}}\right) {\rm L}_\odot \,. \end{aligned} \tag{220}$$

For reference, this yields $L_\nu(h\nu = 13.6 \text{ eV}) \simeq 7.39 \times 10^{27}\, \dot{M}_{\star,\rm gal}$ erg s$^{-1}$ Hz$^{-1}$ (with $\dot{M}_{\star,\rm gal}$ in units of $1\,{\rm M}_\odot$ yr$^{-1}$), consistent with the empirical relation $\sim 8 \times 10^{27}\, \dot{M}_{\star,\rm gal}$ erg s$^{-1}$ Hz$^{-1}$ (Dijkstra *et al.*, 2008). If we insert the result in Eq. (220) into Eq. (216) we arrive at a (normalized) intensity given by:

$$J_{\rm gal,21} \simeq 4900 \left(\frac{\dot{M}_{\star,\rm gal}}{1\,{\rm M}_\odot \text{ yr}^{-1}}\right) \left(\frac{r}{1 \text{ kpc}}\right)^{-2} . \tag{221}$$

To evaluate this we clearly need to know the star formation rate of the galaxy, which is well beyond the scope of Anaxagoras which only deals with short starbursts in low-mass halos. Thus, we can resort to state-of-the-art galaxy formation simulations. The FIRE-2 simulations (Hopkins *et al.*, 2018a), briefly mentioned in the main text, are zoom-in galaxy formation simulations that include a detailed and physically realistic implementation of stellar feedback from radiation pressure, photoheating, SNe, and stellar winds. Ma *et al.* (2018) applied FIRE-2 to study galaxies during the EoR. The authors found that the average star formation rate of galaxies in host halos of mass $M \gtrsim 3 \times 10^7$ M$_\odot$ can be fitted by:

$$\dot{M}_{\star,\rm gal} \simeq 0.081 \text{ M}_\odot \text{ yr}^{-1}\, M_{10}^{1.58} \left(\frac{1+z}{10}\right)^{2.20} , \tag{222}$$

Using this result allow us to estimate the intensity a distance $r$ from a halo of mass $M$:

$$J_{\rm gal,21} \simeq 400\, M_{10}^{1.58} \left(\frac{1+z}{10}\right)^{2.20} \left(\frac{r}{1 \text{ kpc}}\right)^{-2} . \tag{223}$$

## 4.2. Lyman-Werner intensity from the total galaxy population

The mean frequency-integrated background LW intensity $J_{\rm bg,LW}(z)$ at redshift $z$ can be found from the solution of the cosmological radiative transfer equation, and is given by

$$J_{\rm bg,LW}(z) = \frac{(1+z)^3 c}{4\pi} \int_z^{z_{\rm max}} {\rm d}z' \left|\frac{{\rm d}t}{{\rm d}z'}\right| \epsilon_{\rm LW}(z') \,, \tag{224}$$

where $\epsilon_{\rm LW}(z')$ the LW luminosity per comoving volume (in erg s$^{-1}$ cm$^{-3}$) at redshift $z'$. The redshift $z_{\rm max}$ is the maximum redshift an LW photon can be observed from $z$, since LW photons emitted at higher redshifts can be absorbed as Lyman lines before they reach the observer. Numerically we have $1+z_{\rm max} \simeq 1.04\,(1+z)$ (Visbal *et al.*, 2014b). Using this and $t(z') \simeq \frac{2}{3} H_0^{-1} \Omega_{\rm M}^{-1/2} (1+z')^{-3/2}$ yields

$$J_{\rm bg,LW}(z) = \frac{(1+z)^3 c}{4\pi H_0 \sqrt{\Omega_{\rm M}}} \int_z^{1.04\,(1+z)-1} {\rm d}z'\, (1+z')^{-5/2} \epsilon_{\rm LW}(z') \,. \tag{225}$$

The LW specific luminosity can be determined by convolving the mean LW luminosity of a galaxy in a halo of mass $M$ with the halo mass function:

$$\epsilon_{\rm LW}(z') = \int_{M_{\rm min}}^{M_{\rm max}} {\rm d}M\, \frac{\partial n}{\partial M}\, L_{\rm LW}(M, z') \,, \tag{226}$$

where $M_{\rm min}$ and $M_{\rm max}$ are minimum and upper limits to the mass of halos assumed to host stars. Using Eq. (220), we find

$$\epsilon_{\rm LW}(z') \simeq 2.27 \times 10^{-31} \left( \frac{\dot{\rho}_\star}{1\,{\rm M}_\odot\ {\rm yr}^{-1}\ {\rm Mpc}^{-3}} \right)\ {\rm erg\ s}^{-1}\ {\rm cm}^{-3}\,, \tag{227}$$

where $\dot{\rho}_\star = \int_{M_{\rm min}}^{M_{\rm max}} {\rm d}M\,(\partial n/\partial M)\,\dot{M}_{\star,{\rm gal}}$ is the comoving star formation rate density (SFRD). Plugging this into Eq. (225) yields (for the adopted cosmological parameters):

$$J_{\rm bg,LW}(z) \simeq 4.42 \times 10^{-4} (1+z)^3 \int_z^{1.04\,(1+z)-1} {\rm d}z'\,(1+z')^{-5/2} \dot{\rho}_\star(z')\ {\rm erg\ cm}^{-2}\ {\rm s}^{-1}\ {\rm sr}^{-1}\,. \tag{228}$$

Dividing by the LW frequency band $\Delta\nu_{\rm LW} = (13.6\ {\rm eV} - 11.2\ {\rm eV})/h = 5.80 \times 10^{14}$ Hz and then normalizing to $10^{-21}$ erg cm$^{-2}$ s$^{-1}$ Hz$^{-1}$ sr$^{-1}$ yields our final expression for the uniform LW background:

$$J_{\rm bg,21}(z) \simeq 760\,(1+z)^3 \int_z^{1.04\,(1+z)-1} {\rm d}z'\,(1+z')^{-5/2} \dot{\rho}_\star(z')\,, \tag{229}$$

with the SFRD $\dot{\rho}_\star$ in units of ${\rm M}_\odot$ yr$^{-1}$ Mpc$^{-3}$. To apply this result in the thesis in the case of Pop II star formation we fit the SFRD from Ma *et al.* (2018) and evaluate Eq. (229). As shown in Figure 26, the resulting background is well-approximated by

$$J_{\rm bg,21}(z) \simeq 10\, e^{-(z-6)/3.3}\,. \tag{230}$$

The SFRD from Ma *et al.* (2018) is consistent with observations, and the FIRE-2 simulations include a realistic implementation of stellar feedback and produce realistic galaxies. Because of this, Eq. (230) is likely a good estimate of the LW background in ΛCDM.

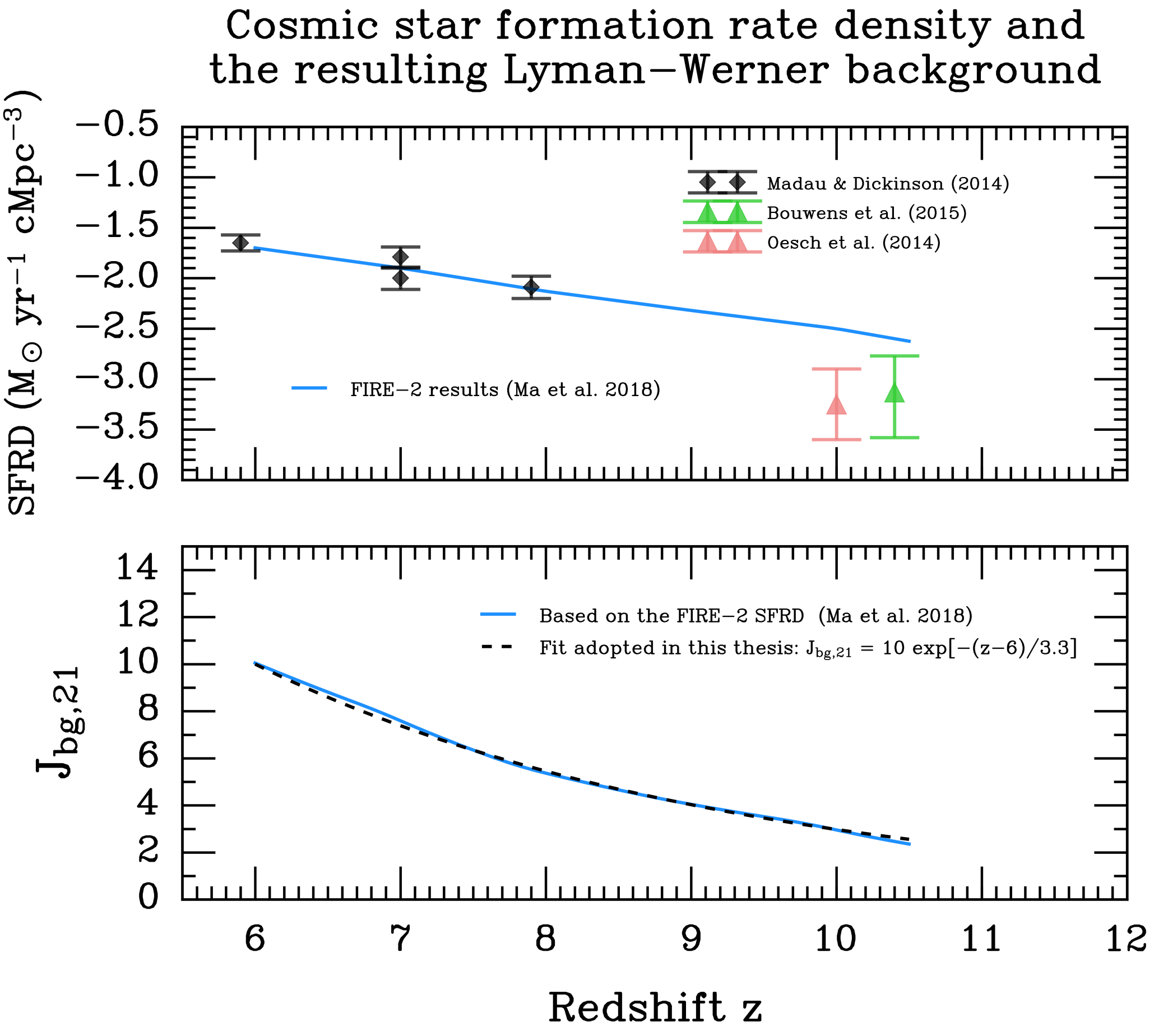

**Figure 26.** *Upper panel:* The comoving star formation rate density (SFRD) at high redshifts predicted by the FIRE-2 simulations compared to observational constraints on the SFRD (Madau & Dickinson, 2014; Oesch *et al.*, 2014; Bouwens *et al.*, 2015). The constraints at $z \sim 10$ are lower limits obtained by integrating the UV luminosity function down to $M_{\rm UV} \sim -17$. Fainter galaxies are not detectable, hence this being a lower bound on the SFRD. *Lower panel:* The resulting Lyman-Werner background obtained by integrating Eq. (229) with the FIRE-2 SFRD. Also shown is the fit in Eq. (230).

# 5. Fluid equations in self-similar form and approximate solutions

## 5.1. Self-similar form of the fluid equations

Let

$$\rho_{\rm gas}\,(r,t) = \frac{\bar\rho(x)}{4\pi G t^2}, \qquad M_{\rm gas}(<r,t) = \frac{c_{\rm s,h}^3 t}{G}\bar m(x), \qquad v(r,t) = c_{\rm s,h}\bar v(x)\,. \tag{231}$$

where $x = r/c_{\rm s,h}t$. For a function $f$, making a change of variables to a set of new variables $x_i$ implies

$$\begin{aligned} \frac{\partial f}{\partial t} &= \frac{\partial f}{\partial x_i}\frac{\partial x_i}{\partial t} \\ \frac{\partial f}{\partial r} &= \frac{\partial f}{\partial x_i}\frac{\partial x_i}{\partial r}\,. \end{aligned} \tag{232}$$

where Einstein summation is implied. We will need the following derivatives:

$$\begin{aligned}\frac{\partial x}{\partial t} &= -\frac{r}{c_{\rm s,h}t^2} = -\frac{x}{t} \\ \frac{\partial x}{\partial r} &= \frac{1}{c_{\rm s,h}t} = \frac{x}{r}\,.\end{aligned} \tag{233}$$

Thus, in our case of only one new variable, for $v(r,t)$ we have

$$\begin{aligned}\frac{\partial v}{\partial t} &= c_{\rm s,h}\frac{{\rm d}\bar{v}}{{\rm d}x}\frac{\partial x}{\partial t} = -\frac{c_{\rm s,h}}{t}\frac{{\rm d}\bar{v}}{{\rm d}x}x \\ \frac{\partial v}{\partial r} &= c_{\rm s,h}\frac{{\rm d}\bar{v}}{{\rm d}x}\frac{\partial x}{\partial r} = \frac{c_{\rm s,h}}{r}\frac{{\rm d}\bar{v}}{{\rm d}x}x\,.\end{aligned} \tag{234}$$

Similarly for $M_{\rm gas}(<r,t)$ we get

$$\begin{aligned}\frac{\partial M_{\rm gas}}{\partial t} &= \frac{\partial}{\partial t}\left(\frac{c_{\rm s,h}^3 t}{G}\bar{m}\right) \\ &= \frac{c_{\rm s,h}^3}{G}\bar{m} + \frac{c_{\rm s,h}^3 t}{G}\frac{\partial \bar{m}}{\partial t} \\ &= \frac{c_{\rm s,h}^3}{G}\bar{m} + \frac{c_{\rm s,h}^3 t}{G}\frac{{\rm d}\bar{m}}{{\rm d}x}\frac{\partial x}{\partial t} \\ &= \frac{c_{\rm s,h}^3}{G}\bar{m} - \frac{c_{\rm s,h}^3}{G}\frac{{\rm d}\bar{m}}{{\rm d}x}x\,.\end{aligned} \tag{235}$$

And for the radial derivative:

$$\frac{\partial M_{\rm gas}}{\partial r} = \frac{c_{\rm s,h}^3 t}{G}\frac{{\rm d}\bar{m}}{{\rm d}x}\frac{\partial x}{\partial r} = \frac{c_{\rm s,h}^2}{G}\frac{{\rm d}\bar{m}}{{\rm d}x}\,. \tag{236}$$

Finally, for $\rho_{\rm gas}(r,t)$ we have

$$\begin{aligned}\frac{\partial \rho_{\rm gas}}{\partial t} &= \frac{\partial}{\partial t}\left(\frac{\bar{\rho}}{4\pi G t^2}\right) \\ &= -\frac{\bar{\rho}}{4\pi G t^3} + \frac{1}{4\pi G t^2}\frac{\partial \bar{\rho}}{\partial t} \\ &= -\frac{\bar{\rho}}{4\pi G t^3} + \frac{1}{4\pi G t^2}\frac{{\rm d}\bar{\rho}}{{\rm d}x}\frac{\partial x}{\partial t} \\ &= -\frac{\bar{\rho}}{4\pi G t^3} - \frac{1}{4\pi G t^3}\frac{{\rm d}\bar{\rho}}{{\rm d}x}x\,.\end{aligned} \tag{237}$$

And for the radial derivative (remember that $x = r/c_{\rm s,h}t$):

$$\frac{\partial \rho_{\rm gas}}{\partial r} = \frac{1}{4\pi G t^2}\frac{{\rm d}\bar{\rho}}{{\rm d}x}\frac{\partial x}{\partial r} = \frac{1}{4\pi G c_{\rm s,h} t^3}\frac{{\rm d}\bar{\rho}}{{\rm d}x}\,. \tag{238}$$

Now we can start to rewrite our PDEs into ODEs. Starting with the equation for $M$:

$$\frac{\partial M_{\rm gas}}{\partial t} + v\frac{\partial M_{\rm gas}}{\partial r} = \frac{c_{\rm s,h}^3}{G}\left\{\bar{m} - \frac{{\rm d}\bar{m}}{{\rm d}x}\left(x - \bar{v}\right)\right\}\,.$$

This should be zero, so we find

$$\frac{{\rm d}\bar{m}}{{\rm d}x}\left(x - \bar{v}\right) = \bar{m}\,, \tag{239}$$

as was stated in Eq. (59) in the main text. Furthermore, $\partial M_{\rm gas}/\partial r = 4\pi r^2 \rho_{\rm gas}$ yields

$$\begin{aligned} \frac{\partial M_{\rm gas}}{\partial r} - 4\pi r^2 \rho_{\rm gas} &= \frac{c_{\rm s,h}^2}{G}\frac{{\rm d}\bar{m}}{{\rm d}x} - \frac{r^2\bar{\rho}}{Gt^2} \\ &= \frac{c_{\rm s,h}^2}{G}\left(\frac{{\rm d}\bar{m}}{{\rm d}x} - x^2\bar{\rho}\right)\,. \end{aligned} \tag{240}$$

Since this is also supposed to equal zero we deduce that

$$\frac{{\rm d}\bar{m}}{{\rm d}x} = x^2\bar{\rho}\,, \tag{241}$$

in agreement with Eq. (60). Next, we focus on the Euler equation:

$$\frac{\partial v}{\partial t} + v\frac{\partial v}{\partial r} = -\frac{c_{\rm s,h}^2}{\rho_{\rm gas}}\frac{\partial \rho_{\rm gas}}{\partial r} - \frac{GM_{\rm gas}}{r^2} - \frac{GM_{\rm DM}}{r^2}\,. \tag{242}$$

For the left-hand side we get

$$\begin{aligned} \frac{\partial v}{\partial t} + v\frac{\partial v}{\partial r} &= -\frac{c_{\rm s,h}}{t}\frac{{\rm d}\bar{v}}{{\rm d}x}x + \bar{v}\frac{c_{\rm s,h}^2}{r}\frac{{\rm d}\bar{v}}{{\rm d}x}x \\ &= \frac{{\rm d}\bar{v}}{{\rm d}x}x\left(-\frac{c_{\rm s,h}}{t} + \bar{v}\frac{c_{\rm s,h}^2}{r}\right)\,. \end{aligned} \tag{243}$$

Since $r = xc_{\rm s,h}t$ this becomes

$$\frac{\partial v}{\partial t} + v\frac{\partial v}{\partial r} = \frac{c_{\rm s,h}}{t}\frac{{\rm d}\bar{v}}{{\rm d}x}\left(-x + \bar{v}\right)\,. \tag{244}$$

And for the right-hand side of the Euler equation, we need

$$\begin{aligned} \frac{c_{\rm s}^2}{\rho}\frac{\partial \rho}{\partial r} &= c_{\rm s,h}^2\left(\frac{\bar{\rho}}{4\pi Gt^2}\right)^{-1} \times \frac{1}{4\pi Gc_{\rm s,h}t^3}\frac{{\rm d}\bar{\rho}}{{\rm d}x} \\ &= \frac{c_{\rm s,h}}{t\bar{\rho}}\frac{{\rm d}\bar{\rho}}{{\rm d}x}\,. \end{aligned} \tag{245}$$

Plugging all of this into the Euler equation yields

$$\frac{c_{\rm s,h}}{t}\frac{{\rm d}\bar{v}}{{\rm d}x}\left(-x + \bar{v}\right) = -\frac{c_{\rm s,h}}{t\bar{\rho}}\frac{{\rm d}\bar{\rho}}{{\rm d}x} - \frac{GM_{\rm gas}}{r^2} - \frac{GM_{\rm DM}}{r^2}\,. \tag{246}$$

Or:

$$\frac{{\rm d}\bar{v}}{{\rm d}x}\left(-x + \bar{v}\right) = -\frac{1}{\bar{\rho}}\frac{{\rm d}\bar{\rho}}{{\rm d}x} - \frac{GM_{\rm gas}t}{c_{\rm s,h}r^2} - \frac{GM_{\rm DM}t}{c_{\rm s}r^2}\,. \tag{247}$$

To proceed we need to eliminate $\bar{\rho}$. Since $\bar{m} = x^2\bar{\rho}\,(x-\bar{v})$ we find

$$\frac{\mathrm{d}\bar{m}}{\mathrm{d}x} = 2x\bar{\rho}\,(x-\bar{v}) + x^2\frac{\mathrm{d}\bar{\rho}}{\mathrm{d}x}\,(x-\bar{v}) + x^2\bar{\rho}\left(1-\frac{\mathrm{d}\bar{v}}{\mathrm{d}x}\right)\,. \tag{248}$$

Furthermore, we had $\mathrm{d}\bar{m}/\mathrm{d}x = x^2\bar{\rho}$ so that

$$x^2\bar{\rho} = 2x\bar{\rho}\,(x-\bar{v}) + x^2\frac{\mathrm{d}\bar{\rho}}{\mathrm{d}x}\,(x-\bar{v}) + x^2\bar{\rho}\left(1-\frac{\mathrm{d}\bar{v}}{\mathrm{d}x}\right)\,. \tag{249}$$

Or:

$$x^2 = 2x\,(x-\bar{v}) + x^2\frac{1}{\bar{\rho}}\frac{\mathrm{d}\bar{\rho}}{\mathrm{d}x}\,(x-\bar{v}) + x^2\left(1-\frac{\mathrm{d}\bar{v}}{\mathrm{d}x}\right)\,. \tag{250}$$

Rearranging a little yields

$$\frac{1}{\bar{\rho}}\frac{\mathrm{d}\bar{\rho}}{\mathrm{d}x} = -\frac{2}{x} + \frac{1}{(x-\bar{v})}\frac{\mathrm{d}\bar{v}}{\mathrm{d}x}\,. \tag{251}$$

Plugging this into the Euler equation gives us

$$\frac{\mathrm{d}\bar{v}}{\mathrm{d}x}\,(-x+\bar{v}) = \frac{2}{x} - \frac{1}{(x-\bar{v})}\frac{\mathrm{d}\bar{v}}{\mathrm{d}x} - \frac{GM_{\mathrm{gas}}t}{c_{\mathrm{s,h}}r^2} - \frac{GM_{\mathrm{DM}}t}{c_{\mathrm{s,h}}r^2}\,. \tag{252}$$

Or, after a little rearrangement:

$$\frac{\mathrm{d}\bar{v}}{\mathrm{d}x}\left\{(x-\bar{v})^2-1\right\} = -\frac{2\,(x-\bar{v})}{x} + \left\{\frac{GM_{\mathrm{gas}}t}{c_{\mathrm{s,h}}r^2} + \frac{GM_{\mathrm{DM}}t}{c_{\mathrm{s,h}}r^2}\right\}(x-\bar{v})\,. \tag{253}$$

Since $M_{\mathrm{gas}} = c_{\mathrm{s,h}}^3 t\bar{m}/G$ we have $GM_{\mathrm{gas}}t/c_{\mathrm{s,h}}r^2 = c_{\mathrm{s,h}}^2 t^2\bar{m}/r^2 = \bar{m}/x^2$. Furthermore, for an isothermal DM halo we have $M_{\mathrm{DM}}(<r) = (1-f_{\mathrm{B}})Mr/R_{\mathrm{vir}}$, so that $GM_{\mathrm{DM}}t/c_{\mathrm{s,h}}r^2 = (1-f_{\mathrm{B}})GMt/rc_{\mathrm{s,h}}R_{\mathrm{vir}} = (1-f_{\mathrm{B}})v_{\mathrm{vir}}^2 t/rc_{\mathrm{s,h}}$, or

$$\frac{GM_{\mathrm{DM}}t}{c_{\mathrm{s,h}}r^2} = \frac{(1-f_{\mathrm{B}})}{x}\left(\frac{v_{\mathrm{vir}}}{c_{\mathrm{s,h}}}\right)^2\,. \tag{254}$$

Using these results in Eq. (253) yields

$$\frac{\mathrm{d}\bar{v}}{\mathrm{d}x}\left\{(x-\bar{v})^2-1\right\} = \beta\frac{(x-\bar{v})}{x} + \frac{\bar{m}}{x^2}\,(x-\bar{v})\,, \tag{255}$$

where $\beta \equiv (1-f_{\mathrm{B}})(v_{\mathrm{vir}}/c_{\mathrm{s,h}})^2 - 2$, as claimed in Eq. (61).

## 5.2. Approximate solutions for large $\beta$

Next we will derive approximate solutions for the gas accretion rate, valid for large $\beta$. This will be particularly useful for understanding the numerical results, and aiding in deriving accurate fits to the exact solutions. First we start by deriving a formal solution for $\bar{m}(x)$. From Eq. (239) we have $\mathrm{d}\bar{m}/\mathrm{d}x = \bar{m}/(x-\bar{v})$, which can be integrated formally to give:

$$\begin{aligned}\bar{m}(x) &= \bar{m}(x=0)\,e^{\int_0^x \mathrm{d}x'/[x'-\bar{v}(x')]} \\ &\equiv \bar{m}(x=0)\,e^{\mathcal{I}(x)}\,.\end{aligned} \tag{256}$$

Recall that the central gas accretion rate is $\dot{M}_{\mathrm{gas}} = \bar{m}(x=0)\,c_{\mathrm{s,h}}^3/G$, so our goal is to find an approximate solution for $\bar{m}(x=0)$. An approximate solution can be found if we have an approximate solution for $\bar{v}(x)$. As $x\to\infty$ our boundary conditions should give $\bar{m}(x\to\infty)\propto x$, so we can write this as $\bar{m}(x\to\infty) = m_\infty x$ where $m_\infty$ is a constant. Since we also want $\bar{v}$ to go to zero as $x \to \infty$, Eq. (255) then yields (since $x \gg \bar{v}$ in this limit):

$$\frac{\mathrm{d}\bar{v}}{\mathrm{d}x} = \frac{\beta + m_\infty}{x^2} \qquad \text{as } x \to \infty\,. \tag{257}$$

Upon integration we find

$$\bar{v}(x \to \infty) = -\frac{\beta + m_\infty}{x}\,, \tag{258}$$

where the integration constant has been set to zero to satisfy our boundary condition for $\bar{v}$. Next we consider the opposite limit, namely $x \to 0$. In this case $x \ll \bar{v}$ and Eq. (255) give

$$\frac{\mathrm{d}\bar{v}}{\mathrm{d}x}\bar{v}^2 = -\beta\frac{\bar{v}}{x} - \frac{\bar{m}(x=0)}{x}\frac{\bar{v}}{x} \qquad \text{as } x \to 0\,. \tag{259}$$

For finite $\beta$ we will necessarily find that the first term on the right becomes subdominant. Thus, in reality we get

$$\frac{\mathrm{d}\bar{v}}{\mathrm{d}x}\bar{v} = -\frac{\bar{m}(x=0)}{x^2} \qquad \text{as } x \to 0\,, \tag{260}$$

which has the solution $\bar{v}$

$$\bar{v}(x \to 0) = -\left\{\frac{2\bar{m}(x=0)}{x}\right\}^{1/2}\,. \tag{261}$$

This has a simple physical interpretation: near the center of the halo the gas falls in at close to the escape speed. Consider now the integral $\mathcal{I}(x)$ for sufficiently large (but finite) $x$. We roughly have:

$$\begin{aligned}
\mathcal{I}(\text{large } x) &= \int_0^x \frac{\mathrm{d}x'}{x' - \bar{v}(x')} \\
&\sim \int_0^{x_{\text{small}}} \frac{\mathrm{d}x'}{x' - \bar{v}(x \to 0)} + \int_{x_{\text{small}}}^x \frac{\mathrm{d}x'}{x' - \bar{v}(x \to \infty)} \qquad (262)\\
&= \int_0^{x_{\text{small}}} \frac{\mathrm{d}x'}{x' + \{2m(x=0)/x'\}^{1/2}} + \int_{x_{\text{small}}}^x \frac{\mathrm{d}x'}{x' + (\beta + m_\infty)/x'}\,,
\end{aligned}$$

where $x_{\text{small}}$ is unknown value of $x$ that we expect to be small. We see that the first integral is subdominant since it is finite and equal to some constant $\mathcal{C}$ whereas the second integral grows with $x$:

$$\int_{x_{\text{small}}}^x \frac{\mathrm{d}x'}{x' + (\beta + m_\infty)/x'} = \frac{1}{2}\ln(\beta + m_\infty + x^2) - \frac{1}{2}\ln(\beta + m_\infty + x_{\text{small}}^2)\,. \tag{263}$$

Consider then the limit of sufficiently large $\beta$ and/or $m_\infty$, so that $\beta + m_\infty \gg x_{\text{small}}^2$. We then get

$$\mathcal{I}(\text{large } x) \simeq \mathcal{C} + \frac{1}{2}\ln\left(1 + \frac{x^2}{\beta + m_\infty}\right)\,. \tag{264}$$

Thus, as $x \to \infty$ we arrive at:

$$\bar{m}(x \to \infty) \simeq e^{\mathcal{C}}\bar{m}(x=0)\,\frac{x}{(\beta + m_\infty)^{1/2}}\,.$$

This should equal $m_\infty x$, and so we find

$$\bar{m}(x=0) \simeq e^{-\mathcal{C}} m_\infty (\beta + m_\infty)^{1/2}\,. \tag{265}$$

As long as $\mathcal{C} \lesssim 1$ we then end up with

$$\bar{m}(x=0) \simeq m_\infty(\beta+m_\infty)^{1/2} \qquad \text{if } \mathcal{C} \lesssim 1\,. \tag{266}$$

For self-consistency we can evaluate the integral for $\mathcal{C}$ using this expression for $\bar{m}(x=0)$. We have:

$$\mathcal{C} = \int_0^{x_{\rm small}} \frac{{\rm d}x'}{x' + \{2m(x=0)/x'\}^{1/2}} = \frac{2}{3}\ln\left[1 + \frac{x_{\rm small}^{3/2}}{\{2m(x=0)\}^{1/2}}\right]\,. \tag{267}$$

This is smaller than unity for $x_{\rm small}^{3/2}/\{2m(x=0)\}^{1/2} \lesssim 3.5$, or:

$$m(x=0) \gtrsim 4.1\times10^{-5}\left(\frac{x_{\rm small}}{0.1}\right)^3\,, \tag{268}$$

and so self-consistency of Eq. (266) yields

$$\bar{m}(x=0) \simeq m_\infty(\beta+m_\infty)^{1/2} \qquad \text{if } m_\infty(\beta+m_\infty)^{1/2} \gtrsim 4.1\times10^{-5}\left(\frac{x_{\rm small}}{0.1}\right)^3\,. \tag{269}$$

We can now apply this to the low and high-mass limits, which have different values for $m_\infty$.

### 5.2.1. Low-mass limit

First we consider halos with masses $M \leqslant 2.92\times10^6\,[(1+z)/10]^{3/2}\,{\rm M}_\odot$ for which Eq. (78) yield

$$m_\infty = 3.14\times10^{-5}\,(\beta+2)\zeta\,. \tag{270}$$

This gives us

$$\begin{aligned} \bar{m}(x=0) \;&\simeq\; 3.14\times10^{-5}\,(\beta+2)\zeta\left\{\beta + 3.14\times10^{-5}\,(\beta+2)\zeta\right\}^{1/2} \\ &\text{if}\quad (\beta+2)(\beta+m_\infty)^{1/2}\zeta \gtrsim 1.3\left(\frac{x_{\rm small}}{0.1}\right)^3 \end{aligned} \tag{271}$$

While $x_{\rm small}$ is not known in our approximate solution, if it is indeed small (say $\lesssim 0.1$), then the inequality is likely to hold for most cases of interest. To see this, note that (as discussed in the main text) the halo mass should exceed the cosmological Jeans mass, giving us $\zeta \gtrsim 32$. Because of this, a sufficient condition for the validity of Eq. (271) for $\bar{m}(x=0)$ is:

$$(\beta+2)\{\beta + 1.0\times10^{-3}\,(\beta+2)\}^{1/2} \gtrsim 0.041\left(\frac{x_{\rm small}}{0.1}\right)^3\,. \tag{272}$$

Thus, if $x_{\rm small} = 0.1$ this is satisfied for all positive $\beta$, which covers most of the interesting parameter space. In summary, we expect Eq. (271) for $\bar{m}(x=0)$ to give a correct solution for $\bar{m}(x=0)$ as long as $\beta$ is sufficiently large (perhaps it only needs to be $\beta \gtrsim 0$). We can determine the corresponding normalized central gas accretion rate

$$\begin{aligned} \frac{\dot{M}_{\rm gas}}{v_{\rm vir}^3/G} \;&=\; \bar{m}(x=0)\left(\frac{1-f_{\rm B}}{\beta+2}\right)^{3/2} \\ &\simeq\; 2.43\times10^{-5}\,\zeta\left\{\frac{\beta}{\beta+2} + 3.14\times10^{-5}\,\zeta\right\}^{1/2}\,. \end{aligned} \tag{273}$$

### 5.2.2. High-mass limit

Next we consider the high-mass limit $M > 2.92 \times 10^6\,[(1+z)/10]^{3/2}\,\mathrm{M_\odot}$. In this case we have $m_\infty = 0.530\,(\beta+2)$, giving us

$$\bar{m}(x=0) \simeq 0.530\,(\beta+2)\{\beta + 0.530\,(\beta+2)\}^{1/2}\,, \tag{274}$$

which should be valid if $0.530\,(\beta+2)\{\beta+0.530\,(\beta+2)\}^{1/2} \gtrsim 4.1 \times 10^{-5}\,(x_{\rm small}/0.1)^3$. For $x_{\rm small} = 0.1$ this is satisfied for all $\beta \gtrsim -0.692$, and so basically the whole range of values of $\beta$ for which gas accretion is possible. The normalized gas accretion rate in this case is given by

$$\begin{aligned} \frac{\dot{M}_{\rm gas}}{v_{\rm vir}^3/G} &= \bar{m}(x=0)\left(\frac{1-f_{\rm B}}{\beta+2}\right)^{3/2} \\ &\simeq 0.410\left\{\frac{\beta}{\beta+2} + 0.530\right\}^{1/2}. \end{aligned} \tag{275}$$

# 6. Understanding the physics of dust radiative feedback

As discussed in the main text, radiation pressure from stars can be boosted by a factor $\simeq 1 + \tau_{\rm IR}$, where $\tau_{\rm IR}$ is the optical depth in the IR due to dust. This factor arises because stellar photons absorbed by dust (mainly in the optical and UV) will be re-radiated in the IR, and perform a random walk before escaping a gas cloud. During each scatter some momentum will be imparted in a random direction, with the net result being a boost $\simeq 1 + \tau_{\rm IR}$. In this Appendix I provide a rigorous derivation of this result. Let us consider a homogeneous spherical gas cloud of radius $R$, density $\rho(r)$, UV optical depth $\tau_{\rm UV}$, and IR optical depth $\tau_{\rm IR}$. At the center of the gas cloud is a star cluster shining with UV luminosity $L_{\rm UV}$. The radiative transfer equations for the IR intensity $I_{\rm IR}$ in spherical coordinates, assuming spherical symmetry reads (see e.g. Chandrasekhar, 1960; Low & Lynden-Bell, 1976; Tomaselli & Ferrara, 2021):

$$\mu \frac{\partial I_{\rm IR}}{\partial r} + \frac{1-\mu^2}{r}\frac{\partial I_{\rm IR}}{\partial \mu} = j_{\rm dust} - \kappa^{\rm IR}_{\rm abs}\rho I_{\rm IR} - \kappa^{\rm IR}_{\rm sc}\rho (I_{\rm IR} - J_{\rm IR})\,.$$

Here $\mu \equiv \cos\theta$ is the cosine of the angle between the ray and the radial unit vector $\hat{\boldsymbol{r}}$ (i.e. $\hat{\boldsymbol{n}}\cdot\hat{\boldsymbol{r}} = \mu$), $j_{\rm dust}$ is the dust source function, $\kappa^{\rm IR}_{\rm abs}$ is the IR mass absorption coefficient (in cm$^2$ g$^{-1}$), $\kappa^{\rm IR}_{\rm sc}$ the IR mass scattering coefficient (in cm$^2$ g$^{-1}$), and $J_{\rm IR} \equiv \frac{1}{2}\int_{-1}^{1} {\rm d}\mu\, I_{\rm IR}$ the mean IR intensity. For simplicity I have assumed that scattering is isotropic.[78] Let each dust grain have radius $a$ and shine with luminosity (mainly in the IR) $L_{\rm gr}$. Then the dust source function is $j_{\rm dust} = n_{\rm dust}L_{\rm gr}/4\pi$, where $n_{\rm dust}$ is the dust number density, and $L_{\rm gr}$ the luminosity of a single grain. The grain luminosity can be determined from energy balance with heating terms. Dust can be heated by radiation from stars (mainly in the UV) and by collisions if the gas temperature $T$ exceed the dust temperature $T_{\rm dust}$ (see e.g. Chapter 24 in Draine, 2011). Thus, assuming energy balance, the grain cooling rate $n_{\rm dust}L_{\rm gr}$ is:

$$n_{\rm dust}L_{\rm gr} = n_{\rm dust}Q^{\rm UV}_{\rm abs}\pi a^2 \times 4\pi J_{\rm UV} + n_{\rm dust}Q^{\rm IR}_{\rm abs}\pi a^2 \times 4\pi J_{\rm IR} + \Lambda_{\rm coll}(T, T_{\rm dust})\,. \tag{276}$$

Here $Q^{\rm UV}_{\rm abs}$ ($Q^{\rm IR}_{\rm abs}$) is the absorption efficiency in the UV (IR), $J_{\rm UV}$ is the mean intensity in the UV, and $\Lambda_{\rm coll}(T, T_{\rm dust})$ is the collisional heating term. At this point we can note that $n_{\rm dust}Q^{\rm UV}_{\rm abs}\pi a^2 = \kappa^{\rm UV}_{\rm abs}\rho$ and $n_{\rm dust}Q^{\rm IR}_{\rm abs}\pi a^2 = \kappa^{\rm IR}_{\rm abs}\rho$, so that

$$j_{\rm dust} = \kappa^{\rm UV}_{\rm abs}\rho J_{\rm UV} + \kappa^{\rm IR}_{\rm abs}\rho J_{\rm IR} + \frac{\Lambda_{\rm coll}}{4\pi}\,. \tag{277}$$

With this result the radiative transfer equation becomes

$$\mu \frac{\partial I_{\rm IR}}{\partial r} + \frac{1-\mu^2}{r}\frac{\partial I_{\rm IR}}{\partial \mu} = \kappa^{\rm UV}_{\rm abs}\rho J_{\rm UV} + \frac{\Lambda_{\rm coll}}{4\pi} - \kappa^{\rm IR}_{\rm ext}\rho (I_{\rm IR} - J_{\rm IR})\,, \tag{278}$$

where $\kappa^{\rm IR}_{\rm ext} \equiv \kappa^{\rm IR}_{\rm sc} + \kappa^{\rm IR}_{\rm abs}$ is the extinction coefficient in the IR. The UV flux from the central star is $4\pi J_{\rm UV} = F_{\rm UV} = L_{\rm UV}e^{-\kappa^{\rm UV}_{\rm abs}\rho r}/4\pi r^2$, and so

$$\mu \frac{\partial I_{\rm IR}}{\partial r} + \frac{1-\mu^2}{r}\frac{\partial I_{\rm IR}}{\partial \mu} = \frac{\kappa^{\rm UV}_{\rm abs}\rho L_{\rm UV}e^{-\kappa^{\rm UV}_{\rm abs}\rho r}}{16\pi^2 r^2} + \frac{\Lambda_{\rm coll}}{4\pi} - \kappa^{\rm IR}\rho (I_{\rm IR} - J_{\rm IR})\,. \tag{279}$$

Let us now take the zeroth moment of this:

$$\frac{1}{2}\int_{-1}^{1}{\rm d}\mu\,\left\{\mu \frac{\partial I_{\rm IR}}{\partial r} + \frac{1-\mu^2}{r}\frac{\partial I_{\rm IR}}{\partial \mu}\right\} = \frac{1}{2}\int_{-1}^{1}{\rm d}\mu\,\left\{\frac{\kappa^{\rm UV}_{\rm abs}\rho L_{\rm UV}e^{-\kappa^{\rm UV}_{\rm abs}\rho r}}{16\pi^2 r^2} + \frac{\Lambda_{\rm coll}}{4\pi} - \kappa^{\rm IR}_{\rm ext}\rho (I_{\rm IR} - J_{\rm IR})\right\}\,. \tag{280}$$

On the right we note that $\frac{1}{2}\int_{-1}^{1}{\rm d}\mu\,(I_{\rm IR} - J_{\rm IR}) = 0$ since $\frac{1}{2}\int_{-1}^{1}{\rm d}\mu\, I_{\rm IR} = J_{\rm IR}$. Thus:

$$\frac{1}{2}\int_{-1}^{1}{\rm d}\mu\,\left\{\mu \frac{\partial I_{\rm IR}}{\partial r} + \frac{1-\mu^2}{r}\frac{\partial I_{\rm IR}}{\partial \mu}\right\} = \frac{\kappa^{\rm UV}_{\rm abs}\rho L_{\rm UV}e^{-\kappa^{\rm UV}_{\rm abs}\rho r}}{16\pi^2 r^2} + \frac{\Lambda_{\rm coll}}{4\pi}\,. \tag{281}$$

[78] In other words, $\langle\mu\rangle = \langle\cos\theta\rangle = 0$.

Using the definition $H_{\rm IR} \equiv \frac{1}{2}\int_{-1}^{1} {\rm d}\mu\, \mu I_{\rm IR}$ (the first moment of the IR intensity), we also have[79]

$$\begin{aligned} \frac{1}{2}\int_{-1}^{1} {\rm d}\mu\, \mu \frac{\partial I_{\rm IR}}{\partial r} &= \frac{\partial H_{\rm IR}}{\partial r} \\ \frac{1}{2}\int_{-1}^{1} {\rm d}\mu\, (1-\mu^2) \frac{\partial I_{\rm IR}}{\partial \mu} &= 2H_{\rm IR}\,. \end{aligned} \tag{282}$$

This gives us

$$\frac{\partial H_{\rm IR}}{\partial r} + \frac{2H_{\rm IR}}{r} = \frac{\kappa_{\rm abs}^{\rm UV}\rho L_{\rm UV} e^{-\kappa_{\rm abs}^{\rm UV}\rho r}}{16\pi^2 r^2} + \frac{\Lambda_{\rm coll}}{4\pi}\,. \tag{283}$$

Or, in terms of the IR flux $F_{\rm IR} = 2\pi \int_{-1}^{1} {\rm d}\mu\, \mu I_{\rm IR} = 4\pi H_{\rm IR}$:

$$\frac{\partial F_{\rm IR}}{\partial r} + \frac{2F_{\rm IR}}{r} = \frac{\kappa_{\rm abs}^{\rm UV}\rho L_{\rm UV} e^{-\kappa_{\rm abs}^{\rm UV}\rho r}}{4\pi r^2} + \Lambda_{\rm coll}\,. \tag{284}$$

The left-hand side can be written as $r^{-2}\partial_r(F_{\rm IR} r^2)$, which can then be integrated to give

$$F_{\rm IR} r^2 = -\frac{L_{\rm UV} e^{-\kappa_{\rm abs}^{\rm UV}\rho r}}{4\pi} + \frac{1}{3}\Lambda_{\rm coll} r^3 + \mathcal{C}\,, \tag{285}$$

where $\mathcal{C}$ is an integration constant. Demanding that $F_{\rm IR} r^2 \to 0$ as $r \to 0$ can be used to determine this constant, giving us our solution for the IR flux:

$$F_{\rm IR}(r) = \frac{L_{\rm UV}}{4\pi r^2}(1 - e^{-\kappa_{\rm abs}^{\rm UV}\rho r}) + \frac{1}{3}\Lambda_{\rm coll} r\,. \tag{286}$$

The acceleration due to radiation pressure is $a_{\rm rad} = \int_0^\infty {\rm d}\nu\, F_\nu \kappa_{{\rm rp},\nu}/c$ (see Eq. 119), where $\kappa_{{\rm rp},\nu}$ is the radiation pressure cross section at frequency $\nu$ (units of ${\rm cm^2\,g^{-1}}$). For isotropic scattering this is equal to the extinction coefficient (so that $\kappa_{\rm rp}^{\rm IR} = \kappa_{\rm ext}^{\rm IR} = \kappa_{\rm sc}^{\rm IR} + \kappa_{\rm abs}^{\rm IR}$).[80] Thus, upon integrating over frequency in our two-band model we find

$$\begin{aligned} a_{\rm rad} &= \frac{1}{c}\kappa_{\rm abs}^{\rm UV} F_{\rm UV} + \frac{1}{c}\kappa_{\rm ext}^{\rm IR} F_{\rm IR} \\ &= \frac{\kappa_{\rm abs}^{\rm UV} L_{\rm UV} e^{-\kappa_{\rm abs}^{\rm UV}\rho r}}{4\pi r^2 c} + \frac{\kappa_{\rm ext}^{\rm IR} L_{\rm UV}}{4\pi r^2 c}(1 - e^{-\kappa_{\rm abs}^{\rm UV}\rho r}) + \frac{1}{3c}\kappa_{\rm ext}^{\rm IR}\Lambda_{\rm coll} r\,. \end{aligned} \tag{287}$$

---

[79] The second integral can be evaluated using partial integration. We have:

$$\frac{1}{2}\int_{-1}^{1} {\rm d}\mu\, (1-\mu^2)\frac{\partial I_{\rm IR}}{\partial \mu} = \frac{1}{2}(1-\mu^2) I_{\rm IR}\,\big|_{-1}^{1} + \int_{-1}^{1} {\rm d}\mu\, \mu I_{\rm IR}\,.$$

The first term on the right is zero, and we also have $\int_{-1}^{1} {\rm d}\mu\, \mu I_{\rm IR} = 2H_{\rm IR}$, and so

$$\frac{1}{2}\int_{-1}^{1} {\rm d}\mu\, (1-\mu^2)\frac{\partial I_{\rm IR}}{\partial \mu} = 2H_{\rm IR}\,.$$

[80] In general the radiation pressure cross section per unit mass would be (see e.g. p. 249 in Draine, 2011)

$$\kappa_{\rm rp} = \kappa_{\rm abs} + (1 - \langle\mu\rangle)\kappa_{\rm sc}\,,$$

where $\langle\mu\rangle = \langle\cos\theta\rangle$ is the average of the cosine of the scattering angle. For isotropic scattering, as assumed in this Appendix, $\langle\mu\rangle = 0$. In the extreme case of perfect backward scattering we would instead have $\langle\mu\rangle = -1$, indicating that each scatter doubles the momentum input since the scattered photon travels backwards.

The total force (i.e. the radiative momentum injection rate) on the spherical cloud is found by multiplying by $\rho$ and integrating over the spherical cloud:

$$\dot{P}_{\rm rad} = \frac{L_{\rm UV}}{c} \int_0^R {\rm d}r \left\{ \rho\kappa_{\rm abs}^{\rm UV} e^{-\kappa_{\rm abs}^{\rm UV}\rho r} + \rho\kappa_{\rm ext}^{\rm IR}(1 - e^{-\kappa_{\rm abs}^{\rm UV}\rho r}) + \frac{4\pi\rho\kappa_{\rm ext}^{\rm IR}\Lambda_{\rm coll}}{3L_{\rm UV}} r^3 \right\} . \tag{288}$$

Evaluating the integral gives us

$$\dot{P}_{\rm rad} = \frac{L_{\rm UV}}{c} \left\{ (1 - e^{-\kappa_{\rm abs}^{\rm UV}\rho R}) + \rho\kappa_{\rm ext}^{\rm IR} R - \frac{\kappa_{\rm ext}^{\rm IR}}{\kappa_{\rm abs}^{\rm UV}}(1 - e^{-\kappa_{\rm abs}^{\rm UV}\rho R}) + \frac{\pi\rho\kappa_{\rm ext}^{\rm IR}\Lambda_{\rm coll}}{3L_{\rm UV}} R^4 \right\} . \tag{289}$$

Noting that $\kappa_{\rm abs}^{\rm UV}\rho R = \tau_{\rm UV}$, $\kappa_{\rm ext}^{\rm IR}\rho R = \tau_{\rm IR}$, and $\kappa_{\rm ext}^{\rm IR}/\kappa_{\rm abs}^{\rm UV} = \tau_{\rm IR}/\tau_{\rm UV}$ yields

$$\begin{aligned} \dot{P}_{\rm rad} &= \frac{L_{\rm UV}}{c} \left\{ 1 - e^{-\tau_{\rm UV}} + \tau_{\rm IR} - \frac{\tau_{\rm IR}}{\tau_{\rm UV}}(1 - e^{-\tau_{\rm UV}}) + \tau_{\rm IR}\frac{\pi\Lambda_{\rm coll}R^3}{3L_{\rm UV}} \right\} \\ &= \frac{L_{\rm UV}}{c} \left\{ 1 + \frac{\tau_{\rm IR}}{1 - e^{-\tau_{\rm UV}}} - \frac{\tau_{\rm IR}}{\tau_{\rm UV}} + \frac{\tau_{\rm IR}}{1 - e^{-\tau_{\rm UV}}}\frac{\pi\Lambda_{\rm coll}R^3}{3L_{\rm UV}} \right\} \left(1 - e^{-\tau_{\rm UV}}\right) . \end{aligned} \tag{290}$$

The dust heating rate is $\Lambda_{\rm coll} = 1.12\times10^{-32}\,(T - T_{\rm dust})T^{1/2}(1 - 0.8\,e^{-75/T})(Z/Z_\odot)n_{\rm H}^2$ erg cm$^{-3}$ s$^{-1}$ (Hopkins *et al.*, 2022), whereas the UV/optical luminosity from a central Pop II star cluster of mass $M_\star$ is $L_{\rm UV} = (\bar{\Psi}_{\rm FUV} + \bar{\Psi}_{\rm NUV} + \bar{\Psi}_{\rm opt})M_\star \simeq 670\,(M_\star/1\,{\rm M}_\odot)\,{\rm L}_\odot = 2.6\times10^{36}\,(M_\star/1\,{\rm M}_\odot)$ erg s$^{-1}$ (using values from Table 2). Thus:

$$\frac{\pi\Lambda_{\rm coll}R^3}{3L_{\rm UV}} = 1.3\times10^{-13}\,\frac{(T - T_{\rm dust})T^{1/2}(1 - 0.8\,e^{-75/T})}{(M_\star/1\,{\rm M}_\odot)} \left(\frac{R}{1\ {\rm pc}}\right)^3 \left(\frac{n_{\rm H}}{1\ {\rm cm}^{-3}}\right)^2 \left(\frac{Z}{Z_\odot}\right) . \tag{291}$$

If we have a star formation efficiency $f_\star$, then $M_\star \simeq f_\star 4\pi R^3 m_{\rm H} n_{\rm H}/3 \simeq 0.1\,f_\star(R/1\ {\rm pc})^3(n_{\rm H}/1\ {\rm cm}^{-3})$ M$_\odot$, and so

$$\frac{\pi\Lambda_{\rm coll}R^3}{3L_{\rm UV}} \simeq 1.3\times10^{-12}\,\frac{(T - T_{\rm dust})T^{1/2}(1 - 0.8\,e^{-75/T})}{f_\star} \left(\frac{n_{\rm H}}{1\ {\rm cm}^{-3}}\right) \left(\frac{Z}{Z_\odot}\right) . \tag{292}$$

Even for $T \gg T_{\rm dust}$ and very large densities (e.g. $n_{\rm H} \sim 10^7$ cm$^{-3}$) this ratio is very small, and so can be neglected in Eq. (290). Thus:

$$\dot{P}_{\rm rad} = \frac{L_{\rm UV}}{c} \left\{ 1 + \frac{\tau_{\rm IR}}{1 - e^{-\tau_{\rm UV}}} - \frac{\tau_{\rm IR}}{\tau_{\rm UV}} \right\} \left(1 - e^{-\tau_{\rm UV}}\right) . \tag{293}$$

Since $\tau_{\rm IR}/\tau_{\rm UV} = \kappa_{\rm ext}^{\rm IR}/\kappa_{\rm abs}^{\rm UV} \sim 3\times10^{-2}$ at most (from Table 2), it follows that if $\tau_{\rm UV}$ is small, so is $\tau_{\rm IR}$. In the two limits we have

$$1 + \frac{\tau_{\rm IR}}{1 - e^{-\tau_{\rm UV}}} - \frac{\tau_{\rm IR}}{\tau_{\rm UV}} \simeq \begin{cases} 1 + \tau_{\rm IR} & \text{if } \tau_{\rm IR} \gg 1 \\ \\ 1 & \text{if } \tau_{\rm IR} \ll 1 \end{cases} . \tag{294}$$

An an approximation, the exact expression can therefore be replaced by $1 + \tau_{\rm IR}$ for all $\tau_{\rm IR}$, as assumed in Eq. (124), giving us:

$$\dot{P}_{\rm rad} \simeq \frac{L_{\rm UV}}{c}(1 + \tau_{\rm IR})\left(1 - e^{-\tau_{\rm UV}}\right) . \tag{295}$$

I derived the exact expression in Eq. (293) well after the application and results from Anaxagoras, and it is unfortunately too late to redo the runs with Eq. (293) instead of Eq. (295). But as demonstrated here, the two expressions give essentially the same results. Future versions of Anaxagoras will make use of the slightly more accurate exact result in Eq. (293).

# 7. Understanding the physics of Ly$\alpha$ radiative feedback

In this Appendix we discuss the physics of Ly$\alpha$ radiative feedback for the curious reader who want a more in-depth understanding of the topic. The derivations here are not exact, but rather approximate order-of-magnitude estimates. For exact derivations, see the lengthy solutions of the radiative transfer equations by Lao & Smith (2020) and Tomaselli & Ferrara (2021). We begin by deriving the expression for the radiation pressure in a gas cloud that is optically thick to Ly$\alpha$ radiation. As in Appendix 6, we start with the radiative transfer equation in spherical coordinates assuming spherical symmetry:

$$\mu \frac{\partial I_\nu}{\partial r} + \frac{1-\mu^2}{r}\frac{\partial I_\nu}{\partial \mu} = -\alpha_\nu I_\nu + j_{\rm source} + j_{\rm scatter}\,,$$

where $\mu \equiv \cos\theta$ is the cosine of the angle between the ray and the radial unit vector $\hat{\boldsymbol{r}}$ (i.e. $\hat{\boldsymbol{n}}\cdot\hat{\boldsymbol{r}} = \mu$), $\alpha_\nu$ is the absorption coefficient, $j_{\rm source}$ the source function, and $j_{\rm scatter}$ the scattering term. At very high optical depths relevant for Ly$\alpha$ scattering we expect the radiation to be nearly isotropic within the cloud (i.e. not depend very much on $\mu$), and so we can use the Eddington approximation $I_\nu(r,\mu) \simeq a_\nu(r) + b_\nu(r)\mu$, with $b_\nu \ll a_\nu$ (i.e. only retaining a linear dependence on $\mu$). The moments of the intensity are then:

$$J_\nu \equiv \frac{1}{2}\int_{-1}^{1} {\rm d}\mu\, I_\nu = a_\nu$$

$$H_\nu \equiv \frac{1}{2}\int_{-1}^{1} {\rm d}\mu\, \mu I_\nu = \frac{b_\nu}{3} \tag{296}$$

$$K_\nu \equiv \frac{1}{2}\int_{-1}^{1} {\rm d}\mu\, \mu^2 I_\nu = \frac{a_\nu}{3}\,,$$

and so $K_\nu = \frac{1}{3}J_\nu$. Using this and taking moments of the radiative transfer equation yields

$$H_\nu = -\frac{1}{3\alpha_\nu}\frac{\partial J_\nu}{\partial r}\,, \tag{297}$$

or, since the energy density is $u_\nu = 4\pi J_\nu/c$ and $4\pi H_\nu = 2\pi\int_{-1}^{1}{\rm d}\mu\,\mu I_\nu = F_\nu$ is the flux:

$$F_\nu = -\frac{c}{3\alpha_\nu}\frac{\partial u_\nu}{\partial r}\,, \tag{298}$$

Thus, the momentum injection rate, assuming a spherical gas cloud, is (see Eq. 120)

$$\dot{P}_{\rm rad} = -\frac{1}{c}\times\frac{4\pi c}{3}\int_0^\infty {\rm d}\nu \int_0^R {\rm d}r\, r^2 \frac{\kappa_\nu \rho_{\rm gas}}{\alpha_\nu}\frac{\partial u_\nu}{\partial r}\,, \tag{299}$$

or since $\kappa_\nu\rho_{\rm gas} = \alpha_\nu$:

$$\dot{P}_{\rm rad} = -\frac{1}{c}\times\frac{4\pi c}{3}\int_0^\infty {\rm d}\nu \int_0^R {\rm d}r\, r^2 \frac{\partial u_\nu}{\partial r}\,. \tag{300}$$

Thus, integrating over frequency, the momentum injection rate into the cloud from Ly$\alpha$ radiation is expected to be:

$$\dot{P}_{{\rm Ly}\alpha} = -\frac{1}{c}\times\frac{4\pi R^2 c}{3}\int_0^1 {\rm d}\bar{r}\,\bar{r}^2 \frac{\partial u_{{\rm Ly}\alpha}}{\partial \bar{r}}\,, \tag{301}$$

where we have introduced $\bar{r} \equiv r/R$. Next we estimate the Ly$\alpha$ energy density at radius $\bar{r}$. Let a central source emit Ly$\alpha$ photons with luminosity $L_{{\rm Ly}\alpha}$. Consider an infinitesimal volume $4\pi r^2\,{\rm d}r$. The energy density in this region is $u_{\rm Ly\alpha} \simeq L_{\rm Ly\alpha} {\rm d}t_{\rm trap}/4\pi r^2\,{\rm d}r = L_{\rm Ly\alpha}{\rm d}t_{\rm trap}/4\pi R^3 \bar{r}^2\,{\rm d}\bar{r}$, where ${\rm d}t_{\rm trap}$ is the infinitesimal time it takes for the Ly$\alpha$ photons to escape this volume. Thus:

$$\dot{P}_{\rm Ly\alpha} \simeq -\frac{1}{c} \times \frac{L_{\rm Ly\alpha} c}{3R} \int_0^1 {\rm d}\bar{r}\,\bar{r}^2 \frac{\partial}{\partial \bar{r}} \left( \frac{1}{\bar{r}^2} \frac{{\rm d}t_{\rm trap}}{{\rm d}\bar{r}} \right) , \tag{302}$$

The trapping time $t_{\rm trap}$ for a cloud of radius $R$ can be derived roughly as follows (e.g. Spaans & Silk, 2006):

- In the wings of the Ly$\alpha$ line, the cross-section is $\sigma(x) = \sigma_0 \phi(x)$ where $\sigma_0$ is the cross-section at line center, $\phi(x) \simeq a_{\rm V}/\sqrt{\pi} x^2$ and $x \equiv (\nu - \nu_0)/\Delta\nu_0$ is the frequency displacement from the line center in units of the thermal width $\Delta\nu_0$ (e.g. Tomaselli & Ferrara, 2021). Each scattering result in a random frequency shift $\Delta x \sim 1$, so after $N$ scatterings the random walk in frequency will give a frequency displacement $x_{\rm tot} \sim N^{1/2}\Delta x \sim N^{1/2}$. The corresponding physical displacement (also a random walk), is $r \sim N^{1/2}\ell$, where $\ell = 1/n_{\rm H}\sigma(x) \simeq \sqrt{\pi} x_{\rm tot}^2/n_{\rm H}\sigma_0 a_{\rm V}$ is the mean-free-path of the photon. This yields (ignoring the factor $\sqrt{\pi}$)

  $$r \sim \frac{x_{\rm tot}^3}{n_{\rm H}\sigma_0 a_{\rm V}} . \tag{303}$$

  The Ly$\alpha$ photon will escape the gas cloud when $r = R$, corresponding to a frequency shift $x_{\rm tot} \sim (a_{\rm V}\tau_0)^{1/3}$, where $\tau_0 = n_{\rm H}\sigma_0 R$ is the optical depth at line center. The corresponding photon escape (or "trapping") time is $t_{\rm trap} = N \times \ell/c \sim x_{\rm tot}^4/n_{\rm H}\sigma_0 a_{\rm V} c \sim (a_{\rm V}\tau_0)^{4/3}/n_{\rm H}\sigma_0 a_{\rm V} c$. Using $n_{\rm H}\sigma_0 = \tau_0/R$ then yields $t_{\rm trap} \sim (a_{\rm V}\tau_0)^{1/3} R/c$.

Thus, we expect that the time it takes to escape out to a radius $r$ is $t_{\rm trap}(r) \simeq (r/c)[a_{\rm V}\tau_0(r)]^{1/3}$ (e.g. Spaans & Silk, 2006; Dijkstra, 2017), where $\tau_0(r) \propto r$ is the optical depth at the center of the Ly$\alpha$ line. Thus, if we let $t_{\rm trap} \equiv t_{\rm trap}(R)$ we have $t_{\rm trap}(\bar{r}) = t_{\rm trap}\bar{r}^{4/3}$, and so ${\rm d}t_{\rm trap}/{\rm d}\bar{r} = \frac{4}{3} t_{\rm trap}\bar{r}^{1/3}$, giving us

$$\begin{aligned}
\dot{P}_{\rm Ly\alpha} &\simeq -\frac{1}{c} \times \frac{L_{\rm Ly\alpha} c}{3R} \frac{4}{3} t_{\rm trap} \int_0^1 {\rm d}\bar{r}\,\bar{r}^2 \frac{\partial}{\partial \bar{r}} \left( \bar{r}^{-5/3} \right) \\
&= \frac{1}{c} \times \frac{L_{\rm Ly\alpha} c}{3R} \frac{4}{3} t_{\rm trap} \frac{5}{3} \int_0^1 {\rm d}\bar{r}\,\bar{r}^{-2/3} \\
&= \frac{20}{9} \frac{t_{\rm trap}}{t_{\rm light}} \frac{L_{\rm Ly\alpha}}{c} \\
&\equiv M_{\rm F} \frac{L_{\rm Ly\alpha}}{c} ,
\end{aligned}$$

where $t_{\rm light} = R/c$ is the light-crossing time for the cloud. Thus, we see that resonant scattering of the Ly$\alpha$ photons amplify the radiation pressure force by a factor $M_{\rm F} \simeq (20/9)(t_{\rm trap}/t_{\rm light}) \sim 2\,(a_{\rm V}\tau_0)^{1/3}$, not too far off from the exact solution of $M_{\rm F} = 3.51\,(a_{\rm V}\tau_0)^{1/3}$ found by Lao & Smith (2020) and Tomaselli & Ferrara (2021). While our order-of-magnitude derivation here has not been exact, it neatly explain the origin of the force multiplication factor without getting bogged down in mathematical details.

# 8. H II feedback: Expansion of H II regions, and disk destruction time

We saw in Section 5.4.1 that the differential equation governing the vertical extent of the ionization front at some radius $R$ in disk reads

$$\frac{\mathrm{d}z_{\mathrm{HII}}}{\mathrm{d}t'} \simeq c_{\mathrm{s,HII}} \left( \frac{\dot{Q}_{\mathrm{ion}} \epsilon_{\star,\mathrm{max}} \Sigma_{\mathrm{disk}} m_{\mathrm{H}}^2}{2 f_Y k_1 \rho_{\mathrm{mid}}^2} \right)^{1/4} \frac{1}{z_{\mathrm{HII}}^{1/4}} \left\{ \frac{\epsilon_{\mathrm{ff}} t'/t_{\mathrm{ff}}}{1 + \epsilon_{\mathrm{ff}} t'/t_{\mathrm{ff}}} \right\}^{1/4} , \tag{304}$$

This differential equation can be separated to get the time $t(H\,|R)$ (after disk formation) for the H II region to reach a height $H$ over the disk midplane:

$$\begin{aligned} \int_0^H \mathrm{d}z_{\mathrm{HII}}\, z_{\mathrm{HII}}^{1/4} &= c_{\mathrm{s,HII}} \left( \frac{\dot{Q}_{\mathrm{ion}} \epsilon_{\star,\mathrm{max}} \Sigma_{\mathrm{disk}} m_{\mathrm{H}}^2}{2 f_Y k_1 \rho_{\mathrm{mid}}^2} \right)^{1/4} \int_0^{t(H|R)-t_{\mathrm{min}}} \mathrm{d}t' \left\{ \frac{\epsilon_{\mathrm{ff}} t'/t_{\mathrm{ff}}}{1 + \epsilon_{\mathrm{ff}} t'/t_{\mathrm{ff}}} \right\}^{1/4} \\ &= c_{\mathrm{s,HII}} \left( \frac{\dot{Q}_{\mathrm{ion}} \epsilon_{\star,\mathrm{max}} \Sigma_{\mathrm{disk}} m_{\mathrm{H}}^2}{2 f_Y k_1 \rho_{\mathrm{mid}}^2} \right)^{1/4} \frac{t_{\mathrm{ff}}}{\epsilon_{\mathrm{ff}}} \int_0^{x=\epsilon_{\mathrm{ff}}[t(H|R)-t_{\mathrm{min}}]/t_{\mathrm{ff}}} \mathrm{d}x' \left\{ \frac{x'}{1+x'} \right\}^{1/4} . \end{aligned} \tag{305}$$

The left-hand side integrates to $\frac{4}{5}H^{5/4}$, and so

$$\frac{4}{5} H^{5/4} = c_{\mathrm{s,HII}} \left( \frac{\dot{Q}_{\mathrm{ion}} \epsilon_{\star,\mathrm{max}} \Sigma_{\mathrm{disk}} m_{\mathrm{H}}^2}{2 f_Y k_1 \rho_{\mathrm{mid}}^2} \right)^{1/4} \frac{t_{\mathrm{ff}}}{\epsilon_{\mathrm{ff}}} \int_0^{x=\epsilon_{\mathrm{ff}}[t(H|R)-t_{\mathrm{min}}]/t_{\mathrm{ff}}} \mathrm{d}x' \left\{ \frac{x'}{1+x'} \right\}^{1/4} . \tag{306}$$

The integral on the right-hand side has a very messy, non-invertable analytical result. To get a simpler, approximate analytical expression we note that for $x \lesssim 1$ we have

$$\begin{aligned} \int_0^x \mathrm{d}x' \left\{ \frac{x'}{1+x'} \right\}^{1/4} &\simeq \int_0^x \mathrm{d}x'\, x'^{1/4} \\ &= \frac{4}{5} x^{5/4} . \end{aligned} \tag{307}$$

In the other case of $x \gtrsim 1$ we instead get

$$\begin{aligned} \int_0^x \mathrm{d}t' \left\{ \frac{x'}{1+x'} \right\}^{1/4} &\simeq \int_0^1 \mathrm{d}x'\, x'^{1/4} + \int_1^x \mathrm{d}x' \\ &= \frac{4}{5} + x - 1 \\ &= -\frac{1}{5} + x . \end{aligned} \tag{308}$$

Thus, to conclude, we expect the integral to have the following properties:

$$\int_0^x \mathrm{d}x' \left\{ \frac{x'}{1+x'} \right\}^{1/4} \simeq \begin{cases} (4/5)\, x^{5/4} & \text{for } x \ll 1 \\ x & \text{for } x \gg 1 \end{cases} . \tag{309}$$

Motivated by this, I compared the following interpolation function with the exact numerical result for the integral:

$$\int_0^x \mathrm{d}x' \left\{ \frac{x'}{1+x'} \right\}^{1/4} \simeq (1+x^{5/4})^{4/5} - 1\,. \tag{310}$$

This approximation is accurate to within 3.2% for all $x$, and so is good enough for our purposes. With this result we have

$$\frac{4}{5} H^{5/4} \simeq c_{\rm s,HII} \left( \frac{\dot{Q}_{\rm ion}\epsilon_{\star,\rm max}\Sigma_{\rm disk} m_{\rm H}^2}{2 f_Y k_1 \rho_{\rm mid}^2} \right)^{1/4} \frac{t_{\rm ff}}{\epsilon_{\rm ff}} \left\{ \left[ 1 + \left( \frac{\epsilon_{\rm ff}[t(H\,|R) - t_{\rm min}]}{t_{\rm ff}} \right)^{5/4} \right]^{4/5} - 1 \right\}. \tag{311}$$

Rearranging a little yields

$$\left[ 1 + \left( \frac{\epsilon_{\rm ff}[t(H\,|R) - t_{\rm min}]}{t_{\rm ff}} \right)^{5/4} \right]^{4/5} - 1 \simeq \frac{4\epsilon_{\rm ff}}{5t_{\rm ff}} \frac{H^{5/4}}{c_{\rm s,HII}} \left( \frac{2 f_Y k_1 \rho_{\rm mid}^2}{\dot{Q}_{\rm ion}\epsilon_{\star,\rm max}\Sigma_{\rm disk} m_{\rm H}^2} \right)^{1/4}. \tag{312}$$

We can simplify this expression further by first noting that $\rho_{\rm mid}^2 = \pi^2 G^2 \Sigma_{\rm disk}^4 / 4\sigma_{\rm gas}^4$ (Eq. 106) so that

$$\left[ 1 + \left( \frac{\epsilon_{\rm ff}[t(H\,|R) - t_{\rm min}]}{t_{\rm ff}} \right)^{5/4} \right]^{4/5} - 1 \simeq \frac{4\epsilon_{\rm ff}}{5t_{\rm ff}} \frac{H^{5/4}}{c_{\rm s,HII}} \left( \frac{f_Y k_1 \pi^2 G^2 \Sigma_{\rm disk}^3}{2\sigma_{\rm gas}^4 \dot{Q}_{\rm ion}\epsilon_{\star,\rm max} m_{\rm H}^2} \right)^{1/4}. \tag{313}$$

The H II region will break out of the disk at time $t_{\rm HII,breakout}$ when $H > H_{\rm disk}$, where $H_{\rm disk}$ is the vertical extent of the disk over the midplane. We can define this quantity via $\Sigma_{\rm disk} = 2\rho_{\rm mid} H_{\rm disk}$, which yields $H_{\rm disk} = 2z_0 = \sigma_{\rm gas}^2/\pi G \Sigma_{\rm disk}$ (from Eqs. 105 and 107). This gives us $H_{\rm disk}^{5/4} = \sigma_{\rm gas}^{10/4}/(\pi G \Sigma_{\rm disk})^{5/4}$, and so

$$\left[ 1 + \left( \frac{\epsilon_{\rm ff}[t_{\rm HII,breakout} - t_{\rm min}]}{t_{\rm ff}} \right)^{5/4} \right]^{4/5} - 1 \simeq \frac{4\epsilon_{\rm ff}}{5c_{\rm s,HII} t_{\rm ff}} \left( \frac{f_Y k_1 \sigma_{\rm gas}^6}{2\pi^3 \dot{Q}_{\rm ion}\epsilon_{\star,\rm max} m_{\rm H}^2 G^3 \Sigma_{\rm disk}^2} \right)^{1/4}. \tag{314}$$

When $t_{\rm HII,breakout} - t_{\rm min}$ becomes comparable to the star formation time-scale $t_{\rm ff}/\epsilon_{\rm ff}$ at the outer edge of the disk (where $R = R_{\rm disk}$ and $t_{\rm min} = t$), we expect H II feedback to become so efficient that the disk is destroyed before any more gas can be turned into stars. Thus, we can define the photoionization feedback time-scale $t_{\rm HII,destr}$ to be the time after disk formation, such that $t_{\rm HII,breakout} - t_{\rm min} \equiv t_{\rm ff}$ at $R = R_{\rm disk}(t_{\rm HII,destr})$. At the outer edge of the disk we have $\Sigma_{\rm disk} = \Sigma_{\rm disk,0}[R_{\rm disk} = R_{\rm disk}(t_{\rm HII,destr})] \propto t_{\rm HII,destr}^{-1}$ (Eq. 101). Furthermore $t_{\rm ff}[R_{\rm disk} = R_{\rm disk}(t_{\rm HII,destr})] \propto t_{\rm HII,destr}$ (Eq. 116), and so the right-hand side of Eq. (314) scale as $t_{\rm HII,destr}^{-1} \times t_{\rm HII,destr}^{1/2} = t_{\rm HII,destr}^{-1/2}$, whereas *the left-hand side is independent of $t_{HII,destr}$* because $\epsilon_{\rm ff}[t_{\rm HII,breakout} - t_{\rm min}]/t_{\rm ff} = 1$ by definition. Using these results, we are free to introduce a completely arbitrary time-scale $t_{\rm arbitrary}$ and scale the quantities accordingly, so that

$$2^{4/5} - 1 \simeq \frac{4\epsilon_{\rm ff}\tilde{t}^{-1/2}}{5c_{\rm s,HII} t_{\rm ff}(t_{\rm arbitrary})} \left\{ \frac{f_Y k_1 \sigma_{\rm gas}^6}{2\pi^3 \dot{Q}_{\rm ion}\epsilon_{\star,\rm max}(t_{\rm HII,destr}) m_{\rm H}^2 G^3 \Sigma_{\rm disk,0}^2(t_{\rm arbitrary})} \right\}^{1/4}, \tag{315}$$

where $\tilde{t} \equiv t_{\rm HII,destr}/t_{\rm arbitrary}$. We can then solve for $t$ to get

$$t_{\rm HII,destr} \simeq \frac{t_{\rm arbitrary}}{(2^{4/5}-1)^2} \left\{ \frac{4\epsilon_{\rm ff}}{5c_{\rm s,HII} t_{\rm ff}(t_{\rm arbitrary})} \right\}^2 \left\{ \frac{f_Y k_1 \sigma_{\rm gas}^6}{2\pi^3 \dot{Q}_{\rm ion}\epsilon_{\star,\rm max}(t_{\rm HII,destr}) m_{\rm H}^2 G^3 \Sigma_{\rm disk,0}^2(t_{\rm arbitrary})} \right\}^{1/2}. \tag{316}$$

This is the result used in the main text. Note that this is not an entirely analytical expression because $\epsilon_{\star,\rm max}$ has a complicated dependence on $t_{\rm HII,destr}$ via its dependence on $\Sigma_{\rm disk,0}(t_{\rm HII,destr})$.

# 9. Fraction of disk volume occupied by supernova remnants

We saw in Section 5.4.2 that the fraction of volume at a given radial position $\bar{R}$ in the disk occupied by SN remnants is approximately given by

$$f_{\rm SN}(t) \simeq \frac{2\pi}{3z_0}\bar{\mathscr{R}}_{\rm SN}\epsilon_{\star,\rm max}\Sigma_{\rm disk}\int_{t_{\rm SN}+t_{\rm min}}^{t}{\rm d}t_{\rm form}\,R_{\rm s}^3(t-t_{\rm form})\,. \tag{317}$$

For convenience, let $x \equiv t - t_{\rm form}$ so that ${\rm d}x = -{\rm d}t_{\rm form}$ and we have

$$f_{\rm SN}(t) \simeq \frac{2\pi}{3z_0}\bar{\mathscr{R}}_{\rm SN}\epsilon_{\star,\rm max}\Sigma_{\rm disk}\int_{0}^{t-(t_{\rm SN}+t_{\rm min})}{\rm d}x\,R_{\rm s}^3(x)\,. \tag{318}$$

Consider times $t-(t_{\rm SN}+t_{\rm min}) > t_{\rm cool}$, which is equivalent to $t \gtrsim t_{\rm SN}+t_{\rm min}$ since $t_{\rm cool} \ll t_{\rm SN} = 3.401$ Myrs for all relevant densities. For $x < t_{\rm cool}$ the shock radius is $R_{\rm s}(x) = R_{\rm cool}(x/t_{\rm cool})^{2/5}$ (the Sedov-Taylor solution), whereas $R_{\rm s}(x) = R_{\rm cool}(x/t_{\rm cool})^{1/4}$ for $x \geqslant t_{\rm cool}$. We then have:

$$\begin{aligned}
\int_{0}^{t-(t_{\rm SN}+t_{\rm min})}{\rm d}x\,R_{\rm s}^3(x) &= R_{\rm cool}^3 t_{\rm cool}^{-6/5}\int_0^{t_{\rm cool}}{\rm d}x\,x^{6/5} + R_{\rm cool}^3 t_{\rm cool}^{-3/4}\int_{t_{\rm cool}}^{t-(t_{\rm SN}+t_{\rm min})}{\rm d}x\,x^{3/4} \\
&= \frac{5}{11}R_{\rm cool}^3 t_{\rm cool} + \frac{4}{7}R_{\rm cool}^3 t_{\rm cool}^{-3/4}\left\{[t-(t_{\rm SN}+t_{\rm min})]^{7/4} - t_{\rm cool}^{7/4}\right\} \\
&= -\frac{9}{77}R_{\rm cool}^3 t_{\rm cool} + \frac{4}{7}R_{\rm cool}^3 t_{\rm cool}^{-3/4}[t-(t_{\rm SN}+t_{\rm min})]^{7/4}\,.
\end{aligned} \tag{319}$$

Thus:

$$f_{\rm SN}(t) \simeq \frac{2\pi}{3z_0}\bar{\mathscr{R}}_{\rm SN}\epsilon_{\star,\rm max}\Sigma_{\rm disk}R_{\rm cool}^3 t_{\rm cool}\left\{-\frac{9}{77}+\frac{4}{7}\left[\frac{t-(t_{\rm SN}+t_{\rm min})}{t_{\rm cool}}\right]^{7/4}\right\}\,. \tag{320}$$

# 10. Halo growth time-scales: Estimates using extended Press-Schechter theory

As discussed in the main text, the typical halo mass accretion rate of a halo of mass $M$ can be estimated analytically using extended Press-Schechter theory (Neistein *et al.*, 2006; Correa *et al.*, 2015a; Salcido *et al.*, 2018):

$$\frac{\dot{M}}{M} = \sqrt{\frac{2}{\pi}} \frac{\delta_{\rm crit}}{\sqrt{S_q - S}} \frac{\dot{\mathfrak{D}}}{\mathfrak{D}^2}\,, \tag{321}$$

where $S \equiv \sigma^2(M)$, $S_q \equiv \sigma^2(M/q)$, $\mathfrak{D}(z)$ is the growth factor of linear density perturbations, $\delta_{\rm crit} = 1.686$ is the critical threshold value of an extrapolated linear density fluctuation at halo formation, and $q$ is to a good approximation a constant (at most it only depends very weakly on halo mass, see Correa *et al.*, 2015a). As shown in Appendix 2, the RMS linear overdensity $\sigma(M)$ in $\Lambda$CDM can be well-approximated by the following formula:

$$\begin{aligned} \sigma\left(M\right) &\simeq \mathcal{N}\left\{\frac{36}{1+3\bar{M}} - \ln^3\left(\frac{\bar{M}}{1+\bar{M}}\right)\right\}^{1/2} \\ \mathcal{N} &= 0.0845\,\sigma_8 \\ \bar{M} &\equiv \frac{8M}{M_{\rm eq}} \\ M_{\rm eq} &= 2.4\times10^{17}\left(\frac{\Omega_{\rm M}h^2}{0.14}\right)^{-1/2}\left(\frac{1+z_{\rm eq}}{3400}\right)^{-3/2}\ {\rm M}_{\odot}\,. \end{aligned} \tag{322}$$

We are focused on low-mass halos for which $\bar{M} \ll 1$, so that,

$$\sqrt{S_q - S} \simeq \mathcal{N}\sqrt{\ln^3\bar{M} - \ln^3(\bar{M}/q)}\,. \tag{323}$$

We can simplify further by making a 1$^{\rm st}$-order Taylor expansion, namely $\ln^3(\bar{M}/q) = (\ln\bar{M} - \ln q)^3 \simeq \ln^3\bar{M} - 3\ln^2\bar{M}\,\ln q$. This approximation is justified — e.g for halos $M = 10^6\,M_\odot$ we have $\left|\ln\bar{M}\right| = 24.1$, which is much larger than $\ln q \simeq \ln 2.38 = 0.867$ using the value of $q$ for $M = 10^6\,M_\odot$ from Correa *et al.* (2015a). Using this approximation we have

$$\sqrt{S_q - S} \simeq (3\ln q)^{1/2}\mathcal{N}\left|\ln\bar{M}\right|\,. \tag{324}$$

Thus, the halo mass accretion rate in Eq. (321) becomes

$$\frac{\dot{M}}{M} \simeq \sqrt{\frac{2}{3\pi\ln q}} \frac{\delta_{\rm crit}}{\mathcal{N}\left|\ln\bar{M}\right|} \frac{\dot{\mathfrak{D}}}{\mathfrak{D}^2}\,. \tag{325}$$

Padmanabhan (2002) (p. 416) gives the following fit to the growth factor in a $\Lambda$CDM Universe:

$$\mathfrak{D}\left(z\right) \simeq \left\{\frac{\Omega_{\rm M} + 0.4545\,\Omega_\Lambda}{\Omega_{\rm M}\left(1+z\right)^3 + 0.4545\,\Omega_\Lambda}\right\}^{1/3}\,. \tag{326}$$

This fit gives the correct asymptotic scalings: In the early matter-dominated epoch we obtain $\mathfrak{D}(z) \propto (1+z)^{-1}$, whereas in the late vacuum-dominated era we get $\mathfrak{D}(z) \to$ constant. Since we are interested in high redshifts (mainly $z > 6$), we can use the results for the former. We find:

$$\begin{aligned} \mathfrak{D}(z \gg 1) &\simeq \left(1 + 0.4545\,\frac{\Omega_\Lambda}{\Omega_\mathrm{M}}\right)^{1/3} \frac{1}{1+z} \\ &= \left(1 + 0.4545\,\frac{\Omega_\Lambda}{\Omega_\mathrm{M}}\right)^{1/3} a\,, \end{aligned} \tag{327}$$

where I have written the growth factor in terms of the scale factor $a = 1/(1+z)$ on the second line. The time derivative of the growth factor then becomes

$$\begin{aligned} \dot{\mathfrak{D}} &= \frac{\mathrm{d}\mathfrak{D}}{\mathrm{d}a}\,\dot{a} \\ &\simeq \left(1 + 0.4545\,\frac{\Omega_\Lambda}{\Omega_\mathrm{M}}\right)^{1/3} Ha \\ &\simeq \mathfrak{D}\,\sqrt{\Omega_\mathrm{M}} H_0 (1+z)^{3/2}\,, \end{aligned} \tag{328}$$

where I have used $H \simeq \sqrt{\Omega_\mathrm{M}} H_0 (1+z)^{3/2}$ for the Hubble parameter during the matter-dominated epoch. Thus, $\dot{\mathfrak{D}}/\mathfrak{D}^2 = \sqrt{\Omega_\mathrm{M}} H_0 (1+z)^{3/2}/\mathfrak{D}$, or:

$$\begin{aligned} \frac{\dot{\mathfrak{D}}}{\mathfrak{D}^2} &\simeq \frac{\sqrt{\Omega_\mathrm{M}} H_0}{(1 + 0.4545\,\Omega_\Lambda/\Omega_\mathrm{M})^{1/3}}\,(1+z)^{5/2} \\ &= 3.08 \times 10^{-16} \left(\frac{1+z}{10}\right)^{5/2} \mathrm{s}^{-1}\,. \end{aligned} \tag{329}$$

Using this in Eq. (325) yields

$$\begin{aligned} \frac{\dot{M}}{M} &\simeq 1.42 \times 10^{-16}\,\frac{\delta_\mathrm{crit}}{(\ln q)^{1/2} \mathcal{N} \left|\ln \bar{M}\right|} \left(\frac{1+z}{10}\right)^{5/2} \mathrm{s}^{-1} \\ &\simeq \frac{3.04 \times 10^{-15}}{\sigma_8 \left|\ln \bar{M}\right|} \left(\frac{1+z}{10}\right)^{5/2} \mathrm{s}^{-1}\,, \end{aligned} \tag{330}$$

where on the second line I have used $\delta_\mathrm{crit} = 1.686$, $q \simeq 2.38$, and $\mathcal{N} = 0.0845\,\sigma_8$.

# 11. Properties of NFW halos

Early cosmological simulations in the 1990's by Navarro *et al.* (1995, 1996, 1997) showed that the density profiles of CDM halos can be well-approximated by the following formula:

$$\begin{aligned}\rho_{\rm NFW}(r) &= \frac{\Delta_{\rm vir}\bar{\rho}_{\rm M}}{3f(c)}\frac{c^3}{(r/R_{\rm s})(1+r/R_{\rm s})^2} \\ &= \frac{4.78\times10^{-25}\,c^3\ {\rm g\ cm^{-3}}}{3f(c)(r/R_{\rm s})(1+r/R_{\rm s})^2}\left(\frac{1+z}{10}\right)^3,\end{aligned} \tag{331}$$

where $R_{\rm s}\equiv R_{\rm vir}/c$ is the scale radius, $c$ is the so-called halo concentration, and $f(c)\equiv\ln(1+c)-c/(1+c)$. In the inner region ($r\ll R_{\rm s}$) the density scales like $\rho_{\rm NFW}(r)\propto r^{-1}$, whereas the density drops approximately like $\rho_{\rm NFW}(r)\propto r^{-3}$ in the outer region ($R_{\rm s}<r\leq R_{\rm vir}$). In writing Eq. (331) I have made the approximation that the DM is gravitationally dominant — in effect ignoring a factor $1-f_{\rm B}\simeq0.843$ in Eq. (331). From Eq. (331) we can determine the enclosed mass $M(<r)$:

$$\begin{aligned}M(<r) &= 4\pi\int_0^r {\rm d}r'\,r'^2\rho_{\rm NFW}(r') \\ &= \frac{4\pi\Delta_{\rm vir}\bar{\rho}_{\rm M}c^3}{3f(c)}\int_0^r {\rm d}r'\,\frac{r'^2}{(r'/R_{\rm s})(1+r'/R_{\rm s})^2} \\ &= \frac{4\pi\Delta_{\rm vir}\bar{\rho}_{\rm M}c^3}{3f(c)}\,R_{\rm s}^3\int_0^{r/R_{\rm s}} {\rm d}x\,\frac{1}{x(1+x)^2} \\ &= \frac{4\pi\Delta_{\rm vir}\bar{\rho}_{\rm M}c^3}{3f(c)}\,R_{\rm s}^3\left\{\frac{1}{1+r/R_{\rm s}}+\ln\left(1+\frac{r}{R_{\rm s}}\right)-1\right\}.\end{aligned} \tag{332}$$

We can simplify this expression by noting that $4\pi\Delta_{\rm vir}\bar{\rho}_{\rm M}c^3R_{\rm s}^3/3=4\pi\Delta_{\rm vir}\bar{\rho}_{\rm M}R_{\rm vir}^3/3=M$ is the total mass of the halo (including baryons). Thus:

$$\begin{aligned}M(<r) &= \frac{M}{f(c)}\left\{\frac{1}{1+r/R_{\rm s}}+\ln\left(1+\frac{r}{R_{\rm s}}\right)-1\right\} \\ &= \frac{M}{f(c)}\left\{\ln\left(1+\frac{r}{R_{\rm s}}\right)-\frac{r/R_{\rm s}}{1+r/R_{\rm s}}\right\} \\ &= M\,\frac{f(r/R_{\rm s})}{f(c)}.\end{aligned} \tag{333}$$

This result can be applied to find the maximum radius a star ejected radially with velocity $v_\star$ from an initial radius $r_0$ in the halo can reach. Assuming that the NFW halo is gravitationally dominant, the equation of motion for the star is simply

$$\ddot{r}=-\frac{GM(<r)}{r^2}. \tag{334}$$

Integrating this once with our initial condition $\dot{r}(r=r_0)=v_\star$ yields

$$\frac{1}{2}\dot{r}^2=\frac{1}{2}v_\star^2-\int_{r_0}^r {\rm d}r'\,\frac{GM(<r')}{r'^2}. \tag{335}$$

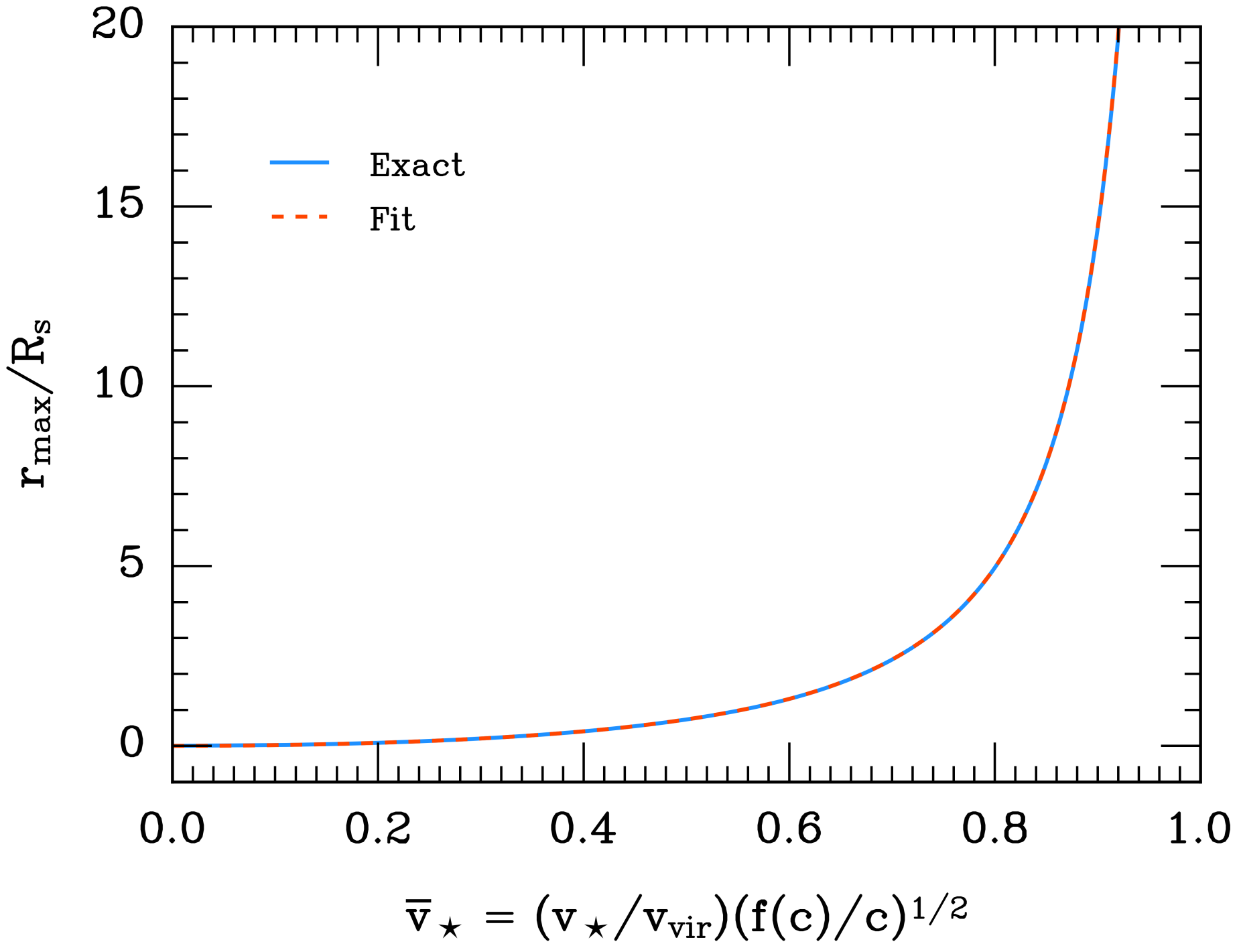

**Figure 27.** A comparison between the exact inverse of Eq. (339) — for which there is no analytical expression — and the fit in Eq. (341).

The maximum radius $r_{\rm max}$ the star will reach is then determined by

$$v_\star^2 = 2\int_{r_0}^{r_{\rm max}} {\rm d}r'\,\frac{GM(<r')}{r'^2}$$

$$= \frac{2GM}{f(c)}\int_{r_0}^{r_{\rm max}} {\rm d}r'\,\frac{f(r'/R_{\rm s})}{r'^2} \qquad (336)$$

$$= \frac{2cGM}{f(c)R_{\rm vir}}\int_{r_0/R_{\rm s}}^{r_{\rm max}/R_{\rm s}} {\rm d}x\,\frac{f(x)}{x^2}\,,$$

or

$$v_\star^2 = \frac{2cv_{\rm vir}^2}{f(c)}\int_{r_0/R_{\rm s}}^{r_{\rm max}/R_{\rm s}} {\rm d}x\,\frac{f(x)}{x^2}\,. \qquad (337)$$

The integral can be evaluated analytically:

$$\int_{r_0/R_{\rm s}}^{r_{\rm max}/R_{\rm s}} {\rm d}x\,\frac{f(x)}{x^2} = \int_{r_0/R_{\rm s}}^{r_{\rm max}/R_{\rm s}} {\rm d}x\,\frac{1}{x^2}\left\{\ln(1+x)-\frac{x}{1+x}\right\}$$

$$= \frac{\ln(1+r_0/R_{\rm s})}{r_0/R_{\rm s}} - \frac{\ln(1+r_{\rm max}/R_{\rm s})}{r_{\rm max}/R_{\rm s}}\,. \qquad (338)$$

Thus, $r_{\rm max}/R_{\rm s}$ is a function of $\bar{v}_\star$:

$$\bar{v}_\star^2 = 1-\frac{\ln(1+r_{\rm max}/R_{\rm s})}{r_{\rm max}/R_{\rm s}}\,, \qquad (339)$$

where

$$\bar{v}_\star^2 \equiv \frac{v_\star^2}{2v_{\rm vir}^2}\frac{f(c)}{c} + 1 - \frac{\ln(1 + r_0/R_{\rm s})}{r_0/R_{\rm s}}\,. \tag{340}$$

We can invert Eq. (339) using a fitting function. Note first that for $r_{\rm max}/R_{\rm s} \ll 1$ a Taylor expansion yields $\bar{v}_\star^2 \simeq \frac{1}{2}(r_{\rm max}/R_{\rm s})$, so that $r_{\rm max}/R_{\rm s} \simeq 2\bar{v}_\star^2$ for $\bar{v}_\star^2 \ll 1$. As shown in Figure 27, for $\bar{v}_\star < 20$ (covering more then enough of the range of values for $r_{\rm max}/R_{\rm s}$ within the halo virial radius given a typical halo concentration $c \sim 3$) we see that $r_{\rm max}/R_{\rm s}$ is reasonably fitted by:[81]

$$\frac{r_{\rm max}}{R_{\rm s}} = \frac{2\bar{v}_\star^2}{(1 - 1.012\,\bar{v}_\star^2 + 0.01815\,\bar{v}_\star^4)^{1.325}}\,. \tag{341}$$

In the limit $\bar{v}_\star \to 0$ this reduces to the exact result $r_{\rm max}/R_{\rm s} \simeq 2\bar{v}_\star^2$ noted earlier. We can also calculate the escape velocity of the halo (i.e. the velocity $v_\star = v_{\star,\rm esc}$ such that $r_{\rm max} = R_{\rm vir} = cR_{\rm s}$). From Eq. (339) we find

$$v_{\star,\rm esc} = \sqrt{2}v_{\rm vir}\left\{\frac{c}{f(c)}\right\}^{1/2}\left\{\frac{\ln(1 + r_0/R_{\rm s})}{r_0/R_{\rm s}} - \frac{\ln(1+c)}{c}\right\}^{1/2}\,. \tag{342}$$

[81]The fitting function adopted was

$$\frac{r_{\rm max}}{R_{\rm s}} = \frac{2\bar{v}_\star^2}{(1 - \alpha\,\bar{v}_\star^2 + \beta\,\bar{v}_\star^4)^{\gamma}}\,,$$

with the best-fit coefficients $\alpha$, $\beta$, and $\gamma$ determined using `curve_fit` from `scipy.optimize`.